# Towards a mathematical formalism for classifying phases of matter

A. Bauer, J. Eisert, C. Wille

Dahlem Center for Complex Quantum Systems, Freie Universität Berlin, Arnimallee 14, 14195 Berlin

11.03.2019

## Abstract

We propose a unified mathematical framework for classifying phases of matter. The framework is based on different types of combinatorial structures with a notion of locality called lattices. A tensor lattice is a local prescription that associates tensor networks to those lattices. Different lattices are related by local operations called moves. Those local operations define consistency conditions for the tensors of the tensor network, the solutions to which yield exactly solvable models for all kinds of phases. To make the connection to physics we provide a unified definition of phases of matter of local classical and quantum models with arbitrary boundaries, defects or anyons by representing them as tensor networks. The main part of this work consists of different implementations of the framework, reproducing many models of non-chiral topological order in a systematic way. Among those are symmetry-breaking and topological phases in up to three space-time dimensions, their boundaries, defects, domain walls and symmetries, as well as their anyons for 2+1-dimensional systems. Along the way, we also obtain combinatorial geometric interpretations to algebraic structures closely related to *-algebras, their representations, unitary fusion categories, weak Hopf algebras, and the representations of their quantum doubles. We also deliver ideas of how other kinds of phases, like SPT/SET, fermionic, free-fermionic, chiral, and critical phases, can be described within our framework. We also define another structure called contracted tensor lattices which generalize tensor lattices: The former associate tensors instead of tensor networks to lattices, and the consistency conditions for those tensors are defined by another kind of local operation called gluings. Using this generalization, our framework also covers mathematical structures like axiomatic (non-fully extended or defective) TQFTs, that do not directly describe phases on a microscopic physical level, but formalize certain aspects of potential phases, like the anyon statistics of 2+1-dimensional phases. We also introduce the very powerful concept of (contracted) tensor lattice mapping, unifying a lots of different operations, such as stacking, anyon fusion, anyon condensation, equivalence of different fixed point models, taking the Drinfel'd centre, trivial defects or interpreting a bosonic model as a fermionic model.

# Contents

# Chapter 1

# Introduction

## 1.1 Why we should care about phases of matter

Classifying phases of matter means dividing the space of models into families with the same qualitative properties. By *model* we roughly mean the mathematical description of a physical system with many degrees of freedom. A classical example of such a model would be "the 2-dimensional classical Ising model on square lattices". Slightly more precisely a model is a thermodynamic-limit sequence of systems (i.e., a sequence of systems with more and more degrees of freedom) which is homogeneous (i.e., it looks the same at different places and for different systems in the sequence). The key property that such a model needs to have in order to talk about phases is *locality*, i.e., the interactions between the degrees of freedom have to be local. All first-principle models, in particular models in condensed-matter physics such as the Hubbard model, but even the standard model of high energy physics, are local models of this type. In contrast to how the word is often used, for us a "model" has no free parameters, but represents a single point in parameter space. So in the example above, we should have fixed the parameters and talk about e.g. "the Ising model at $\beta = 0.5$ and $h = 0.1$". In Section (4) we will present a more detailed discussion of the required properties of models such that their classification into different phases is well defined.

Two models belong to the same *phase* if they can be mapped onto each other under *local* deformations of the microscopic degrees of freedom that do not change global properties. In many formulations such a deformation consists of a constant-depth circuit of local operations, each of which does not affect the system outside of a small region. For example, this could be a constant-depth circuit of local unitaries acting on the ground state of a quantum many-body system [1]. For example such local deformations include basis changes in the Hilbert space of the local degrees of freedom. We follow a simpler, more general, and more constructive approach by describing models via tensor networks, see Section (4.1). Our framework unifies and includes several tensor network based classifications put forth in the literature [2, 3, 4, 5, 6], but at the same time goes significantly beyond known approaches based on representing quantum states as tensor networks. For quantum systems our formulation in terms of tensor networks is done via a Trotterization of the imaginary time evolution. The local operations acting on the tensor networks are called tensor-network moves, see Section (2.3). A comparison of the resulting phase definition to other notions of phases is presented in Section (4.2).

A *physical theory* consists of a mathematical model together with a *dictionary* that tells us what part of the real world the model describes and which observables in the model correspond to which observations or experiments in the real world. When condensed matter physicists write down a mathematical model they usually implicitly give this dictionary by the names they choose for observables and terms in the Hamiltonian. For example, if some observable in the definition of a Hamiltonian is called $\sigma_x$ it is commonly assumed that a measurement of this observable can be determined by a magnetic field in the $x$-direction in the real world. In the general case, however, we may have to choose such a dictionary explicitly. A fixed model can correspond to different theories when choosing different dictionaries, and conversely two different models can correspond to the same theory when we choose the dictionaries accordingly. In particular the same model can correspond to a classical statistical system for one and to a quantum mechanical ground state for another dictionary as is illustrated in the examples provided in Section (5). The only restriction that we have for choosing a dictionary is due to locality: Local observables in the model should correspond to local measurements in the real world. In fact all measurements we can perform eventually in the real world are measurements corresponding to local observables.

Consider a fixed dictionary and the set of all models that formally fit to this dictionary (e.g., all models with fixed Hilbert space dimension of the local degrees of freedom in quantum mechanics). Now consider the local observable corresponding to a fixed measurement (e.g., the single-qubit observable $\sigma_z$, for all different qubit models). As the deformations that do not change the phase essentially allow for arbitrary local changes in the degree of freedom (such as, e.g., arbitrary local basis changes), the phase of the model tells us literally nothing about the specific measurement outcome of such a fixed measurement. In that sense, as predictions for the outcomes of local measurements are the only thing that we are interested in practically, the phase of a model is a property of no practical relevance to predict expectation values. More pedantically, the entire idea of having different models with the same dictionary does not fit into the mindset of phases of matter in practice: A dictionary can only be chosen depending on a single model, and there is no canonical way to state when two different models use the same dictionary (e.g., in quantum mechanics one should not think of two different models being embedded into the same Hilbert space, even they have the same dimension).

However, such a god-given fixed dictionary that is automatically forced onto us when writing down a mathematical model does not exist. Two models (without dictionary) can represent the same physical theory (after an according choice of dictio-nary) exactly if they are in the same phase. So in that sense, the phase of a model is the *only* property of practical relevance, as long as we do not fix the dictionary connecting it to the real world.

Although the phase does not say anything about the measurement outcome for fixed observables, it does tell us something about the qualitative properties that certain (yet to be determined) observables of the model have. E.g., in the case of *symmetry-breaking phases* [7], the phase tells us that there exits some local observables (known as order parameters) whose 2-point correlations do not decay (and how many independent ones there are). How those observables are microscopically implemented though depends on the microscopic model, and is not a property of the phase. In a limited few-parameter family of models, like "the 2-dimensional Ising model" there might be a fixed candidate for such an order parameter, like "the local magnetization". However, in the space of all 2-dimensional lattice models such a choice is not possible any more.

According to the previous paragraph, the phase is the only important property of a model concerning all questions that are independent of its microscopic realization. One such question for example concerns the simulation of one physical system via another: If two models are in the same phase, then we can locally encode one into the other. Thus, one can efficiently simulate the other up to a constant overhead (apart from some factor which depends on how we compare the two space(-time) volumes with each other). This implies that if an $(n + 1)$-dimensional quantum system is in the same phase as an $(n+1)$-dimensional classical system, then the ground state properties of the former can be efficiently calculated by a classical statistical device. Another example is fault-tolerant quantum computation [8]. The question of whether a model allows for a noise threshold below which quantum error correction is possible in principle is only a property of the phase.

## 1.2   What this work is about

The main purpose of this work is three-fold:

1. We aim at providing a new viewpoint on phases of matter by representing all different kinds of physical many-body models as tensor networks, and classifying those tensor networks.

2. We give a unified mathematical framework that allows us to construct fixed point models for all different sorts of phases of matter.

3. We introduce a simpler presentation of specific known fixed point models that we find is more natural and that eliminates various technical details.

While the study of quantum phases of matter using tensor networks has a solid tradition in the recent literature [1, 2, 9, 3, 5, 10, 11], our approach is significantly different in spirit: We do not use tensor networks to represent ground states of quantum models, but to represent the imaginary time evolution. So though we use the language of tensor networks, it is much closer to the approach via state-sum constructions [12, 13, 14, 15]. However, our framework works on a much more general and abstract level, using different lattice types and tensor types: This way it includes e.g. fermionic phases or all sorts of defects. We also believe that taking this generalized viewpoint is the key to classifying chiral topological phases, a task that has not been fully accomplished so far.

## A unifying viewpoint on phases of matter

In Chapter 4, we put our mathematical framework for constructing fixed point models into the context of phases of matter. We see that various classical or quantum local many-body models can be described in a unified language via different types of tensor networks living on different types of lattices. Consider the following examples:

● Classical statistical models in $d$ spatial dimensions are tensor networks of real tensors on $d$-dimensional lattices.

● Ground state properties of quantum models in $d$ spatial dimensions are described by $(d + 1)$-dimensional (complex or real) tensor networks via the Trotterization of their imaginary time evolution. Unlike working with a Hilbert space associated to the space only, this viewpoint yields a simple all-in-one picture for the whole situation.

● Classical or quantum models with a boundary are tensor networks on lattices with a boundary.

● Classical or quantum models with any sorts of defects or domain walls are tensor networks on lattices with special lower-dimensional embedded sub-lattices.

The list of examples can be continued if we consider different types of tensors. As we point out in the outlook of this work in Section 6) we can extend our formalism to include

● Symmetries can be implemented in three different ways: 1) Space-group symmetries are encoded by using a lattice type that has the desired symmetries, e.g., if a model is defined on a square lattice without any arrows specifying a direction, the associated tensor-network automatically has to inherit the symmetries. 2) Local symmetries (i.e., ones representable by a *tensor-network operator*) in quantum systems can be seen as co-dimension 1 defects perpendicular to the time direction, by inserting the tensor-network operator into the Trotterized imaginary time evolution. 2) On-site symmetries can be modelled by another tensor type: Symmetric tensors are ones where each index carries a group representation, such that the total tensor is in the trivial symmetry sector. (This is the formulation of symmetry that has to be used if we want to classify *SPT phases* where the symmetry itself is trivial as a topological defect, see Section 6.1.3).

● Fermionic quantum models are also modelled by tensor networks. To this end we have to use fermionic tensors instead of standard (bosonic) ones.

● If we want to restrict to quadratic fermionic Hamiltonians, one can do this by working with *Gaussian tensors*.

Typically physical models describing condensed matter or similar systems are translation invariant, i.e., the lattices they are defined on are regular grids. The same is true for models arising from field theories on a flat background after discretization. We can now introduce moves that locally deform the local grids, yielding e.g., what we call topological lattice types. For tensor networks on such lattice types the moves define consistency conditions for the tensors, that make them what is often called *fixed point models*. The more restrictive those consistency conditions are the easier it is to classify those tensor networks and to analytically compute properties of the models. Often, tensor networks on regular grids can be extended to lattice types with very powerful moves, such as topological lattice types. Some examples for this are:

• Topologically ordered quantum models in 2 space dimensions can be extended from $2+1$-dimensional regular grids (on which their discretized imaginary time evolution is defined) to a lattice type on 3-dimensional topological manifolds, e.g., simplicial complexes with Pachner moves.

• Classical models in a symmetry breaking or trivial phase can be extended to a topological lattice type.

• Anyons can be seen as 0-dimensional defects in 2-dimensional space. They can be extended from $2+1$-dimensional regular grids with 1-dimensional special lines running in time direction to topological 3-dimensional lattice types with 1-dimensional sub-regions that can be arbitrarily curved and run in any space-time direction.

• Arbitrary topological defects can be extended to lattices on topological manifolds with topological sub-manifolds.

• Symmetries implemented locally in topological models can be extended to lattices on $n$-dimensional manifolds with an $(n-1)$-dimensional sub-manifold.

• Fermionic topologically ordered models can be extended to topological lattices with a combinatorial representation of a *spin structure*.

• We believe that 2-dimensional critical classical systems can be extended to a conformal lattice type, however, we do not know any entirely combinatorial structure modelling conformal transformations, yet.

The examples above illustrate in which sense our framework constitutes a unifying viewpoint on phases of matter, but the list might be far from exhaustive. In the last Section of this work (Sec (6.2) we hint at several further conceptual insights provided by our framework regarding the robustness of phases (Section (6.2.1) accessible via the structure of the solutions of the CTL axioms and possible notions of a 'phase' of real time evolutions and dissipative models (Section 6.2.2). In addition we present instructive and practical ideas, i.e., in Section (6.2.4) we propose that one can find concrete fixed point models of a certain type by numerically solving the consistency conditions for the tensors and in Section (6.2.5), we propose an idea for an algorithm that variationally determines whether two models are in the same phase, based on the idea of tensor-network moves.

## What the general framework captures

In Chapter (2) we give an extremely versatile framework to construct families of fixed point models for lots of different phases. In the following we summarizes the main strengths of our formalism:

• Our formalism is mathematically very simple and involves nothing but basic linear algebra (tensor products and index contractions of vector spaces) and combinatorics paired with a lot of geometric intuition. Because of that this work is completely self-contained. Therefore we hope that it is also understandable for readers with no knowledge of category theory or algebra. However, we do give the relation between our formalism and the usual categorial or algebraic language, see, e.g., Section (5.4.6, 5.8.7, 5.9.6, 5.11.5).

• We model the complex numbers in an unconventional way (see Section (2.2.3)): Even though they are not explicitly part of our framework, complex tensors emerge starting from real tensors, via their realifications. This has several consequences: 1) The two seemingly different or unrelated concepts of the absence of an orientation and the unitarity of quantum mechanics are one and the same thing in our formalism (see Section (2.4.5)). 2) Anti-unitary symmetries are automatically contained in our framework without having to include them explicitly.

• Our framework unifies fixed point constructions for various different situations by taking different lattice types and tensor types, as already outlined in the paragraph before. In particular, it works in any dimension by taking lattices of that dimension. It works for classifying topological phases in the same same way as for classifying topological sectors of lower-dimensional parts of a model, such as anyons (see Section (5.11)), topological boundaries, domain walls and co-dimension defects (see Section (5.5, 5.10)), twists (see Section (6.1.10)), global symmetries, or ground states (which are nothing but 0-dimensional defects, see Section (5.1, 5.6, 5.7, 6.1.10)). To this end we simply need to consider lattices including lower dimensional parts that corresponds to the aforementioned objects. Our formalism captures SPT/SET phases by either using symmetric tensors, or lattices with a combinatorial representation of a group homology class (see Section (6.1.3)). It captures fermionic topological models by taking fermionic tensors and lattices with a combinatorial representation of a spin structure (see Section (6.1.1)). We believe also critical/gapless/conformal models can be classified by our framework by using a conformal lattice type (see Section (6.1.6)).

• We introduce the concept of universality (which is not related to the notion of "universality" in the context of statistical physics, see Section (2.3.2)) for a type of fixed point models: Such a type is universal if any other type of fixed point model on the same type of lattice can be encoded in this type (usually by blocking). All the concrete implementations in Section (5) in more than 1 dimensions are too simple to be universal. It seems very plausible that there are phases that are captured by a universal type of fixed point model, but that cannot be described by the models in Section (5). We believe that this is the case for chiral topological phases in $2+1$ dimensions, such as in the fractional quantum Hall effect: A universal type of fixed point model for $2+1$-dimensional topological lattices naturally allows for many of the properties of chiral phases, such as the absence of a commuting-projector Hamiltonian [16] or the absence of a gapped boundary (see Section (6.1.5)). It is not hard to write down a universal type of fixed point model, though the arising consistency conditions are relatively complex and it is not easy to find solutions.

• Our framework allows to systematically construct many different types of microscopic fix point models for various phase. For example for topological phases in $2 + 1$ dimensions we give 1) one CTL type yielding fixed point models similar to the Levin-Wen string-net models [17] (see Section (5.8)), 2) one CTL type yielding fixed point models similar to the Kitaev quantum double models for Hopf algebras [18] (see Section (5.9)) and 3) one CTL type that contains the colour code [19] as one specific realization (see Section (6.1.11)). Even though for the pure task of classifying phases of matter having different types of fixed point models is not necessary, constructing different microscopic models for certain phases is of great practical importance. One type of microscopic model may be suited better for a specific practical purpose than another model, even though they both correspond to the same phase. Such purposes could be quantum computation, error correction, or engineering of specific phases of matter using certain components. E.g., the colour code allows for better transversal gates for topological quantum computation than two copies of the toric code string-net [19] even though they are in the same phase [20]. On the other hand encoding the toric code phase into the colour-code CTL type yields a model that is much more complicated than the string-net representation of the toric code. Another example is the implementation of topological Hamiltonians via mesoscopic circuits [21], where it is of great importance to pick the right microscopic realization of the phase.

• Our notion of tensor lattice (short TL) (see Section (2.3)) formalizes tensor networks on lattices describing fixed point models of microscopic physical systems. However, the mathematical literature on topological phases is dominated by effective descriptions of physical systems that do not directly correspond to a local microscopic realization. Those are so-called axiomatic TQFTs [22] and variants thereof, such as spin TQFTs, non-fully extended TQFTs, open-closed TQFTs, or TQFTs with defects. In order to also capture those constructions we define the more general notion of a contracted tensor lattice (short CTL) (see Section (2.4)). Contracting cut-out patches of a TL yields a CTL, but not every CTL can be obtained from a TL by contraction of patches (see Section (2.4.6)).

• We introduce the concept of mapping (see Section (2.3.2, 2.4.3)) between different types of tensors, lattices, and (C)TLs, which unifies many different operations on models and their phases. Examples are of mappings are fusion of anyons, fusion of domain walls or other types of defects, condensation of defects/anyons at a boundary, stacking multiple layers of a model, taking the Drinfel'd centre to obtain the anyon theory, taking the overlap of a ground state PEPS with itself, explicitly breaking symmetries, interpreting free-fermion phases as many-body phases, or equivalence of different types of fixed point models. The concept of mapping also allows us to define different classes of lattices and (C)TLs: Lattice classes roughly correspond to different kinds of deformability, such as "2-dimensional topological", "3-dimensional topological with topological boundaries and spin structure", or "2-dimensional regular grids without any deformability". TL classes correspond to different classes of order, such as "$2 + 1$-dimensional topological order", "$3 + 1$-dimensional free-fermionic topological order", "anyons in $2 + 1$-dimensional topological order"

• In the conventional approach to fixed point models one usually starts with some categorial or algebraic structure and then shows that if we use this structure as input for a certain construction, then this yields fixed point models with certain desired properties. We present the story in a different order: (Contracted) tensor lattices are arbitrary tensors that fulfill consistency conditions directly connected to lattice moves, such as Pachner moves for topological order. It then turns out that the tensors fulfilling the consistency conditions are very similar to known algebraic structures like fusion categories, weak Hopf algebras, or *-algebras. However, for us this similarity is only a side remark, and the objects that classify different phases are the tensors themselves. A similar viewpoint can be found in Ref. [23].

## Concrete implementations of the framework

In Chapter (5) and Section (6.1) we discuss concrete implementations of the general framework, their properties, and connection to existing models.

• Section (5.1, 5.2, 5.3) cover rather trivial fixed point models in 0 and 1 space-time dimensions, for the sake of completeness and for a simple demonstration of the formalism.

• In Section (5.4) we construct fixed point models for $(1 + 1)$-dimensional topological models. In the complex-real case, the corresponding phases are known as symmetry-broken phases (otherwise topological phases protected by time-reversal symmetry). The fixed point models are very similar to those in Ref. [24].

• In Section (5.5) we study topological boundaries, codimension 1 defects or domain walls of 2-dimensional topological models. In the complex-real case those are projections onto one particular symmetry broken sector (and direct sums thereof).

• In Section (5.6, 5.7) we give simple examples for CTLs that do not come from TLs: Physically those correspond to collections of all ground states, or of all independent local observables.

• In Section (5.8) we give a type of fixed point model for topological (or symmetry breaking) order in $2 + 1$ dimensions, that essentially correspond to the Turaev-Viro model [12, 13] in the state-sum language, or the Levin-Wen model [17] in the Hamiltonian language. Our formulations makes the single translation between both languages quite obvious. Also we resolve various technical restrictions, as, e.g., the tetrahedral symmetry of the $F$-symbols in Ref. [17], or the trivalence of the underlying graphs.

• In Section (5.9) we give another type of fixed point model for topological order in $2 + 1$ dimensions. Those essentially correspond to the Kuperberg invariant [15] in the state-sum language, or the quantum double models [8] in the Hamiltonian language (or more precisely their generalizations to (weak) Hopf *-algebras [18, 25]). Our formalism provides a simple geometric picture for the equivalence between Levin-Wen models and quantum double models.

• In Section (5.10) we obtain fixed point models for gapped boundaries, co-dimension 1 defects or domain walls of the

models from Section (5.9) that to our knowledge have no analogue in the literature. In Ref. [26] the authors give models for gapped boundaries and domain walls in the language of fusion categories. Our models cover equivalent phases, yet the microscopic implementation is quite different.

We also make clear the connection of our formalism to the description of topologically ordered systems with *projected entangled pair states (PEPS)* and *virtual matrix product operator (MPO) symmetries* [3, 4]: Fixed point models for topological boundaries (like the ones in Section (5.10)) yield PEPS representations for certain ground states (see Section (4.4)). This yields a very direct physical interpretation of the virtual MPO symmetries: Those are simply topological defects living within the physical boundary, see Section (6.1.10). From this viewpoint it is also easy to see that there are topologically distinct PEPS representations for one and the same fixed point model, corresponding to the topologically distinct boundaries. Those can have different virtual symmetries, corresponding to different boundary defects (though the fusion categories that capture their statistics are *Morita equivalent*).

• In Section (5.11) we obtain a type of fixed point model for anyons in the models in Section (5.9) (or defects projecting onto the different sectors of a symmetry-broken phase). Again we do not know of any analogue construction in the literature. However, it is well-known that anyons in quantum double models correspond to irreducible representations of the *Drinfel'd quantum double* [27] of the input Hopf algebra, which we can recover from our formalism. Yet the tensors consistency conditions describing our fixed model for the anyon are simpler than those for representations of the quantum double.

• In Section (6.1) we sketch implementations of the framework for various other situations: In Section (6.1.1) we define fermionic tensors and lattices with combinatorial spin structure, which is all we need to define fermionic fixed point models. In Section (6.1.2) we sketch the definition of Gaussian tensors and how they can be mapped to fermionic tensors. This way one can build a bridge between classifications of phases for free fermions [28], and for quantum spin systems [1]. In Section (6.1.3) we define symmetric tensors, which is all we need to describe SPT phases. For certain SPT/SET phases it also suffices to take another type of lattice, namely homology lattices. In Section (6.1.4) we sketch the definition of lattices with combinatorial orientation. TLs on such lattices do not have the unitarity properties that are present in standard quantum mechanics, so they describe non-Hermitian quantum phases. In Section (6.1.5) we conjecture that if we take a universal state-sum construction, then also chiral topological phases in $2+1$ dimensions can be described by such a fixed point model. In Section (6.1.6) we dream of a 2-dimensional lattice type with moves corresponding to conformal transformations. This would yield an algebraic classification of critical 2-dimensional phases on a direct physical microscopic lattice level. In Section (6.1.7) we demonstrate how axiomatic TQFTs can be formalized as CTLs. In Section (6.1.8) we sketch the same thing for extended TQFTs. In Section (6.1.10) we briefly sketch fixed point models for other types of defects, such as twists, defects in the boundary of $2+1$-dimensional systems, or general 0-dimensional defects. In Section (6.1.11) we define yet another microscopic representation of topological phases in $2+1$ dimensions inspired by the colour code, in order to demonstrate how easy it is to get new types of microscopic fixed point models from some geometric intuition using our framework.

## 1.3 How to read this work

To keep the structure of this work as clear as possible, we separated the general formalism from the concrete implementations, as well as the mathematical structures from their physical interpretation.

The general mathematical formalism is presented in Chapter (2). Chapter (3) is basically a list of many important concrete lattice types, and the intuitive continuum picture that they describe. Chapter (5) is a list of concrete (C)TL types. Both Chapter have a heavy encyclopedic character: Apart from a few cross-references, the single sections are independent implementations of the basic formalism in Chapter (2. Section (6.1) contains various ideas for other (C)TL types that describe other sorts of phases.

One of the main purposes of this work is to provide a clean, simple, unified and self-contained language for talking about phases and their fixed point models. In order to keep this work self-contained and to avoid confusion we put relations to other approaches from the literature and to the more physical language only to certain specified places: The different physical contexts in which the general formalism can be applied for classifying phases of matter given in Chapter (4). Each of the sections in Chapter (5) contains one subsection with giving the physical context and the relation to known fixed point models, as well as one subsection giving the relation to conventional algebraic structures. All other statements that connect the formalism to the conventional language or that are not self-contained are marked as "comment". Only in Chapter (6) we loosen this strict separation.

The paper is not designed for a linear read-through: For the understanding of the later sections it is not always necessary to understand everything before in detail. In Chapter (2) and Chapter (3) we payed a lot of effort to give or at least sketch rigorous combinatorial definitions for all the concepts we introduce. Those combinatorial definitions can also be regarded as a proof of principle, and their detailed understanding will not be important for the physical context in Chapter (4) and the concrete (C)TL types in Chapter (5): In those section we will work mostly with a lot of geometric intuition and pictures. To facilitate this intuitive understanding we added many informal definitions and examples to the Chapters (2, 3).

We suggest the following strategy for reading this work: Go over the informal definitions and examples in Chapter (2) to get a rough feeling for the important vocabulary (listed in the beginning of that Chapter). Then take a look into Chapter (4) to until the physical interpretation becomes sufficiently clear. Then pick one or two of the concrete (C)TL types in Chapter (5) (e.g., 2E-CTLs and 3E-CTLs) and try to understand those fixed point constructions within the general framework. All the vocabulary introduced in this work will be hyperlinked to the place where it is defined, so if unfamiliar notions occur one can easily follow the links till everything is clear. After understanding the main notions of the formalism, the rest of this work can be regarded as an encyclopedia of fixed point models, all of them presented in a simple unified language.

This paper is thought as a long-term project. In future versions we will expand the encyclopedia by adding new sections presenting new types of fixed point models for all different kinds of situations. In particular we will work out properly the ideas in Section (6.1). We will also strive to simplify and beautify the main formalism as much as possible, and make it more accessible by adding many more examples. Also we will improve the presentation for the (C)TL types in Section (5) and add more comments and concrete examples. In particular we will rewrite the CTL types that come from TLs in the language of TLs which makes the way we obtain the basic tensors and axioms conceptually much easier.

## 1.4 Notational conventions

### Lattices

**Remark 1.** We will draw backgrounds of a CTL in three different ways:

1. We draw one complete lattice representative of the background.

2. We draw the index lattice and for the rest of the background draw the corresponding continuum picture. E.g., if the extended backgrounds are boundary (higher order) manifolds, we will draw this manifold and draw the index lattice onto its boundary.

3. We draw only the index lattice and assume that the rest of the background is clear from the context, or just describe in words what it looks like.

**Remark 2.** When we draw a higher order manifold we will often use different colours for the different regions, such that we can see when multiple disconnected pieces belong to the same region. We might use the same colour for different regions if they can be distinguished by their dimension. If there aren't enough distinguishable colours we will put labels next to the different connected components of the regions instead.

When we draw a higher order manifold we cannot in general tell of what type it is, as we cannot tell whether there is another empty region or not, and what the upper link of potential empty regions is. If this is the case we will just say so.

**Remark 3.** Truly generic 3-dimensional objects cannot be drawn on a sheet of paper without drawing multiple 2-dimensional slices. We will thus not try to draw 3-dimensional things, but instead 1) take the skeleton consisting only of the 0-, 1-, and 2-dimensional parts and 2) project this skeleton onto the paper plain, i.e., "press it flat". Thereby we will often end up with multiple layers stacked on top of each other. We will draw 2-dimensional parts of those layers with a semi-transparent filling, such that they do not hide away the back layers. More specifically, for drawing higher order manifolds, we will only draw the 0-, 1- and 2-regions, and draw them on top of each other by using semi-transparent colours for the 2-regions.

If there are (higher dimensional) parts of the object that cannot be captured by the picture, we will simply describe them in words.

**Remark 4.** For drawing lattices living on 2-manifolds according to Remark (3), two layers (the **front** and **back** layer) will always suffice. Most of the times we will only draw the point-like and line-like objects (i.e., edges and vertices) of the lattice, and assume that the area-like objects (i.e., faces) are everywhere in between. When the 2-manifold on which the lattice lives is not a sphere but e.g., a torus, we will shade the layers in order to indicate where the area-like objects (i.e., faces) are located.

In the back layer we will draw line-like objects dashed, and point-like objects hollow, and labels grayed out.

**Remark 5.** For drawing lattices in 3 dimensions we will either draw the 1-skeleton or 2-skeleton of the lattice, or draw single local components and indicate by labels how they are connected with each other.

**Remark 6.** In order to mark some lattice elements, we overlay them semi-transparently with the following colour:

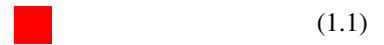

$$(1.1)$$

More precisely those markings look like:

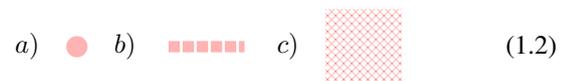

$$(1.2)$$

a) is for point-like, b) for line-like and c) for area-like objects of a lattice.

**Remark 7.** In order to indicate that two components a lattice are glued, we will mark them as in Remark (6). If more than one pair of components are glued we will try to make clear which component is glued to which other component by placing the pairs of glued component near each other. If the latter is not possible or unclear, we will put labels next to all glued components instead of marking them, and use the same label for two components that are glued.

**Remark 8.** When a (part of a) lattice is empty (i.e., the trivial lattice) we will literally draw nothing and just leave some blank space.

**Remark 9.** In this remark we summarize how we will draw different components of lattices in general. Vertices will be drawn as small dots,

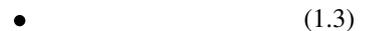

$$(1.3)$$

Edges will be drawn as lines connecting the vertices,

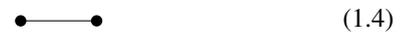

$$(1.4)$$

When we cut out a patch of some lattice we will indicate this by continuing the cut-off edges by three dots,

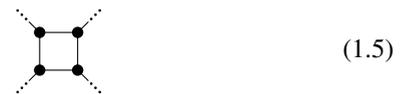

$$(1.5)$$

Faces will usually not be drawn at all. Whenever some region is encircled by edges we will assume that there is a face in between (if the lattice type has faces). When there are regions where there is no face, we will shade all regions that have faces, see also Remark (4).

An orientation of an edge is indicated by a simple arrow, e.g., the following edge is oriented towards the vertex on the right:

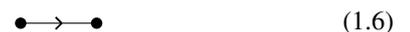

$$(1.6)$$

If we additionally have a dual orientation we will use an arrow that is only on one side, e.g., the following edge is oriented towards the vertex on the right and towards the face on the top:

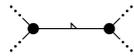

$$(1.7)$$

To indicate a favorite connected edge of a face we will place a small half-circle at the favorite edge. E.g., in the following example the edge on the right is the favorite edge of the 4-gon face in the middle:

$$(1.8)$$

## Tensors

We will use different notations for tensors throughout this work.

**Remark 10.** In **sub- or superscript notation**, a tensor is denoted by a letter or symbol with a list of sub- or superscript labels that represent the indices. More precisely each index corresponds to a position in the sub- or superscript. The letters or symbols used as label are arbitrary and only needed to specify how to match up the indices when we equate two tensors. E.g., a tensor $T$ with 3 indices is denoted by

$$T_{a,b,c} \quad \text{, or} \quad T_{abc}. \quad (1.9)$$

The tensor product is denoted by placing two tensor symbols next to each other. E.g., the tensor product $A$ of two tensors $B$ and $C$ with 3 and 2 indices

$$A_{a,b,c,d,e} = B_{a,b,c} C_{d,e}. \quad (1.10)$$

Contraction of two indices is denoted by using the same letter for the corresponding labels. E.g., the contraction $D$ of the first and third index of a tensor $E$ with 4 indices is

$$D_{a,b} = E_{x,a,x,b}. \quad (1.11)$$

**Remark 11.** In **tensor-network notation** a tensor is denoted by a labeled box or any other shape with lines emanating from the boundary whose endpoints carry labels. So each of those lines corresponds to one index. E.g., for a 3-index tensor $T$, we have

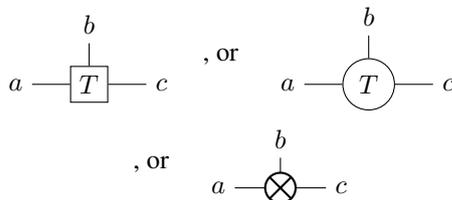

$$(1.12)$$

Again the labels tell us how to match up indices when we equate two tensors. Often we will omit the labels and instead use the position of the endpoints and/or the directions they are pointing at to indicate the latter. Which index is which is determined by where exactly the corresponding line sticks out of the shape/box. Shapes/boxes might be drawn rotated but never reflected. If the shape is round there might be no way to distinguish the indices by where they stick out. In this case the notation is only consistent when the tensor is invariant under cyclic permutation of its indices. Different index types are often reflected by using different line-styles for the corresponding lines.

As for sub- or superscript notation, the tensor product is denoted by placing two tensors next to each other, e.g.,

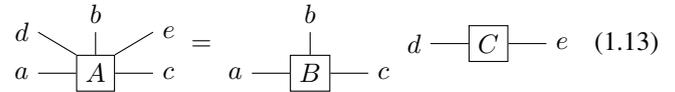

$$(1.13)$$

Contraction of two indices is denoted by connecting the corresponding lines. E.g.,

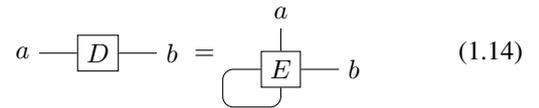

$$(1.14)$$

In this notations, tensor networks are represented by graph-like pictures. E.g.

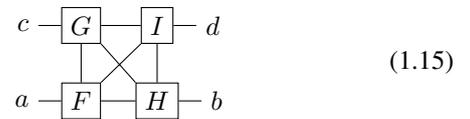

$$(1.15)$$

Two lines crossing does not have any effect. When too many connections would clutter the diagram, contraction might also be denoted by using the same label for the contracted indices instead, as for subscript notation. E.g.,

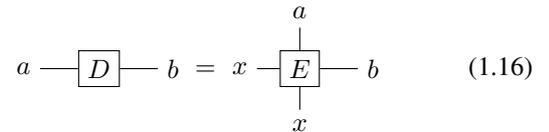

$$(1.16)$$

Note that the empty tensor-network evaluates to the number 1, so the latter is represented by drawing literally nothing. In this case we will just leave some blank space.

The identity tensor is consistently denoted by a free line, where the two endpoints of the line represent the two indices. E.g., for real tensors we have

$$a \text{———} b = \delta_{a,b} = \begin{cases} 1 \text{ if } a = b \\ 0 \text{ otherwise} \end{cases} \quad (1.17)$$

**Remark 12.** For CTLs there will be a very natural notation for drawing the tensors associated to lattices: We can simply draw the lattice (or only the index lattice, see Remark (1)). If we want to equate the tensors from the CTL with tensors in the above notations, we have to put labels. As indices of the CTL tensors will be associated to places of the lattice, we can simply draw the labels near those places in the drawn lattice. E.g., if we have a CTL which associates indices to every edge of a 1-dimensional simplicial complex we denote the tensor associated to the triangle as

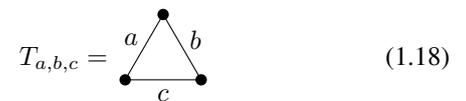

$$(1.18)$$

**Remark 13.** The three notations are actually not too different from each other and will often be combined, e.g., a tensor in subscript notation can be contracted with a tensor in tensor network notation by using the same letter for the corresponding index labels.

**Remark 14.** We will use the above notations also if we have tensors in the generalized sense of Remark (32). In this case we will simply state the basis set of which indices depends on the value of which other indices.

**Remark 15.** We will use the following ways of representing concrete real/complex tensors:

First, in **array notation** we will arrange all tensor entries in an array which has as many dimensions as indices. So for one index this is just a sequence, for two indices we get a 2-dimensional array with rows and columns, for three indices we have a sequence of blocks that are 2-dimensional arrays, and so on. Thereby the row, column, block, and so on of each entry corresponds to an element of the basis set of the first, second, third, and so on index of the tensor.

Second, **sparse notation** will be useful when many entries of a tensor are 0. Then we will write down a comma-separated list, with entries like $a_1|a_2|a_3| \ldots : x$, where $a_1$, $a_2$, $a_3$, $\ldots$ are elements of the basis sets of the first, second, third, and so on index, and $x$ is the tensor entry for this index configuration. All basis set element configurations that do not appear in this list have entry 0.

We will sometimes combine the two notations, and use array notation for some of the indices, and sparse form for the remaining indices. In this case the entries of the array form are given as tensors in sparse form or vice versa.

## Miscellaneous

**Remark 16.** Elements of lattices or (C)TLs that do already exist for a sub-type of the lattice or (C)TL type will be often referred to by giving the name of the sub-type as a prefix.

**Remark 17.** Many of the definitions will use "commutative diagrams" (not always in the strict mathematical sense) as a pictorial illustration. Thereby we will make use of two different line styles for the relations between the objects to distinguish whether there exists a transformation relating two objects, or whether the diagram has to hold for all transformations relating two objects:

$$\begin{array}{ccc} \text{-----} & : & \text{exists} \\ \text{wwwww} & : & \text{forall} \end{array} \tag{1.19}$$

# Chapter 2

# General formalism

In this section, we define the general mathematical structures underlying this work: A lattice type is a combinatorial data structure with a particular notion of locality, and a lattice is a particular instance of such a data structure. Those data structures will usually come with nice intuitive geometric pictures. A TL is a local prescription that associates tensor networks to all lattices of a fixed type. A CTL is a map that associates tensors to lattices such that the indices of the tensors correspond to different local components of the lattice. For both TLs and CTLs those maps have to be such that certain local changes or global operations in the lattice correspond to certain operations on the tensor networks or tensors:

$$
\begin{array}{ccc}
\textbf{Lattices:} & & \textbf{Tensor networks:} \\
\text{Symmetries,} & \xrightarrow{\text{ TL }} & \text{Index permutation,} \\
\text{Moves} & & \text{Tensor-network moves} \\
& & \\
\textbf{Lattices:} & & \textbf{Tensors:} \\
\text{Disjoint union,} & & \text{Tensor product,} \\
\text{Symmetries,} & \xrightarrow{\text{ CTL }} & \text{Index permutation,} \\
\text{Moves,} & & \text{Invariance,} \\
\text{Gluings} & & \text{Contraction}
\end{array}
\tag{2.1}
$$

There are different types of TLs and CTLs depending on which lattice type we use and how the geometry of the tensor network or the index prescription of the tensors look like. In the Section (2.1) we will explain the general notions of lattice, move and gluing. In Section (2.2) we define tensors in a rather abstract way and then, and discuss different tensor types, most importantly real and complex tensors. Finally we show how to put the lattice and the tensor side together and define TLs and CTLs in Section (2.3) and Section (2.4).

**Index of important definitions**

The following definitions in this will be important for understanding the other chapters. Note, however, that it is by no means necessary to understand them on a complete formal level, in most cases a rough intuitive understanding will suffice.

- lattice, (type)

- place, location (type)

- lattice mapping

- restriction

- lattice class

- disjoint union

- product type, sub-type, joint type

- decoration

- move

- moved lattice type, basic move

- background

- circuit move

- gluing

- glued lattice type, basic gluing

- background gluing

- finitely generated, basic lattice

- history, history move

- defining history

- finitely axiomatized, basic history move

- tensor (type)

- index (type)

- basis (type)

- tensor product, contraction

- tensor network, evaluation

- gauge transformation

- tensor mapping

- real tensor

- complex number tensor

- complex tensor

- realification

- delta tensor

- preferred basis real tensor

- tensor-network move

## 2.1 Lattices, moves and gluings

In contrast to how the word is often used, by lattice we do not refer to a regular grid but to something far more general: Lattices are a very fundamental concept of combinatorial data structures that have a notion of locality. Because of its generality there are many different ways to formalize this concept. For every concept that we introduce we will first give an intuitive, informal definition, and then sketch a formal definition within one particular **universal formalization**. The situation is a bit similar to what one has in compatibility theory: Computation is a very fundamental concept and everyone who has programmed something has a very clear intuition about what a computer can do. In order to make it rigorous we have to choose a particular formalization, as for example Turing machines, lambda calculus, or any Turing complete programming language.

### 2.1.1 Lattices

**Informal definition 1.** We can picture a lattice as a collection of small objects called cells covering some sort of underlying space called background. This background can be something like a topological manifold (which will be the case for large parts of this work), a manifold with boundary, a manifold with a geometric, conformal or spin structure, something that we will call higher order manifold, or even something completely different. The so-called connections provide a combinatorial data structure that describes how cells near each other on the background are located with respect to each other. There are different lattice types consisting of cells and connections of different types. The types can be associated to different lattices classes, corresponding to different sorts of backgrounds.

The actual lattice will be just the combinatorics of the cells and connections. The background will not be an explicit part of the description but rather emerges from the combinatorics.

**Informal definition 2.** A **patch** of a lattice is given by all cells of the lattice inside a certain part of the background. Whenever we call something **small**, we want to indicate that a it is of a fixed constant size (but in principle arbitrary) size, relative to a fixed lattice type, but that it cannot depend on a specific lattice of that type. I.e., a **small patch** of a lattice is a patch that could be regarded as a patch of different lattices of a fixed type, that contains a constant number of cells, independent of the lattice. Two cells in a lattice are called **near** each other if their distance measured in the number of connections needed to get from one to the other is small.

### Pre-types, locations, places and mappings

**Informal definition 3.** A lattice pre-type refers to the information of which different cells types and connection types there are. A lattice of a pre-type is given by sets of cells with arbitrary connections.

**Definition 1.** A **lattice pre-type** is defined by a set of **cell types** and a set of **connection types**, where to each connection type we associate a **source cell type** and a **target cell type**. In other words, it is a directed graph, whose vertices are the cell types and whose links are the connection types.

**Definition 2.** The **trivial lattice pre-type** is the lattice pre-type with the empty set of cell types and connection types.

**Example 1.** Consider the following (very simple) lattice pre-type with only one cell type $\mathcal{C}$ and one connection type $\mathcal{D}$ whose source and target is both that cell type:

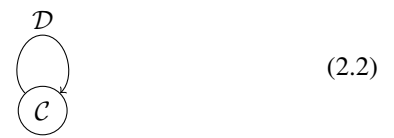

$$(2.2)$$

**Definition 3.** A **lattice** of a certain pre-type consists of one finite set for each cell type (whose elements will be called **cells** of the given type) and one finite set for each connection type (whose elements will be called **connections** of the given type). On those sets we have the following data structure: For each connection type there is a **source map** and a **target map** mapping from the corresponding set of connections to the set of the source cell type and to the set of the target cell type. In other words, we can identify the connections with pairs of source and target cell, however, each pair can occur multiple times. If two cells are source and target of a common connection we say that they are **connected**.

**Remark 18.** Two lattices of a pre-type are called **equivalent** if there is a bijection between their cells and connections of different type that leaves the source and target maps invariant. We will think of equivalent lattices as the same lattice. In particular when we denote lattices by drawing them graphically there will be no way to distinguish different but equivalent lattices.

**Remark 19.** For every lattice pre-type there is the **trivial lattice**, with all sets of cells and connections being empty.

**Example 2.** Consider the lattice pre-type from Eq. (2.2). A lattice of this pre-type is a set of cells, and a set of connections to each of which a source and a target cell is associated. So if we draw the cells as vertices and the connections as directed edges, we can represent a lattice of this pre-type as a directed graph. E.g.,

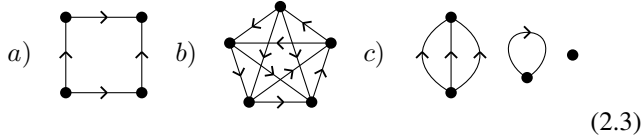

$$(2.3)$$

**Informal definition 4.** A location of a specific type refers to a collection of cells which are connected among each other in a specific way. A location (type) is called place (type) if it only involves cells in a small connected patch.

**Definition 4.** A **location type** for a lattice pre-type is a directed graph where each vertex is labeled by a cell type and each link is labeled by a connection type, such that: The labels of the source and target vertex of a link are the source and target cell type of the associated connection type. In other words, it is a second lattice pre-type with a homomorphism to the given lattice pre-type. A location type is called $n$-**location type** if the directed graph has $n$ connected components. 1-location types are called **place types**.

**Definition 5.** A **location** (of given type) in a lattice (of the according pre-type) is an injective map from the vertices and links of the location type to the cells and connections of the according type, such that: The source and target of the connection associated to a link are the cells that are associated to the source and target vertex of the link. A **place** is a location for a place type.

**Remark 20.** Each cell type defines a place type consisting of a single vertex labeled by this cell type. Each cell of the type in a lattice is a place of this type.

**Remark 21.** Each connection type defines a place type consisting of a single link between two vertices. The link is labeled by the connection type and the vertices by its source and target cell types. Each connection of the type is a place of this type.

**Example 3.** Consider the following place type for the lattice pre-type in Eq. (2.2):

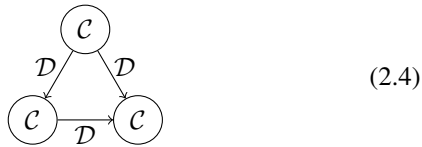

$$(2.4)$$

Consider the following place of this type in the lattice b) in Eq. (2.3) (the corresponding cells and connections have been marked according to Remark (6)):

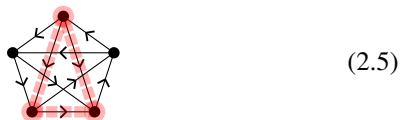

$$(2.5)$$

**Definition 6.** A **sub-type** of a place type $\mathcal{P}$ is another place type $\mathcal{S}$ together with an injective homomorphism from the labeled graph given by $\mathcal{S}$ to the labeled graph given by $\mathcal{P}$. In other words, the latter graph contains the former as a labeled sub-graph.

**Definition 7.** Consider a fixed place type $\mathcal{P}$ together with another place type $\mathcal{S}$ as a sub-type, and a place of type $\mathcal{P}$ in a lattice. The **sub place** of the place with respect to the sub-type is the place of type $\mathcal{S}$ that is obtained after restricting the place to the sub-graph given by the sub-type.

**Example 4.** Consider the following sub-type of the place type in Eq. (2.4):

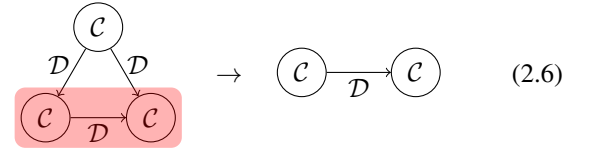

$$(2.6)$$

The according sub place of thee place in Eq. (2.5) is given by

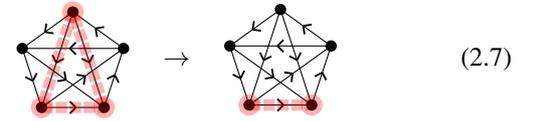

$$(2.7)$$

**Informal definition 5.** A lattice mapping is a locality-preserving prescription that transforms lattices of one pre-type into lattices of another pre-type. I.e., how the lattice of the second type looks like at one spot can only depend on how the first type looks like at the same spot.

**Definition 8.** Let us consider two lattice pre-types $\mathcal{A}$ and $\mathcal{B}$. A **cell mapping prescription** is a pair of a cell type of $\mathcal{B}$ and place type of $\mathcal{A}$. A **connection mapping prescription** with respect to two cell mapping prescriptions $\mathcal{X}$ and $\mathcal{Y}$ is a pair of connection type of $\mathcal{B}$ and place type of $\mathcal{A}$ containing place types of $\mathcal{X}$ and $\mathcal{Y}$ as sub-types. The cell types of $\mathcal{X}$ and $\mathcal{Y}$ have to be the source and target cell type of the connection type of $\mathcal{B}$.

A **lattice mapping** (or just **mapping** if the context is clear) from $\mathcal{A}$ to $\mathcal{B}$ consists of

- a set of cell mapping prescriptions, and

- a set of connection mapping prescriptions with respect to pairs of the cell mapping prescriptions.

**Definition 9.** Given a lattice mapping, for every lattice $A$ of pre-type $\mathcal{A}$, the **mapping of** $A$ is the lattice $B$ of pre-type $\mathcal{B}$ constructed in the following way:

- For each cell mapping prescription, $B$ has one cell of the according type for every place of $A$ of the according type.

- For each connection mapping prescription, $B$ has one connection of the according type for every place of $A$ of the according type. This place of $B$ has two subplaces corresponding to the two sub-types. The cells associated to those subplaces are the source and target cell of the connection.

**Definition 10.** For every lattice pre-type, the **identity mapping** is the lattice mapping from this pre-type to the pre-type again that has one cell mapping prescription for every cell type and one connection mapping prescription for every connection type, such that the corresponding place types are given by the same cell type/connection types. I.e., the mapping of every lattice is the lattice itself.

**Definition 11.** For every pair of lattice pre-types $\mathcal{A}$ and $\mathcal{B}$, the **trivial mapping** from $\mathcal{A}$ to $\mathcal{B}$ is the lattice mapping without any cell mapping prescription of connection mapping prescription. I.e., every lattice of type $\mathcal{A}$ is mapped to the trivial lattice of type $\mathcal{B}$.

**Example 5.** Consider the following lattice mapping from the lattice pre-type in Eq. (2.2) to that same pre-type again: There is one cell mapping prescription:

$$\mathcal{C} \quad : \quad \boxed{\mathcal{C}} \!-\!\!\underset{\mathcal{D}}{}\!\!-\! \boxed{\mathcal{C}} \tag{2.8}$$

and one connection mapping prescription:

$$\mathcal{D} \quad : \quad \mathcal{C} \!-\!\underset{\mathcal{D}}{}\!-\! \mathcal{C} \!-\!\underset{\mathcal{D}}{}\!-\! \mathcal{C}$$

Target: $\mathcal{C} \!-\!\underset{\mathcal{D}}{}\!-\! \mathcal{C} \!-\!\underset{\mathcal{D}}{}\!-\! \mathcal{C}$ \qquad (2.9)

Source: $\mathcal{C} \!-\!\underset{\mathcal{D}}{}\!-\! \mathcal{C} \!-\!\underset{\mathcal{D}}{}\!-\! \mathcal{C}$

i.e., in the graphical representation with edges and vertices this mapping is the following operation: Replace place of type given in Eq. (2.8) by a cell of type $\mathcal{C}$, replace every edge by a vertex, and connect two vertices if the corresponding edges share a common vertex and their directions are aligned. E.g.,

$$\tag{2.10}$$

**Restrictions**

**Informal definition 6.** A feature type refers to one particular aspect of how a lattice can look like locally in a small patch around places of a given type. The feature (of this type) around some place in the lattice describes how this aspect is realized around a given place in a lattice.

**Definition 12.** A **feature type** for a central place type $\mathcal{P}$ in a lattice pre-type $\mathcal{L}$ is a mapping to another lattice pre-type $\mathcal{X}$ such that every place type of all cell mapping prescriptions and connection mapping prescriptions contains $\mathcal{P}$ as a sub-type.

**Definition 13.** The **feature** of a given type around a place $P$ (of type $\mathcal{P}$ above) in a lattice $L$ (of pre-type $\mathcal{L}$ above) is the mapping of $L$ by the feature type, where we only take cells and connections for places that have $P$ as their sub place.

**Example 6.** For the cells of a fixed type $\mathcal{C}$, the number of connections of a fixed type $\mathcal{D}$ that have this cell as source (target) cell is a feature type: The corresponding lattice pre-type $\mathcal{X}$ consists of one cell type, and no connection types. The central place type consists of a single vertex labeled by $\mathcal{C}$. There is one single cell mapping prescription that associates to the only cell type of $\mathcal{X}$ the place type given by a single connection of type $\mathcal{D}$. This place type contains the central place type as a sub-type by taking only the source (target) cell.

**Informal definition 7.** For our final definition of a lattice we will introduce certain restrictions that restrict how the lattice can look like around different places. The key concept of locality will be imprinted by imposing that there can only be a finite number of possibilities of how a lattice looks like around a each place.

**Definition 14.** A **restriction** for a lattice pre-type is a feature type together with a set of features of that type. A **restricted lattice pre-type** is a lattice pre-type together with a set of restrictions.

**Definition 15.** A restriction is **fulfilled** at a place in a lattice (of a certain pre-type) if the feature (of type given by the restriction) around the place equals one of the features in the set of features of the restriction. A **lattice** of a given (restricted pre-)type is a lattice of the corresponding pre-type that fulfills all restrictions around every according place.

**Definition 16.** A **lattice type** is a restricted lattice pre-type such that for all lattices of that restricted pre-type and every feature type there can only be a finite number of features of this type.

**Remark 22.** A restricted lattice pre-type is a lattice type if and only if the number of connections connected to each cell type can only take values from a finite set. So in order to get a proper lattice type it suffices to add restrictions limiting this number of connections.

**Example 7.** Consider the lattice pre-type Eq. (2.2) together with the two restrictions that every cell has to be the source and the target of exactly one connection. This directly defines a lattice type. Lattices of this type are not arbitrary directed graphs any more, but rather oriented loops of edges. E.g.,

$$\tag{2.11}$$

**Definition 17.** A **mapping** from a lattice type $\mathcal{A}$ to a lattice type $\mathcal{B}$ is a mapping from the pre-type $\mathcal{A}$ to the pre-type $\mathcal{B}$ such that for every lattice $A$ of type $\mathcal{A}$, the mapping of $A$ is a lattice of type $\mathcal{B}$. I.e., for restricted lattice pre-types, a mapping has to preserve the restrictions.

**Informal definition 8.** Two lattice types are in the same class if they model the same kind of structure in a combinatorially different way.

**Definition 18.** Two lattice types $\mathcal{A}$ and $\mathcal{B}$ are said to be in the **same class** if there is a mapping $\mathcal{T}_{\mathcal{A}}$ from $\mathcal{A}$ to $\mathcal{B}$ and a mapping $\mathcal{T}_{\mathcal{B}}$ from $\mathcal{B}$ to $\mathcal{A}$ such that $\mathcal{T}_{\mathcal{A}} \circ \mathcal{T}_{\mathcal{B}} = \mathrm{id}$ and $\mathcal{T}_{\mathcal{B}} \circ \mathcal{T}_{\mathcal{A}} = \mathrm{id}$, where $\mathrm{id}$ is the trivial lattice mapping on $\mathcal{A}$ and $\mathcal{B}$, respectively. As a diagram:

$$\tag{2.12}$$

**Example 8.** Consider one the one hand the lattice type with one cell type (called vertices) and no connections, such that lattices of this type consist of an arbitrary number of vertices. On the other hand consider the lattice type with one cell type (called vertices) and one connection type (called edges) with this cell type as source and target, with the following restriction: Every vertex is either source or target of exactly one edge. Lattices of this type consist of an arbitrary number of pairs of connected vertices.

Those two lattice types are obviously equivalent: Replacing every pair of connected vertices by a single vertex and vice versa yields a one-to-one correspondence between the lattices. The corresponding lattice mappings act as, e.g.,

$$\bullet \quad \overset{\bullet}{\bullet} \quad \longleftrightarrow \quad \overset{\bullet}{\bullet} \quad \overset{\bullet}{\bullet} \tag{2.13}$$

**Disjoint union, product, sub-types and joint types**

**Informal definition 9.** The disjoint union of two lattices is given by putting two lattices next to each other and interpreting them as one single lattice.

**Definition 19.** The **disjoint union** of two lattices $A$ and $B$ (of the same type) is the lattice (of again the same type) whose sets of cells and connections are the disjoint union of the sets of cells and connections of $A$ and $B$ (for each cell type and connection type separately).

**Informal definition 10.** For two lattice types we can construct the product type: A lattice of this type is just a pair of one lattice for each of the two types. If there is a subset of lattices that only uses a subset of cells and connections, this subset forms a sub-type of a lattice type. When we have two such sub-types that correspond to equivalent lattice types, we can build the joint type for those two sub-types by identifying the two subsets of cells and connections.

**Definition 20.** The **product** of two lattice pre-types is the disjoint union of the two directed graphs. The **product** of two lattice types is the product of the two pre-types with the disjoint union of the two sets of restrictions.

**Definition 21.** A **sub-type** of a lattice pre-type is another lattice pre-type together with an identification with a sub-graph of the former lattice pre-type.

A **sub-type** of a lattice type is a sub-type of the corresponding pre-type, together with all the restrictions of the original lattice type whose place types involve only cells and connections of the sub-type.

**Observation 1.** Consider a lattice type with a sub-type. Every lattice of the sub-type can be interpreted as a lattice of the original type.

**Observation 2.** The product of two types canonically contains both original lattice types as sub-types. Every lattice of the product type is a disjoint union of two lattices of each of the original types, interpreted as lattice of the product type.

**Definition 22.** Consider a lattice pre-type containing another lattice pre-type twice as a sub-type. The **joint lattice pre-type** of the lattice pre-type is the lattice pre-type obtained by identifying the corresponding sub-graphs. The **joint type** of a lattice types with a double sub-type is the corresponding joint pre-type together with the disjoint union of all restrictions. Thereby restrictions that are completely contained in the sub-types can be equal, and in this case we can eliminate one of the copies of a restriction.

**Example 9.** Consider the product of twice the same lattice type. This contains the original lattice type twice as a sub-type. The joint type for this is the original lattice type again.

**Example 10.** Consider blue 1-dimensional simplicial complexes with green boundary together with purple 1-dimensional simplicial complexes with orange boundary as a lattice type. 1-dimensional simplicial complexes (without boundary) are contained twice as a sub-type by restricting to the blue/purple simplicial complexes with empty boundary. The joint type for those two sub-types is given by (black) 1-dimensional simplicial complexes with two sorts of (green or orange) boundary vertices. Consider the following examples of lattices of the original type in a) and b) and lattices of the joint type in c) and d):

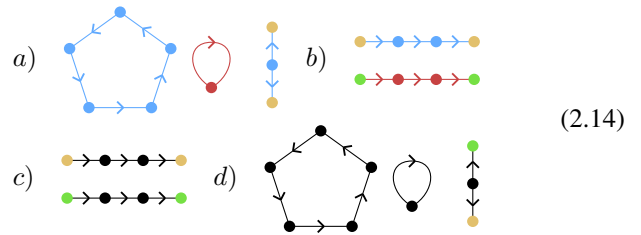

$$\tag{2.14}$$

**Informal definition 11.** In Section (3) we define a few very basic lattice types. The lattice types that will be used in Section (5) build up on those basic lattice types but will be slightly refined by adding additional combinatorial data. We will refer to such additional data **decorations**.

**Example 11.** A typical example for a decoration is adding an orientation to all cells of a type that have a geometric interpretation as edges. Or more generally, for each cell of a certain type that is connected to multiple other cells with the same kind of connection, specifying a favorite connected cell.

**Examples**

**Example 12.** The non-negative integers define a lattice type, with only one cell type and no connections: The different lattices are given by different numbers of cells.

**Example 13.** Directed graphs can be seen as a lattice type if we add some restrictions, say, set the maximum number of links to which a vertex can be connected to some $n$. There is one cell type, corresponding to the vertices. There is one connection type with the vertices as its source and target, corresponding to the links. There is one restriction: For every vertex the number of connections to this vertex (as source or target) has to be at most $n$.

**Example 14.** Also undirected graphs form a lattice type: To this end we use two cell types, corresponding to the vertices and to the links. There is one connection type corresponding to the pairs of link and connected vertex. There are two restrictions: 1) the number of vertices connected to every link is exactly 2. 2) the number of links connected to every vertex is a most $n$ (as in the directed case).

**Example 15.** Square grids with oriented edges can be seen as a lattice type: Just as for directed graphs, there is one cell type corresponding to the vertices, and one connection type corresponding to the edges of the grid. But now we have the following restrictions: 1) The number of edges connected to every vertex is exactly $4$. 2) Both the plaquettes and the edges in the grid are a place type. So we can look at the feature around a vertex that maps the plaquettes around the vertex to cells of one type, and the edges shared between two such plaquettes to connections between those cells. Then we restrict the features of this type to be a $4$-gon.

The lattices of this type are $m \times n$ square grids with periodic, but potentially twisted boundary conditions.

**Example 16.** Trees (restricted by some maximum number of children) cannot be seen as lattices: Trees by construction must not have cycles, however, the acyclicity is not a local property, as cycles can have arbitrary size. Thus, we cannot formulate acyclicity as a restriction. One can of course exclude cycles up to a fixed maximum size for any maximum size, but one cannot exclude cycles of all possible sizes all-together.

**Example 17.** The paradigmatic example for a lattice type are triangulations of $n$-manifolds (or $n$-manifolds with boundary), for a fixed $n$ also known as *simplicial complexes*. An according combinatorial structure will be defined in Section ($3$).

**Example 18.** Blocking/fine-graining is a typical example of a mapping. E.g., there is a mapping from $2$-dimensional square grids to $2$-dimensional square grids by replacing each plaquette of the square grid by $4$ plaquettes:

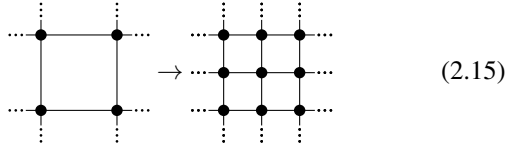

$$(2.15)$$

### 2.1.2 Moves

In this section we will introduce local operations called moves that connect different lattices. Moves will be important for defining consistency conditions for TLs and CTLs later.

**Moved types and backgrounds**

**Informal definition 12.** A move is a prescription of how to exchange a small patch of a lattice with another patch (or vice versa).

**Definition 23.** A **move** for a given lattice pre-type consists of

• two place types $\mathcal{P}_1$ and $\mathcal{P}_2$. Places of those types will be referred to as **move places**,

• a map that associates to every vertex of $\mathcal{P}_1$ a set of connection types (the **dangling connections**) that have the cell associated to the vertex as their source or target. The same for $\mathcal{P}_2$, and

• a map from the set of connection types from the previous item for $\mathcal{P}_1$ to the vertices of $\mathcal{P}_2$. The same for $\mathcal{P}_1$ and $\mathcal{P}_2$ interchanged.

A move can be **applied** to every place $P_1$ of type $\mathcal{P}_1$ such that for every cell of the place, the connections to this cell that are not part of $P_1$ are of a type in the set of dangling connection types. Thereby 1) remove all cells and connections of $P_1$. 2) paste the cells and connections of $\mathcal{P}_2$. 3) After the removal of $P_1$ there are dangling connections that do not have a source/target any more. The third item above tells which cells of $\mathcal{P}_2$ to use as source/target of those dangling connections.

Conversely, a move can be applied to every place $P_2$ of type $\mathcal{P}_2$.

**Definition 24.** A **moved lattice type** is a lattice type equipped with a set of moves that are consistent with the restrictions. Those moves are called the **basic moves**.

**Remark 23.** Every lattice type can be made a moved lattice type by choosing the empty set of basic moves.

**Example 19.** Consider the lattice type given by the pre-type in Eq. (2.2) together with the restrictions in example (7). Consider the following move: The pair of place types together with dangling connections (dashed lines) to other parts of the lattice is given by

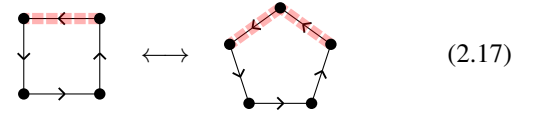

$$(2.16)$$

The dangling connections of $\mathcal{P}_2$ on the left (right) are mapped to the left (right) vertex of $\mathcal{P}_1$ and vice versa.

Applying the move means replacing a single edge in the lattice with two edges and vice versa. E.g., (see Remark (6)):

$$(2.17)$$

**Informal definition 13.** Intuitively the background can be imagined as some sort of underlying space into which the lattice is embedded. The basic moves change the lattice but not this underlying space.

**Definition 25.** Two lattices $A$ and $B$ of the same moved type are said to **have the same background** if $B$ is connected to $A$ by a sequence of basic moves. I.e., as a diagram:

$$A \xleftarrow{\quad \text{basic moves} \quad} B \qquad (2.18)$$

For a given moved lattice type the equivalence classes of lattices under having the same background are called **backgrounds**.

**Example 20.** Consider the moved lattice type given by the move in example (19) and the according lattice type. Lattices of this type are given by a set of loops with different numbers of edges. For each loop we can arbitrarily change the number of edges by the basic move, however, we can never remove or add a loop. Thus, there is exactly one background for each number of loops.

**Definition 26.** A **sub-type** of a moved lattice type is a sub-type of the corresponding lattice type together with all the moves of the original moved type whose place types $\mathcal{P}_1$ and $\mathcal{P}_2$ is contained in the sub-type.

**Mappings and classes**

**Definition 27.** A **(lattice) mapping** from a moved lattice type $\mathcal{A}$ to a moved lattice type $\mathcal{B}$ is a mapping between the two lattice types such that for every basic move relating two lattices $A$ and $A'$ of type $\mathcal{A}$ there is a basic move (or a sequence of basic moves) relating the mappings $B$ and $B'$ of $A$ and $A'$. I.e., the following diagram commutes:

$$
\begin{array}{ccc}
A & \xrightarrow{\text{mapping}} & B \\
\Big\updownarrow \text{\scriptsize basic} & & \Big\updownarrow \text{\scriptsize basic} \\
\text{\scriptsize moves} & & \text{\scriptsize moves} \\
A' & \xrightarrow{\text{mapping}} & B'
\end{array}
\tag{2.19}
$$

**Definition 28.** A **parallel move** is a map that associates different basic moves to all places of different types, such that those moves act on non-overlapping patches and thus can be applied in parallel. A **circuit move** is a finite sequence of parallel moves. Thereby the places where the moves are associated to in later steps of the sequence are places of the original lattice.

A **circuit move** with respect to a lattice mapping is a circuit move applied to the result of that lattice mapping, such that the places where the moves are associated to are places in the original lattice before the mapping. Applying such a circuit move after a lattice mapping defines another lattice mapping. Circuit moves (with respect to the trivial lattice mapping) define lattice mappings between twice the same lattice type.

**Definition 29.** Two moved lattice types $\mathcal{A}$ and $\mathcal{B}$ are said to be **in the same class** if there is a mapping $\mathcal{T}_\mathcal{A}$ from $\mathcal{A}$ to $\mathcal{B}$ and a mapping $\mathcal{T}_\mathcal{B}$ from $\mathcal{B}$ to $\mathcal{A}$, as well as circuit moves $\mathcal{C}_\mathcal{A}$ and $\mathcal{C}_\mathcal{B}$ for $\mathcal{A}$ and $\mathcal{B}$ such that both $\mathcal{C}_\mathcal{A} = \mathcal{T}_\mathcal{B} \circ \mathcal{T}_\mathcal{A}$ and $\mathcal{C}_\mathcal{B} = \mathcal{T}_\mathcal{A} \circ \mathcal{T}_\mathcal{B}$. As diagrams:

$$
\begin{array}{ccc}
 & \mathcal{A} & \\
 & \xrightarrow{\mathcal{T}_\mathcal{A}} & \\
\mathcal{C}_\mathcal{A} \Big\downarrow & & \mathcal{B} \\
 & \mathcal{A} & \\
 & \xrightarrow{\mathcal{T}_\mathcal{B}} &
\end{array}
\qquad
\begin{array}{ccc}
 & \mathcal{T}_\mathcal{B} & \mathcal{B} \\
 & & \\
\mathcal{A} & & \Big\downarrow \mathcal{C}_\mathcal{B} \\
 & \xrightarrow{\mathcal{T}_\mathcal{A}} & \mathcal{B}
\end{array}
\tag{2.20}
$$

Equivalence classes of moved lattice types under being in the same class are called **lattice classes**.

**Observation 3.** If two moved lattice types are in the same class then there is a one-to-one correspondence between their backgrounds.

**Comment 1.** Different lattice classes often correspond to well known continuum structures, such as "topological manifolds", "manifolds with boundary" or "manifold with orientation", i.e., the backgrounds for those classes are in one-to-one correspondence with equivalence classes of such continuum structures under isomorphism. Such structures can in principle be defined rigorously but their continuum nature makes dealing with them very unconstructive. In our approach we will go a much more convenient way: We define them in a purely constructive, combinatorial and finite way via lattices and moves, avoiding continuum technicalities.

However, we will still sometimes go the reverse way: We start with an intuitive picture for some continuum structure we want to describe, and use that intuition as a guide to define a concrete moved lattice types in the corresponding lattice class.

**Comment 2.** The mathematical framework described so far has some categorial structure that we will briefly outline for the reader familiar with those concepts. Note, however, that the categorial structure does not capture the most important concept, namely locality.

A moved lattice type defines a symmetric monoidal category: The lattices are the objects, and sequences of basic moves connecting them are the morphisms. Thereby two sequences represent the same morphism if they are related by 1) changing the order of moves that act on non-overlapping patches and 2) if a move is directly followed by its inverse, removing those from the sequence. The tensor product is given by the disjoint union. All morphisms are invertible, by taking the sequence of inverse moves with the reverse order. The backgrounds are isomorphism classes of objects. lattice mappings are functors between the categories given by different lattice types. Circuit moves for lattice mappings are natural transformations between the lattice mappings. All those natural transformations are invertible, and thus natural isomorphisms. The diagrams Eq. (2.20) correspond to $\mathcal{T}_\mathcal{A}$ being the weak inverse of $\mathcal{T}_\mathcal{B}$ and vice versa.

All moved lattice types together are the objects of a (symmetric monoidal) strict 2-category: The 1-morphisms are given by lattice mappings. The 2-morphisms are given by circuit moves for those lattice mappings. All 2-morphisms are invertible, as circuit moves can be undone. The lattice classes are the equivalence classes of objects in this 2-category.

However, the categorial structure does not go much further: We will not view the linear algebra side as a linear category. This would be unnatural for classical or quantum mechanical systems, as all quantities are tensors, which do not have a canonical bi-partition of their indices as "target" or "source" of a linear operator. In particular the mathematical structures we define will not be functors to the category of vector spaces.

### Examples

**Example 21.** Square grids do not allow for moves (at least not for ones that can be applied to lattices of arbitrary size). This is because the restrictions make it impossible to locally change the lattice.

**Example 22.** Consider trivalent graphs as lattice type. We can add the following basic move:

$$\tag{2.21}$$

With this basic move we can shrink any cycle of the graph to a one-edge bubble, e.g.:

$$\tag{2.22}$$

So we can transform any lattice into a cycle-free graph with additional one-edge bubbles. All such graphs are equivalent under basic moves, so each background for this moved lattice type is uniquely specified by giving this number of one-edge loops, which corresponds to the number of independent cycles of the graphs corresponding to those backgrounds.

**Example 23.** Consider triangulations of $n$-manifolds (with boundary) as in example (17). The canonical choice of a set of basic moves are local changes of the triangulation that do not change the topology of the underlying manifold. Then by construction, the backgrounds are in one-to-one correspondence to $n$-manifolds (up to homeomorphism). The according combinatorial operations will be precisely defined in Section (3) and are known as *Pachner moves*.

### 2.1.3 Gluings

For defining CTLs we need another kind of local operations relating different lattices that we will call gluings.

**Glued types and mappings**

**Informal definition 14.** A gluing is a prescription of how to remove two small patches from a lattice and replace it by one single patch. I.e., parts of the lattice that were far apart before the move can be near each other after the gluing.

**Definition 30.** A **gluing** is the same as a move, apart from the following differences:

- The place types $\mathcal{P}_1$ and $\mathcal{P}_2$ can also be location types (i.e., consisting of two or more place types instead of one). We will refer to locations of type $\mathcal{P}_1$ as **gluing locations**.

- There is only a map from the dangling connections of $\mathcal{P}_1$ to the cells of $\mathcal{P}_2$, but not vice versa.

A gluing can only be applied in one direction, by replacing a location of type $\mathcal{P}_1$ with a location of type $\mathcal{P}_2$.

When pairs of places on multiple disconnected components of a lattice are glued such that we get one connected component, we will say that the components have been **glued together**.

**Definition 31.** A lattice type together with a set of gluings that are consistent with the restrictions is called a **glued lattice type**. We refer to those gluings as **basic gluings**.

**Example 24.** Consider triangulations of $n$-manifolds with boundary as described in example (17). We can add the following basic gluing: Take two cells of the triangulation in the boundary and attaching them, such that those two cells become one single cell in the interior. We will give the combinatorial definition of such gluings in Section (3).

**Definition 32.** For a glued, moved lattice type the basic gluings can change the background of a lattice. We will call those operations changing the background the **background gluings**.

**Definition 33.** A **sub-type** of a glued lattice type is a sub-type of the corresponding lattice type together with all the gluings of the original glued type whose location type is contained in the sub-type.

**Definition 34.** A **(lattice) mapping** from a glued lattice type $\mathcal{A}$ to a glued lattice type $\mathcal{B}$ is a mapping between the two lattice types such that for every basic gluing relating two lattices $A$ and $A'$ of type $\mathcal{A}$ there is a basic gluing relating the mappings

$B$ and $B'$ of $A$ and $A'$. I.e., the following diagram commutes:

$$
\begin{array}{ccc}
A & \xrightarrow{\text{mapping}} & B \\
\text{basic} \Big\updownarrow & & \Big\updownarrow \text{basic} \\
\text{gluings} & & \text{gluings} \\
A' & \xrightarrow{\text{mapping}} & B'
\end{array}
\tag{2.23}
$$

**Basic lattices and histories**

For CTLs to be a useful framework we want the glued, moved lattice types to be finite in certain ways. In this section we will describe two such finiteness conditions, namely being finitely generated and finitely axiomatized.

**Definition 35.** We will call a glued, moved lattice type **finitely generated**, if there is a finite set of lattices, called **basic lattices**, such that: All lattices can be obtained by basic moves and basic gluings from the disjoint union of many copies of basic lattices. The backgrounds corresponding to the basic lattices will be called the **basic backgrounds**.

Usually we are interested in finding sets of basic lattices that are as small as possible, so normally we would impose that no basic lattice can be obtained from the other basic lattices by basic moves and basic gluings.

**Definition 36.** Consider a glued, moved lattice type that is finitely generated. A **history** for a lattice is a prescription of how a lattice with the same background can be obtained from the disjoint union of many basic lattices by basic gluings. We can take **disjoint union** of histories, yielding a history for the disjoint union of the lattices. We can add gluings to a history, yielding a history for the glued lattice.

A **defining history** is an algorithm that takes an arbitrary lattice and yields a history for that lattice. The algorithm may also be non-deterministic, i.e., the resulting history might depend on non-canonical choices taken during the algorithm. Giving a history mapping is the constructive way of proving that a lattice type is finitely generated.

A **history mapping** is a local prescription that maps every lattice to a history for that lattice. I.e., such a history mapping associates basic lattices and basic gluings to all places of a finite set of different types in the lattice. A history mapping is a special case of a defining history.

**Definition 37.** A **history move** is a pair of histories for the same lattice. A history (as a set of basic lattices and gluings) can occur as part of another history. A history move $M$ can be applied to a history $H$ by replacing one history of $M$ as part of $H$ with the other history of $M$.

**Definition 38.** Consider a glued, moved lattice type that is finitely generated with a defining history. Such a type is called **finitely axiomatized** if there is a finite set of history moves, called **basic history moves**, such that:

- All defining histories for a fixed lattice are connected by basic history moves (if the defining history is non-deterministic).

- The defining history for a disjoint union is connected to the disjoint union of the defining histories by basic history moves (this is usually trivially the case).

- The defining history after a basic gluing is connected to the history where we add the corresponding gluing by basic history moves (this is often trivial for CTLs coming from TLs).

- The defining history after a basic move is connected to the history before the move by basic history moves.

**Remark 24.** In this remark we give a heuristic procedure to obtain a set of basic history moves.

Given a glued moved lattice type that is finitely generated we can construct another lattice type. We will call lattices of this new type the **history lattices**. The cells of a history lattice are the (copies of) basic lattices, the basic moves and basic gluings. Two basic moves/gluings are connected if they act right after another in an overlapping manner. A basic move/gluing is connected with a copy of a basic lattice if it acts on that basic lattice. History lattices have to be closed, in the sense that every basic move has to be connected to basic moves before and after, or to basic lattices. Although history lattices consist of basic moves, there is in general no order of the moves, such that they yield a valid sequence of moves that can be applied to a lattice.

One can define a canonical gluing for the history lattices, which we will refer to as **history gluing**: The gluing locations are given by the occurrence of two copies of the same basic lattice in the lattice. When a history lattice is glued at those two copies, we remove them both from the lattice, and connect everything that was connected to the one copy with everything that was connected to the other copy.

Every patch of a history lattice defines a history by only taking the basic lattices and gluings. A bi-partition of a history lattice into two parts yields two histories for the same lattice, which often (but not always) defines a history move.

One good strategy to search for basic history moves is the following: Find a set of **basic history lattices**, from which any history lattice can be obtained from disjoint union of many copies and history gluing. Then take all history moves arising from bi-partitions of those basic history lattices as basic history moves.

## 2.2 Tensors and contraction

### 2.2.1 Tensors: General concepts

In this section we will define tensors of different type in an abstract language.

**Example 25.** To already give the reader a feeling for what we are aiming for, let us quickly describe real tensors (which are defined within the abstract framework in Section (2.2.2)). By a real tensor we simply mean an array of real numbers, labeled by a collection of coefficients called indices. All indices have the same range (called basis), at least if they are all of the same type. Indices of different type can have different range. E.g., tensors with one or two indices are known as vectors or matrices. Contracting two indices means forcing them to be equal and summing over all values, as e.g., in taking the trace of a matrix, or multiplying a vector with a matrix.

Note that this notion of tensor is different (simpler and more fundamental) from the one used in differential geometry in that we do not care for how a tensor transforms under coordinate transformations (as we do not even have coordinates or a manifold).

**Definition 39.** A **tensor type** consists of

● a set of **index types** to each of which we associate a data structure called the **basis type**. This basis type has to posses a notion of trivial element and product (that makes it a symmetric monoidal category). A particular instance of a basis type will be called a **basis**,

● a data structure defined with respect to a set of **indices** for each index type and one basis of the corresponding type for each index type. One particular instance of this data structure is called a **tensor**. The set of indices (i.e., the number of indices) of each type will be referred to as the **shape** of the tensor,

● an operation called the **tensor product** that maps two tensors to one tensor whose shape is the disjoint union of the two shapes (i.e., the disjoint union of the sets of indices for each index type separately), and

● an operation called the **contraction** that takes two indices of the same type of a tensor and yields another tensor with the two indices removed. We then say that the two indices have been **contracted**.

The tensor product must be commutative and associative and if we contract multiple pairs of indices of one tensor, the order of contraction must not matter.

**Definition 40.** A **tensor network** is a set of tensors together with a set of contractions between their indices. Indices that are not contracted are referred to as **open indices**. The **evaluation** of a tensor network is the tensor resulting from taking the tensor product of all involved tensors and then performing all contractions. The indices of the evaluation are the open indices of the tensor network.

**Remark 25.** For some special tensor types we might also accept slightly generalized notions of contraction that allow e.g., contracting groups of more than two indices. See, e.g., Section (2.2.4).

**Definition 41.** Consider a tensor type and choose one index type with a given basis. There are tensors $G$ with two indices of this type (i.e., "matrices"), such that: For each tensor $T$ we can take the tensor product with one copy of this matrix for each index of the chosen type, and then contract each such index with the first index of the associated copy of this matrix. I.e., in tensor-network notation (see Remark (11)):

$$(2.24)$$

This operation has to be such that it commutes with contraction. We will refer to such a matrix as **gauge transformation**. I.e., for the usual notion of contraction it has to satisfy:

$$(2.25)$$

For slightly generalized notions of contraction we can get different conditions. See, e.g., Section (2.2.4).

**Definition 42.** A **tensor mapping** from a tensor type $\mathcal{A}$ to a type $\mathcal{B}$ is a way of blocking or reinterpreting the indices of the tensors of type $\mathcal{A}$ to obtain tensors of type $\mathcal{B}$, that is compatible with tensor product and contraction. More precisely, it is a map $\mathcal{M}$ that associates

- to every index type of $\mathcal{B}$ one set of index types of $\mathcal{A}$. This set can be empty or contain the same index type multiple times. It can be used to map every tensor shape $I$ of $\mathcal{B}$ to a tensor shape $\mathcal{M}[I]$ of $\mathcal{A}$ by replacing each index by the corresponding set of indices,

- to every set of bases (for the different index types) of $\mathcal{A}$ a set of bases of $\mathcal{B}$,

- for every tensor shape $I$ of $\mathcal{B}$, to every tensor of type $\mathcal{A}$ with shape $\mathcal{M}[I]$ a tensor of type $\mathcal{B}$ of shape $I$. The bases of this tensor are given by the item before.

The following diagrams have to commute for all tensors $T$ of the according types and shapes:

$$
\begin{array}{ccc}
T_{\mathcal{A}} & \xrightarrow{\;\mathcal{M}\;} & T_{\mathcal{B}} \\
\text{contraction} \downarrow & & \downarrow \text{contraction} \\
T'_{\mathcal{A}} & \xrightarrow{\;\mathcal{M}\;} & T'_{\mathcal{B}}
\end{array}
\tag{2.26}
$$

$$
\begin{array}{ccc}
(T^1_{\mathcal{A}}, T^2_{\mathcal{A}}) & \xrightarrow{\;\mathcal{M} \times \mathcal{M}\;} & (T^1_{\mathcal{B}}, T^2_{\mathcal{B}}) \\
\substack{\text{tensor} \\ \text{product}} \downarrow & & \downarrow \substack{\text{tensor} \\ \text{product}} \\
T_{\mathcal{A}} & \xrightarrow{\;\mathcal{M}\;} & T_{\mathcal{B}}
\end{array}
\tag{2.27}
$$

**Definition 43.** For every tensor type there is the **identity tensor mapping** from this type to itself given by taking the same tensor again. The **trivial tensor mapping** from a tensor type $\mathcal{A}$ to a tensor type $\mathcal{B}$ associates 1) to each index type of $\mathcal{B}$ the empty set of index types of $\mathcal{A}$, 2) the trivial basis to all index types of $\mathcal{A}$, independent on the bases of $\mathcal{B}$ 3) the trivial tensor of type $\mathcal{A}$ to the trivial tensor of $\mathcal{B}$.

**Example 26.** Consider real tensors with two index types. There is the following tensor mapping to real tensors with one index type: It associates to the one index type a pair of the two index types. The basis set of the one index is the cartesian product of the basis sets of the two indices. The mapping takes a tensor with two index types that occur in pairs, and maps it to a tensor with only one index type by blocking each pair.

**Definition 44.** The **product** of two tensor types $\mathcal{A}$ and $\mathcal{B}$ is the tensor type whose set of index types is the disjoint union of the sets of index types of $\mathcal{A}$ and $\mathcal{B}$, and whose data structure is the cartesian product of the two data structures.

We will now introduce some specific tensor types. We will restrict to the very basic types only, however, in Section (6) we will sketch many more, such as fermionic tensors, Gaussian tensors or symmetric tensors.

### 2.2.2 Real tensors

**Definition 45. Real tensors** are a tensor type with one index type, whose basis type are just finite sets (referred to as **basis sets**. A real tensor with indices $i, j, \ldots$ with basis $B$ is a map that associates a real number to each element of $B \times B \times \ldots$. In other words an element of the vector space $\mathbb{R}^{B \times B \times \cdots}$. The tensors product is just the mutual product of all the real numbers, and contraction of two indices is given by forcing them to the same value and summing over this value.

**Observation 4.** The gauge transformations for real tensors are just orthogonal matrices, as the matrix $G$ from Eq. (2.25) is a real matrix in the usual sense and has to fulfill

$$
GG^T = \mathbb{1}.
\tag{2.28}
$$

**Remark 26.** We can also define real tensors with more than one index type. The only difference is that the indices of different type can have different sets as basis $B_i, B_j, \ldots$ and the tensor is a map that associates real numbers to each element of $B_i \times B_j \times \ldots$. Also the gauge transformations consist of applying independent orthogonal matrices for each index type.

### 2.2.3 Complex tensors

**Definition 46.** For every string $s \in \{\text{in}, \text{out}\}^n$ (called the **complex arrow orientations**, define the **complex number tensor** as the following real tensor $\mathbb{C}$ with $n$ indices with basis set $\{\mathbf{1}, \mathbf{i}\}$:

$$
\mathbb{C}^s_{a_1, \ldots, a_n} = \chi \left( \sum_{i=1}^n c(s_i, a_i) \mod 4 \right) 2^{1-n/2},
$$
$$
c(\text{in}, \mathbf{1}) = 0, \quad c(\text{out}, \mathbf{1}) = 0,
$$
$$
c(\text{in}, \mathbf{i}) = 1, \quad c(\text{out}, \mathbf{i}) = 3,
$$
$$
\chi(0) = 1 \quad \chi(1) = 0, \quad \chi(2) = -1 \quad \chi(3) = 0.
\tag{2.29}
$$

In tensor network notation we will denote the string $s$ by putting arrows to each index that point inwards for in and outwards for out. E.g., for $s = (\text{in}, \text{in}, \text{out}, \text{out})$ we write

$$
\mathbb{C}^s_{abcd} = \;
\begin{array}{c}
b \\
\downarrow \\
c \dashv\!\!\!\bigcirc\!\!\!\vdash a \\
\downarrow \\
d
\end{array}
\tag{2.30}
$$

This notation is consistent because of the permutation property from Observation (5). The normalization factor $2^{1-n/2}$ is included to cancel normalizations occurring from contracting closed loops of complex number tensors.

**Remark 27.** One can view complex numbers as vectors in $\mathbb{R}^2$. Then the multiplication, complex conjugation and the number $1 + 0i$ become real-linear maps from $\mathbb{R}^2 \otimes \mathbb{R}^2$, $\mathbb{R}^2$ or $\{0\}$ to $\mathbb{R}^2$. The structure coefficients of those real-linear maps are given by the complex number tensors (just that we have to revert the normalizations in their definition):

$$\langle ab, c \rangle = \left( \begin{pmatrix} 1 & 0 \\ 0 & -1 \end{pmatrix}, \begin{pmatrix} 0 & 1 \\ 1 & 0 \end{pmatrix} \right) = \left( 2^{1/2} \right) \overset{a}{\underset{b}{\textcircled{C}}} \!\!> c \; ,$$

$$\langle a^*, b \rangle = \begin{pmatrix} 1 & 0 \\ 0 & -1 \end{pmatrix} = a \; \triangleleft\!\textcircled{C}\!\!> b \; ,$$

$$\langle \mathbf{1}\mathbf{1} + 0\mathbf{i}, a \rangle = (1, 0) = \left( 2^{-1/2} \right) \textcircled{C}\!\!> a \; . \tag{2.31}$$

where the indices $a$, $b$ and $c$ correspond to the row, column and block, and the $\mathbf{1}$ and $\mathbf{i}$ component correspond first and second entry, respectively.

**Observation 5.** The complex number tensors obey the following **permutation property**: They are invariant under arbitrary index permutations that leave the arrow directions invariant. E.g.

$$c \!>\! \overset{b}{\underset{d}{\textcircled{C}}}\!\!< a \; = \; b \!>\! \overset{d}{\underset{a}{\textcircled{C}}}\!\!< c \tag{2.32}$$

**Observation 6.** The complex number tensors obey the following **arrow reversal property**: They are invariant under inverting all arrow directions. E.g.,

$$\overset{b}{\underset{c}{\textcircled{C}}}\overset{a}{\underset{d}{}} \; = \; \overset{b}{\underset{c}{\textcircled{C}}}\overset{a}{\underset{d}{}} \tag{2.33}$$

**Observation 7.** The complex number tensors obey the following **fusion property**: Any connected tensor network formed by complex number tensors that has matching arrow orientations for every contraction evaluates to the complex number tensor over the open indices. E.g.,

$$a) \quad \overset{a}{\underset{b}{\textcircled{C}}}\!\!>\!\!> \textcircled{C}\!<\!< \textcircled{C} \; = \; a \!>\!\textcircled{C}\!<\! b$$

$$b) \quad \textcircled{C}\!< \; = \; \textcircled{C}\!\!>\!\!> \textcircled{C}\!<$$

$$c) \quad \overset{\ulcorner}{\underset{\llcorner}{\textcircled{C}}}\!\! = \overset{\ulcorner}{\underset{\llcorner}{\textcircled{C}}}\!\!>\!\!> \overset{\ulcorner}{\underset{\llcorner}{\textcircled{C}}} \tag{2.34}$$

$$d) \quad \overset{\textcircled{C}\!<\!\!<}{\underset{\textcircled{C}\!<\!\!<}{}}\overset{<\!\textcircled{C}}{\underset{<\!\textcircled{C}}{}} \; = \; \overset{\ulcorner}{\underset{\llcorner}{\textcircled{C}}}$$

**Observation 8.** The complex number tensors obey the following **arrow obstruction property**: When we contract two indices of a complex number tensor whose arrow directions do not match, we get 0. E.g.

$$\overset{a}{\underset{b}{\textcircled{C}}} \; = \; 0 \tag{2.35}$$

Together with the fusion property we get the more general statement: Every tensor network formed by complex number tensors that has a cycle of contractions along which the number of contractions with non-matching arrow orientations is odd evaluates to 0. E.g.;

$$\overset{a}{>\!\textcircled{C}}\!\!> \; \overset{b}{>\!\textcircled{C}}\!< \; \overset{c}{>\!\textcircled{C}}\!\!> \; \overset{d}{<\!\textcircled{C}}\!< \; = \; a \; \overset{b\;\;c}{<\!\textcircled{C}}\!< \; d \; = 0 \tag{2.36}$$

**Comment 3.** The permutation, arrow reversal and fusion property of the complex number tensors correspond to the fact that the real-linear maps over $\mathbb{R}^2$ given by the multiplication, complex conjugation and number 1 form a unital, associative, commutative real *-algebra. E.g., the associativity for the multiplication follows as

$$(ab)c = a(bc)$$

$$\overset{a}{\underset{}{\textcircled{C}}}\overset{b}{\underset{\textcircled{C}}{}}\;c \; = \; a \; \overset{b}{\underset{\textcircled{C}}{\textcircled{C}}}\overset{c}{\underset{}{}} \tag{2.37}$$

**Definition 47.** **Complex tensors** are a tensor type with one index type whose basis type is a finite set, just as for real tensors. A complex tensor $T$ with indices $i, j, \ldots$ with basis $B$ is a map that associates a real number to each element of the set $B \times B \times \ldots \times \{1, i\}$. In other words, it is a real tensor $T_{\mathrm{real}}$ with an additional index (the **complex index**) of another type whose basis is fixed to the 2-element set $\{1, i\}$. E.g.,

$$a \!-\!\boxed{T}\!\overset{b}{}\!-\!c \quad \leftrightarrow \quad a \!-\!\boxed{T_{\mathrm{real}}}\!\overset{b\quad x}{}\!-\!c \; , \tag{2.38}$$

where $a, b, c \in B$ and $x \in \{1, i\}$. We cannot simply define the tensor product of two complex tensors as the tensor product of the associated real tensors, as we would end up with two instead of one complex index. Instead we have to use the complex number multiplication tensor from Eq. (2.31) to fuse the two complex indices into one. E.g.,

$$a \!-\!\boxed{T_1 \otimes T_2}\!\overset{d\;b\;e}{}\!-\!c \quad \leftrightarrow \quad \overset{e}{\underset{d\!-\!\boxed{T_2}}{}} \; \overset{}{\underset{a\!-\!\boxed{T_1}\overset{b}{}\!\!>\!\textcircled{C}\!\!>\;x}{}}\!-\!c \tag{2.39}$$

The contraction of two indices of a complex tensor in turn is just the contraction of those two indices of the associated real tensor. I.e., real and imaginary part are contracted separately.

**Remark 28.** Just as for real tensors, we can also think of complex tensors with more **index types**.

**Definition 48.** The **realification** of a complex tensor $T$ with $n$ indices and basis $B$ with respect to a string $s \in \{\mathrm{in}, \mathrm{out}\}^n$ is the following real tensor $T_{\mathrm{real}}$: It has the same $n$ indices but the set $B \times \{1, i\}$ as basis. To obtain $T_{\mathrm{real}}$ we take the tensor product with the complex number tensor with $n + 1$ indices, $n$ of them associated to the indices of $T$ and one special index. The arrow direction of the special index is set to "in" and the directions of the remaining indices are given by $s$. Now we contract the special index of the complex number tensor with the complex index of $T$ (when interpreted as a real tensor). $T_{\text{real}}$ is obtained by blocking each index of $T$ with the associated complex number tensor index. E.g.,

$$a \longrightarrow \boxed{T_{\text{real}}} \overset{b}{\Longrightarrow} c \quad = \quad a \longrightarrow \boxed{T} \overset{b}{\longrightarrow} c \qquad (2.40)$$

**Comment 4.** The realification of complex tensors is closely related to Dirac notation. The arrow directions in the realification correspond to weather each index is represented as ket or bra.

**Observation 9.** The tensor product of the realifications of two complex tensors is in general not equal to the realification of the tensor product of the complex tensors. However, a tensor product followed by a contraction of a pair of indices on each component commutes with taking the realification. E.g.,

$$a \longrightarrow \boxed{T_1} \overset{b}{\Longrightarrow} \boxed{T_2} \overset{d}{\Longrightarrow} c$$

$$= a \longrightarrow \boxed{T_1} \overset{b}{\qquad} \boxed{T_2} \overset{d}{\qquad} c \qquad (2.41)$$

$$= a \longrightarrow \boxed{T_1} \boxed{T_2} \overset{b \quad d}{\qquad} c$$

**Remark 29.** Consider the realification of a 2-index complex tensor $T$ with the following arrow directions:

$$\begin{aligned} a) \quad & a \longrightarrow \boxed{T_{\text{real}}} \longrightarrow b \\ b) \quad & a \longrightarrow \boxed{T_{\text{real}}} \longleftarrow b \end{aligned} \qquad (2.42)$$

Both realifications can be interpreted as real-linear maps from the index labeled by $a$ to the index labeled by $b$. However, we can also interpret a) as a complex-linear map, and b) as a complex anti-linear map.

**Remark 30.** The Hermitian conjugate of a complex-linear map written as the realification of a complex tensor as in Remark (29) is given by exchanging the two indices. This is because reversing the arrow orientations automatically includes complex conjugation:

$$b \longrightarrow \boxed{L} \longrightarrow a \quad = \quad b \overset{\boxed{L}}{\qquad} a$$

$$= \quad b \overset{\boxed{L}}{\qquad} a \quad = \quad b \overset{\boxed{L^*}}{\qquad} a \qquad (2.43)$$

$$= \quad a \longrightarrow \boxed{L^\dagger} \longrightarrow b$$

### 2.2.4 Preferred basis tensors

**Definition 49.** The $n$-index **delta tensor** for a basis set $B$ is the following tensor with $n$ indices of type $B$:

$$\delta_{a,b,c,\dots} = \overset{b}{\underset{c}{\searrow}} \cdots a = \begin{cases} 1 \text{ if } a = b = c = \dots \\ 0 \text{ otherwise} \end{cases}. \qquad (2.44)$$

So it is just a multi-variable generalization of the Kronecker-delta symbol.

**Observation 10.** The delta tensors obey the following **fusion property**: The evaluation of any connected tensor network formed by delta tensors (for the same basis set) with $n$ open indices is equal to the $n$-index delta tensor. E.g.

$$\begin{aligned} a) \quad & \overset{b \quad a}{\underset{c \quad d}{\bullet - \bullet}} e \quad = \quad \overset{b \quad a}{\underset{c \quad d}{\searchls}} e \\ b) \quad & a \bullet\!\!\!\bigcirc\!\!\!\bullet b \quad = \quad a \bullet\!\!-\!\!\bullet b \end{aligned} \qquad (2.45)$$

They also obey the **permutation property**: All delta tensors are invariant under all permutations of their indices. E.g.

$$c \overset{b}{\underset{d}{\bullet}} a \quad = \quad c \overset{a}{\underset{d}{\bullet}} b \qquad (2.46)$$

**Comment 5.** The fusion property and the permutation property are equivalent to saying that the 1-index, 2-index and 3-index delta tensors are the structure coefficients of the unit, involution and product of a unital real *-algebra, namely the algebra of real functions over the basis set $B$, under pointwise multiplication.

**Definition 50.** A **preferred basis real tensor** is a real tensor, just that we additionally are allowed to use the delta tensors for contracting indices. E.g., we can contract three indices like:

$$c \overset{b \quad a}{\underset{d \quad e}{\bullet\boxed{T}\bullet}} \qquad (2.47)$$

Therefore, a gauge transformation $G$ for this kind of index have to obey by e.g.

$$\overset{b}{\underset{c}{\bullet}} a \quad = \quad \overset{b}{\underset{c}{\diamond\!\!\bullet\boxed{G}}} a \qquad (2.48)$$

So $G$ has to be a permutation matrix.

**Remark 31.** There is a natural tensor mapping from preferred basis real tensors to real tensors by simply forgetting about the delta tensors (i.e., preferred basis). As every usual contraction for indices of real tensors is also a allowed contraction for indices of preferred basis real tensors, this defines a tensor mapping.

Note that the reverse operation, i.e., adding a preferred basis, is not a tensor mapping, as a contraction via delta tensors is not a usual contraction of real tensors.

**Remark 32.** One can also consider mixed types of real tensors and preferred basis real tensors. I.e., there is one (or more) index type that are equipped with delta tensors, and another (or more) index type that are not.

In this case one can consider a slight generalization of the above notion of tensor: We can make the basis of the indices of some type dependent on the value of one or more indices of a type that is equipped with a preferred basis. I.e., the basis of the former index type is a set $B_i$ that depends on the value $i$ of some index of the latter type whose basis set $I$ is equipped with a preferred basis. E.g., if a tensor has one index of each of the two types then such a tensor associates one real number to every element of $\bigoplus_{i \in I} B_i$, where the sum denotes a disjoint union of sets.

For tensors in those generalized sense we have to make sure that whenever we contract two indices with variable basis set we also contract the indices on whose value their basis set depends on.

## 2.3 Tensor lattices

So far we have defined lattices and tensors/tensor networks. In this section, we will show how to put together lattices and tensor networks, and how the basic moves of a moved lattice type naturally yield consistency conditions for the tensors of the tensor network.

### 2.3.1 Definition

**Definition 51.** Consider a tensor network without any open indices and a subset of its tensors together with a subset of the contractions among those tensors. Such a subset defines another smaller tensor network, but this time with open indices. A **tensor-network move** is a procedure that replaces this smaller tensor network with another tensor network (with possibly different geometry) that evaluates to the same tensor. E.g., consider the following tensor-network move replacing a patch of four tensors with a patch of four new tensors with a different geometry:

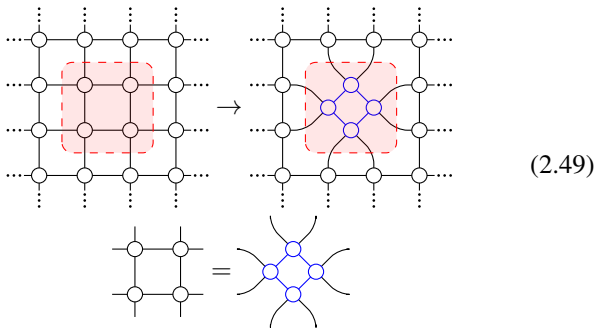

$$(2.49)$$

**Informal definition 15.** A tensor lattice is a prescription that associates a tensor network (without open indices) with each lattice of a given moved type. The geometry of the tensor network and the tensors themselves at a given place can only depend on how the lattice looks like locally in a small patch around that place. The prescription has to be such that the tensor networks on a lattice before and after a basic move at some place only differ by a local tensor network move around this place.

**Definition 52.** A **tensor lattice type** (or short **TL type**) consists of

- a moved lattice type, whose lattices will referred to as **the lattices**,

- a mapping from the underlying lattices to another lattice type whose lattices will be called **tensor network geometries**,

- for every basic move of the underlying lattices, a prescription to select a patch of the tensor network geometries before and after the move, such that the complement of this patch remains unchanged. We will refer to this as the **tensor-network move prescription**,

- a map that associates to every cell type of the tensor network geometries a tensor type, and

- a map that associates to every connection type of the tensor network geometries an index type.

**Definition 53.** A **tensor lattice** (or short **TL**) of a given type is a map $T$ that associates

- One basis (of the according type) to every connection type of the tensor network geometries.

- Consider a cell type and all possible tuples of numbers that state how many connections of each type a cell of this type can have. To each such type associate one tensor which has as many indices of each according type as the tuple states. The basis sets of the different index types are those associated to the according connection types.

For a given lattice, a TL yields a tensor network by replacing each cell of each type by one copy of the associated tensor, and connections between cells by contractions of the associated indices.

A TL has to fulfill the following axioms:

- **Symmetry axiom**: If a TL tensor has multiple indices corresponding to the same connection type, then the tensor has to be invariant under arbitrary permutations of those indices.

- **Move axiom**: For every basic move transforming a lattice $L_a$ into another lattice $L_b$, the tensor networks corresponding to the two patches of the tensor-network move prescription before and after the move have to evaluate to the same tensor. In other words, the two associated tensor networks $T_a$ and $T_b$ are related by a tensor-network move.

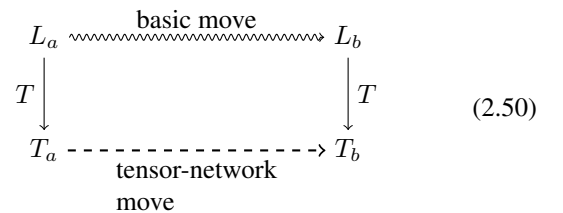

$$(2.50)$$

**Remark 33.** The following picture illustrates a basic move and the corresponding tensor-network move of a TL on a 2-dimensional lattice:

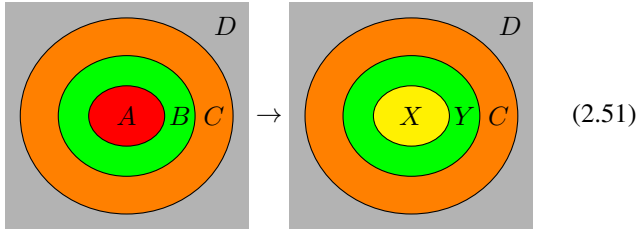

(2.51)

The smallest region represents the patch of the underlying lattice that the basic move acts on. The change of lattice is represented by changing the colour from red to yellow. This change of lattice implies a change of the tensor network inside the second smallest region, represented by a change in labels from $A$ and $B$ to $X$ and $Y$. The tensor network move acts on the third smallest region, so the evaluation of the patch $A$, $B$ and $C$ equals the evaluation of the patch $X$, $Y$ and $C$.

**Remark 34.** It might seem like having the distinction between the underlying lattices and the tensor network geometries is unnecessary, and one could simply work with the latter alone. However, the tensor network geometries often have much less combinatorial information than the underlying lattices, thus neither restrictions nor basic moves can be formulated in terms of the tensor-network geometries.

Also it is often nice have a simple type of underlying lattices and to compare different (more or less complicated) TL types on the same underlying lattice type. For instance this allows one to define the notion of universality for TL classes on a fixed lattice type.

**Definition 54.** A **normalization TL** is a TL where the bases of all index types are trivial. E.g., if we are dealing with real tensors this means that all tensors are just numbers, and the tensor network is just a product of real numbers.

### 2.3.2 Mappings and phases

**Definition 55.** A **parallel tensor-network move** consists of many local tensor-network moves that are done in parallel on subsets that do not overlap. E.g., for a tensor network on a 2-dimensional lattice the following picture illustrates a parallel tensor-network move: The grid in the background represents a tensor network and the red shaded regions are the patches on which the single tensor network moves happen:

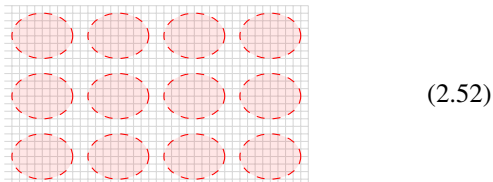

(2.52)

**Definition 56.** A **circuit tensor-network move** consists of a set of parallel tensor-network moves that are applied in sequence. E.g., the following picture of a circuit tensor-network move in 2 dimensions is a sequence of the red, followed by the blue, followed by the green parallel tensor-network move.

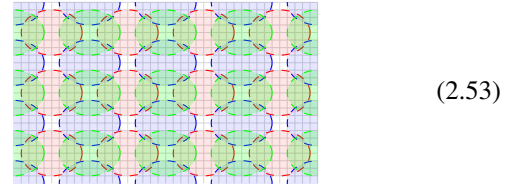

(2.53)

**Definition 57.** Two TLs $A$ and $B$ of the same type said to be **in the same phase** if there is a circuit tensor-network move transforming them into each other. The equivalence classes of TLs of a fixed type under being in the same phase are called **TL phases**. As a diagram:

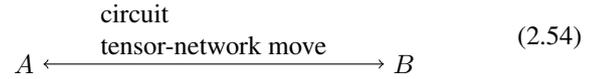

(2.54)

**Definition 58.** A **tensor-network mapping** from one type of tensor-network geometry to another is a prescription of reshaping and blocking the contractions of a tensor network with a geometry of the first type to a tensor network with a geometry of the second type. A **TL mapping** from a TL type $\mathcal{A}$ to a TL type $\mathcal{B}$ consists of 1) a mapping from the underlying lattices of $\mathcal{B}$ to the underlying lattices of $\mathcal{A}$ and 2) A tensor-network mapping from the tensor-network geometries associated to the underlying lattices of $\mathcal{A}$ to the tensor-network geometries associated to the mappings of those. This tensor-network mapping has to be such that a tensor-network move of the TL of type $A$ is automatically a tensor-network move of the TL of type $B$.

A TL mapping is called **exact** if the corresponding lattice mapping is a circuit move. Two TL mappings are **equivalent** if they differ by an exact TL mapping.

**Definition 59.** Two TL types $\mathcal{A}$ and $\mathcal{B}$ are said to be **in the same class** if there is a mapping $\mathcal{T}_{\mathcal{A}}$ from $\mathcal{A}$ to $\mathcal{B}$ and a mapping $\mathcal{T}_{\mathcal{B}}$ from $\mathcal{B}$ to $\mathcal{A}$, such that both $\mathcal{T}_{\mathcal{B}} \circ \mathcal{T}_{\mathcal{A}}$ as well as $\mathcal{T}_{\mathcal{A}} \circ \mathcal{T}_{\mathcal{B}}$ are exact. As a diagram:

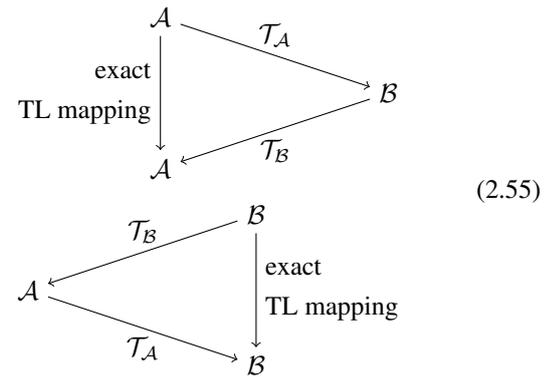

(2.55)

**Definition 60.** For every TL type there is the **identity TL mapping** from this type to itself given by the identity lattice mapping and the identity tensor-network mapping. The **trivial TL mapping** from any TL type to any other TL type is the TL mapping given by the trivial lattice mapping and the trivial tensor-network mapping.

**Definition 61.** Two TLs $A$ and $B$ of types $\mathcal{A}$ and $\mathcal{B}$ are said to be **in the same phase** if:

- $\mathcal{A}$ and $\mathcal{B}$ are in the same class. I.e., there are mappings from $\mathcal{A}$ to $\mathcal{B}$ and vice versa.

• The TL mapping (from the previous point) of $A$ is in the same phase as $B$ (and vice versa).

**Definition 62.** We say that a TL $A$ of type $\mathcal{A}$ **extends** to another type $\mathcal{B}$, if there is a TL $B$ of type $\mathcal{B}$ together with a TL mapping from $\mathcal{B}$ to $\mathcal{A}$ that maps $B$ to $A$.

**Observation 11.** Consider a TL and a circuit move on the underlying lattices. When we replace every basic move of the circuit move by the corresponding tensor-network move of the TL we get a circuit tensor-network move relating the TL on the lattice before and after the circuit move.

This means that each TL is in the same phase as itself even if we regard it as a TL of another type, as in Definition (61). So our definition of TLs of different types being in the same phase is consistent and does not depend on which of two equivalent TL mappings between the two types we choose.

**Definition 63.** The **product** of two TL types is a TL type for the product of the two underlying lattices types. The lattice mapping to the tensor-network geometry is just given by the disjoint union of the two lattice mappings.

The **product** of two TLs is the TL of the product type, whose tensors and bases are just the tensors and bases of the original TLs together.

**Definition 64.** A **sub-type** of a TL type is a TL type defined for a sub-type of the underlying lattices. The mapping to the tensor-network geometries is the same, just that we only map the place types that are contained in the lattice sub-type.

The **sub TL** for a sub-type of its type, is the TL of the sub-type that uses the same tensors, just that we restrict to those associated to cell types that occur in the tensor-network geometries.

**Definition 65.** Consider a TL type $\mathcal{A}$ that contains another TL type $\mathcal{B}$ twice as a sub-type. The **joint type** of the TL type with the two sub-types is a TL whose underlying lattices are the corresponding joint type of the underlying lattices of $\mathcal{A}$. Thereby we identify all pairs of place types in the lattice mapping to tensor network geometry that in occur in the sub-type with each other.

For a TL of type $\mathcal{A}$ such that the two sub TLs are the equal, the **joint TL** is the TL of the sub-type using the same tensors. Because the two sub TLs are equal, it does not matter which tensors we take for the identified place types.

**Informal definition 16.** Ideally we would like to be able to classify TLs on a particular lattice type/class. It would be cumbersome if to this end we had to look at an infinite set of more and more complex TL types for the same lattice type. Luckily it turns out that this is not the case: There are usually universal TL types for a given lattice type that cover all possible other TL types. Other TL types whose tensor network geometries depend on the lattices in a very complicated way can be reduced to such a universal type by a mapping that usually involves a lot of blocking. So TLs of such complicated type will be contained in a universal type with large bases.

**Definition 66.** A TL type $\mathcal{A}$ on a moved lattice type $\mathcal{X}$ is **universal** if every other TL type for a moved lattice type that is in the same class as $\mathcal{X}$ is in the same class as $\mathcal{A}$.

**Comment 6.** The TL types corresponding to the CTL types in the concrete implementations section Section (5) are in general not universal. This leaves hope that there might be more general TL types that allow for new ways of representing phases by fixed point models that are yet unknown. For more details see Section (6.1.5).

## 2.4 Contracted tensor lattices

In this section we will describe how one can associate tensors to lattices, and how the gluings and moves of the lattices naturally yield consistency conditions for the associated tensors. Note that in contrast to TLs, 1) we do directly associate tensors, and not tensor networks to lattices and 2) gluing is an essential lattice operation for CTLs.

Every TL yields a CTL whose lattices are cut-out patches of the TL lattices, and whose basic gluings consist in reconnecting the cut-out patches. The CTL tensors are simply the evaluations of the TL tensor network on the cut-out patches. For more details see Section (2.4.6).

We will introduce CTLs in addition to TLs because of two main reasons: 1) CTLs are more general than TLs: Whereas TLs only represent truly microscopic local physical models, CTLs can also describe effective theories for certain aspects of a model. Examples for this are topological invariants, or (non-fully extended) axiomatic TQFTs (with defects or boundaries), which can e.g., describe the anyon statistics of a model without referring to a potential microscopic realization. 2) CTLs often provide a more practical and condensed language for TLs: If we want to talk about physical quantities in a TL we always have to specify on which patch we evaluate the TL to actually get numbers. In the CTL language we can directly talk about tensors. Also for topological lattice types (which we will mostly deal with) the CTL tensor only depends on the boundary of the cut-out patch (apart from the topological properties of the interior), so we can restrict our drawings to this boundary which is one dimension lower.

### 2.4.1 Definition

**Definition 67.** An **index prescription** for a lattice type and a tensor type is a map that associates to every index type one place type. Given a index prescription and a lattice, for each index type, we can associate one index of this type to each place of the associated type.

**Example 27.** The easiest index prescription for a given lattice type is to simply associate indices of a fixed type to the cells of a fixed type. E.g., one could associate indices to the vertices in a 2-dimensional square grid. Or one could associate one index to each edge of a 2-dimensional simplicial complex.

**Example 28.** A slightly more non-trivial example for an index prescription would be to associate one index to each corner of a 2-dimensional cell complex, i.e., each pair of face and vertex such that the vertex is part of the face.

**Definition 68.** Consider a glued lattice type and a tensor type together with an index prescription, and associate one index of each type to each place of the according type. Imagine a gluing location in the lattice and the set of surrounding indices. If we perform the corresponding basic gluing this set of surrounding indices changes. A **contraction type** is a prescription of how to contract the indices before the gluing to obtain the indices after the gluing. A **contraction prescription** is a map that associates to every basic gluing a contraction type.

**Remark 35.** We sometimes allow for a slight generalization of the formalism: In many of the CTL types presented in Section (5) we allow the contraction types for certain gluing types to involve some external parameters, that we will call the **normalizations**. Those normalizations are tensors that vary for different CTLs of the same type. In particular an CTL in this generalized sense is not determined by the CTL tensors alone but one also has to specify the normalizations.

However, we can always include the normalizations into the CTL tensors at the expense of adding some decorations to the index lattices, which tell whether a normalization has been included yet.

**Definition 69.** A **CTL type** consists of the following:

• A glued, moved lattice type. Lattices of this type will be referred to as the **CTL lattices** (or sometimes just **the lattices**).

• A lattice mapping from the lattices to another lattice type with no basic moves. Lattices of the latter type will be referred to as **index lattices**. We will refer to the result of this mapping for a certain lattice as **the index lattice** of that lattice. The gluing locations of the lattices have to correspond to location types of the index lattices.

• A tensor type. Tensors of this type will be simply referred to as **the tensors**.

• An index prescription for this tensor type and the index lattices. Under the lattice mapping this index prescription is mapped to an index prescription on the lattices.

• A contraction prescription for the lattices and the index prescription on the lattices.

To summarize:

**CTL type**

$$
\begin{array}{ll}
\textbf{lattice type} & \textbf{tensor type} \\
\quad\Big\downarrow \text{mapping} & \\
\textbf{index lattices} & \\
\text{place type} & \longleftarrow \quad \text{index type} \\
\text{basic gluing} & \longrightarrow \quad \text{contraction type}
\end{array}
\tag{2.56}
$$

**Remark 36.** As the index lattices do not have any basic moves, the basic moves of the lattice cannot change its index lattice. This will later be important in order to define the move axiom.

**Remark 37.** Gluing will usually involve a 2-location type consisting of two copies of the same place type. In this case there is a very natural choice for a contraction type: Simply contract all indices surrounding the one place with the corresponding indices surrounding the other place.

**Example 29.** Consider the glued, moved lattice type given by triangulations of $n$-dimensional manifolds with boundary with the basic moves and gluings described in examples (23, 24).

However, restrict to only basic moves that do not change the triangulation of the boundary, but only of the interior. The index lattice is given by taking the triangulation of the boundary only. Take an index prescription that associates one index of one fixed type to each boundary $(n-1)$-simplex. When we glue two boundary $(n-1)$-simplices they disappear from the boundary. So the two surrounding indices before the gluing become zero indices after the gluing. Thus, a consistent contraction type for that basic gluing is to simply contract the two indices. CTL types like this will be defined more precisely in Section (5).

**Informal definition 17.** For many CTL types there is a natural way to use the basic gluings to define another set of moves (in addition to the basic moves) that do change the index lattice, in the following way: 1) Take the disjoint union with some fixed small lattice and 2) glue the move place with a place on this small lattice. We will call such moves the **index moves** and refer to the backgrounds with respect to the basic moves together with the index moves as the **extended backgrounds**.

**Definition 70.** A **contracted tensor lattice** (short **CTL**) $T$ (of a given type) is a map that associates

• to every index type one basis of the corresponding type.

• to every lattice $L$ a tensor $T[L]$. Those tensors will be referred to as the **CTL tensors**. The bases for the different index types of the tensor are the bases from the previous item. $T[L]$ has one index for each place of $L$ of each type in the index prescription.

To summarize:

$$
\begin{array}{lll}
\text{index type} & \longrightarrow & \text{basis} \\
\text{place} & \longrightarrow & \text{index} \\
\text{lattice} & \longrightarrow & \text{tensor}
\end{array}
\tag{2.57}
$$

The map has to obey the following axioms:

1. **Symmetry axiom:** If a lattice $L$ has some symmetry, i.e., the source and target maps are invariant under a permutation of cells and connections, then $T[L]$ has to have the same symmetry acting as the corresponding permutation of indices. As a commuting diagram:

$$
\begin{array}{ccc}
L_A & \xrightarrow{\;\;T\;\;} & T_A \\
\text{symmetry}\Big\downarrow & & \Big\downarrow \substack{\text{index} \\ \text{permutation}} \\
L_B & \xrightarrow{\;\;T\;\;} & T_B
\end{array}
\tag{2.58}
$$

2. **Disjoint union axiom:** The disjoint union of two lattices $L_A$ and $L_B$ has to be consistent with the tensor product of the associated tensors. I.e., $T[L_A \otimes L_B] = T[L_A] \otimes T[L_B]$, or as a commuting diagram:

$$
\begin{array}{ccc}
(L_A, L_B) & \xrightarrow{\;\;T\;\;} & (T_A, T_B) \\
\substack{\text{disjoint} \\ \text{union}}\Big\downarrow & & \Big\downarrow \substack{\text{tensor} \\ \text{product}} \\
L_C & \xrightarrow{\;\;T\;\;} & T_C
\end{array}
\tag{2.59}
$$

3. **Move axiom**: Applying a basic move to a lattice $L_A$ does not change the associated CTL tensor $T_A$. I.e., :

$$
\begin{array}{ccc}
L_A & & \\
 & \searrow^{T} & \\
\text{basic} & & \rightarrow T_A \\
\text{move} & \nearrow^{T} & \\
L_B & &
\end{array}
\qquad (2.60)
$$

4. **Gluing axiom**: Applying a basic gluing to a lattice $L_A$ has to be consistent with the contraction of the corresponding type of indices of $T[L_A]$, i.e., the following diagram commutes:

$$
\begin{array}{ccc}
L_A & \xrightarrow{\;T\;} & T_A \\
{\scriptstyle\text{basic}}\Big\downarrow {\scriptstyle\text{gluing}} & & \Big\downarrow {\scriptstyle\text{contraction}} \\
L_B & \xrightarrow{\;T\;} & T_B
\end{array}
\qquad (2.61)
$$

**Comment 7.** Note that the symmetry axiom is actually already implicitly enforced by the way we think about lattices: If two sets of source and target maps are equivalent up to a bijection between their cells and connections, they are equivalent and we regard them as the same lattice. In particular when we draw a lattice there is no way to distinguish two such equivalent lattices.

## 2.4.2 Basic tensors and axioms

**Definition 71.** We will call a CTL type **finitely generated** if its glued, moved lattice type is finitely generated. In that case we will refer to the CTL tensors associated to the basic lattices as the **basic tensors**.

**Example 30.** All CTL types that we will deal with in this work are finitely generated. It is of course easy to construct counter examples though: Take the CTL type with real tensors on $n$-dimensional simplicial complexes with Pachner moves, but no basic gluing and the empty index prescription. A CTL of this type associates one number to every (triangulable) $n$-manifold. This number can be arbitrary, except for the fact that disjoint union has to be consistent with multiplication. So a set of "basic tensors" is given by the numbers associated to all connected $n$-manifolds, but there is infinitely many of them (except $n = 0$ or $n = 1$).

**Observation 12.** It might seem like CTLs involve dealing with an infinite set of tensors with arbitrarily big number of indices all together, which would be quite cumbersome. However, for finitely generated CTL types, all lattices can be obtained by disjoint union, basic moves and basic gluings from the basic lattices. So according to the CTL axioms all CTL tensors can be obtained from the basic tensors by tensor product and contractions. So in that case a CTL becomes something very finite. E.g., if the tensors of the CTL are real tensors then it is described by a finite set of real numbers, namely the entries of the basic tensors.

**Definition 72.** Consider a CTL type that is finitely generated. Every history defines a tensor network by replacing every basic lattice by the corresponding basic tensor and every basic gluing by the corresponding contraction. Every history move defines an equation between the two corresponding tensor networks. We will refer to such equations as **axioms** of the CTL type. We find that:

- Consider a history move $A$ that is applied to a history move $B$ yielding a history move $C$. Then the corresponding axiom $C$ follows from the axioms $A$ and $B$.

- The axioms corresponding to two history moves $A$ and $B$ imply the axiom corresponding to the disjoint union of $A$ and $B$.

- The axiom corresponding to a history move $A$ implies the axiom corresponding to a history move $A'$ if $A'$ is obtained from $A$ by adding a basic gluing.

So if the glued, moved lattice type is **finitely axiomatized** all axioms follow from the axioms corresponding to the basic history moves. In this case we will refer to the CTL as **finitely axiomatized** and to the latter axioms as the **basic axioms**.

**Observation 13.** Consider a finitely generated and finitely axiomatized CTL type with real tensors. The basic axioms are equations between tensor networks formed from copies of the basic tensors. These equations are linear in each individual copy of each basic tensor, so they are polynomials in the entries of the tensors. So in this case a CTL can be boiled down to a finite set of finite tensors whose entries obey a finite set of polynomial equations.

## 2.4.3 Gauge transformations, phases and mappings

Just as for TLs we can define gauge transformations, phases and mappings for CTLs. For CTLs coming from TLs those are equivalent to the same notions for TLs.

**Observation 14.** Applying the same gauge transformation to all CTL tensors simultaneously preserves all the CTL axioms, and thus yields another CTL. We will refer to equivalence classes of CTLs under gauge transformations as **gauge families** of CTLs.

**Observation 15.** Consider a background with empty index lattice. The associated CTL tensor does not have any indices, so it is does not change under gauge transformations and thus defines an invariant. For CTL types with real tensors those invariants are simply real numbers.

**Definition 73.** A **TL operator** is a slight generalization of a TL with additional open indices. I.e., the tensor network geometries have special cell types that are only the target of exactly one special connection type. To those special cell types there are no associated tensors. Instead they represent the open indices of the type associated to the according special connection type. Those special cell types representing open indices are divided into two groups named **input indices** and **output indices**. (Note that "operator", "input" and "output" seem to suggest some directionality, however, there is none).

**Comment 8.** A TL operator for real/complex tensors is a generalization to what is known as *tensor network operator* (with small bond dimension).

In this case a TL operator can also be thought of as a circuit of local linear maps of finite depth.

**Definition 74.** Roughly, two CTLs $A$ and $B$ of the same type $\mathcal{T}$ are **in the same phase** if there is a TL operator $C$ on the index lattices, such that

- both the input indices and the output indices of $C$ match the index distribution of $\mathcal{T}$.

- Contracting all indices of $A[L]$ with the input indices of $C[L]$ evaluates to $B[L]$, for every lattice $L$.

- Contracting all indices of $B[L]$ with the output indices of $C[L]$ evaluates to $A[L]$, for every lattice $L$.

In pictures:

$$
\begin{array}{ccc}
 & \overset{A}{\nearrow} & T_A \\
L & & \Big\updownarrow C \\
 & \underset{B}{\searrow} & T_B
\end{array}
\tag{2.62}
$$

More precisely, for every bi-partition of a fixed index lattice $I$, consider the bi-partition $C[I]$ into two tensor networks $C_1$ and $C_2$. Those two tensor networks have all the input and output indices from their part of the bi-partition, together with additional open indices arising from cutting $C[I]$ into two parts. E.g., for an interval-like part of a 1-dimensional index lattice we can illustrate $C_1$ like the following thin brown stripe:

$$
x \overline{\phantom{xxx}}\overset{b}{\phantom{xx}} x
\tag{2.63}
$$

where the input (output) indices are summarized by $a$ ($b$), and $x$ is the collective label for the additional indices arising from the bi-partition.

Now for every lattice $L$ with such a bi-partition of its index lattice $I_L$ the following has to hold: Contracting the input indices of $C_1$ with the according indices of $A[L]$ evaluates to the same tensor as contracting the output indices of $C_2$ with the according indices of $B[L]$.

e.g., consider a CTL on lattices whose extended backgrounds are 2-manifolds with boundary, and the index backgrounds correspond to the boundary. The following picture illustrates the condition above for a lattice with 2-disk extended background which has a circle as index background, and the bi-partition of this circle into two intervals and their complement:

$$
\tag{2.64}
$$

**Comment 9.** Being in the same phase is closely related to the categorial concept of *Morita equivalence*. E.g., a certain CTL type on 3-dimensional simplicial complexes are closely related to unitary fusion categories. Then there is a type of TL operator such that such an operator between two CTLs in the same phase is closely related to an invertible bimodule between the corresponding fusion categories [26], which is the notion of Morita equivalence for fusion categories [29].

**Observation 16.** Gauge transformations are TL operators, so CTLs related by gauge transformations are always in the same phase.

**Remark 38.** The more powerful the basic moves/gluings of a lattice type of a CTL type are, the more axioms the CTL tensors have to fulfill, the more rare are solutions to those axioms. For types with very powerful moves we will find that the equivalence classes of CTLs under gauge transformations form discrete sets (up to maybe a few continuous parameters). "Classifying" the phases of the CTL type is then doable as we only have to identify which elements of the discrete set are in the same phase.

**Definition 75.** A **CTL mapping** from one CTL type $\mathcal{A}$ to another type $\mathcal{B}$ consists of

- a lattice mapping $\mathcal{X}$ from the lattices of $\mathcal{B}$ to the lattices of $\mathcal{A}$.

- a tensor mapping $\mathcal{Y}$ from the tensors of $\mathcal{A}$ to the tensors of $\mathcal{B}$. This tensor mapping has to be local and formally consistent with the lattice mapping, i.e., it has to map the index distribution of $\mathcal{A}$ to the index distribution of $\mathcal{B}$. Usually this involves dividing the indices of $\mathcal{A}$ into local groups and blocking those groups of indices.

$\mathcal{X}$ and $\mathcal{Y}$ have to be consistent in the following sense: Performing a contraction associated to a basic gluing of $\mathcal{A}$ and then applying $\mathcal{Y}$ is equal to first applying $\mathcal{Y}$ and then the contraction associated to the corresponding basic gluing of $\mathcal{B}$. The whole situation as a commuting diagram (where by $I$ we denote tensor shape corresponding to the lattices):

$$
\tag{2.65}
$$

**The mapping** of a CTL $A$ of type $\mathcal{A}$ is the CTL $B$ of type $\mathcal{B}$ given by $B = \mathcal{Y} \circ A \circ \mathcal{X}$. As a diagram:

$$
\begin{array}{ccc}
L_\mathcal{B} & \overset{\mathcal{X}}{\longrightarrow} & L_\mathcal{A} \\
B \Big\downarrow & & \Big\downarrow A \\
T_\mathcal{B} & \underset{\mathcal{Y}}{\longleftarrow} & T_\mathcal{A}
\end{array}
\tag{2.66}
$$

**Definition 76.** For every CTL type there is the **identity CTL mapping** from this type to itself given by the identity lattice mapping and the identity tensor mapping. The **trivial mapping** from any CTL type to any other CTL type is the CTL mapping given by the trivial lattice mapping and the trivial tensor mapping.

**Definition 77.** A CTL mapping is called **exact** if the corresponding lattice mapping is a circuit move.

**Definition 78.** Two CTL types $\mathcal{A}$ and $\mathcal{B}$ are said to be **in the same class** if there is a mapping $\mathcal{M}_\mathcal{A}$ from $\mathcal{A}$ to $\mathcal{B}$ and a mapping $\mathcal{M}_\mathcal{B}$ from $\mathcal{B}$ to $\mathcal{A}$ such that both $\mathcal{M}_\mathcal{B} \circ \mathcal{M}_\mathcal{A}$ and $\mathcal{M}_\mathcal{A} \circ \mathcal{M}_\mathcal{B}$

are exact. As a picture:

$$
\begin{array}{c}
\mathcal{A} \\
\downarrow \text{exact} \\
\text{CTL mapping} \quad \xrightarrow{\mathcal{M}_\mathcal{A}} \quad \mathcal{B} \\
\mathcal{A} \quad \xrightarrow{\mathcal{M}_\mathcal{B}}
\end{array}
\qquad (2.67)
$$

$$
\begin{array}{c}
\mathcal{B} \\
\mathcal{A} \quad \xleftarrow{\mathcal{M}_\mathcal{B}} \quad \downarrow \text{exact} \\
\quad \text{CTL mapping} \\
\xrightarrow{\mathcal{M}_\mathcal{A}} \quad \mathcal{B}
\end{array}
$$

**Definition 79.** Two CTLs $A$ and $B$ of type $\mathcal{A}$ and $\mathcal{B}$ are said to be **in the same phase** if the following holds:

- $\mathcal{A}$ and $\mathcal{B}$ are in the same class.

- $B$ is in the same phase as the according mapping of $A$.

### 2.4.4 Product, sub-types and fusion

**Definition 80.** The **product** of two CTL types $\mathcal{A}$ and $\mathcal{B}$ is the CTL type whose lattice type is the product of the lattice types of $\mathcal{A}$ and $\mathcal{B}$, whose tensor type is the product type of the tensor types of $\mathcal{A}$ and $\mathcal{B}$, and whose index distribution prescriptions and contraction prescriptions are given by the disjoint union of the index distribution prescriptions and contraction prescriptions of $\mathcal{A}$ and $\mathcal{B}$.

The **product** of two CTLs $A$ and $B$ of type $\mathcal{A}$ and $\mathcal{B}$ is the CTL that associates to a pair of a lattice of $\mathcal{A}$ and a lattice of $\mathcal{B}$ the product of the CTL tensors associated to the two lattices.

**Definition 81.** A **sub-type** of a CTL type is a sub-type of the corresponding glued, moved lattice type with the subset of the index prescription whose place types are contained in the lattice sub-type, and the subset of the contraction prescriptions whose basic gluings are contained in the lattice sub-type.

The **sub CTL** of a CTL of a type with a sub-type is the CTL of the sub-type obtained by restricting to lattices of the sub-type.

**Definition 82.** Consider a CTL type $\mathcal{A}$ together with another type $\mathcal{B}$ that occurs twice as a sub-type of $\mathcal{A}$. For this situation the **joint type** is obtained from $\mathcal{A}$ by identifying the two occurrences of $\mathcal{B}$. I.e., the lattices are of the joint type, and the index distribution prescription and contraction prescription that were previously defined for the different occurrences act now on the unified occurrence.

Every lattice of $\mathcal{A}$ yields a lattice of the joint type. So given a CTL of type $\mathcal{A}$ we can associate CTL tensors to all lattices of the joint type arising this way. However, not every lattice of the joint type arises this way and if it does that way might not be unique. Thus, in order to define the **joint CTL** for a CTL of type $\mathcal{A}$, the CTL tensors associated to lattices arising from lattices of type $\mathcal{A}$ have to consistently determine the CTL tensors associated to all other lattices of the joint type via the CTL axioms. For this to be possible, at least the two sub CTLs have to be equal. But even then the joint CTL might not defined for every CTL type $\mathcal{A}$. However, for CTLs arising from TLs it is defined, and consistent with the notion of joint TL.

**Definition 83.** We say that there is a **fusion** from a set of CTL types $\mathcal{A}_i$ to another CTL type $\mathcal{B}$ if there is a sequence of taking CTL mappings and joint CTLs that yields $\mathcal{B}$ from the product of all the $\mathcal{A}_i$.

**Observation 17.** The **stacking** of two CTLs $A$ and $B$ of same type is one of the simplest examples of fusion from twice the same type to this type again. The corresponding lattice mapping is just duplication. The tensor mapping is a simple blocking of the two indices. So the CTL tensor associated to a lattice $X$ is the tensor product of the CTL tensors that $A$ and $B$ associate to $X$.

$$
(T_1 \otimes T_2)[X] := T_1[X] \otimes T_2[X] \qquad (2.68)
$$

**Definition 84.** Two CTLs of the same type are said to be **in the same phase relative to** a sub-type, if 1) the CTLs corresponding to the sub-type are equal 2) the two CTLs are in the same phase and 3) the TL operator connecting them is trivial when restricted to the sub-type.

### 2.4.5 Special solutions to the CTL axioms

**Observation 18.** For each CTL type we can choose the trivial basis for each index type (a one-element set in the case of real tensors), and all CTL tensors to be the trivial tensor (the number 1 in the case of real tensors). This obviously satisfies all CTL axioms. We will call this CTL the **trivial CTL**, and the corresponding phase the **trivial phase**.

**Observation 19.** For CTL types whose basis types of the tensor type are given by sets (as for real or complex or fermionic or similar tensor types), there is the **empty CTL**, which has the empty set as basis for all index types. For real, complex, or similar tensor types, and every choice of basis, there is the **zero CTL**, where all entries of all CTL tensors are 0.

**Observation 20.** Consider some CTL type whose tensor type is given by real tensors or similar. For each such type the delta tensors for any basis set $B$ define a CTL in the following way: Set the basis for one selected index type to $B$ and choose the trivial basis set for all other index types. As CTL tensor for a connected lattice (i.e., one that is not a disjoint union of other lattices) take the delta tensor with as many indices as needed. For a disjoint union of many disconnected components take the tensor product of one delta tensor for each component.

We still need to check that this assignment fulfills the CTL axioms: The symmetry axiom is follows from the permutation property of the delta tensors. The internal move and disjoint union axiom are fulfilled by construction. The gluing axiom follows from the fusion property of the delta tensors.

We will call such a CTL coming from the delta tensors a **delta CTL**.

**Definition 85.** Consider a CTL type with real (or similar) tensors CTLs $T_1$ and $T_2$ of this type. The **direct sum** of $T_1$ and $T_2$ defined in the following way:

$$
(T_1 \oplus T_2)[X] := T_1[X] \oplus T_2[X], \qquad (2.69)
$$

where basis set of each index type of $T_1 \oplus T_2$ is the disjoint union of the basis sets of the two corresponding types of indices of $T_1$ and $T_2$. Its entries are equal to the entries of $T_1$ when all index values are from the $T_1$ part of the disjoint union of basis sets, equal to the entries of $T_2$ if all index values are from the $T_2$ part, and 0 otherwise. Alternatively the direct sum can be viewed as a "controlled tensor product" with the delta CTL for a 2-element basis set. E.g.,

$$
\left( - \boxed{T_1} - , \ - \boxed{T_2} - \right) \quad \rightarrow \quad - \boxed{T_{1,2}} -
$$

$$
d - \boxed{T_1 \oplus T_2} - b \quad = \quad d - \boxed{T_{1,2}} - b \tag{2.70}
$$

where $T_1$, $T_2$ and $T_1 \oplus T_2$ denote the corresponding CTL tensors associated to the same lattice.

In the same way we can define the direct sum of a set of $n$ different CTLs and view it as some kind of controlled tensor product with the delta CTL for a $n$-element basis set. If two CTLs share a common sub CTL, one can also build the **direct sum relative** to this sub CTL. Thereby the index types corresponding to the sub-type do not contain the direct sum component.

**Definition 86.** We will call a CTL phase **reducible** if it is in the same phase as a non-trivial direct sum of CTLs. Otherwise we will call it **irreducible**.

**Remark 39.** For every CTL of a type with real (or similar) tensors, we can always take a direct sum with a zero CTL. The corresponding CTL tensors are then not fully supported on the local indices (for which the zero CTL has non-empty basis set), and if we restrict to the locally supported subspace we get back the original CTL. So this embedding of a CTL in a larger space filled up with zeros does not bring anything new. Thus, it makes sense to impose the **local support convention** by restricting to fully supported CTLs only.

Usually we find for a CTL type that there are very simple lattices that can be glued to lattices of certain types of any lattice without changing the background of the lattice. The corresponding CTL tensors can be interpreted as linear maps acting on the indices (usually only a single one) that are involved in the contraction corresponding to the gluing. Those linear maps are just the projectors onto the locally supported subspace. So we can enforce the local support convention by imposing that those linear maps are the identity.

**Definition 87.** Consider a fusion from a set of CTL types $\mathcal{A}_i$ to $\mathcal{B}$. Now for each set of irreducible CTLs $A_i$ of type $\mathcal{A}_i$ and $B$ of type $\mathcal{B}$, the corresponding **irrep coefficient** is the non-negative integer that states how often $B$ occurs as component in the fusion of the $A_i$. We will refer to the collection of the coefficients for all different irreducible $A_i$ and $B$ as **irrep coefficients** of the fusion.

**Definition 88.** Consider a CTL type with real tensors or similar with one selected index type. Imagine choosing some kind of orientation for the background of the lattices that allows to label every place where this index type occurs by a chirality "left handed" or "right handed". Combinatorially we can model this by a **chirality rule** that asserts whether two places of such type near each other have the same chirality or not. The chirality rule has to be locally consistent, but it does not need to be possible to find a global consistent assignment of chiralities for every lattice. If such a global assignment is possible, we call the lattice **orientable**, otherwise it is **non-orientable**. The chirality rule has to obey the following properties: 1) Gluing two places of a connected orientable lattice yields an orientable lattice if and only if the two places have opposite chirality. 2) Gluing two places on two different connected components of an orientable lattice always yields an orientable lattice. 3) Gluing two places of a non-orientable lattice yields a non-orientable lattice.

With such a rule, an orientation for an orientable lattice is a consistent choice of chirality for all places on the lattice. This choice of chirality is determined by choosing the chirality of one single place for each connected component.

Now the **complex number CTL** is the CTL defined in the following way: The basis set of the selected index type is the 2-element set $\{1, \mathbf{i}\}$ and all other basis sets are trivial. The CTL tensor associated to a connected lattice is 0 if the lattice is non-orientable, and the complex tensor with as many indices as needed otherwise. The complex arrow orientations depend on the chirality of each place for a fixed orientation. Due to the arrow reversal property of the complex number tensors, the choice of orientation does not matter.

We still need to check that this assignment fulfills all the CTL axioms: The tensor product axiom and internal move axiom are fulfilled by construction. The symmetry axiom is fulfilled due to the permutation property and the arrow reversal property of the complex tensors. The gluing axiom holds due to the fusion property of the complex tensors together with the chirality rule properties 1) and 2), and the arrow obstruction property of the complex tensors together with the chirality rule property 1) and 3).

**Observation 21.** For a CTL on manifolds with boundary and gluing at the boundary, the orientation defined by the chirality rule as above really corresponds to what is known as an orientation of the manifold. If we calculate the invariant the complex number CTL associates to a closed manifold, we get 0 if the manifold is non-orientable and $2^n$ otherwise, where $n$ is the number of connected components (note that the complex number tensor without any indices is the number 2). So apparently the invariant just counts the number of possible orientations of a closed manifold.

**Definition 89.** A real CTL (one whose CTL tensors are real tensors) is called **complex-real** if it is gauge equivalent to a CTL whose tensors for connected lattices are all realifications of some complex tensor.

**Remark 40.** Consider a complex CTL (i.e., a CTL whose tensors are complex tensors) on lattices that have an orientation (defined by some chirality rule), with the following **unitarity property**: There is a basis such that reversal of the orientation of a connected lattice equals complex conjugation of the CTL tensors in that basis. Each such complex CTL defines a complex-real CTL on the same lattice type without the orientations, in the following way: In order to get the CTL tensor for a connected (non-oriented) lattice pick one orientation and take the realification of the complex CTL tensor, where the arrow directions of all indices are chosen according to the chirality of the corresponding places. Because of the unitarity property of the complex CTL the choice of orientation does not matter. For non-orientable lattices the CTL tensor of the complex-real CTL is 0.

Conversely if we are given a complex-real CTL on lattices without orientation, we can get a complex CTL on the same lattice type where we add an orientation. The CTL tensors of this complex CTL are obtained from the CTL tensors of the complex-real CTL by reverting the realification. For this inversion to be well-defined we need to choose the arrow directions of all indices. Those can be chosen depending on the chirality of the associated places with respect to the orientation of the lattice. This complex CTL automatically has the unitarity property as inverting the arrow directions in the inverse realification equals a complex conjugation.

So we see that complex-real CTLs are the much more natural formulation of complex CTLs on oriented lattice with the unitarity property, because it is simpler in three ways: 1) We only need to work with real tensors. 2) We do not need an orientation. 3) The unitarity property is automatically inbuilt.

**Definition 90.** The tensor product of any real CTL with the complex number CTL yields a complex-real CTL, which we call **complexification** of the original CTL. Note that the complexifications of different gauge-inequivalent real CTLs might lead to gauge-equivalent complex-real CTLs.

**Remark 41.** CTLs provide a very simple mathematical framework that yields very rich algebraic structures. For CTLs with real tensors the basic tensors can often be identified with the structure coefficients of known algebraic structures. Examples that we find in Section (5) are *-algebras, their representations, (weak) Hopf algebras, representations of their Drinfel'd doubles, or unitary fusion categories. In general the algebraic structures emerging from CTLs on $n$-manifolds with boundary and gluing at the boundary seem to be related to $n$-categories.

Linear algebraic structures can usually be written down in terms of a set of linear or anti-linear maps of the form

$$T : V^{\otimes n} \longrightarrow V^{\otimes m},$$

with $V = \mathbb{C}^B$ for some finite basis set $B$. We can represent such a linear or anti-linear map $T$ as a complex tensor with $n + m$ indices simply by its coefficients with respect to the canonical basis $\{|i\rangle\}_{i \in B}$ of $V$:

$$T^{abc...}_{xyz...} = \langle x, y, z, \dots | T | a, b, c, \dots \rangle , \qquad (2.71)$$

with $a, b, \dots \in B$. Conversely, each complex tensor $T$ defines a linear map by taking

$$T |a, b, c, \dots \rangle = \sum_{x, y, z, \dots} T^{abc...}_{xyz...} |x, y, z, \dots \rangle \qquad (2.72)$$

and also an anti-linear map by taking

$$\tilde{T} = K \circ T \qquad (2.73)$$

where $K$ is the complex conjugation in the canonical basis. The complex tensor $T$ itself does not contain any information about whether the according map was linear or anti-linear. So instead we will work with the realification of the complex tensor $T$, to which we feed the information about linearity or anti-linearity via the complex arrow orientations: For a linear map, the ingoing indices $a, \dots$ get ingoing arrow directions and the outgoing indices $x, \dots$ get outgoing arrow directions, whereas for anti-linear maps, all indices get ingoing arrow directions. For complex-real CTLs the CTL basic tensors will correspond to the structure coefficients of the algebraic structures in this realified form.

The axioms of known algebraic structures are usually given in terms of commutative diagrams, i.e., equations between different ways of composing (tensor products of) the linear or anti-linear maps. On the level of the realified structure coefficients, such a composition is nothing but a contraction of the corresponding indices. E.g.

$$T(V(a)) = U(a) \forall a \quad \longleftrightarrow \quad T^i_j V^k_i = U^k_j, \qquad (2.74)$$

where the indices $i$, $j$ or $k$ also carry a label in $\{\mathbf{1}, \mathbf{i}\}$ from the realification which is important when anti-linear maps are involved. Thus, the axioms for known algebraic structures have the same form as the CTL axioms, namely equations between different tensor networks formed by the realified structure coefficients.

### 2.4.6 Relation to TLs

**Observation 22.** Each TL yields a CTL by the following identification:

1. The CTL lattices will be a lattice type obtained from the TL lattices by adding a boundary, which we will call the **cutting boundary**. I.e., the CTL lattices are patches that have been cut out from the TL lattices. E.g., for TLs on $n$-dimensional simplicial complexes the CTL lattices would be $n$-dimensional simplicial complexes with boundary. Or for TLs on $n$-dimensional simplicial complexes with boundary, this would yield $n$-dimensional complexes with two sorts of boundary separated by an $(n-2)$-dimensional simplicial complex. One of the sorts of boundary is the old physical boundary whereas the other sort of boundary and the $(n-2)$-dimensional simplicial complex form the cutting boundary.

2. The index lattice is given by the cutting boundary.

3. All basic moves of the TL become the basic moves of the CTL lattices, just that they are not allowed to change the cutting boundary.

4. The cut-out patches of the TL lattices, i.e., the CTL lattices, can be glued together again at their cutting boundary. This defines a basic gluing for the CTL lattices.

5. When we put the TL tensor network on a CTL lattice (with cutting boundary) get open indices at the cutting boundary. The distribution of those indices is the index distribution prescription of the CTL. This is consistent with the fact that the index lattice is the cutting boundary.

6. When we join two patches of TL lattice together we can also joint together the patches of tensor network on them. Thereby we contract the indices across the cutting boundary pieces where the two patches are glued together. This contraction is the contraction prescription of the CTL associated to the gluing moves that glues the two cutting boundary pieces.

7. The CTL tensor associated to a CTL lattice is the evaluation of the patch of tensor network given by the TL on this lattice with cutting boundary.

The symmetry axiom of the CTL follows from the symmetry axiom of the TL. The disjoint union axiom and gluing axiom of the CTL are fulfilled by construction. The move axiom of the CTL follows from the move axiom of the TL.

**Observation 23.** For every TL on higher order manifolds (see Section (3.1.3)) the procedure above yields a CTL on higher order manifolds with boundary. The central boundary link of the latter equals the central link of the former higher order manifold type. Gluing happens at all boundary regions.

**Observation 24.** Each CTL that is finitely generated and finitely axiomatized defines a TL on the history lattices: At every cell of the history lattice coming from a basic lattice of the CTL put the corresponding basic tensor, and replace every cell of the history lattice coming from a basic gluing by the corresponding contraction.

# Chapter 3

# Concrete lattice types

In this section we will define different lattice types with different moves and gluings that will be important for the concrete CTL constructions in Section (5). In the first part we will give informal definitions and intuition about the continuum structures that the backgrounds of those lattice types will correspond to, and how the backgrounds are affected by different types of gluing. In the second part we will give precise combinatorial definitions for lattices having these kinds of backgrounds.

**Overview**

| lattice type | geometric picture | moves | gluings | continuum picture |
|---|---|---|---|---|
| $n$SC-lattices | simplicial complexes | Pachner moves | simplex gluing | $n$-manifolds |
| $n$CC-lattices | cell complexes | bi-stellar flips | cell gluing | $n$-manifolds |
| $n$SCb-lattices | simplicial complexes with boundary | Pachner moves, boundary attachment | internal gluing, boundary gluing | $n$-manifolds with boundary |
| $n$CCb-lattices | cell complexes with boundary | bi-stellar flips, boundary flips | internal gluing, boundary gluing | $n$-manifolds with boundary |
| $n/o$CC-lattices | higher order cell complexes | region bi-stellar flips | region gluing | $n/o$-manifolds |
| $n/o$CCf-lattices | thickened higher order cell complexes | thickened bi-stellar flips | thickened gluing | framed $n/o$-manifolds |

## 3.1 Continuum picture for different moved lattice classes

The continuum structures introduced in this section will only serve as a guide to construct combinatorial structures representing them as mentioned in Remark (1). Therefore the definition in this section are not to be thought as rigorous definitions but should rather give the reader an as precise as possible picture for those continuum structures.

### 3.1.1 Manifolds

**Definition 91.** By $n$-**manifold** we mean a compact topological $n$-manifold. Thereby we consider two $n$-manifolds as equal if they are homeomorphic.

To be a bit more precise, the lattice types described in Section (3.2) seem to rather correspond to piece-wise linear $n$-manifolds than topological ones. (It is known that there are topological manifolds that to not allow for (unique) piece-wise linear triangulations, and the latter are more or less the ones that our lattice types describe. However, two notions are only different for $n \geq 4$.)

For reasons that will become clear later we also define $-1$-manifolds. There is only one $-1$-manifold, namely the empty one.

For every fixed $n$, the $n$-manifolds are the continuum picture for the backgrounds of one moved lattice class.

**Background gluings**

**Definition 92.** A 0-**surgery gluing** is the following operation that can be applied to any pair of points on a $n$-manifold: Cut out a small $n$-ball around each of the two points. Each of those $n$-balls has the $(n-1)$-sphere as its boundary. Then identify those two boundary parts. Note that for a fixed pair of points there might be different identifications that yield different 0-surgery gluings. Two such operations applied to homeomorphic $n$-manifolds with homotopically equivalent pairs of points and identifications are considered to be equal.

For every fixed $n$, 0-surgery gluing is the continuum picture for a background gluing for $n$-manifolds.

**Example 31.** Some examples of 0-surgery gluing for $n =$

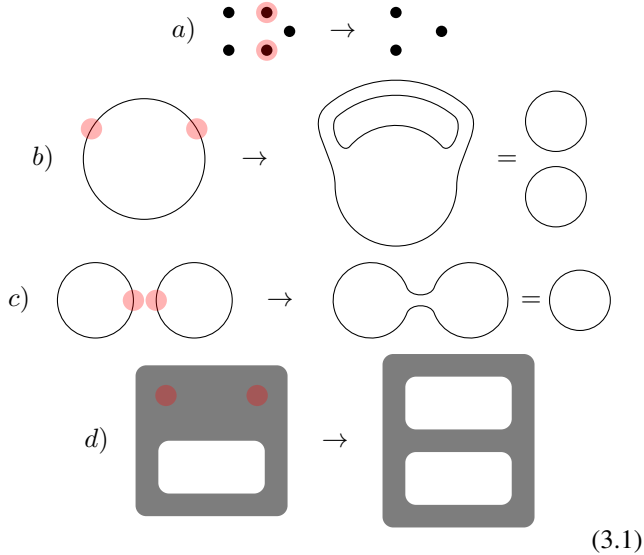

(3.1)

As a) shows that gluing of two points of a 0-manifold makes them disappear. b) and c) show gluing of two points on a 1-manifold. d) shows gluing of two points on a 2-torus, yielding a genus 2 surface. Note again that the right hand side is not uniquely determined from the left hand side as we still need to specify how the two boundaries are identified. E.g., the gluing in b) could equivalently yield a single circle again, or the gluing in d) could also yield a Klein bottle with an additional handle if we glue in an orientation-reversing manner.

**Definition 93.** More generally, for every $0 \leq d \leq n$ we can define $d$-**surgery gluing** as the following operation that can be applied to every embedding of $S_d$ on a $n$-manifold: Cut out a tubular neighbourhood of the embedded $S_d$, identify it with $S_d \times B_{n-d}$ and paste $B_{d+1} \times S_{n-d-1}$. To this end note that there is a canonical identification between the boundaries of $S_d \times B_{n-d}$ and $B_{d+1} \times S_{n-d-1}$. Thereby different identifications of the tubular neighbourhood with $S_d \times B_{n-d}$ correspond to different gluings. Two $d$-surgery gluings acting on homeomorphic $n$-manifolds with homotopically equivalent embeddings of $S_d$ and identifications of the tubular neighbourhood with $S_d \times B_{n-d}$ will be considered equal.

For every fixed $n$ and $d$, the $d$-surgery gluing is the continuum picture for a background gluing for $n$-manifolds.

**Observation 25.** For $n = d$, a $d$-surgery gluing simply removes a $n$-sphere as from the $n$-manifold. One could also define $-1$-surgery gluing as adding a $n$-sphere, however, this is not a typical background gluing for a typical glued lattice type.

**Example 32.** Consider the following $d$-surgery gluings:

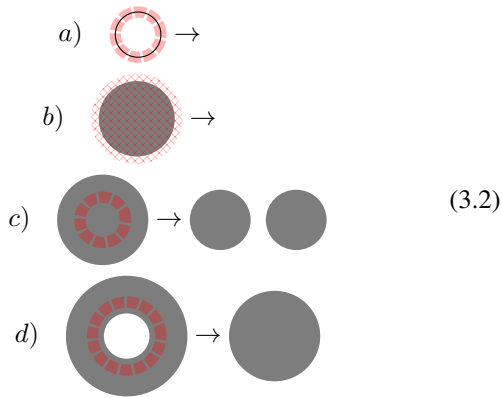

(3.2)

a) shows a 1-surgery gluing removing the circle times the point and adding the disk times the empty $-1$-manifold. b) shows the equivalent in one dimension higher, namely a 2-surgery gluing removing a sphere. c) shows a 1-surgery gluing acting on a sphere removing a stripe (the circle times the interval) and pasting two disks (the 2-ball times the 0-sphere), yielding two spheres. d) shows the same 1-surgery gluing applied to a torus, for the circle winding around a non-contractible loop, yielding a sphere.

### 3.1.2 Manifolds with boundary

**Definition 94.** By a **boundary $n$-manifold** we mean a compact topological $n$-manifold with boundary. Boundary manifolds are considered equal if they are homeomorphic.

Again, technically we should rather talk about piece-wise linear instead of topological manifolds.

For every fixed $n$, the boundary $n$-manifolds are the continuum structure that is in one-to-one correspondence to the backgrounds of a moved lattice class.

**Background gluings**

There are two kinds of gluings for lattice classes described by boundary $n$-manifolds:

**Definition 95. Interiour $d$-surgery gluings** act on a boundary $n$-manifold with a $d$-sphere embedded into its interior. They are the same as the $d$-surgery gluings for $n$-manifolds (without boundary). The boundary is not affected by such a gluing.

For every fixed $d$ and $n$ the interior $d$-surgery gluings are the continuum picture for one background gluing for boundary $n$-manifolds.

**Example 33.** Consider the following interior surgery gluings for boundary 2-manifolds:

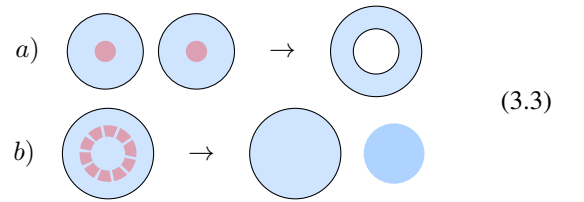

(3.3)

a) shows an interior 0-surgery gluing of two points on two disks yielding an annulus. b) shows an interior 1-surgery gluing of a circle on a disk yielding a disk and a sphere.

**Definition 96.** If we cut out pieces of boundary $n$-manifold we get "manifolds" whose boundary consists of two parts: pieces of the boundary of the original boundary manifold to which we will refer to as **boundary**, and new boundary components arising from the cutting to which we will refer to as **cutting boundary**. The boundary and the cutting boundary are separated by embedded $n-2$-manifolds which we can view as the cutting boundary of the boundary. Now let us define the following three cut-out pieces:

- By $d$-**ball** ($B_d$) we mean the $d$-ball with no cutting boundary.

- By **cutting $d$-ball** ($B_d^c$) we mean the $d$-ball with no boundary but only cutting boundary.

- By **mixed $d$-ball** ($B_d^m$) we mean the $d$-ball that has one $d-1$-ball as boundary and one $d-1$-ball as cutting boundary, meeting at an embedded $d-2$-sphere.

A **boundary $d$-surgery gluing** for $0 \leq d \leq n-1$ is the following operation acting on boundary $n$-manifolds with a $d$-sphere embedded into their boundary: Cut out a tubular neighbourhood of the embedded $d$-sphere. This yields a cut-out piece that can be identified with $S_d \times B_{n-d}^m$. Now there is a canonical identification between the cutting boundaries of $S_d \times B_{n-d}^m$ and $B_{d+1}^c \times B_{n-d-1}$. Thus, cutting out the tubular neighbourhood and filling in $B_{d+1}^c \times B_{n-d-1}$ according to this identification yields a boundary $n$-manifold again. Note that different identifications of the cut-out piece with $S_d \times B_{n-d}^m$ yield different boundary surgery gluings. Two boundary surgery gluings are considered equal if they act on homeomorphic boundary manifolds, and the embedded $d$-spheres and the identification of their tubular neighbourhood with are homotopically equal, and the identifications are equal.

Again, for each $d$ and each $n$ the boundary $d$-surgery gluings acting on $n$-manifolds are the continuum picture for one specific background gluing.

**Example 34.** Consider the following boundary surgery gluings:

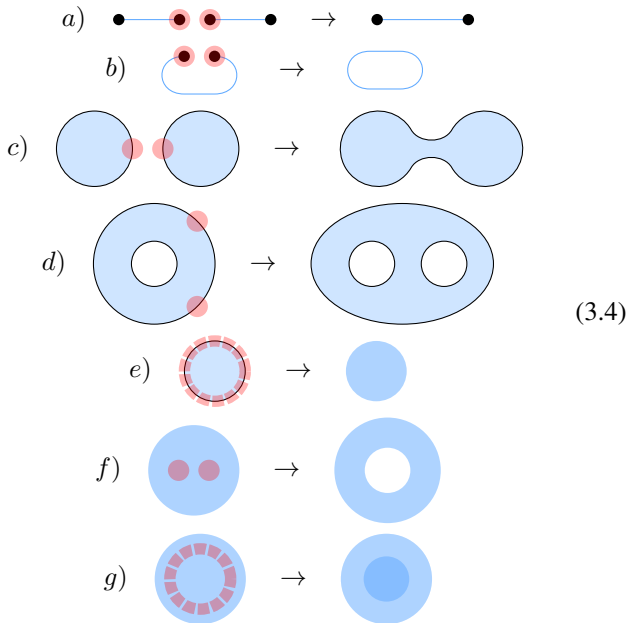

$(3.4)$

a) to d) show boundary 0-surgery gluings on boundary 1-manifolds and boundary 2-manifolds that should be self-explanatory. e) shows a boundary 1-surgery gluing transforming a disk into a sphere. f) shows a boundary 0-surgery gluing transforming a 3-ball into a solid 3-torus. g) shows a boundary 1-surgery gluing transforming a 3-ball into a hollow 3-ball (i.e., the 2-sphere times the interval).

**Proposition 1.** Every boundary $n$-manifold can be obtained from a disjoint union of $n$-balls by boundary surgery gluings (being known as a *handle decomposition* of the manifold).

### 3.1.3 Higher order manifolds

**Informal definition 18.** Higher order manifolds can be imagined as a space composed of several manifolds with boundary of different dimensions. E.g., one can think of manifolds that have different embedded sub-manifolds, as for example surfaces with pattern of lines that meet at points on them.

Each higher order manifold is of a specific type that says what kinds of sub-manifolds of what dimensions can occur and in which way they are put together.

**Definition 97.** A **pre higher order $n$-manifold** is a compact topological space composed of (possibly non-compact or empty) $d$-manifolds ($0 \leq d \leq n$) called $d$-**regions**. Two pre higher order $n$-manifolds are called **equivalent** if they are related by a homeomorphism that leaves the regions invariant.

**Example 35.** Consider the following examples of pre higher order manifolds (which are not higher order manifolds as we will see later) (note Remark (2)):

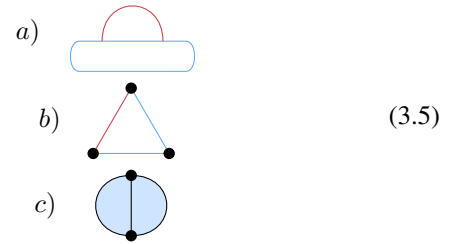

$(3.5)$

a) shows a pre higher order 1-manifold composed of two 1-manifolds (one circle and one open interval). b) shows a pre higher order 1-manifold composed of one 0-manifold (three points) and two 1-manifolds (two and one open intervals). c) shows a pre higher order 2-manifold composed of one 0-manifold (two points), one 1-manifold (three open intervals) and one 2-manifold (two open disks).

**Definition 98.** A **higher order $n$-manifold** $M$ is a pre higher order manifold such that: For every $d$-region there is a higher order $(n-1)$-manifold $C$ whose $x$-regions are in one-to-one correspondence to the $(x+1)$-regions of $M$, such that: For every point of the $d$-region there are arbitrarily small neighbourhoods whose boundary is homeomorphic to $C$.

Note that this is actually a recursive definition as the definition of an $n$-manifold relies on the definition fo an $(n-1)$-manifold. The recursion is terminated by stating that a higher order $-1$-manifold type has no regions at all, and that higher order manifolds of this type are the empty set.

**Remark 42.** The examples for pre higher order manifolds above are all not higher order manifolds: In a) the blue circle has two points (where the red interval joins) where the boundary of each (small enough) neighbourhood includes a red point. In b) the boundary of a small neighbourhood of two of the three points consists of a blue and a red point, whereas the boundary of a small neighbourhood of the third points consists of two blue points. In c) the neighbourhood for the two intervals at the boundary of the disk looks different from the neighbourhood of the interval in the interior, though all intervals correspond to the same region.

Let us consider a few examples of higher order manifolds.

**Example 36.** Higher order 0-manifolds are sets containing different 0-manifolds, i.e., sets of sets of vertices. E.g., see Remark (2):

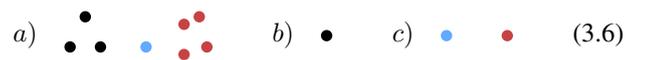

$(3.6)$

a) shows a higher order 0-manifold with 3 0-regions, consisting of 3, 1 and 4 points each. b) and c) show higher order 0-manifolds with one or two regions each consisting of one point, respectively.

**Example 37.** Higher order 1-manifolds consist of a set of 1-regions which have 0-regions as their boundary or meet at 0-regions. Each vertex of a fixed 0-region has to connect the same 1-regions. E.g., (see Remark (2):

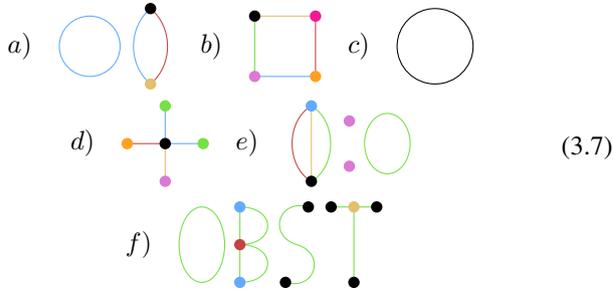

(3.7)

a) shows a higher order 1-manifold with two 1-regions and two 0-regions. b) shows an example with four 1-regions and four 0-regions. c) shows an example with a single 1-region. d) shows a higher order 1-manifold with four 0-regions, and three 1-regions. e) shows another example with three 0-regions and three 1-regions. f) has one 1-region (consisting of one circle and 8 intervals) and 4 0-regions (consisting of 1, 1, 2 and 5 points, respectively).

**Example 38.** Consider the following higher order 2-manifolds (see Remark (3)):

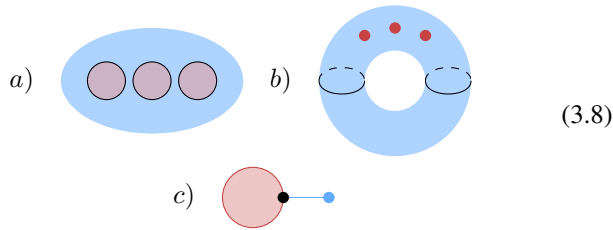

(3.8)

a) shows a 2-sphere decomposed into a purple 2-region consisting of 3 disks, and a blue 2-region covering the rest, separated by a single black 1-region consisting of three circles. b) shows a 2-torus covered by a single blue 2-region consisting of two annuli, separated by a black 1-region consisting of two circles. One of the annuli has three punctures coming from an embedded purple 0-region consisting of three points. c) consists of two 0-regions (one point each), two 1-regions (one interval each) and one 2-region (a disk).

**Example 39.** Every 3-dimensional cell complex defines a higher order 2-manifold: replace each face of the cell complex by a disk-like 2-region, each edge by an interval-like 1-region and each vertex by a 0-region consisting of one point.

**Definition 99.** The *d*-ball completion of a higher order *n*-manifold $X$ is the following higher order manifold: 1) take $X \times B_d$ ($B_d$ is the *d*-ball), 2) add a $(d-1)$-region $S$ consisting of the $(d-1)$-sphere and 3) for every point $s$ in the boundary $(d-1)$-sphere of $B_d$, shrink $X \times s$ to a single point that is identified with the corresponding point of $S$.

**Remark 43.** The 0-ball completion is a bit special: It takes the higher order manifold times the point, yielding the same higher order manifold. As the point has no boundary there is nothing to identify. So the 0-ball completion is the higher order manifold itself.

**Example 40.** Consider the following higher order manifolds and their ball completions:

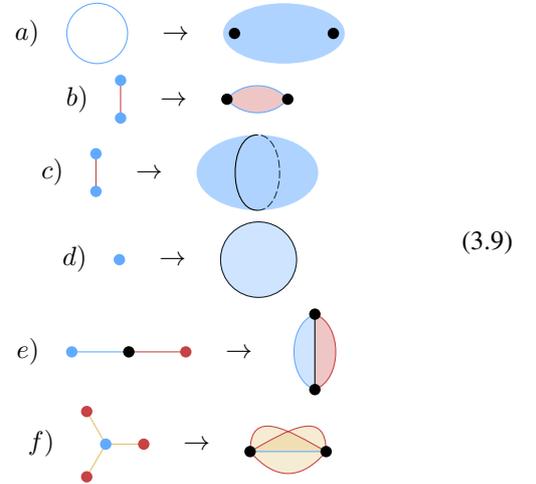

(3.9)

a) shows the 1-ball completion of a circle yielding a sphere with two embedded points. b) shows the 1-ball completion of an interval yielding a disk with two embedded points in the boundary. c) shows the 2-ball completion of an interval yielding a 3-ball whose boundary 2-sphere is separated into two parts by an embedded circle. d) shows the 2-ball completion of a point yielding a disk. e) shows the 1-ball completion of two intervals of different regions joined together. f) shows the 1-ball completion of three intervals meeting a one point, yielding three disks whose boundary is divided into two intervals, with one of the intervals of every disk being identified.

**Definition 100.** For any *d*-region of a higher order *n*-manifold the boundary of a small neighbourhood has to be equal to the *d*-ball completion of a higher order $(n-d-1)$-manifold, which we will call the **upper link** of that region.

**Remark 44.** The *x*-regions of the upper link of a *d*-region are one-to-one correspondence to the $(d+x+1)$-regions of the original higher order manifold. The upper link of a region in the upper link of another region equals the upper link of the equivalent region in the original higher order manifold. E.g., in example (41) the upper link of $g$ in $w$ equals the upper link of $g$ in the overall higher order manifold.

**Example 41.** Consider the following higher order 2-manifold (see Remark (2)):

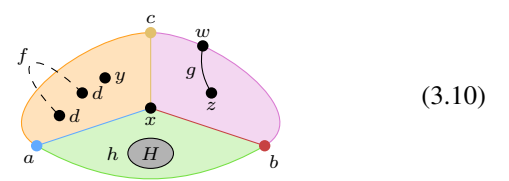

(3.10)

The upper links of some of the regions are:

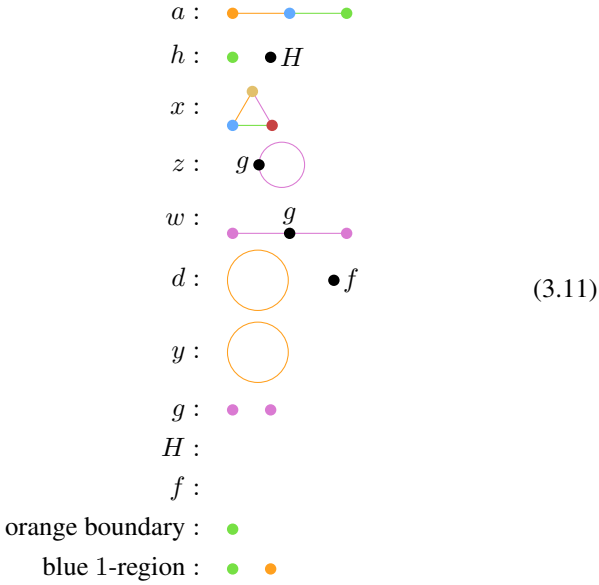

$$\begin{array}{rl} a: \\ h: \\ x: \\ z: \\ w: \\ d: \\ y: \\ g: \\ H: \\ f: \\ \text{orange boundary}: \\ \text{blue 1-region}: \end{array} \qquad (3.11)$$

**Definition 101.** The **type** of a higher order manifold refers to the information of what regions there are and what the upper links of those regions are.

For every fixed type the higher order $n$-manifolds of this type are the continuum picture for one specific moved lattice class. If the class has no $d$-regions for $d < n-o$ then we will also refer to higher order $n$-manifolds of this type as $n/o$-**manifolds**.

**Definition 102.** A higher order manifold type is called **central** if there is one region such that all regions in its upper link are non-empty. This region is then called the **central region**. In this case the class is determined by drawing the upper link of this central region, which we will call the **central link**.

**Definition 103.** The **disjoint union** of two higher order manifolds of the same type is obtained by taking for each $d$-region of the disjoint union the disjoint union of the two corresponding $d$-manifolds. I.e., intuitively taking the disjoint union means placing the two higher order manifolds next to each other.

**Observation 26.** $n$-manifolds form a (central) $n/0$-manifold type consisting of only one single $n$-region, with empty upper link.

Now let's consider some examples of different higher order manifolds of specific types:

**Example 42.** Consider a central $2/2$-manifold type with the following upper link of the central 0-region:

black 0-region: 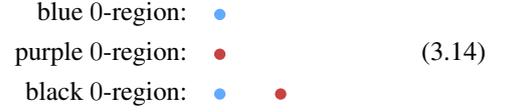 $\qquad (3.12)$

$2/2$-manifolds of this type are e.g.:

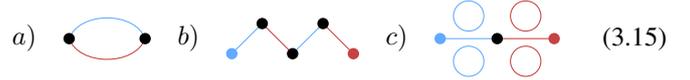
$$(3.13)$$

The 1-regions and 2-regions of the $2/2$-manifolds of this type correspond to the 0-regions and 1-regions of the central upper link. The central 0-region (or any other region) can also be empty as shown in d).

**Example 43.** Consider the following non-central $1/1$-manifold type: There are three 0-regions with the following upper links:

blue 0-region: 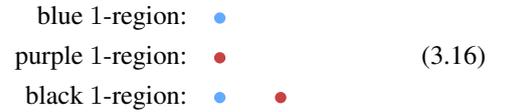
purple 0-region: $\qquad (3.14)$
black 0-region:

Together with the according blue and purple 1-regions. $1/1$-manifolds of this type are e.g.:

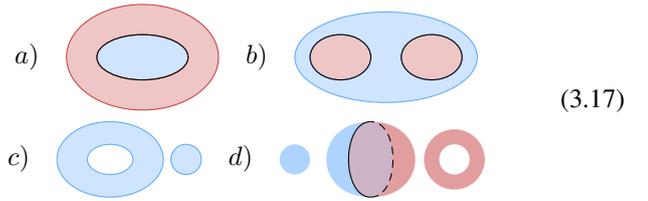
$$(3.15)$$

**Example 44.** Consider the following non-central $2/1$-manifold type: There are three 1-regions with the following upper links:

blue 1-region: 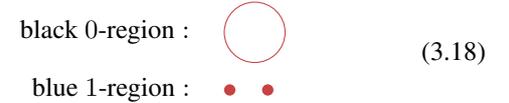
purple 1-region: $\qquad (3.16)$
black 1-region:

Together with the according blue and purple 2-regions. $2/1$-manifolds of this type are e.g.:

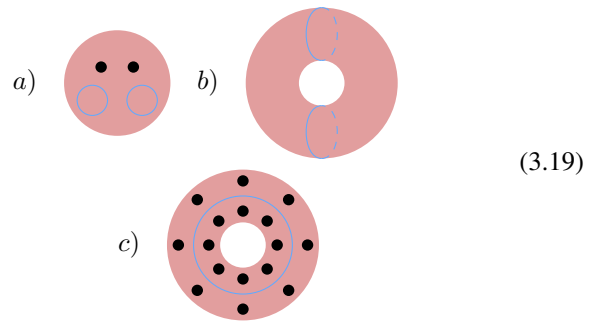
$$(3.17)$$

a), b) and c) show are $2/1$-manifolds that can be embedded into the plain and should be self-explanatory. d) consists of a sphere decomposed into a blue and purple half-sphere, separated by a black circle. The blue and purple 2-regions also have a second connected component, namely a sphere and a torus, respectively.

**Example 45.** Consider the $2/2$-manifold type with one 0-region, one 1-region and one 2-region, and the following upper links:

black 0-region: 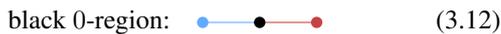
blue 1-region: $\qquad (3.18)$

$2/2$-manifolds of this type are 2-manifolds with embedded points and lines that do not interact with each other:

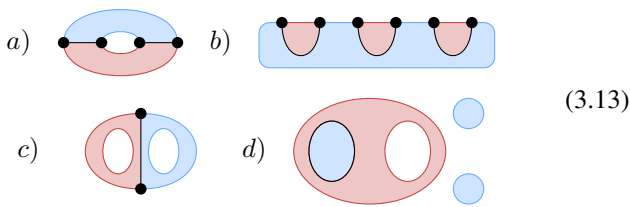
$$(3.19)$$

a) shows a sphere with two embedded circles and two embedded points. b) shows a torus with two embedded circles winding around a non-contractible loop. c) shows a torus with one embedded circle and various embedded points.

**Remark 45.** If we want to constructively build a higher order manifold of a specific type we have to start with the lowest-dimensional regions. I.e., for a $n/o$-manifold type, we start by choosing an $(n-o)$-manifold for each $(n-o)$-region of the type. Then we add an $(n-o+1)$-manifold with boundary for each $(n-o+1)$-region of the type, such that their boundary consists of the $(n-o)$-manifolds in a way that is consistent with their upper links, and so on.

**Remark 46.** Consider a $n/o$-manifold type and one $d$-region of that type. Take the central $n/(n-d)$-manifold type with the upper link of the $d$-region as the central upper link. Every $n/(n-d)$-manifold of this type is also a $n/o$-manifold of the original type. E.g., we can take the blue 1-region in example (44): The higher order manifold c) is of the according central type.

**Definition 104.** The **stellar cone** of a higher order $n$-manifold $X$ is the higher order $n+1$-manifold $Y$ with the following properties:

• For every $d$-region of $X$ there is one $d$-region of $Y$ and one $d+1$-region of $Y$.

• The $d$-regions of $Y$ corresponding the $d$-regions of $X$ form a higher order $n$-manifold which is just $X$.

• There is one additional 0-region consisting of a single vertex (the **central vertex**) and for every $d$-region $B$ of $X$ the corresponding $d+1$-region $C$ of $Y$ (together with its boundary) is homeomorphic to $B \times [0,1]$ where $B \times 0$ is identified with the $d$-region corresponding to $B$ and $B \times 1$ is contracted to a single vertex that is identified with the central vertex. The boundary of $C$ consists of the $d$-region of $Y$ corresponding to $B$ and the $x$-regions of $Y$ corresponding to the $x$-regions of $X$ that form the boundary of $B$, and the $x+1$-regions of $Y$ corresponding to the $x$-regions of $X$ that form the boundary of $B$.

**Example 46.** Consider the following higher order manifolds and their stellar cones:

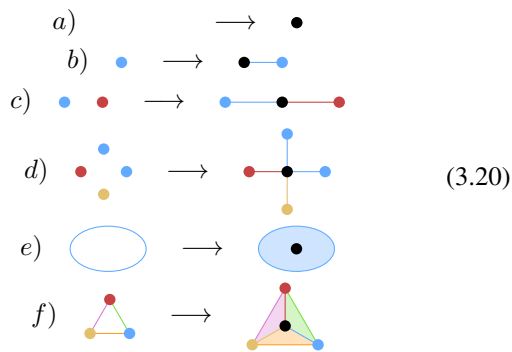

(3.20)

**Definition 105.** We will refer to a specific $n/(o-1)$-manifold type as **boundary** $n/o$-**manifolds** if its regions are separated into two sets called the **bulk** $d$-**regions** and the **boundary** $d$-**regions**, such that:

• There is a one-to-one correspondence between bulk $d$-regions and boundary $(d-1)$-regions.

• The upper link of each boundary region is the stellar cone of the upper link of the corresponding bulk region. The central vertex of that stellar cone corresponds to that bulk region itself.

**Definition 106.** The boundary regions of a boundary $n/o$-manifold form a boundary $(n-1)/o$-manifold, that we will call the **boundary** of the boundary $n/o$-manifold.

**Definition 107.** A boundary $n/o$-manifold type is called **central** if there is one bulk $(n-o)$-region such that all other regions are non-empty in the upper link of this $(n-o)$-region. This upper link is then called the **boundary central link**.

**Remark 47.** Boundary $n$-manifolds are a boundary $n/0$-manifold type consisting of only one bulk $n$-region and one boundary $(n-1)$-region.

**Observation 27.** The stellar cone of a $n/o$-manifold is a boundary $(n+1)/o$-manifold.

Let us consider a few examples of higher order manifolds with boundary.

**Example 47.** Boundary higher order 1-manifolds look like higher order 1-manifolds (without boundary) just that for every 1-region there is one 0-region whose vertices are endpoints of exactly those 1-regions.

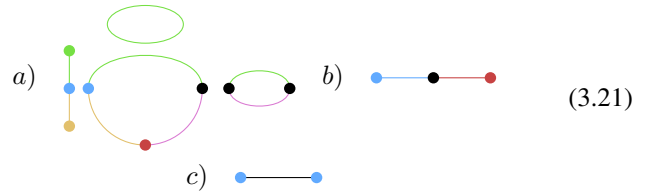

(3.21)

a) shows a boundary higher order 1-manifold involving three bulk 1-regions and three bulk 0-regions. There are three boundary 0-regions, but only two of them are non-empty though. b) shows the stellar cone of a higher order 0-manifold consisting of two points. c) shows a higher order boundary 1-manifold consisting of one bulk 1-region and one boundary 0-region.

**Example 48.** Consider the following boundary higher order 2-manifolds:

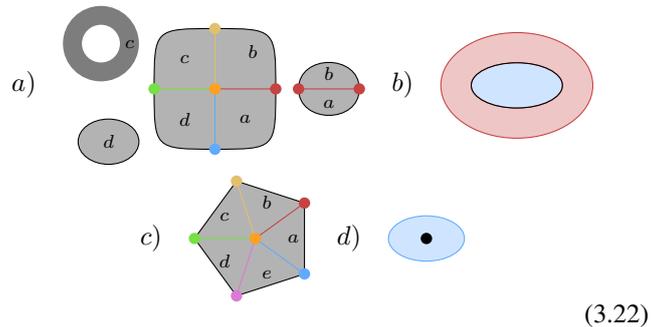

(3.22)

a) shows a boundary higher order 2-manifold where we labelled the bulk 2-regions together with their corresponding boundary 1-regions by the letters $a$, $b$, $c$ and $d$, and there are four bulk 1-regions with the corresponding boundary 0-regions. The bulk 2-region $c$ is the disjoint union of a torus and an (open) disk. The bulk 2-region $d$ is the disjoint union of two (open) disks. b) shows an example with two bulk 2-regions and one bulk 1-region. Only the boundary region corresponding to the purple bulk 2-region is non-empty. c) and d) show boundary higher order 2-manifolds that are stellar cones of a higher order 1-manifold.

**Example 49.** Consider the following boundary higher order 3-manifolds (see Remark (3)):

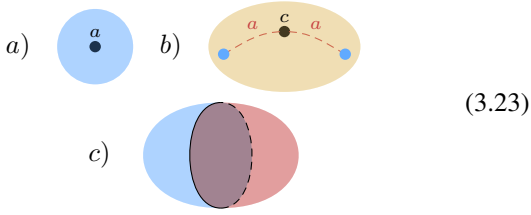

$$(3.23)$$

a) consists of a bulk 0-region directly embedded into a ball-like bulk 3-region which has a sphere-like boundary 2-region as boundary. I.e., the 3-region is an open ball with a puncture b) a bulk 3-region (a 3-ball) with the according boundary 2-region (a 2-sphere), one bulk 0-region (a point in the middle of the ball), one boundary 0-region (two points at the poles of the sphere) and the according bulk 1-region (two intervals connecting the poles with the point in the middle). c) has two (blue and purple) bulk 3-regions separated by a (black) bulk 2-region. The corresponding boundary regions are two disk boundary 2-regions separated by a circle boundary 1-region.

**Observation 28.** For every central $n/o$-manifold type, the $(n - o + 1)$-ball completion of the central upper link is a $n/o$-manifold of that type. It can be somehow viewed as the canonical representative of this type. Also for every $d$-region of a $n/o$-manifold type, the $(d + 1)$-ball completion of the upper link of this $d$-region is a $n/o$-manifold of that type.

**Background gluings**

**Definition 108.** Consider a $n/o$-manifold type, one of its $d$-regions and an $0 \leq x \leq d$. Consider the following two boundary higher order manifolds:

- The $(d - x)$-ball completion $Z$ of the upper link of the $d$-region. $Z$ is a boundary higher order $(n - x)$-manifold.

- The $(d - x)$-ball completion $Y$ of the stellar cone of the upper link of the $d$-region. $Y$ is a boundary higher order $(n - x - 1)$-manifold.

There is a canonical identification of the boundaries of the boundary higher order manifolds $S_x \times Y$ and $B_{x+1} \times Z$.

A $x$**-surgery gluing** for the selected $d$-region and $x$ is the following operation that acts on a $n/o$-manifold of this type with a $x$-sphere embedded into the selected $d$-region: The tubular neighbourhood of the embedded $x$-sphere can be identified with $S_x \times Y$. Cut out this $S_x \times Y$ and replace it with $B_{x+1} \times Z$ according to the canonical identification of their boundaries. Note that different identifications of the tubular neighbourhood with $S_x \times Y$ can yield different surgery gluings. Two surgery gluings are considered equal if their higher order $n$-manifolds are equal, the embedded $x$-spheres are homotopically equivalent and the identifications of their tubular neighbourhood with $S_x \times Y$ are equivalent.

For a fixed class of $n/o$-manifolds, a fixed $d$-region and a fixed $0 \leq x \leq d$, the $x$-surgery gluings are the continuum picture for a background gluing for a moved lattice class.

**Example 50.** Consider the following examples of surgery gluings for higher order manifolds:

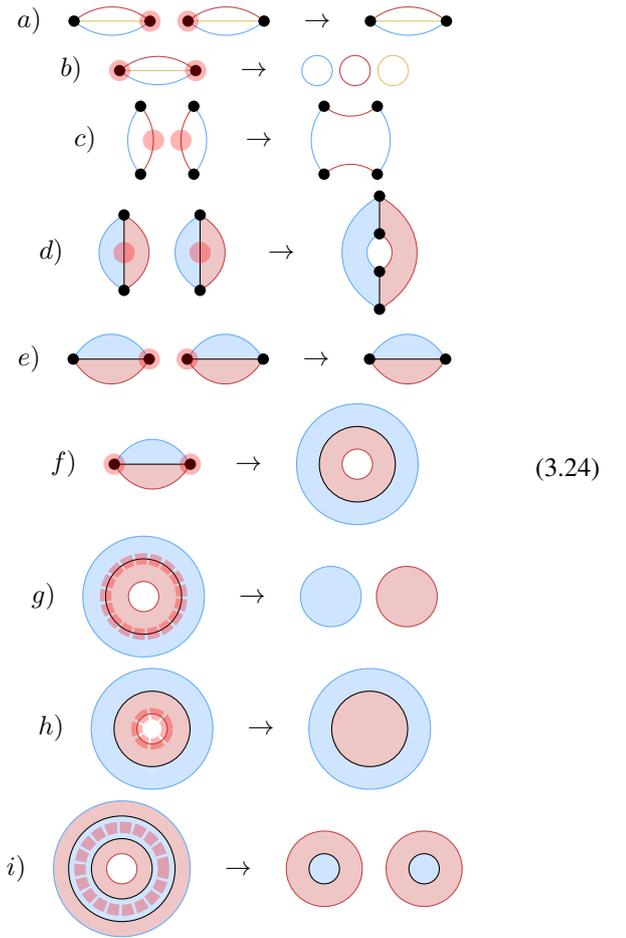

$$(3.24)$$

a) to f) are 0-surgery gluings that should be self-explanatory. g) shows a 1-surgery gluing of the 1-region that divides an annulus into two 2-regions, yielding one disk for each of the 2-regions. h) shows a 1-surgery gluing acting on the boundary 1-region of some $2/1$-manifold. i) shows a 1-surgery gluing acting on a 2-region.

**Remark 48.** $d$-surgery gluings of manifolds and interiour/boundary $d$-surgery gluings for boundary manifolds are a special case of $d$-surgery gluings of higher order manifolds.

**Proposition 2.** Every boundary $n/o$-manifold of a fixed type can be obtained from a disjoint union of copies of the $d+1$-ball completion of the upper link of different $d$-regions by surgery gluings for the boundary regions.

### 3.1.4 Singular manifolds

**Definition 109.** A **singular $n$-manifold type** is a higher order $n$-manifold type such that 1) there are no regions that have the same upper link and 2) the upper link of any $d$-region as an overall topological space must never be the $(n - d - 1)$-sphere. A **singular $n$-manifold** is a higher order manifold of such a type.

For each fixed $n$ and each singular $n$-manifold type, the singular $n$-manifolds of this type are the continuum picture for one moved lattice class.

**Example 51.** Consider the following examples for singular manifolds:

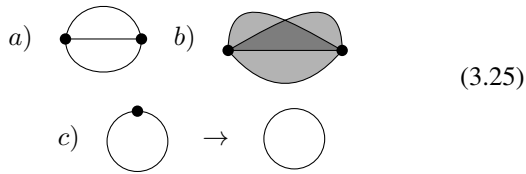

$$(3.25)$$

a) shows a singular 1-manifold with one 1-region consisting of two open intervals and a 0-region consisting of two points. The upper link of the 0-region consists of three points. b) shows a singular 2-region consisting of three disks that meet at an interval. There is one 1-region, namely the interior of the three disks. The central interval where the three disks meet is a 1-region, whose upper link consists of three points. The other three intervals form a 1-region with a single point as upper link. The two points form a 0-region with the stellar cone of three points as upper link. c) shows an example of a higher order manifold that is not a singular manifold: The upper link of the 0-region is a 0-sphere, so if we compose all regions together we get an ordinary manifold.

**Remark 49.** One can also define **singular higher order manifolds** where each region itself can be a singular manifold.

### 3.1.5 Framed higher order manifolds

**Definition 110.** A **framing type** for a fixed $n/o$-manifold type is a set of pairs of its regions $(A, B)$ where the dimension of $A$ is smaller than the dimension of $B$. When a pair $(A, B)$ is part of the subset we will say that $A$ **has a framing** with respect to $B$.

Concretely, consider a region $A$ of a $n/o$-manifold of a fixed type, and the manifold $B_l$ representing a region $B$ in the upper link of $A$. We can equip each point of $A$ with its own copy of the region $B_l$ of the upper link, and identify it with the corresponding intersection of the boundary of a small perpendicular disk around that point. The space of all such copies of $B_l$ forms a fiber bundle of $B_l$ over $A$. A **framing** of $A$ with respect to $B$ is a section of this fiber bundle, i.e., a map that associates to every point of $A$ a point in the region $B_l$ of the upper link at that point, such that this map continuously varies as we vary the point in $A$.

A **framed $n/o$-manifold** for a given framing type for a given $n/o$-manifold type is a $n/o$-manifold of this type, together with one framing for each pair of regions in the framing type. Two framed $n/o$-manifolds are considered equal if the corresponding $n/o$-manifolds are equal and their framings are equivalent up to homotopy.

The framed $n/o$-manifolds for a fixed type and framing type are the continuum picture for the backgrounds of a moved lattice class.

**Observation 29.** A framing of $A$ with respect to $B$ is only sensible if $B_l$ is non-empty in the upper link of $A$, otherwise no framed $n/o$-manifold of a type with such a type of framing exists, unless the region $A$ is empty.

**Observation 30.** A framing of $A$ with respect to $B$ can be trivial, in the sense that for every $n/o$-manifold of the corresponding type there is exactly one possible framing. In this case removing the framing of $A$ with respect to $B$ from the framing type does not change anything. E.g., this is the case when $B_l$ has the topology of an (open) $n$-ball.

**Example 52.** Consider a type of $3/2$-manifolds with the central upper link given by a circle, i.e., . 3-manifolds with embedded lines. Now introduce a framing of the 1-region with respect to the 3-region. Such a framing can be pictured by replacing the 1-region by a thin stripe (or "ribbon") whose two margins are coloured differently. So for the $3/2$-manifold consisting of a 3-ball with an embedded loop, there is one framing for every non-negative integer, corresponding to how many full twists the ribbon along the loop has.

**Example 53.** Consider a $3/1$-manifold type with the central upper link given by a single 0-region with two points:

$$(3.26)$$

i.e., 3-manifolds with embedded 2-manifolds. Now introduce a framing of the 2-region with respect to the 3-region. Such a framing can be pictured by consistently adding to each point of the 2-region an arrow that points perpendicular to the 2-region, i.e., an orientation of the embedded 2-manifold within the 3-manifold. For every connected component there are two such choices of orientation, corresponding to different framed $3/1$-manifolds.

**Example 54.** If a framing is non-trivial, we will usually find that there are more than one different framings for a fixed non-trivial $n/o$-manifold. But there can also be no framing at all: Consider for example $2/1$-manifolds with the central upper link given in Eq. (3.26). I.e., 2-manifolds with special embedded lines, and consider a framing of the central 1-region with respect to the 2-region, i.e., an orientation of the embedded 1-manifold within the 2-manifold. Now consider the $2/1$-manifold that is a 2-sphere with $n$ embedded circles, then for every circle we can choose the orientation independently, so we get $2^n$ different framings for this $2/1$-manifold. On the other hand consider a Klein bottle with an embedded loop winding around the orientation-reversing cycle. When going once around the loop the orientation changes, so there is no consistent choice of arrow directions for the loop. So there is no framing for this $2/1$-manifold.

## 3.2 Combinatorial constructions

In this section we will consider concrete combinatorial structures corresponding to different lattice types. For each construction we will first give some geometric intuition, then define the combinatorial structure, then see whether and how it fits into our universal formalization, and finally give different moves and gluings for it.

### 3.2.1 Simplicial complexes

Simplicial complexes are a quick and simple way to formulate a moved lattice type whose backgrounds are described by $n$-manifolds.

#### Geometric intuition

**Informal definition 19.** A simplicial complex can be imagined as a decomposition of a $n$-manifold into $n$-simplices. All edges of the simplices carry orientations (known as *branching structure*) such that there is no 2-simplex (triangle) where the three edges are oriented in a cyclic manner (all pointing clockwise or counter-clockwise).

**Example 55.** Consider the following examples of 1-dimensional simplicial complexes:

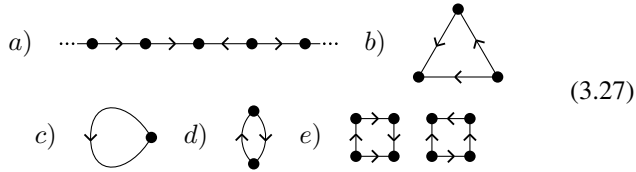

$$(3.27)$$

a) shows a patch of a simplicial complex. b), c) and d) show complexes that are decompositions of a circle. As shown in c) and d), simplicial complexes need not be drawable with straight lines when embedded in the plain. e) is a simplicial complex on the disjoint union of two circles.

**Example 56.** Consider the following 2-dimensional simplicial complexes (see Remark (4)):

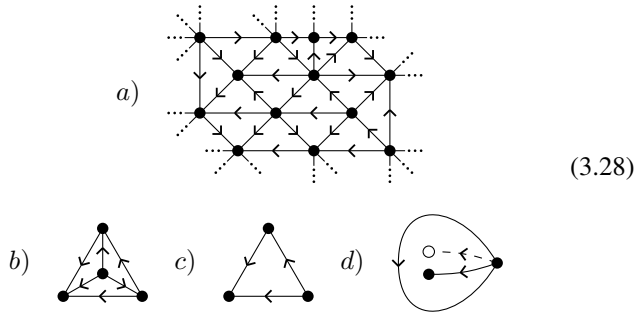

$$(3.28)$$

a) shows a patch of a complex. b) shows the boundary of a 3-simplex as 2-dimensional simplicial complex consisting of 4 2-simplices. c) shows a complex consisting of only two 2-simplices (both the back and front layer consist of one 2-simplex only). d) also shows a complex with two 2-simplices just that now there are two 1-simplices that are connected to the same 2-simplex twice.

#### Combinatorial definition

**Definition 111.** A **$n$-dimensional simplicial complex** is a set whose elements we call $n$-simplices together with a set of tuples, each consisting of two pairs of one $n$-simplex and one number between 0 and $n$. Those tuples are called $(n-1)$-simplices. There is the combinatorial constraint that every pair of $n$-simplex and number occurs exactly once in a tuple.

Obviously the $n$-simplices ($(n-1)$-simplices) in the combinatorial definition correspond to the $n$-simplices ($(n-1)$-simplices) in the intuitive description. The two $n$-simplices in a tuple indicate part of which the $n$-simplices the corresponding $(n-1)$-simplex is. The branching structure allows us to uniquely label the $(n-1)$-simplices that form the boundary of a $n$-simplex by numbers from 0 to $n$. This way the numbers in the tuple specify at which position the $(n-1)$-simplex occurs within the corresponding $n$-simplex. The combinatorial constraint models the face that every $n$-simplex is connected to $n+1$ $(n-1)$-simplices at all different positions.

**Remark 50.** $x$-simplices of dimension $x < n-1$ are not explicitly part of our combinatorial definition of a simplicial complex. Though we can define them implicitly: A $x$-**subsimplex** of an $n$-simplex can be represented as that $n$-simplex together with an ordered set of $x+1$ different numbers from 0 to $n$ specifying the vertices of the $n$-simplex forming that subsimplex. Now when two $n$-simplices are connected via an $(n-1)$-simplex we can determine which of their subsimplexes should correspond to the same simplex in the overall simplicial complex. This defines an equivalence relation on the $x$-subsimplices, and the equivalence classes can be defined as the $x$-**simplices**.

**Definition 112.** For a fixed $x$-simplex in a $n$-dimensional simplicial complex, consider all $(n-x-1)$-simplices such that the vertices of the $(n-x-1)$-simplex together with the vertices of the $x$-simplex are all vertices of one common $n$-simplex in the lattice. Those form an $(n-x-1)$-dimensional simplicial complex: The connections between the $(n-x-1)$-simplices are simply inherited by the connections between the $n$-simplices spanned by them and the $x$-simplex. We call this $(n-x-1)$-dimensional simplicial complex the **link** the $x$-simplex.

**Remark 51.** The link of a $n$-simplex in a $n$-dimensional simplicial complex is the empty set (the only $-1$-dimensional simplicial complex). The link of an $(n-1)$-simplex is by definition a pair of vertices. The link of a $x$-simplex with $x < n-1$ is always a connected simplicial complex (one where every $n$-simplex can be reached from every other via connections): The $(n-x-1)$-simplices in the link of the $x$-simplex correspond to the $n$-simplices of which the $x$-simplex is a subsimplex. The $x$-simplex is the equivalence class formed by those subsimplices, and two subsimplices are in the same equivalence class exactly when the corresponding $(n-x-1)$-simplices are connected in the link. Apart from those constraints it is easy to construct simplicial complexes with $x$-simplices with arbitrary link.

#### Formalization

**Definition 113.** $n$-dimensional simplicial complexes can be formulated as a restricted lattice pre-type in our universal formulation: Take two cell types, corresponding to the $n$-simplices and the $(n-1)$-simplices of the complex. Take $n+1$ connection types between the $n$-simplices and the $(n-1)$-simplices that indicate which of the $(n-1)$-simplices are part of which $n$-simplex in which way. There are two local restrictions: 1) Each $n$-simplex has to be connected to exactly one $(n-1)$-simplex via each of the $n+1$ connections, corresponding to the fact that each $n$-simplex has $n+1$ boundary $(n-1)$-simplices labelled from 0 to $n$. 2) Each $(n-1)$-simplex is connected to exactly two $n$-simplices via all the $n+1$ connections together, corresponding to the fact that each $(n-1)$-simplex is in between two $n$-simplices.

This restricted pre-type is already a lattice type as the number of connections connected to each cell is restricted, see Remark (22).

**Remark 52.** The link of a $x$-simplex (apart from the trivial cases $x = n$ and $x = n-1$) is itself not a feature type in the restricted lattice pre-type given by simplicial complexes. This is because for its determination the number of connections we have to take from a starting subsimplex can be arbitrarily high. So we cannot directly add restrictions that control the link. However, we can implicitly restrict the link to a finite number of possibilities by imposing how local patches of the link can look like.

For each finite set of possible links we get another lattice type. We commonly refer to lattices of those types as $n$**SC-lattices**.

### Moves

**Definition 114.** Consider the $(n + 1)$-simplex. Each (non-empty and proper) subset of its $n$-simplices forms a patch with the topology of a $n$-ball in a $n$-dimensional simplicial complex. A $x$-**Pachner move** is a move on $n$SC-lattices that replaces the patch formed by $x$ of the $n$-simplices of the $(n + 1)$-simplex by the patch formed by the remaining $n + 2 - x$ $n$-simplices. The $x$-Pachner moves are inverses to the $(n + 2 - x)$-Pachner moves. Due to the branching structure, every subset yields a different $x$-Pachner move.

**Example 57.** Consider the following Pachner moves:

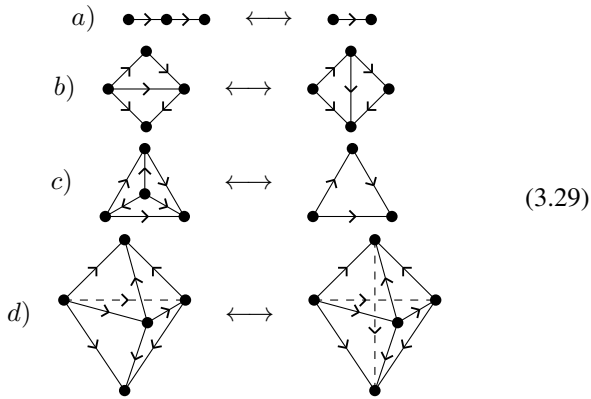

$$(3.29)$$

a) shows a 1-Pachner move for 1SC-lattices, b) and c) show 2-Pachner moves and 3-Pachner moves for 2SC-lattices, and d) shows a 2-Pachner move for 3SC-lattices. The left picture shows two tetrahedra separated by one horizontal face, whereas the right picture consists of three tetrahedra, meeting at one vertical edge in the middle, and separated by three vertical faces.

Let us consider the moved lattice type consisting of $n$SC-lattices with all Pachner moves.

**Definition 115.** The $n + 1$-simplex defines a $n$SC-lattice with $n + 2$ $n$-simplices and $(n + 2)(n + 1)/2$ $n - 1$-simplices. We will call the background of this $n$SC-lattice the **sphere**.

**Definition 116.** A $x$-simplex in a $n$-dimensional simplicial complex is called **non-singular** if its link has a sphere background. Otherwise we call it **singular**. A $n$SC-lattice type is called **non-singular** if the allowed links for $x$-simplices have $(n - x - 1)$-sphere background, for all $X$. Unless stated otherwise, $n$SC-lattice will refer to a lattice of such a type.

**Remark 53.** 1-dimensional simplicial complexes are automatically non-singular, as the link of each 0-simplex is fixed to the 0-sphere background by construction. 2-dimensional simplicial complexes are automatically non-singular as the link has to be connected (Remark (51)) and connected 1-dimensional simplicial complexes always have a circle background.

**Observation 31.** A Pachner move acts on the links of the $x$-simplices nearby by Pachner moves as well. Also a Pachner move inserts or removes simplices in the lattice, those have always a simplex as link, i.e., they are non-singular. So singular simplices can never be removed by Pachner moves, and moreover the background of their link can never change.

**Remark 54.** We can build a continuum space from a $n$SC-complex by replacing every $n$-simplex with the corresponding boundary $n$-manifold, and gluing all those manifolds according to the combinatorics. This space looks like a $n$-manifold within every $n$-simplex, but at the $x$-simplices there can be $x$-singularities. Indeed each singular $x$-simplex leads to a $x$-singularity whose upper link is exactly the background of the link of the simplex. So the continuum picture for the backgrounds of a non-singular $n$SC-lattice type is given by $n$-manifolds, whereas for a singular $n$SC-lattice type it is given by manifolds with singularities with the corresponding upper links.

### Gluings

**Definition 117. Simplex gluing** is the following operation that can be applied to any pair of $n$-simplices in a $n$SC-lattice that do not share a common $(n - 1)$-simplex: Remove both $n$-simplices from the set of simplices. Each of the $n$-simplices is connected to $n + 1$ $(n - 1)$-simplices numbered from 0 to $n$. Now we replace every pair of $(n - 1)$-simplices that have the same position within the two glued $n$-simplices by a single $(n - 1)$-simplex that connects the two $n$-simplices that were connected to the glued $n$-simplices via those $(n - 1)$-simplices before.

**Observation 32.** On the level of the $n$-manifolds representing the backgrounds, simplex gluing corresponds to a 0-surgery gluing.

**Example 58.** Consider the following simplex gluings of $n$SC-lattices (see Remark (7, 4)):

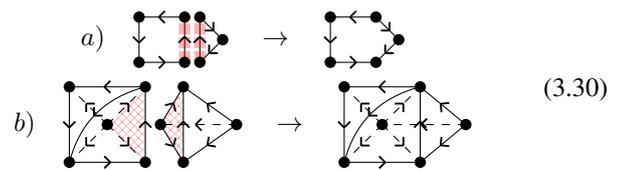

$$(3.30)$$

a) shows a gluing of two edges of a 1SC-lattice with background consisting of two circles yielding a 1SC-lattice with background consisting of a single circle. b) shows a gluing of two triangles of a 2SC-lattice with background consisting of two 2-spheres changing its background to a single 2-sphere.

**Definition 118.** More generally, **level $d$ simplex gluing** is an operation that can be applied to any pair of $n$-simplices in a $n$SC-lattice that share $d$ common $(n - 1)$-simplices. The only difference to the ordinary simplex gluing is that the common $(n - 1)$-simplices are removed together with the glued $n$-simplices.

**Example 59.** Consider the following level $d$ simplex gluings:

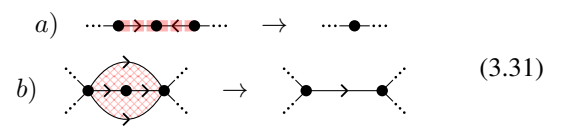

$$(3.31)$$

a) shows a level 1 simplex gluing applied to a 1SC-lattice. b) shows a level 2 simplex gluing applied to a 2SC-lattice.

### 3.2.2 Cell complexes

Cell complexes are another (moved) lattice type whose continuum picture for the backgrounds is described by manifolds. Opposed to simplicial complexes they are more flexible and natural at the expense of being a bit harder to define. Singularities can be dealt with in a more direct way, and also non-connected upper links are possible. Many lattice types in this work will build up on cell complexes, just that there will be additional restrictions or decorations.

**Geometric intuition**

**Informal definition 20.** A non-singular $n$-dimensional cell complex can be pictured as a decomposition of a $n$-manifold into cells, which are polytopes of dimension up to $n$, i.e., vertices, edges, faces, volumes, and so on.

**Example 60.** Non-singular 0-dimensional cell complexes consist of an arbitrary number of vertices. E.g.

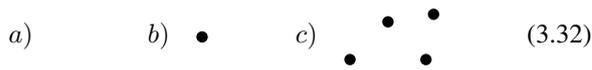

$$(3.32)$$

This lattice with an empty set of cells shown in a) is a valid cell complex in every dimension.

**Example 61.** Non-singular 1-dimensional cell complexes look like a disjoint union of arbitrary polygons. E.g.

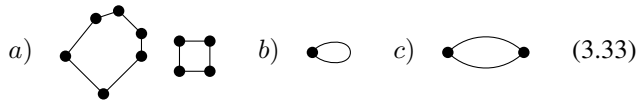

$$(3.33)$$

The polygons with one and two edges/vertices in b) and c) are also allowed even if they cannot be drawn on a piece of paper with straight edges.

**Example 62.** Non-singular 2-dimensional cell complexes look like planar graphs embedded in some closed surface. E.g., (see Remark (4)):

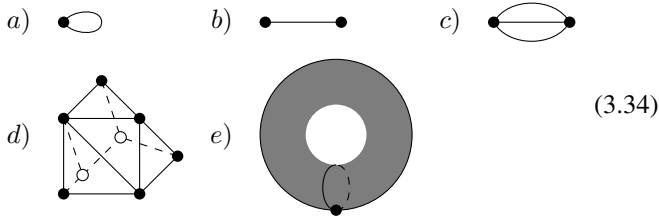

$$(3.34)$$

a) is a complex with two 1-gon faces, one edge and one vertex, and c) consists of three 2-gon faces with three edges and two vertices. Drawing b) is a bit special as there is only one face, which is shared by the front and back layer. The face has the shape of a 2-gon and is "wrapped" around a sphere such that the two edges are identified. d) is is a rather generic cell complex on a sphere. e) is a complex on a torus that consists of one 4-gon face. It is obtained by gluing opposite edges together. When drawing it in the plain, the front and back layer have the shape of an annulus.

**Example 63.** Non-singular 3-dimensional cell complexes describe decompositions of 3-manifolds into volumes, faces, edges and vertices. Consider the following example (see Remark (5)):

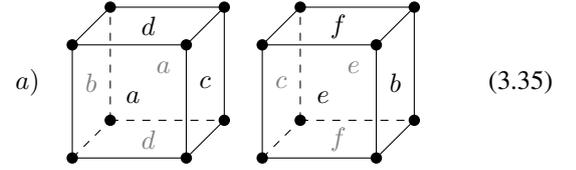

$$(3.35)$$

a) shows a cell complex that is a decomposition of a 3-torus into two cubes, separated by 6 4-gon faces. Note that for the above drawing to uniquely specify the cell complex one should indicate in which way the identically labelled faces are identified. Here we assumed an identification "with the opposite side".

**Remark 55.** Non-singular $n$-dimensional cell complexes and higher order $n$-manifolds might seem like very related concepts. Indeed, every non-singular cell complex yields a higher order manifold by replacing each $d$-cell by an open $d$-ball representing its own $d$-region. However, the two notions are conceptually very different: Whereas cell complexes are a rigorous combinatorial structure defining a lattice type, higher order $n$-manifolds are only an intuitive continuum picture for the backgrounds of a moved lattice class. Also the mapping from cell complexes to higher order manifolds does not work in the opposite direction: The regions of a higher order manifold can themselves have arbitrarily complicated topology, whereas the cells of a cell complex are always represented by a $d$-ball. Also the $d$-regions of a higher $n$-manifold alltogether do not have to form a $n$-manifold, whereas the cells of a non-singular cell complex by definition represent a decomposition of a $n$-manifold.

**Combinatorial definition**

**Definition 119.** A $n$-**dimensional cell complex** consists of sets $C_i$ for $0 \leq i \leq n$ and $R_{ij}$ for $0 \leq i < j \leq n$. We call the elements of the sets $C_i$ $i$-**cells**, and we also call 0-cells **vertices**, 1-cells **edges**, 2-cells **faces** and 3-cells **volumes**. We will call the elements of the sets $R_{ij}$ $i,j$-**relation-cells**.

On those sets we have the following data structures:

1. Two maps $S$ and $T$ from the relation-cells to the cells

$$S_{ij} : R_{ij} \to C_i \qquad T_{ij} : R_{ij} \to C_j. \quad (3.36)$$

where $S$ stands for **source** and $T$ for **target**.

2. A map $\circ$ called **junction** taking two relations into one relation

$$\circ : R_{ij} \times R_{jk} \to R_{ik} \quad (3.37)$$

such that $a \circ b$ is only defined when $T(a) = S(b)$ and such that $S(a \circ b) = S(a)$ and $T(a \circ b) = T(b)$.

The junction has to obey the following associativity constraint:

$$a \circ (b \circ c) = (a \circ b) \circ c \quad (3.38)$$

**Remark 56.** Let us see how the intuitive notion of cell complexes above is described by this combinatorial structure: The $n$-cells are just the $n$-dimensional polytopes of the decomposition. Each of the relation-cells tells that the source cell is contained in the boundary of the target cell. A source and target cell can also be connected multiple times, as in the example of an edge ending two times at the same vertex. The junctions describe the following observation: If one cell is contained in the boundary of a second cell and the second cell is in turn contained in the boundary of a third cell, then the first cell is also contained in the boundary of the third cell. Thus, if there are relations between cell $a$ and $b$ and between cells $b$ and $c$, then there is a corresponding relation between cells $a$ and $c$.

Here is an example of an intuitive 2-dimensional cell complex that is a decomposition of a sphere where we labelled all cells and relation-cells and give the sources, targets and junctions:

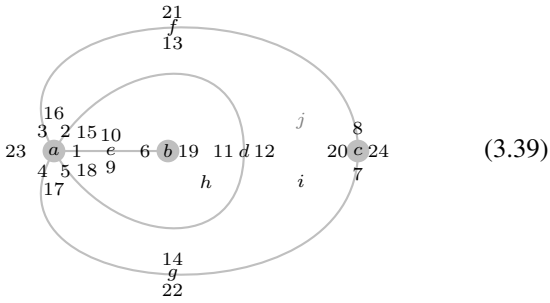

(3.39)

The cells of different dimensions are

$$
\begin{array}{c|c}
\text{dimension} & \text{cells} \\
\hline
0 & a, b, c \\
1 & d, e, f, g \\
2 & h, i, j
\end{array}
\tag{3.40}
$$

The relation-cells, their sources and targets are

| | | |
|---|---|---|
| $1: a \to e$ | $2: a \to d$ | $3: a \to f$ |
| $4: a \to g$ | $5: a \to d$ | $6: b \to e$ |
| $7: c \to g$ | $8: c \to f$ | $9: e \to h$ |
| $10: e \to h$ | $11: d \to h$ | $12: d \to i$ |
| $13: f \to i$ | $14: g \to i$ | $15: a \to h$ |
| $16: a \to i$ | $17: a \to i$ | $18: a \to h$ |
| $19: b \to h$ | $20: c \to i$ | $21: f \to j$ |
| $22: g \to j$ | $23: a \to j$ | $24: c \to j$ |

(3.41)

All possible junctions are given by

| | | |
|---|---|---|
| $1 \circ 9 = 18$ | $1 \circ 10 = 15$ | $2 \circ 11 = 15$ |
| $2 \circ 12 = 16$ | $3 \circ 13 = 16$ | $4 \circ 14 = 17$ |
| $5 \circ 12 = 17$ | $5 \circ 11 = 18$ | $6 \circ 10 = 19$ |
| $6 \circ 9 = 19$ | $7 \circ 14 = 20$ | $8 \circ 13 = 20$ |
| $8 \circ 21 = 24$ | $7 \circ 22 = 24$ | $3 \circ 21 = 23$ |
| $4 \circ 22 = 23$ | | |

(3.42)

**Remark 57.** For $n \leq 0$ there are no relation-cells at all. For $n \leq 1$ there are no junctions as all relation cells go from vertices to edges. For $n \leq 2$ there is no associativity constraint as there is no possibility how three relation-cells could be joined together.

Cell complexes as combinatorially defined above are more general than the non-singular cell complexes defined informally before. This is because there can be arbitrary singularities. Consider the following examples:

**Example 64.** 1-dimensional cell complexes are similar to *hypergraphs*, i.e., graphs where each edge is not nessecarily connected to two but to an arbitrary number of vertices (and slightly more general as an edge can also be connected multiple times to the same vertex). E.g.

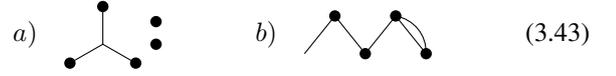

(3.43)

a) shows an edge connected to three vertices, and two additional vertices connected to no edge. b) shows a cell complex containing a vertex that is connected to three edges, and an edge connected to only one vertex.

**Example 65.** For 2-dimensional cell complexes the associativity constraint is still trivial, so they are something like "hypergraphs generalized to three kinds of objects".

**Definition 120.** When we flip all dimensions of the cells of a cell complex $x \to n - x$, exchange source and target for all relation-cells, and change the order of the junction, we get another cell complex. This operation is known as *Poincaré duality*, and we will call the cell complex after application of this operation the **dual lattice**.

**Example 66.** In two dimensions, Poincaré duality exchanges vertices and faces, e.g., consider the following patch of a 2-dimensional cell complex (in black) and its dual (in red):

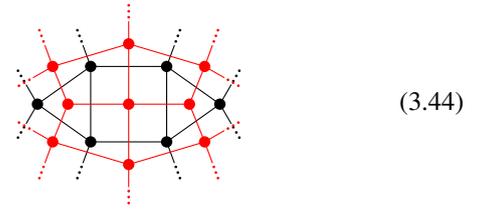

(3.44)

**Informal definition 21.** The upper link of a $x$-cell in a $n$-dimensional cell complex can be obtained by 1) put a small $(n-x)$-ball around a point in the cell perpendicular to the cell. 2) consider the intersection of the $(n-x-1)$-sphere that is the boundary of this small ball with the sorrounding cell complex. This yields an $(n-x-1)$-dimensional cell complex that is the upper link. The lower link of a $x$-cell is the $(x-1)$-dimensional cell complex that forms the boundary of this cell.

**Definition 121.** The **upper link** of a $x$-cell $X$ in a $n$-dimensional cell complex $N$ is an $(n - x - 1)$-dimensional cell complex $N_X$ defined in the following way:

1. The $y$-cells of $N_X$ are given by all $x, x+y+1$-relation-cells of $N$ whose source is $X$.

2. Consider a $y$-cell $Y$ and a $z$-cell $Z$ of $N_X$ with $y < z$, such that $y$ corresponds to the $x, (x + y + 1)$-relation-cell $Y_N$ of $N$ and $z$ corresponds to the $x, (x + z + 1)$-relation-cell $Z_N$ of $N$. The set of $y, z$-relation-cells of $N_X$ with source $Y$ and target $Z$ is given by the set of $(x + y + 1, x + z + 1)$-relation-cells $r$ of $N$ such that $Y_N \circ r = Z_N$. Note that the total set of all $x, y$-relation-cells of $N_X$ is then given by the disjoint union of the sets for all possible sources and targets $Y$ and $Z$, which might contain some relations-cells of $N$ multiple times.

3. Consider a $y$-cell $Y$, $z$-cell $Z$ and $w$-cell $W$ of $N_X$ with $y < z < w$, corresponding to a $x, (x+y+1)$-relation-cell $Y_N$, a

$x, (x+z+1)$-relation-cell $Z_N$ and a $x, (x+w+1)$-relation-cell $W_N$ of $N$. Further consider a $y, z$-relation-cell $R$ with source $Y$ and target $Z$ and a $z, w$-relation cell $S$ with source $Z$ and target $W$ of $N_X$. Those correspond to $(x+y+1), (x+z+1)$- and $(x+z+1), (x+w+1)$-relation-cells $R_N$ and $S_N$ of $N$ with the property that $Y_N \circ R_N = Z_N$ and $Z_N \circ S_N = W_N$. By associativity in $N$ we get $Y_N \circ (R_N \circ S_N) = W_N$. So $R_N \circ S_N$ is a $(x+y+1), (x+w+1)$-relation-cell of $N$, which corresponds to a $y, w$-relation-cell of $N_X$ with source $Y$ and target $W$. We define this relation-cell as the outcome of $R \circ S$.

Let us try to illustrate the situation in a picture:

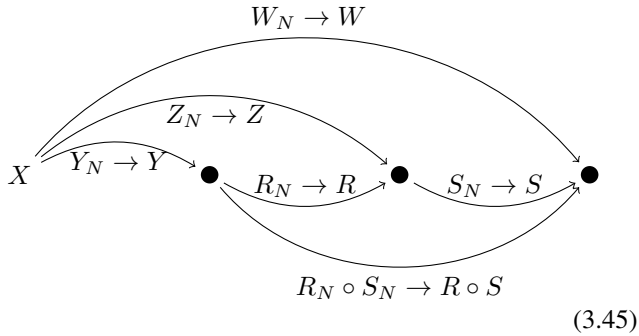

$$(3.45)$$

Analogously define the **lower link** of a $x$-cell $X$ as the $(x-1)$-dimensional cell complex whose cells are formed by the relation-cells from cells of lower dimension to $X$. It is the dual notion to the upper link.

**Comment 10.** What the above definition of a upper (lower) link roughly does is to restrict the lattice to only cells and relations that are connected to the cell $X$ that have higher (lower) dimension. Just with the subtlety that e.g., if a cell is connected twice to $X$ the upper (lower) link will contain two copies of that cell.

**Example 67.** Consider the following examples of cell complexes (black) and the and upper links of specific cells (marked red).

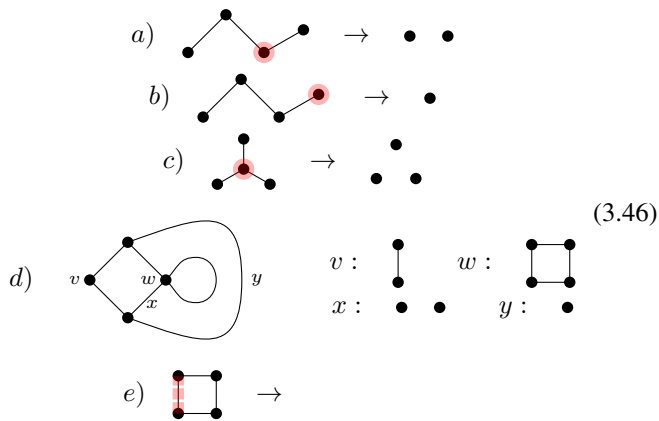

$$(3.46)$$

a), b) and c) show the upper links of vertices in 1-dimensional cell complexes. d) shows a 2-dimensional cell complex that is a decomposition of a disk into a 1-gon and two 4-gons. The right side depicts the upper links of two vertices $v$ and $w$ as well as two edges $x$ and $y$. e) shows the upper link of an edge in a 1-dimensional cell complex which is always empty (i.e., the only $-1$-dimensional cell complex).

**Example 68.** Consider the following lower links of cells in cell complexes:

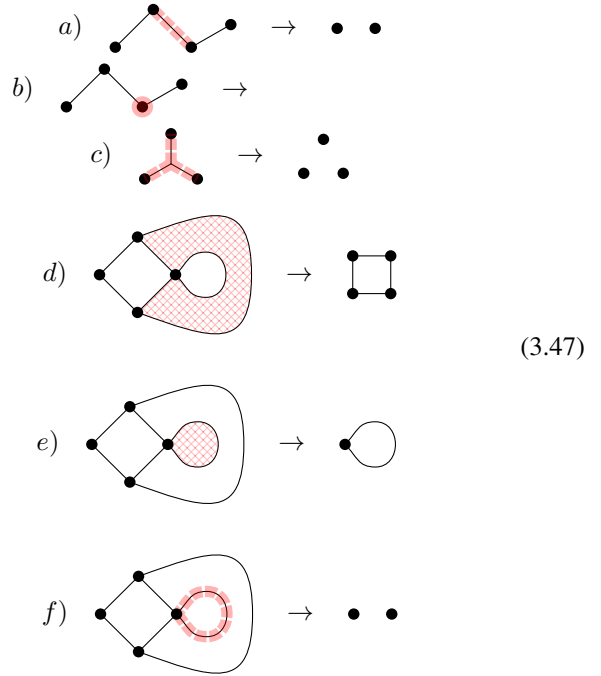

$$(3.47)$$

a) shows the lower link of an edge of a 1-dimensional cell complex consisting of two vertices. The lower link of vertices is always empty (i.e., the only $-1$-dimensional cell complex), as shown in b). c) shows an edge that is connected to three vertices with the according lower link. d) and e) show the lower link of faces in a 2-dimensional cell complex. As shown in d) a cell of the original complex can occur multiple times in the lower link when there are multiple relation-cells connecting the two. The same situation arises in f) where the concerned edge is connected twice to one vertex that thus appears twice in the lower link.

### Formalization

**Remark 58.** One can make $n$-dimensional cell complexes a restricted lattice pre-type in the following way: The cells of different type are formed by the $i$-cells and $i, j$-relation-cells as well as triples of relation-cells $a$, $b$ and $c$ such that $a \circ b = c$. We call the latter cell types the $i, j, k$-triple-cells. The source (target) maps become different types of connections between the $i, j$-relation-cells and the $i$-cells ($j$-cells). In order to match the original combinatorial definition we have to add the restriction that each relation-cell is connected exactly once to a source (target) cell. Additionally there are three other kinds of types of connections between the triple-cells and the relation-cells that are part of the triple. We have to add another local restriction that for every pair of $i, j$-relation-cell and $j, k$-relation-cell that are connected to the same $j$-cell as source or target there is exactly one triple-cell to which both are connected exactly once. The associativity constraint is another restriction.

However, a cell can be connected to arbitrarily many relation-cells, so restricted pre-type described above does not yet define a lattice type.

**Observation 33.** The upper and lower links of $x$-cells are feature types (for every $x$) of the restricted lattice pre-type defined by cell complexes.

**Definition 122.** Using the feature types given by the upper and lower link, we can add further restrictions to the restricted pretype to make cell complexes a lattice type. I.e., we can restrict the upper (lower) link of all cell types to a finite set of possible $(n-x-1)$-dimensional cell complexes $((x-1)$-dimensional cell complexes). To such lattice types we will commonly refer to $n$**CC-lattices**. Though for a specific type we will have to state what precisely the possible upper and lower links are.

**Moves**

**Definition 123.** The **cone product** of two cell complexes $X$ and $Y$ of dimension $x$ and $y$ is a cell complex $Z$ of dimension $x+y+1$ with the following property: For every $a$-cell of $X$ $(Y)$ there is an $a$-cell of $Z$. Those cells are connected among each other like the cells of $X$ $(Y)$. For every pair of $a$-cell of $X$ and $b$-cell of $Y$ there is a $a+b+1$-cell of $Z$. Every cell of $Z$ associated to such a pair is connected to the cells associated to the first and second cell in the pair. All cells associated to pairs with a fixed first (second) cell of $X$ $(Y)$ are connected among each other like the cells of $Y$ $(X)$.

**Example 69.** Consider the following pairs of cell complexes and their cone product:

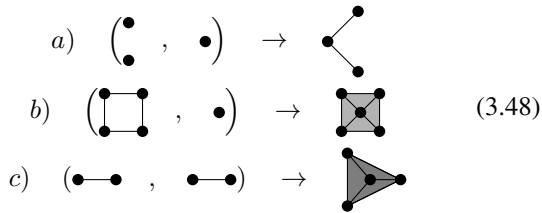

$$(3.48)$$

a) shows the cone product of a pair of vertices with a single vertex, yielding one vertex connected to two other vertices by two edges. b) shows the cone product of a 4-gon with a single vertex yielding a 2-dimensional cell complex consisting of four triangles. c) shows the cone product of two single edges with vertices as endpoints, yielding a tetrahedron with 1 volume, 4 faces, 6 edges and 4 vertices.

**Definition 124.** The **filling** of a $n$CC-lattice is the $(n+1)$CC-lattice obtained from adding a single $(n+1)$-cell that has the $n$CC-lattice as its lower link.

**Definition 125.** For every pair of $x$-dimensional cell complex $X$ and $y$-dimensional cell complex $Y$ with $x+y<n$, the **bi-stellar flip** with respect to $X$ and $Y$ is the following move acting on $n$CC-lattices: The move can be applied to places where the cone product of $X$ with the filling of $Y$ occurs as a patch of cells in the lattice. This patch of cells is then replaced by the cone product of $Y$ with the filling of $X$.

**Example 70.** Consider the following bi-stellar flips:

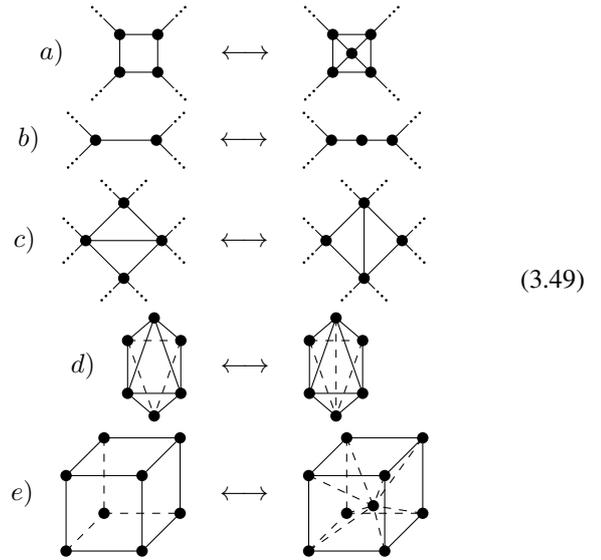

$$(3.49)$$

a) shows a bi-stellar flip in 2CC-lattices, where $X$ is the empty $-1$CC-lattice, and $Y$ is the 4-gon 1CC-lattice. b) shows a bi-stellar flip on 2CC-lattices where $X$ is the empty $-1$CC-lattice and $Y$ is the 0CC-lattice consisting of two vertices. For c) both $X$ and $Y$ are equal to the 0CC-lattice consisting of two vertices. d) shows a bi-stellar flip on 3CC-lattices that removes two pyramides attached at their base with 4 tetrahedra sharing one common central edge. $X$ is the 0CC-lattice consisting of two points and $Y$ is the 4-gon. e) shows a cube volume of a 3CC-lattice that is decomposed into 6 tetrahedra by a bi-stellar flip with $X$ the empty $-1$CC-lattice and $Y$ the 2CC-lattice formed by the boundary faces of the cube.

Let us consider the moved lattice type consisting of $n$CC-lattices with all bi-stellar flips that are consistent with the allowed lower and upper links of the precise type of $n$CC-lattice.

**Definition 126.** Define the **standard $n$-sphere** as the $n$-dimensional cell complex consisting of 2 $x$-cells for all $0 \le x \le n$ such that each two cells of different dimension have exactly one relation between them. So the junction of two relation-cells is already determined by what the sources and targets are. We will call the background of the standard $n$-sphere $n$**-sphere**.

**Example 71.** Standard spheres in 0, 1 and 2 dimensions look like:

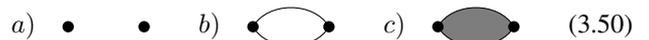

$$(3.50)$$

The standard 0-sphere a) consists of two vertices, the standard 1-sphere b) is a loop of two edges, and the standard 2-sphere c) consists of two 2-gon faces where one is in the front and one in the back layer.

**Definition 127.** A $x$-cell is **lower non-singular** if its lower link has sphere background, and **upper non-singular** if the upper link has sphere background. Otherwise it is called **lower singular** or **upper singular**. A specific $n$CC-lattice type is called **non-singular** if all the allowed lower and upper links have sphere background. Those are the types that we will usually have in mind.

**Observation 34.** The dual of a non-singular $n$CC-lattice is again a non-singular $n$CC-lattice.

**Proposition 3.** If a non-singular $n$CC-lattice type allows for enough possible upper and lower links such that there is a sufficiently large set of consistent bi-stellar flips, then the backgrounds really correspond to $n$-manifolds. So any two $n$CC-lattice types with this property are in the same class.

**Observation 35.** A non-singular $n$SC-lattice type and a non-singular $n$CC-lattice type are in the same class if they both allow sufficiently many possible (upper or lower) links. To demonstrate this we give two lattice mappings between them:

The first lattice mapping from the $n$SC-lattices to the $n$CC-lattices is given by simply removing the edge orientations.

The second lattice mapping from $n$CC-lattices to $n$SC-lattices is known as a **barycentric subdivision**: We replace each cell (of any dimension) of the cell complex by a vertex of the corresponding simplicial complex. In general we take one $x$-simplex for each set $R_0, \ldots, R_x$ of $d_i, d_{i+1}$-relation-cells $R_i$ with $0 \leq d_0 < \ldots < d_x \leq n$ such that $T_{d_i, d_{i+1}}(R_i) = S_{d_{i+1}, d_{i+2}}(R_{i+1})$ (i.e., a path of relation cells going through $x+1$ cells). In particular, $n$-simplices correspond to full paths with $d_i = i$.

Finally, consider the lattice mapping from $n$CC-lattices to $n$CC-lattices given by a barycentric sibdivision followed by removing the orientations. It equals a circuit move of bi-stellar flips. Also the lattice mapping from $n$SC-lattices to $n$SC-lattices given by removing the edge orientations and then performing a barycentric subdivision equals a circuit move of Pachner moves. Thus, the axiom Eq. (2.20) is fulfilled.

**Example 72.** Consider the following example of a barycentric subdivision on a patch of a 2CC-lattices:

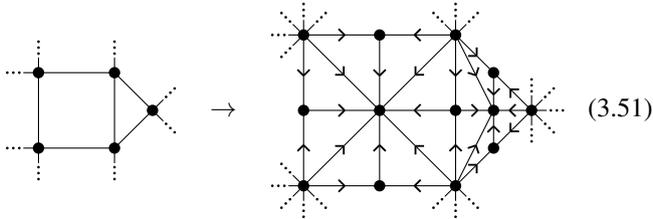

$$(3.51)$$

**Remark 59.** A barycentric subdivision can also be defined as a map from cell complexes to cell complexes by not taking the arrow orientations into account. Conjugating this variant of the barycentric subdivision with taking the Poincaré dual yields a mapping that we will refer to as **thickening** of a cell complex.

**Gluings**

**Definition 128.** Consider an $(n-1)$CC-lattice $L_l$. **Cell gluing** with respect to $L_l$ is the following operation that can be applied to every pair of $n$-cells of a $n$CC-lattice that do not have any cells of their lower link in common, with an identification of their lower links with $L_l$. First the two glued $n$-cells are removed. Now the identification of the two lower links with $L_l$ yields for every cell connected to the first glued $n$-cell a corresponding cell connected to the second glued $n$-cell. Each of those pairs of connected cells is replaced by a single cell that inherits all the relations from both those cells (ecxept for the relations to the glued $n$-cells that disappear).

**Example 73.** Consider the following examples of cell gluing for $n$CC-lattices, for dfferent $n$ (see Remark (7)):

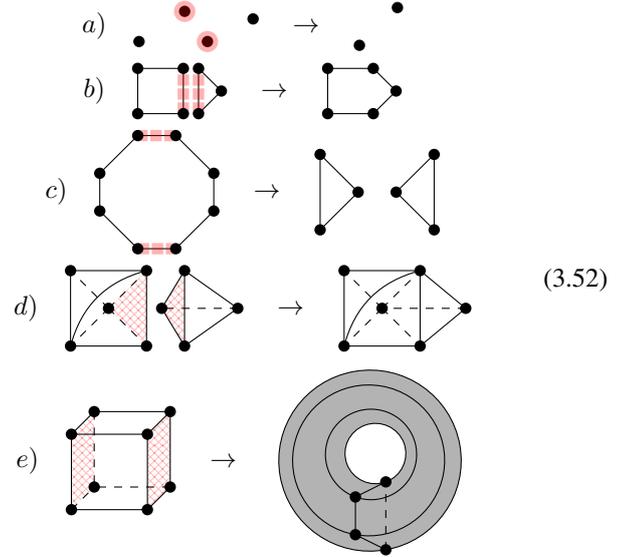

$$(3.52)$$

As a) shows, gluing two vertices of an 0CC-lattice just makes them disappear. b) and c) show gluings of two edges of a 1CC-lattice. The two glued edges are removed and their adjacent vertices are identified pairwise. d) and e) show 2CC-lattices glued at two faces. This works because both faces have the shape of a 4-gon. Note that in all examples the left side does not specify the identification, so the results on the right side could be different.

**Definition 129.** More generally, **level $d$ cell gluing** is an operation that can be applied to two $n$-cells of a $n$CC-lattice with equal lower link that are connected to $d$ common $(n-1)$-cells. The operation is the same as for ordinary cell gluing, just that the $(n-1)$-cells to which both the $n$-cells are connected also disappear.

**Example 74.** Consider the following examples for higher level cell gluing:

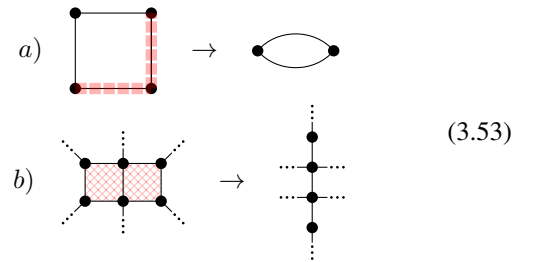

$$(3.53)$$

a) shows level 1 gluing for a 1CC-lattice. b) shows level 1 gluing for patch of a 2CC-lattice.

### 3.2.3 Simplicial complexes with boundary

**Geometric intuition**

**Informal definition 22.** A (non-singular) $n$-dimensional simplicial complex with boundary can be pictured as a decomposition of a $n$-manifold with boundary into $n$-simplices, such that the decomposition restricted to the boundary yields a decomposition of the corresponding $(n-1)$-manifold (without boundary) into $(n-1)$-simplices.

**Example 75.** Consider the following (non-singular) simplicial complexes with boundary (see Remark (2)):

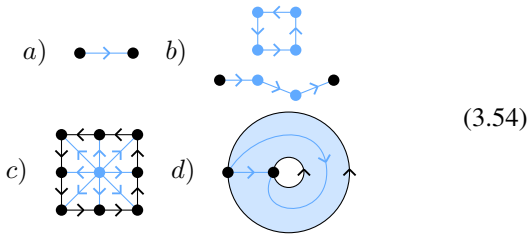

$$(3.54)$$

a) shows a 1-dimensional simplicial complex with boundary on an interval, b) shows one on the disjoint union of an interval and a circle. c) shows a 2-dimensional simplicial complex with boundary on a disk whereas d) is a decomposition of an annulus.

**Combinatorial definition**

**Definition 130.** To model simplicial complexes with boundary combinatorially, we can separate a set of special $(n-1)$-simplices that we will call **boundary $(n-1)$-simplices** from the rest in its own set. Those boundary $(n-1)$-simplices are also targets of the maps that associate to each $n$-simplex the $(n-1)$-simplices of their boundary. The combinatorial constraint that each $(n-1)$-simplex has to be connected to two $n$-simplices still holds in the interior. However, the boundary $(n-1)$-simplices must only be connected to one single $n$-simplex.

**Definition 131.** $x$-simplices of a simplicial complex with boundary can be defined analogously to the case without boundary. The **link** of a $x$-simplex can also be defined analogously. Note, however, that for $x$-simplices in the boundary, the link is not a simplicial complex, but a simplicial complex with boundary.

**Formalization**

**Remark 60.** Simplicial complexes with boundary can be interpreted as a restricted lattice pre-type just in the same way as simplicial complexes without boundary. The restriction for the boundary $(n-1)$-simplices has to be adapted accordingly. Again this restricted pre-type already forms a proper lattice type.

**Definition 132.** The link is again not a feature type, however, we can again implicitly restrict it to a finite set of possiblilities by imposing that there are no patches of the link of a certain size. We will refer to lattices of such a type with finitely many possible links as $n$**SCb-lattices**.

**Moves**

**Definition 133.** A **boundary attachment move** is the following operation that can be applied to $n$SC-lattices: Take a $n$-simplex that has 1 to $n$ of its $n+1$ $(n-1)$-simplices in the boundary and remove it, or reversely, attach an $n$-simplex to the lattice by gluing it to the complex at 1 to $n$ of its boundary $(n-1)$-simplices.

**Example 76.** Consider the following boundary attachment moves:

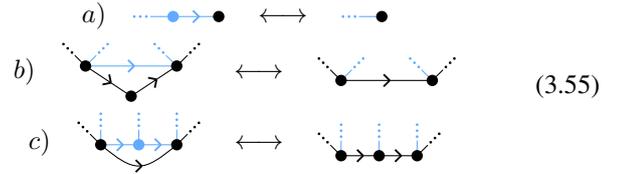

$$(3.55)$$

**Remark 61.** The Pachner moves defined for $n$SC-lattices can also be applied to $n$SCb-lattices. They act only in the interiour and leave the boundary unchanged.

Let us consider the moved lattice type consisting of $n$SCb-lattices together with Pachner moves in the interiour and boundary attachment moves.

**Observation 36.** $n$SC-lattices are a sub-type of $n$SCb-lattices.

**Gluings**

**Definition 134. Boundary simplex gluing** is the following operation that can be applied to every pair of boundary $(n-1)$-simplices of a $n$SCb-lattice that do not share a common $(n-2)$-simplex: Replace the two glued boundary $(n-1)$-simplices by a single bulk $(n-1)$-simplex that is connected to the bulk $n$-simplices to which the glued boundary $(n-1)$-simplices were connected to.

**Example 77.** Consider the following examples of boundary simplex gluing:

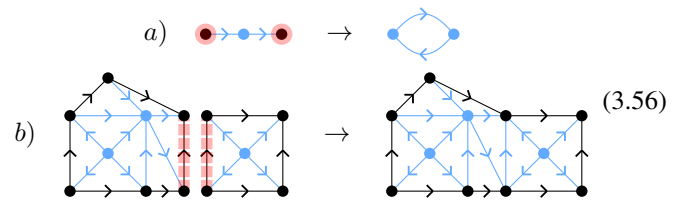

$$(3.56)$$

**Definition 135. Level $d$ boundary simplex gluing** is an operation that can be applied to every pair of boundary $(n-1)$-simplices of a $n$SCb-lattice that are connected to $d$ common boundary $(n-2)$-simplices. The only difference to ordinary boundary simplex gluing is that the common $(n-2)$-simplices become part of the bulk during the gluing.

**Example 78.** Consider the following examples of higher level boundary simplex gluing:

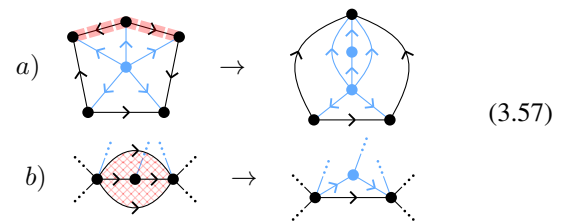

$$(3.57)$$

a) shows level 1 boundary simplex gluing of a 2SCb-lattice. b) shows level 2 boundary simplex gluing of a patch in a 3SCb-lattice. Note that the blue parts of the lattice correspond to the interiour.

### 3.2.4 Cell complexes with boundary

**Geometric intuition**

**Informal definition 23.** Similar to simplicial complexes with boundary, (non-singular) cell complexes with boundary are decompositions of a boundary $n$-manifold into polytopes of different dimension, such that the decomposition in the interior matches the decomposition of the boundary.

**Example 79.** Consider the following cell complexes with boundary:

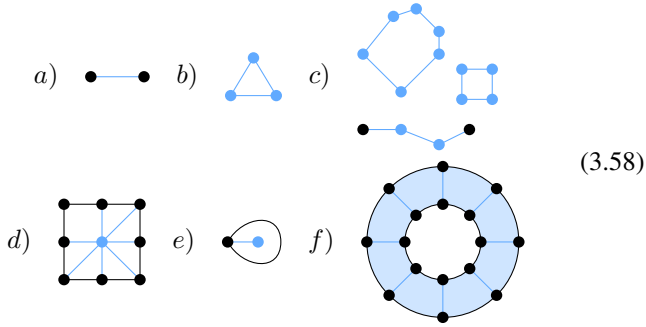

$$(3.58)$$

where we draw the boundary part in black and the bulk part in blue, see Remark (2). a), b) and c) are 1-dimensional cell complexes with boundary, where for b) the boundary is actually empty. d) and e) show 2-dimensional cell complexes with boundary on a disk, whereas f) is a decomposition of an annulus (where we also coloured the faces to make this clear).

**Combinatorial definition**

**Remark 62.** In fact cell complexes can already contain boundaries when we allow for singularities: Cell complexes with boundary are cell complexes whose cells are lower non-singular everywhere and upper non-singular everywhere exept for the set of cells that form the boundary. Those boundary $x$-cells have an upper link with the background of an $(n-x-1)$-ball instead of an $(n-x-1)$-sphere.

**Definition 136.** We will treat cell complexes with boundary as cell complexes with singularities, but model them more explicitly. To this end we simply add sets of **boundary $x$-cells** (for $x < n$) in addition to the **bulk $x$-cells** (for $x \le n$). There are also three kinds of relation-cells, namely those between the bulk cells, between the boundary cells, and between bulk and boundary cells. All the cells and relation-cells together form an $n$-dimensional cell complex (with singularities). The boundary cells alone form an $(n-1)$-dimensional cell complex.

**Definition 137.** The **upper link** of a cell in a cell complex with boundary is just the upper link in the overall (singular) cell complex, just that there are now different kinds of $x$-cells in the upper link because there are different kinds of relation-cells. The upper link of a bulk cell only consists of cells coming from one kind of relation-cell, hence it is an ordinary cell complex. The upper link of the boundary cells However, consists of two kinds of cells coming from two kinds of relation-cells, i.e., it is a cell complex with boundary itself.

The **lower link** of a $x$-cell in a cell complex with boundary is just the lower link in the overall cell complex. In contrast to the upper link, we do not distinguish between different kinds of cells for the lower link.

**Example 80.** Consider the following examples for upper links of cells in a cell complex with boundary:

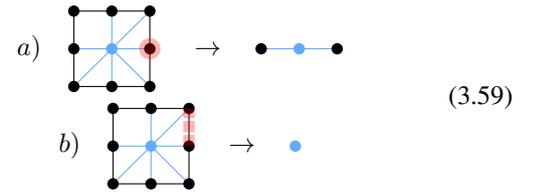

$$(3.59)$$

a) shows the lower link of a boundary vertex being a 1-dimensional cell complex with boundary. b) shows the lower link of a boundary edge being a single vertex.

**Example 81.** Consider the following examples for lower links of cells in a cell complex with boundary:

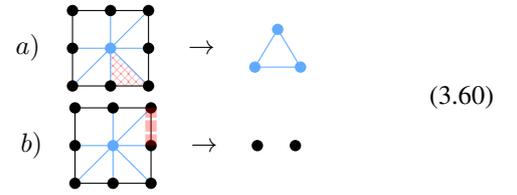

$$(3.60)$$

a) shows the upper link of a bulk face of a 2-dimensional cell complex with boundary being a triangle. Even though the face is connected to a boundary edge, the edges and vertices in the resulting triangles are all coloured the same. b) shows the upper link of a boundary edge consisting of two vertices.

**Formalization**

**Definition 138.** Cell complexes with boundary can be modelled as a restricted lattice pre-type, just as ordinary cell complexes can. The only difference is that for every dimension (except for $x = n$) we have two types of $x$-cells, namely the boundary and bulk cells, and three types of relation-cells.

**Definition 139.** The upper and lower link are still featuer types in cell complexes with boundary. We can restrict the upper and lower link for each type of cell to a finite set of possible cell complexes/cell complexes with boundary. This way we get a proper lattice type. We will commonly refer to lattices of such types as $n$**CCb-lattices**.

**Moves**

**Definition 140.** For every pair of $x$-dimensional cell complex $X$ and $y$-dimensional cell complex $Y$ with $x + y \le n - 3$, the **boundary bi-stellar flip** is the following move: It can be applied to every patch of cells in a $n$CCb-lattice that looks like the cone product of $X$, the filling of $Y$, and a single vertex. Thereby the cone product of $X$ with the filling of $Y$ alone has to be part of the boundary. The patch of cells is removed and replaced by the cone product of the filling of $X$ with $Y$ with the single vertex.

Applying a boundary bi-stellar flip modifies the boundary by an ordinary bi-stellar flip.

**Example 82.** Consider the following examples of boundary bi-

stellar flips:

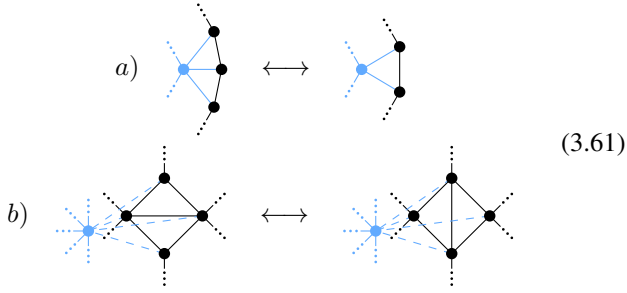

(3.61)

a) shows a boundary bi-stellar flip of a 2CCb-lattice exchanging two neighbouring triangles at the boundary bi a single one. b) shows a boundary bi-stellar flip of a 3CCb-lattice exchanging two neighbouring tetrahedra at the boundary by two other neighbouring tetrahedra.

**Remark 63.** Ordinary bi-stellar flips for $n$CC-lattices can also be applied to $n$CCb-lattices. Thereby the boundary remains unchanged.

Let us consider the moved lattice type consisting of $n$CCb-lattices together with all the bi-stellar flips and boundary bi-stellar flips that are consistent with the allowed upper and lower links.

**Definition 141.** The **standard $n$-ball** is the filling of the standard $(n-1)$-sphere. In other words, it is the standard $n$-sphere where we only take a single $n$-cell and interpret all other cells as boundary cells. We will refer to the background of the standard $n$-ball as the **$n$-ball**.

**Definition 142.** A boundary $x$-cell in a $n$CCb-lattice is called **non-singular** if its lower link has $(x-1)$-sphere background and its upper link has $(n-x-1)$-ball background. Otherwise it is called **singular**. For bulk cells, singularity is analogous to singularity for cells in $n$CC-lattices.

A specific $n$CCb-lattice type is called **non-singular** if the allowed lower and upper links for bulk cell types have sphere background, the lower links for all boundary cell types have sphere background, and the upper links of all boundary cell types have ball background. When talking abound $n$CCb-lattices we will mostly have in mind such types.

**Gluings**

**Definition 143.** For every $(n-2)$-dimensional cell complex $L$, **boundary cell gluing** is the following operation that can be applied to any pair of boundary $(n-1)$-cells of a $n$CCb-lattice $X$ that are not connected to any common $(n-2)$-cells, together with an identification of their lower links with $L$: First, remove the two glued boundary $(n-1)$-cells from the complex. Then replace every pair of boundary cells of $X$ connected to one of the glued cells, that are identified with the same cell of $L$, by a single boundary cell that interits the connections from both (except for the connections to the two glued cells). Third, add a bulk $(n-1)$-cell whose upper link equals $L$, and that is connected to all the cells that the glued cells were connected to before.

**Example 83.** Consider the following examples for boundary cell gluing (see Remark (4)):

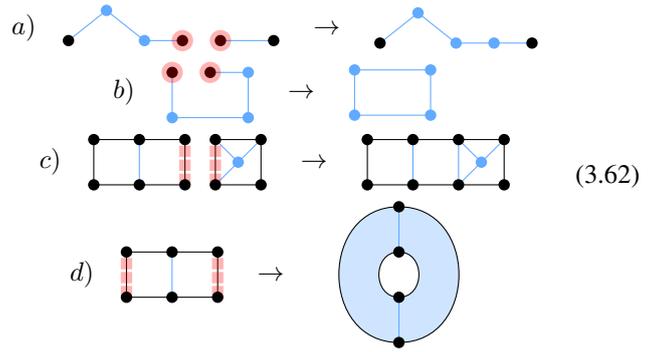

(3.62)

a) shows gluing two vertices of a 1CCb-lattices on two intervals yielding a 1CCb-lattice on a single interval. b) shows gluing two boundary vertices of a 1CCb-lattice on an interval yielding a 1CCb-lattice on a circle (with empty boundary). c) shows gluing two boundary edges of a 2CCb-lattice on two disks yielding a 2CCb-lattice on a single disk. d) shows gluing two boundary edges of a 2CCb-lattice on a disk yielding a 2CCb-lattice on an annulus. Note that different identifications of the upper links (which we didn't draw here) yield different gluings. E.g., in the example d) one could also choose an orientation reversing identification yielding a 2CCb-lattice on a Möbius strip instead of an annulus.

**Remark 64.** Similar to $n$SCb-lattices or $n$CC-lattices one can also define **level $d$ boundary cell gluing**. In contrast to the ordinary boundary cell gluing, the glued boundary cells are connected to $d$ common boundary $(n-2)$-cells. Those common connected boundary $n-2$-cells become bulk $n-2$-cells during the gluing.

**Example 84.** Consider the following higher level boundary cell gluings:

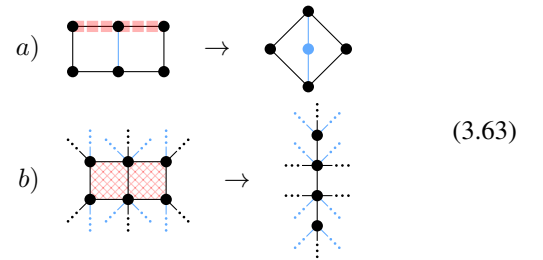

(3.63)

a) shows level 1 gluing of two neighbouring boundary edges of a 2CCb-lattice. b) shows level 1 gluing of two adjacent boundary faces of some patch of a 3CCb-lattice (note that the blue lines are edges in the bulk of the lattice).

**Remark 65.** Every (non-singular) $n$CCb-lattice type is finitely generated. The basic lattices are given by the fillings of all allowed lower links of $n$-cells, interpreted as cell complex with boundary with only one bulk cell. This is easily shown by giving an according history mapping: In a given $n$CCb-lattice associate to each $n$-cell the corresponding filling of the upper link. When two $n$-cells in the original $n$CCb-lattice share an $(n-1)$-cell, glue those boundary $(n-1)$-cells of the basic lattices accordingly.

**Example 85.** Consider the history mapping above for a 2CCb-lattice (see Remark (7)):

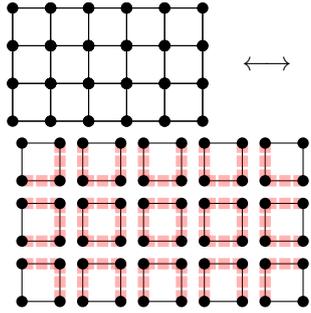

$$\longleftrightarrow \tag{3.64}$$

### 3.2.5 Higher order cell complexes

**Geometric intuition**

**Informal definition 24.** A higher order cell complex is a decomposition of a higher order manifold into polytopes of different dimension. Thereby the decomposition of each regions have to match the decomposition of the regions in the lower link of that region. I.e., we could get such a higher order cell complex constructively by choosing a cellulation for the lowest dimensional regions, and then step by step adding cellulations for the higher dimensional regions that are consistent with the previous cellulations.

**Example 86.** Consider the following examples of higher order cell complexes. The corresponding higher order manifolds of which they are decompositions are drawn on the right:

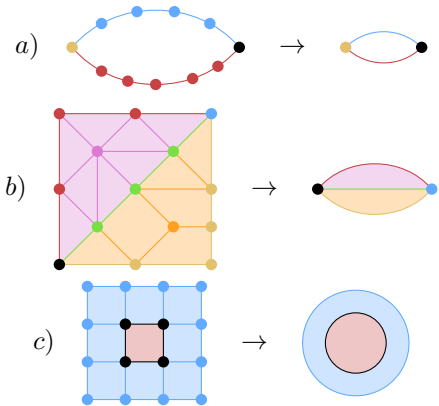

$$\tag{3.65}$$

**Combinatorial definition**

**Definition 144.** A $n$-dimensional higher order cell complex type consists of a set of $d$-**regions**, for different $d \leq n$. In a $n$-**dimensional higher order cell complex** of a given type, each $d$-region has its own sets of $x$-cells for all $0 \leq x \leq d$. There are relation-cells between the $x$-cells of all pairs of regions, with its target and source cells corresponding to those regions. There are junction maps that join together pairs of relation-cells where the target cell of the first equals the source cell of the second. The cells, relation-cells and junctions altogether (forgetting that they were associated to different regions) have to form a $n$-dimensional cell complex.

**Definition 145.** The **upper link** of a $x$-cell in a $d$-region of a $n$-dimensional higher order cell complex is an $(n-x-1)$-dimensional higher order cell complex. It is defined as for ordinary cell complexes, just that now the $x$-cell is connected to relation-cells of different types, that correspond to the different

$y$-regions for $y > d$. As the cells of the upper link correspond to the connected relation-cells, they naturally form an $(n-x-1)$-dimensional higher order cell complex whose $y$-regions are in one-to-one correspondence with the $y + x + 1$-regions of the original higher order cell complex type.

The **lower link** of a $x$-cell in a $d$-region is just the lower link of that $x$-cell in the overall $n$-dimensional cell complex. I.e., for the lower link we do not distinguish between cells of different regions.

**Formalization**

**Remark 66.** $n$-dimensional higher order cell complexes (of a fixed type) can be formulated as a restricted lattice pre-type in the same way as ordinary cell complexes, just that there are even more kinds of cells for distinguishing the different regions.

**Remark 67.** As for ordinary cell complexes, the upper and lower links are feature types for the lattice pre-type given by higher order cell complexes.

**Definition 146.** One can make $n$-dimensional higher order cell complexes (of a fixed type) a lattice type by allowing only a finite set of possible upper and lower links for the $x$-cells, for all different $x$ and regions. When $d$ is the smallest dimension of a $d$-region in the type, we will commonly refer to lattices of such a type as $n/(n-d)$CC-lattices. Note that the precise type However, depends on which regions of which dimensions are there and what the possible upper and lower links are.

**Moves**

**Definition 147.** The **cone product** of two higher order cell complexes $X$ and $Y$ is just the cone product of the underlying cell complexes of $X$ and $Y$. The only difference is that the cells are associated to different regions. As the cells of the cone product correspond to cells of $X$ and $Y$ as well as pairs of cells of $X$ and $Y$, the cone product has one region for every region of the types of $X$ and $Y$, and for every pair of such regions.

**Definition 148.** For every $d$-region in a $n$-dimensional higher order cell complex and every triple of $x$-dimensional (ordinary) cell complex $X$, $y$-dimensional cell complex $Y$ (with $x + y = d - 2$) and $(n-d-1)$-dimensional higher order cell complex $Z$, define the **higher order bi-stellar flip** with respect to $X$, $Y$, and $Z$ as the following operation that can be applied to a higher order cell complex: Consider a $(x+1)$-cell of the $d$-region whose neighbourhood looks like the cone product of the filling of $X$ times $Y$ times $Z$. Remove this part and replace it by the cone product of $X$ times the filling of $Y$ times $Z$. When performing the cone product with $Z$, the cells of the cone product that are related to a cell of $Z$ are associated to the same region as this cell of $Z$.

In other words, within the $d$-region the higher order bi-stellar flip is just an ordinary bi-stellar flip with respect to $X$ and $Z$.

**Example 87.** Consider the following higher order bi-stellar

flips:

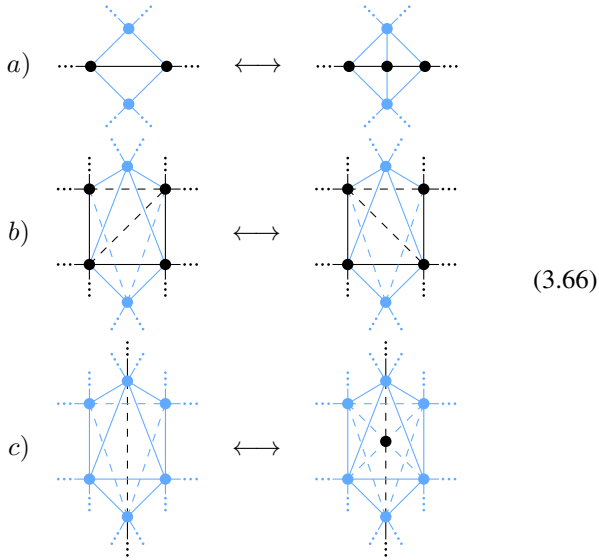

a) shows a higher order bi-stellar flip for a 1-region in a 2-dimensional higher order cell complex, with $X$ being two vertices, $Y$ being empty, and $Z$ being two (blue) vertices. b) shows a higher order bi-stellar flip for a 2-region in a 3-dimensional higher order cell complex, with $X$ being two vertices, $Y$ being two vertices, and $Z$ being two (blue) vertices. c) shows a higher order bi-stellar flip for a 1-region in a 3-dimensional higher order cell complex, with $X$ being two vertices, $Y$ being empty and $Z$ being the (blue) 4-gon.

Let us consider the moved lattice type consisting of $n/o$CC-lattices together with all higher order bi-stellar flips that are consistent with the possible upper and lower links.

**Definition 149.** The $d$-**ball completion** of a $n/o$CC-lattice $X$ is the $(n+d)/(n+o)$CC-lattice obtained from taking $X$ times the standard $d$-ball and identifying $X$ times the boundary of the standard $d$-ball with the standard $d-1$-sphere.

**Example 88.** Consider the following examples of ball completions:

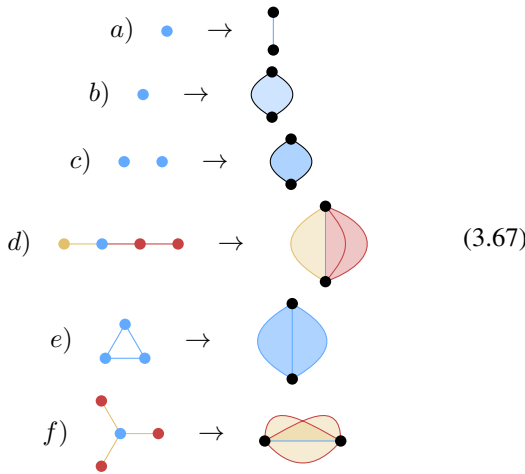

a) shows the 1-ball completion of a single vertex yielding a single edge. b) shows the 2-ball completion of a single vertex yielding a single 2-gon face. c) shows a 2-ball completion of two vertices yielding two 2-gon faces that share their two edges. d) shows the 1-ball completion of a 1/1CC-lattice formed by 4 edges yielding a 2/2CC-lattice formed by 4 2-gon faces. e) shows the 1-ball completion of a triangle yielding three 2-gon faces each pair of which shares one edge. f) shows a 1-ball completion of three edges meeting at one point yielding a 2/2CC-lattice formed by three 2-gons meeting at one edge.

**Definition 150.** A $d$-region in a specific type of $n/o$CC-lattice is called **non-singular** if 1) there is one $(n-d-1)$-dimensional higher order cell complex $L$ such that for every $x \leq d$, the allowed upper links for the $x$-cells in the $d$-region can be obtained from the $(d-x)$-ball completion of $L$ by higher order bi-stellar flips. In this case we will call the background of $L$ the **upper link** of the $d$-region. 2) for every $x \leq d$, the allowed lower links of the $x$-cells can be obtained from the standard $x$-sphere by bi-stellar flips.

A type of $n/o$CC-lattice is called non-singular if all of its regions are non-singular.

**Remark 68.** In every non-singular $n/o$CC-lattice, we can "fill in" the $x$-cells by $x$-balls, and get a higher order manifold. The $d$-regions of the type of $n/o$CC-lattice are in one-to-one correspondence to the $d$-regions of the higher order manifold type. The upper link of a $d$-region of the higher order manifold type is the background of the upper links of the corresponding $d$-region of the non-singular $n/o$CC-lattice type.

**Definition 151.** We will refer to $n/(o-1)$CC-lattices of a type, whose corresponding $n/(o-1)$-manifold type is a boundary $n/o$-manifold type, as $n/o$**CCb-lattices**.

**Gluings**

**Definition 152.** For every $d$-region in a $n/o$CC-lattice type, and every possible upper link $U$ and lower link $L$ for the $d$-cells of the $d$-region, **higher order cell gluing** with respect to $U$ and $L$ is the following operation that can be applied to a pair of $d$-cells in the $d$-region together with an identification of their upper and lower link with $U$ and $L$. The two glued cells should have non-overlapping lower links.

1. Remove the two glued $d$-cells.

2. For every cell $l$ of $L$, replace the cell connected to the first glued cell identified with $l$ and the cell connected to the second glued cell identified with $l$ by one single cell. Those cells are connected among each other in the same way as before, and inherit all connections to other cells from the two cells that were identified (except for the connections to the glued cells).

3. For each $x$-cell of $U$ add one $(d+x)$-cell, such that those cells are connected among each other just like the cells of $U$. Consider the $(d+x+1)$-cell in the original $n/o$CC-lattice corresponding to the $x$-cell of the upper link. The new $(d+x)$-cell is connected to this original $(d+x+1)$-cell. Finally, each of those new $(d+x)$-cells is connected to all the identified cells associated to cells of $L$ from point 2.

**Remark 69.** Cell gluing and boundary cell gluing are just special cases of higher order cell gluing, where $U$ is empty, or the single vertex, respectively.

(3.66)

(3.67)

**Example 89.** Consider the following examples for higher order cell gluings:

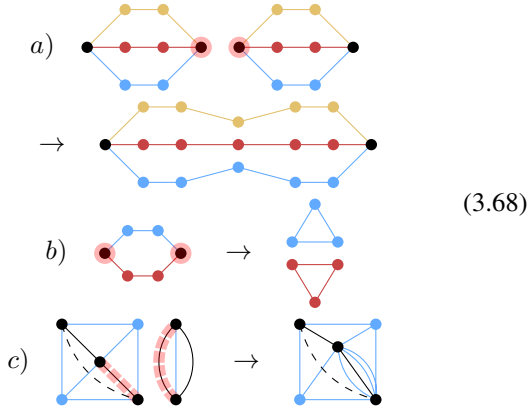

$$(3.68)$$

a) and b) show gluings of two vertices of a 0-region of a 1/1CC-lattice. c) shows gluing of two edges of a 1-region in a 2/1CC-lattice.

**Remark 70.** One can also define **higher level higher order cell gluing** where the two glued cells are allowed to have cells of their lower link in common. E.g., consider the following level 1 gluing of two neighbouring edges of a 2/1CC-lattices:

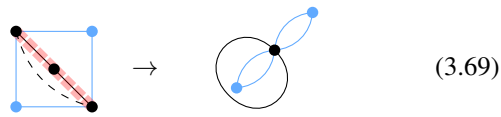

$$(3.69)$$

### 3.2.6 Thickened higher order cell complexes

**Geometric intuition**

**Informal definition 25.** A Thickened higher order cell complex is a higher order cell complex where some of some $d$-regions are thickened, in the following sense: Instead of $x$-cells from $x = 0$ to $x = d$ it consists of $x$-cells from $x = i$ to $x = d + i$, for some fixed $d$. Apart from this the complex has similar structure: E.g. imagine a 1-region embedded into a 2-region, and $i = 1$. Then this 1-region is represented by a sequence of faces instead of a sequence of edges. In contrast to ordinary higher order cell complexes two such sequences can be directly parallel to each other, only separated by a sequence of edges of the 2-region.

**Example 90.** Consider the following examples for thickened higher order cell complexes, together with the corresponding backgrounds (under all higher order bi-stellar flips from the next section):

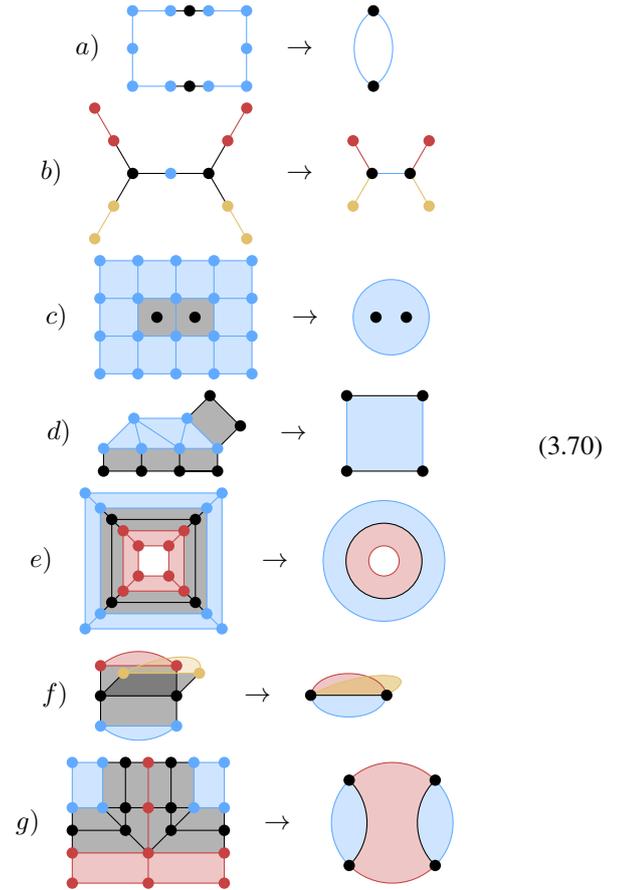

$$(3.70)$$

a) and b) are examples for a 1-thickened 0-region. c) has a 2-thickened 0-region. e) has a 1-thickened 1-region. d), f) and g) have a 1-thickened 1-region and a 1-thickened 0-region.

**Combinatorial definition**

**Definition 153.** A higher order cell complex is a cell complex whose cells are divided into different regions, where a $d$-region contains $x$-cells for $0 \leq x \leq d$. In a **thickened higher order cell complex** there can also be $i$-**thickened $d$-regions**. In this case the region contains $x$-cells for $i \leq x \leq d + i$ instead (such that ordinary regions in ordinary higher order cell complexes are 0-thickened).

**Definition 154.** The **unthickening** of a thickened higher order cell complex with respect to a $i$-thickened $d$-region is a thickened higher order cell complex where this region is 0-thickened. It is obtained by letting $x$ run from $i$ to $d + i$ and replacing every $x$-cell of the $d$-region by the stellar cone of its lower link. Thereby the central vertex is assumed to belong to the $d$-region itself. The unthickening is a lattice mapping. When applied to a cell complex (with a single region), the unthickening of that region equals the barycentric subdivision. The unthickening of all (thickened) regions alltogether yields an ordinary higher order cell complex.

**Definition 155.** A **thickened higher order cell complex type** is the same as a higher order cell complex type. A thickened higher cell complex of a given type is one whose unthickening is a higher order cell complex of the according type. Thickened $o$th order $n$-dimensional cell complex of a fixed type are commonly referred to as $n/o$**CCt-lattices**.

**Definition 156.** The **thickening** of a $n/o$CCt-lattice with respect to a $0$-thickened $d$-region is a $n/o$CCt-lattice where this $d$-region is $(n-d)$-thickened. To this end we let $x$ run from $d$ to $0$ and replace every $x$-cell of the $d$-region by the cone product of its upper and lower link. The thickening is a lattice mapping. When applied to a cell complex (with a single region), the thickening equals the thickening of the cell complex.

### Moves and gluings

Higher order bi-stellar flips can be defined analogously to the non-thickened case. Also higher order cell gluing can be defined analogously: For a $i$-thickened $x$-region this gluing happens at the $x + i$-cells.

**Example 91.** Consider the following examples for gluings of cells of thickened regions:

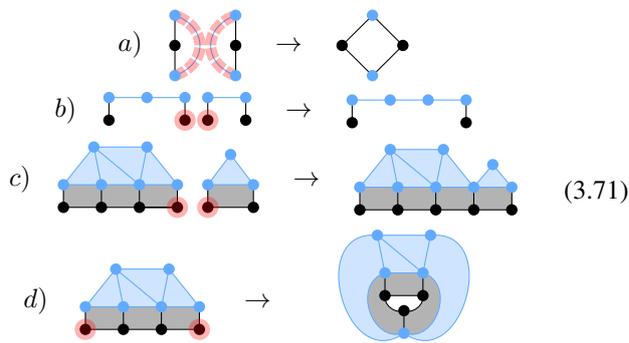

$(3.71)$

# Chapter 4

# Phases of matter

Phases of matter are families of models with the same qualitative properties. For the notion of phase that we aim for, those models need to have the following properties:

1. They must feature a thermodynamic limit, a
2. statistical description,
3. locality, and
4. homogeneity.

1. Phases are all about scaling properties, so a model cannot be just a single physical system, but is has to be a sequence of systems of increasing size, i.e., in this sequence there need to be models with an arbitrarily high number of degrees of freedom.

2. Different phases correspond to different global pattern of how the individual degrees of freedom are correlated with each other. In order to be able to talk about correlation, the kind of predictions that we make with our models need to be of a statistical nature. This is the case for classical statistical physics, as well as quantum physics.

3. Phases are equivalence classes of models under local changes on the microscopic degrees of freedom that leave certain global properties invariant. So locality is a key ingredient for the definition. To this end, one usually distributes degrees of freedom over some regular grid, and regards something as local if it only involves degrees of freedom inside a patch of the grid, whose size is independent of an eventual thermodynamic limit. Then for classical statistical or quantum mechanical systems to be local, the classical/quantum Hamiltonian has to be a sum of local terms.

4. The most general way in which one could constructively write down a thermodynamic limit sequence of physical systems would be an algorithm that constructs systems of larger and larger size. However, for such an algorithm there is no guarantee that the systems at different sizes have anything to do with each other. One could construct things that look like a rectangle with boundary which looks like a toric code on the left half and like a trivial phase on the other half at one size, and at the next size like the transverse-field Ising model with periodic boundary conditions, and so on. To rule out such things we need some notion of homogeneity, i.e., the systems have to locally look alike at different sizes, and at different places within every system of a fixed size. The most common way of having homogeneity is translation invariance: The systems are defined on regular periodic grids of arbitrary size and the physical laws (i.e., Hamiltonian terms) are just repeated everywhere in the grid. We will generalize this by imposing that the systems live on some local combinatorial structure (i.e., a lattice type) and that the physical laws can only depend on how this combinatorial structure looks like locally.

In Section (4.1) we will demonstrate how different kinds of models with the above properties can be represented by tensor networks. Examples include (trotterized) real and imaginary time evolutions of local quantum Hamiltonians with or without a physical boundary, classical partition functions and correlation functions. In Section (4.2) we adapt the notion of TL phase developed in Section (2.3) such that it can be used for actual realistic physical models, i.e. models with finite correlation length, and compare it to other definitions of phases of matter known in the literature. The tensor-network constructions from Section (4.1) as such corespond to TLs without any basic moves, but in many instances they can be extended to TLs with non-trivial basic moves. This concept of extendibility introduced in Section (4.3) makes our formalism applicable to the classification of phases of matter. In Section (4.4) we show that TLs on different backgrounds can model various physical situations, e.g., topological or symmetry-breaking fixed point models with boundaries, domain walls and defects. In addition they can be used to represent and/or calculate various characteristic quantities of a phase as, e.g., local order parameters or modular matrices. At last we consider how the concept of (C)TL mappings can be used to describe various constructions of interest such as the fusion of defects or taking the Drinfel'd double.

## 4.1 Many-body systems as tensor networks

**Observation 37.** The imaginary time evolution of any $(d + 1)$-dimensional translation-invariant many-body local quantum Hamiltonian defines a TL on $(d+1)$-dimensional regular grids. At least this is the case when we perform a Trotter decomposition of the corresponding exponential of the Hamiltonian and are willing to neglect a small error arising from this Trotterization. E.g., for a nearest-neighbour Hamiltonian in $1 + 1$ dimen-

sions, we get

(4.1)

where on the left side there is the imaginary time evolution operator as a tensor and the right side is the corresponding tensor network.

Note, however, that the tensor network on the right as such does not converge to any meaningful TL on a square lattice: First of all the tensors $e^{-\epsilon H}$ all converge to the identity operator for $\epsilon \to 0$. Second, the correlation length in imaginary time direction is a constant when measured in the actual time, so it increases like $1/\epsilon$ when measured in units of the tensor network. So simply taking the limit of the tensor network on the right side is clearly not a sensible way to discretize imaginary time.

A procedure that does converge to a valid tensor network discretization is the following: Consider a fixed finite (small) time interval, and some lattice sites in spatial direction as a unit cell. Within this unit cell consider the Trotterized imaginary time evolution with smaller and smaller time steps $\epsilon$, and block this patch of tensor network into one tensor. With a naive blocking we have to cut $\mathcal{O}(1/\epsilon)$ tensors/indices along the time-like boundary of the tensor network patch in the unit cell, so the bond dimension will grow exponentially in $1/\epsilon$. However, what one typically observes is that one can massively truncate those indices to a much smaller bond dimension $D$. Most importantly the error of the truncation does not depend on $\epsilon$, or more precisely it converges to a finite value $\eta$ when we decrease $\epsilon$. Additionally $\eta$ decreases exponentially with increasing $D$. (This is just what we observe in numerical algorithms performing imaginary or real time evolution of a state with a tensor-network parametrization, such as *iTEBD* [30, 31, 32, 33]: The growth of entanglement entropy does not essentially depend on the chosen Trotterization step.)

**Observation 38.** Also the real time evolution defines a tensor network in the same way. However, classifying quantum phases of matter corresponds to classifying the imaginary time evolution, not the real-time evolution.

**Observation 39.** A time evolution under a non-Hermitian Hamiltonian also defines a tensor network like above. Such a tensor network is an interesting mixture between real-time and imaginary time evolution: The tensors in the Trotterization are neither Hermitian nor unitary. Also, the evolution of density matrices under a local Lindbladian yields a tensor network in the same form [34]. The only difference is that all indices are composites of a ket and a bra part.

**Observation 40.** The partition function of every $n$-dimensional translation-invariant many-body classical statistical system defines a TL on $n$-dimensional regular grids [35, 36]. In conventional language the partition function is given by

$$Z = \sum_c W(c) = \sum_c e^{-\beta H(c)} =$$
$$\sum_c e^{-\beta \sum_i H_i(c)} = \sum_c \prod_i e^{-\beta H_i(c)}, \quad (4.2)$$

where the sum over $c$ is a sum over all configurations of classical statistical degrees of freedom on a regular grid. The $H_i$ are real functions for different places $i$ that only depend on the configuration of the degrees of freedom inside a small patch around $i$. We can rewrite the above expression as a tensor network in the following way: First replace every degree of freedom by a delta tensor with the basis set equal to the set of its configurations and with one index for each $H_i$ involving this degree of freedom. Second replace all of the places $i$ with the Boltzmann weight tensor $e^{-\beta H_i}$ that has one index for every degree of freedom it depends on. Now we have one Boltzmann weight tensor index and one delta tensor index for every pair of place $i$ and degree of freedom that is involved in the corresponding $H_i$. Finally, we contract all those pairs of indices.

E.g., for a classical model in two spatial dimensions with degrees of freedom on all vertices and one Hamiltonian term for each plaquette that only depends on the degrees of freedom at the four corners of the plaquette, we get the following tensor network:

(4.3)

**Observation 41.** The discretized imaginary-time path integral of a local quantum $(d+1)$-dimensional Lagrangian is formally the same as a $(d+1)$-dimensional classical statistical system. So according to Observation (40) this yields yet another way to represent a quantum system by a TL.

Again one might think there are problems with the discretization as the correlation length stays constant when measured in the distance and thus increases with the discretization when measured in the distance within the discretized tensor network. We can apply the same argument as in Observation (37), just now in all directions and not only time direction: Instead of considering smaller and smaller unit cells we fix a size of the unit cell and then make smaller and smaller discretizations within that unit cell. Then we block the tensor network within this unit cell and find that we can truncate it with an error that is independent of the time/space step of the discretization, and decreases exponentially in the chosen bond dimension.

**Observation 42.** Consider a translation-invariant local quantum Hamiltonian with a physical boundary. The imaginary time evolution of such a system defines a TL on regular grids with boundary. E.g., in $1 + 1$ dimensions with a nearest-neighbour bulk Hamiltonian and a 2-body boundary term, we get:

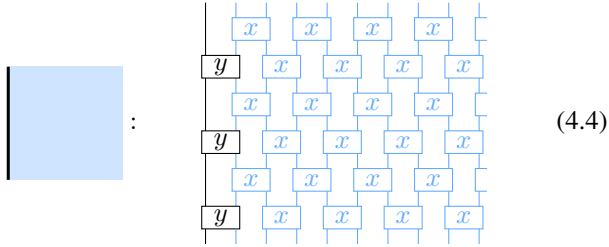

(4.4)

where $x$ and $y$ are a local imaginary time evolution Trotter step of the bulk Hamiltonian and the boundary Hamiltonian, respectively. Note that in the depicted tensor network, the left side is a real physical boundary of the tensor network, whereas the other sides are cut off and thus have open indices. The left side shows a sketch of the geometry of the tensor network: in each differently coloured region the tensor network can look different.

**Observation 43.** Another configuration is given by two different bulk Hamiltonians separated by a domain wall. Here the geometry of the tensor network in, e.g., $1 + 1$ dimensions looks like:

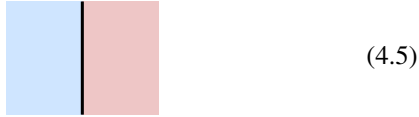

(4.5)

**Observation 44.** Another example is an imaginary time evolution starting from some tensor network state. Here the geometry in $1 + 1$ dimensions looks like:

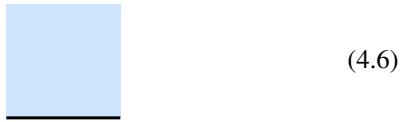

(4.6)

**Observation 45.** Yet another geometry of tensor network is given by the following physical situation: Consider a $2 + 1$-dimensional system with a point-like defect. Then the 3-dimensional imaginary time evolution tensor network is different from the rest around the line that the defect point defines along time direction.

**Observation 46.** If we want to obtain the ground state expectation value of a local observable $O$ in a quantum system we have to calculate $\mathrm{Tr}(\rho O)$, where $\rho$ is the ground state projector. This amounts to inserting the operator $O$ into the imaginary time evolution tensor network. E.g., in $1 + 1$ dimensions the situation could look like:

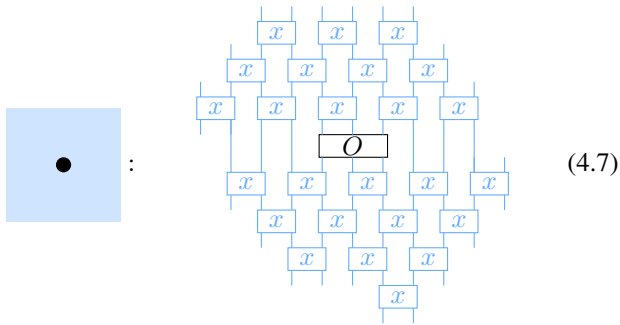

(4.7)

To actually get the expectation value we still have to normalize by the evaluation of the tensor network without $O$ inserted.

If we want to calculate the probability vector of a POVM instead of an expectation value we can insert the POVM projections which would result in evaluating a tensor network with one open index.

**Observation 47.** Also the imaginary time $n$-point correlation functions (including ground state correlators if they have the same time coordinates) correspond to the evaluation of a tensor network, just that we have to insert local operators at $n$ different space-time coordinates instead of a single one as for local expectation values. E.g., in $1 + 1$ dimensions with a 2-point function the geometry of the tensor network would look like:

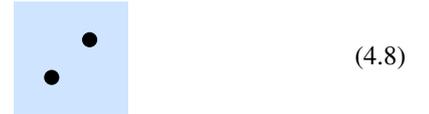

(4.8)

**Remark 71.** Tensor networks of the above geometries can also describe boundaries, domain walls, local expectation values or correlation functions of classical statistical models. In this case there is no distinction between the space and time directions.

## 4.2 Physical TL phases

### Definition

In the previous section, we have recapitulated notions of phases of matter and have described several important physical situations in terms of tensor networks. What is more, in Section (2.3), we have already defined a natural equivalence relation as "being in the same phase" for TLs. For realistic physical models, however, this is not a meaningful equivalence relation, as it would be unlikely for two models to be in the same phase. In order to get a sensible definition we take into account the following two points:

1. Circuit tensor-network moves can only correlate/decorrelate the microscopic degrees of freedom arbitrarily within a fixed sharp locality. But for realistic physical models we typically observe the following: In addition to the long-distance behaviour (which is constant or zero for gapped/topological systems), we have a part of the 2-point correlations that decays exponentially with the distance, the decay rate being known as *correlation length*. Two realistic physical models can only be mapped onto each other by circuit tensor-network moves with fixed locality if we cut off those exponentially small tails. So for two realistic physical models to be in the same phase we should also be allowed to truncate a small part of the tensors, such that the effect of the truncation becomes exponentially small in the size of the patches on which we compare the two TLs.

2. We have said that the models we look at yield statistical predictions, and found in Observation (46) that the probabilities or expectation values for such predictions are given by the evaluation of a tensor network. However, in order to get actual probabilities we still have to normalize by the evaluation of the tensor network without probabilities or expectation values. Thus, if two tensor networks only differ by a normalization they make the same statistical predictions and should be regarded as the same model.

In order to make sense of the "exponential tail" we need a quantitative measure of *locality*. Note that this is not a quantitative measure in the sense that we can compare different localities in different situations, but only in the sense that we can look at the scaling properties of other quantities depending on the locality.

**Definition 157.** The **locality** of a tensor-network move is the maximum distance of two tensors in the subset, measured in the number of contractions it takes to get from one to another. The **locality** of a TL is the maximum locality of the tensor-network moves that correspond to its basic move. The **locality** of a parallel tensor-network move is the maximal locality of the single tensor-network moves. The **locality** of a circuit tensor-network move is the sum of the localities of the contained parallel tensor-network moves.

**Definition 158.** A **physical tensor-network move** of error $\epsilon$ is a tensor-network move where Eq. (2.49) holds only up to normalization and an error $\epsilon$. More precisely, the evaluations $A$ and $B$ of the cut-out and pasted patches of a tensor network fulfil

$$d\left(\frac{A}{\|A\|}, \frac{B}{\|B\|}\right) \leq \epsilon. \tag{4.9}$$

Here $\|\cdot\|$ is some norm and $d(\cdot, \cdot)$ is some measure of distance that we do not specify here. A **physical circuit tensor network move** is a circuit tensor network move consisting of physical tensor-network moves of error $\epsilon$.

**Comment 11.** Let us have one more word on the norm and distance function used in Definition (158). In the case of classical statistical physics or quantum physics (when we work on the density matrix level) every index of the TL is equipped with a *-algebra structure, and all TL tensors are positive elements with respect to this *-algebra structure. Then a natural choice would be to take for $\|\cdot\|$ the overlap with the *-algebra unit element, and for $d(\cdot, \cdot)$ the relative entropy for the *-algebra structure, as the latter precisely measures how well two models can be distinguished by statistical observations. On the other hand it might be the case that the choice of norm and distance above does not matter to some extent, as 1) we only care for the very rough scaling properties of the distance and 2) the norm and distance are only applied to the vector space associated to the boundary of local patches (of increasing size though), and never to the whole many-body Hilbert space.

**Definition 159.** Two TLs $A$ and $B$ are said to be **in the same physical phase** if there is a sequence $M_l$ of physical circuit tensor-network moves of locality $l$ and error $\epsilon_l$ mapping $A$ into $B$, such that:

$$\exists \quad \alpha, \beta \in \mathbb{R}^+: \qquad \epsilon_l \leq \alpha e^{-\beta l}. \tag{4.10}$$

The equivalence classes of TLs under being in the same physical phase are called **physical TL phases**.

**Example 92.** Consider a TL type on a square lattice with a tensor on each plaquette, such that the tensors on neighbouring plaquettes are contracted. Take the tensors of the following two

TLs $A$ and $B$ with a 2-element basis set:

$$
\begin{array}{c}
b \!-\!\! \boxed{A} \!\!-\! d \quad = \quad 0|0|0|0 : 1 \\
\end{array}
$$

with vertical labels $a$ (top) and $c$ (bottom), and

$$
\begin{array}{c}
b \!-\!\! \boxed{B} \!\!-\! d \quad = \quad \begin{aligned} &0|0|0|0 : 1, \\ &1|1|1|1 : \tfrac{1}{2} \end{aligned}
\end{array}
\tag{4.11}
$$

Physically those correspond to the trivial 2-dimensional classical model and the classical Ising model with non-zero local field at zero temperature. Consider cutting out a patch $P$ of $l \times l$ tensors. The evaluations of the TLs on such a patch are:

$$
\begin{aligned}
P[A] &= 0|0|\ldots|0 : 1, \\
P[B] &= 0|0|\ldots|0 : 1, \quad 1|1|\ldots|1 : 2^{-l^2}.
\end{aligned}
\tag{4.12}
$$

So we have

$$d\left(\frac{P[A]}{\|P[A]\|}, \frac{P[B]}{\|P[B]\|}\right) \lesssim e^{-l}. \tag{4.13}$$

Thus, replacing one patch by the other defines a physical tensor-network move of an error that vanishes exponentially with $l$. We can build a physical circuit tensor-network move out of those, mapping $A$ to $B$. So $A$ and $B$ are in the same physical TL phase. This matches conventional definitions of phases, due to which the Ising model becomes trivial immediately after adding a perturbation that explicitly breaks the symmetry.

**Remark 72.** The exponential scaling in Eq. (4.9) might seem a bit ad hoc. Our motivation behind it is the following: With $C_E$ measurements we can extract the probability or expectation value of an observable in a real-world experiment to an accuracy $\epsilon \sim 1/\sqrt{C_E}$. I.e., the experimental complexity $C_E$ for extracting the probability/expectation value is $C_E \sim \text{poly}(1/\epsilon) \sim \exp(n_d)$ where $n_d$ is the number of digits of the probability/expectation value.

On the other hand, the vector space corresponding to the boundary of a patch of locality $l$ has dimension $D \sim \exp(l)$. Any operation that involves evaluating such a patch necessarily needs a computational complexity $C_C \sim D \sim \exp(l)$. So if we set $1/\epsilon \sim \exp(l)$ we get $C_E \sim C_C$. So with the exponential scaling we find that the computational/numerical effort for testing if two models are physically equivalent is of the same order as the experimental effort it takes to distinguish the two models by measurements.

### Comparison to other definitions

Let us now see how the above definition of a physical TL phase relates to common definitions of phases of matter. There are several such definitions that are all more or less equivalent. Like our definition, all of them define a phase of matter as an equivalence class of physical models. Different notions of what a model is and different equivalence relations yield different definitions of phases of matter. A number of conditions for two models being in the same phase have been considered in the literature. They follow largely the logic of the subsequent list. The following discussion is deliberately kept not entirely precise; it will be made precise again when we connect these notions of phases with our formalism of TL phases.

1. There is a path (in the space of all models) connecting the two models along which the expectation values of all local observables change analytically.

2. There is a path along which the gap in the Hamiltonian does not close.

3. There is a constant-time evolution under a local Hamiltonian connecting the ground states.

4. There is a constant-depth local generalized unitary circuit connecting the ground states.

1. This is the notion of phase commonly used for classical statistical systems. This definition is very phenomenological and does not reveal why phases of matter are such a fundamental concept. Also it is very unconstructive, i.e., there is no way to make this definition a numerical algorithm deciding whether two models are in the same phase. In order to determine this one would have to check for all paths in the huge space of all models, and check all local observables along this path. And even then, analyticity is algorithmically not a very natural property, and it is not so easy to decide whether a sequence of curves at larger and larger finite system sizes, with only a finite resolution, describes something analytic or not.

2. This is the definition most commonly used in quantum mechanical systems. One could also use it for classical statistical systems by looking at the gap of the transfer matrix instead of the Hamiltonian. It is much more restricted than our definition of a phase: First, it only works for gapped models. A serious obstacle here is that one cannot decide whether a given model has a Hamiltonian gap in the first place [37]. Specifically, there only a few tools known that lower bound gaps for frustration-free models [38, 39], and for generic models one would be unable to provide such a bound, This is also why the discussion in Ref. [2] is restricted to frustration-free parent Hamiltonians of matrix product states. Second, to distinguish symmetry breaking phases we will need to add further technicalities, as in a first order phase transition to a symmetry-breaking phase there is a gap all along the path, but the ground state degeneracy changes abruptly. Also we cannot distinguish $0 + 1$-dimensional phases of matter, including $0 + 1$-dimensional topological defects, as those cannot be scaled in the spatial dimension and so we cannot define a gap. Also this definition hides away that phases are all about local operations preserving global properties.

3. This definition makes the importance of locality for phases very apparent: The constant-time evolutions under a local Hamiltonian are exactly the local changes in the microscopic degrees of freedom that preserve global properties. However, it has the following caveat: If we view phases as equivalence classes of ground states, we have to define what "a ground state" could be. Taking any thermodynamic limit sequence of translation-invariant states does not make any sense, as the vast majority of the states in an exponentially growing many-body Hilbert space are completely unphysical. The only real solution to this is to say that "a ground state" is by definition the ground state of a local translation-invariant Hamiltonian. So we can only seemingly forget about the Hamiltonian and work on the level of ground states. Also, if we have a non-trivial ground state degeneracy we have to specify which ground states we mean as for topologically ordered systems not even the different ground states of the same model are related by a constant-time evolution of a local operator. Also, for models that are not gapped, the "ground state in the thermodynamic limit" is not a very well-defined object.

4. This is the discrete-time version of the previous definition. It provides an even clearer picture for the role of locality by disregarding continuum technicalities. Though it suffers from the same problems as the definition before.

The different definitions are all more or less equivalent. It is common knowledge that all local observables behave in an analytic way if the gap does not close, so 2) implies 1). Furthermore, if two models have the same ground state, then the linear interpolation $\lambda H_1 + (1 - \lambda) H_2$ provides such a gapped path, so gapped quantum phases can in principle be viewed as equivalence classes of ground state spaces. It was shown via the quasi-adiabatic continuation [40] that a gapped path between $H_1$ and $H_2$ corresponds to a constant-time evolution under a local (time-dependent) Hamiltonian, so 2) implies 3). By Trotterizing the time evolution we can approximate it by a short range local unitary circuit [1, 41], so 3) implies 4). Also, if we have a local unitary circuit or a local-Hamiltonian time evolution between two models, we can perform it slowly to get a path on which the gap does not close, so 3) or 4) implies 2). Let us now see how our definition of physical TL phases compares to those conventional definitions of phases of matter. Let us start with a small subtlety that we need to take care of:

**Remark 73.** The above construction for tensors networks for quantum mechanical and classical systems does not actually take the quantum/classical Hamiltonian as an input, but a decomposition of this Hamiltonian as a sum of local parts. Different such decompositions will yield different tensor networks. However, the conventional definitions of phases above are a property of the Hamiltonian alone and do not depend on its decomposition into local parts. So in order to agree with conventional definitions of phases, the different tensor networks arising from different decompositions of the Hamiltonian should be in the same phase according to our definition. This is indeed the case.

In the classical case this can be seen in the following way: Each two decompositions of a classical Hamiltonian can be connected by 1) adding and subtracting the same local term $H_i$: $H = H + H_i - H_i$ 2) combining two neighbouring local terms $H_x$ and $H_y$ into $H_z = H_x + H_y$: $\tilde{H} + H_x + H_y = \tilde{H} + H_z$. Both operations do not change the TL evaluated on a patch when performed inside this patch, so they correspond to tensor-network moves.

In the quantum case, the same is true up to a small error connected to the Trotterization. We have

$$e^{-\epsilon(H_1 + H_2)} = e^{-\epsilon H_1} e^{-\epsilon H_2} + \mathcal{O}(\epsilon^2) \qquad (4.14)$$

so local Hamiltonian terms can be freely combined in the limit of small $\epsilon$.

**Relation to the short-range circuit definition**

The constant-depth local generalized unitary circuit definition of phases of matter is the closest to our formulation. We will first give a quick recap of the former and then show how it is related to the latter.

**Definition 160.** Consider a sequence of ground state vectors $|\phi\rangle_s$ of a translation-invariant local gapped quantum Hamiltonian for larger and larger system sizes $s$ (with periodic boundary conditions). Each of the ground states is a large complex tensor with one index for every site of the system. A **generalized local unitary** of locality $l$ is an operator $U$ acting on the indices of the ground states on this grid inside a patch of size $l$, with the following property:

$$U^\dagger U |\phi\rangle_s = |\phi\rangle_s \qquad \forall s, \qquad (4.15)$$

e.g., in $1+1$ dimensions the above equation looks like

 (4.16)

As shown, $U$ can change the number of indices and their basis sets. For an **approximate generalized local unitary**, Eq. (4.16) holds only up to an error $\epsilon$ that is independent of $s$ (we do not specify the measure of distance here).

**Remark 74.** The generalized local unitaries differ from ordinary local unitaries in two ways: First, the unitarity is only probed in one direction: There is no restriction to $UU^\dagger$, only to $U^\dagger U$. So a generalized local unitary can isometrically embed the space it acts on into an arbitrary larger space. This way it can, e.g., add arbitrary unentangled degrees of freedom to the state. Second, the unitarity is only probed acting on the state vector $|\phi\rangle$, i.e., $U^\dagger U = \mathbb{1}$ only has to hold in the subspace of the indices of the patch that is locally supported by $|\phi\rangle$. This way, we can, e.g., get rid of arbitrary unentangled degrees of freedom of $|\phi\rangle$.

**Definition 161.** A (approximate) **parallel generalized local unitary** of locality $l$ (and error $\epsilon$) is a product of many (approximate) generalized local unitaries of locality $l$ (and error $\epsilon$) whose support regions do not overlap such that they can be applied in parallel.

**Definition 162.** A (approximate) **circuit generalized local unitary** (of error $\epsilon$) is a product of a small number of (approximate) parallel generalized local unitaries (of error $\epsilon$). Its locality $l$ is given by the sum of the localities of the individual parallel generalized local unitaries. It can also be interpreted as a finite-depth local circuit made from generalized unitaries. E.g., a circuit generalized local unitary in $1+1$ dimensions can look like:

 (4.17)

where each single $U$ is a generalized local unitary.

**Definition 163.** Two sequences of ground state vectors of gapped local Hamiltonians $|\phi\rangle_s$ and $|\psi\rangle_s$ are said to be **in the same phase**, if there is a sequence of approximate circuit generalized local unitaries of locality $l$ and error $\epsilon(l)$ that map $|\phi\rangle_s$ onto $|\psi\rangle_s$, such that $\epsilon$ decreases exponentially with $l$, independent of $s$. The equivalence classes of sequences of ground states under being in the same phase are called **gapped quantum phases of matter**.

**Remark 75.** Consider a local gapped Hamiltonian, i.e., with a constant number of eigenstates (called the **thermodynamic limit ground states**) that are separated by a gap from the rest of the spectrum that stays uniformly constant in the thermodynamic limit. Further assume that the energy difference among the thermodynamic limit ground states vanishes exponentially in the system size $s$, which is the generic behaviour for such Hamiltonians. Classifying gapped quantum phases of matter means classifying the thermodynamic limit ground state spaces of such gapped local Hamiltonian.

In principle, we can get the projector onto the thermodynamic limit ground state space by taking the (normalized) imaginary time evolution operator $e^{-\beta H}$ for $\beta \to \infty$. This will converge exponentially quickly to the thermodynamic limit ground state space, as the latter is separated from the rest of the spectrum by a constant gap. However, at each fixed system size the thermodynamic limit ground states are generically not exactly degenerate. So the imaginary time evolution operator will converge to the projector onto one unique thermodynamic limit ground state that happens to be the exact lowest energy eigenstate at this system size.

However, for below some threshold value of $\beta$ we will not resolve the separation between the different thermodynamic limit ground states. As this separation is assumed to vanish exponentially with $s$, this threshold value increases linearly with $s$. So if we scale both imaginary time $\beta$ and system size $s$ simultanously in a fixed ratio $s/\beta$ (that has to be high enough) we get the thermodynamic limit ground state projector.

So exactly the fact that we are looking at ground states (and not finite-temperature density matrices) means that we have to scale both the time and spatial directions in the imaginary time evolution. This is exactly what we do in our TL definition of phases when scaling up the locality of circuit tensor-network moves.

We will now demonstrate that applying a circuit generalized local unitary to the ground states can be implemented via tensor-network moves in the imaginary time evolution tensor network.

**Observation 48.** By performing a tensor-network move, we can insert a pair of local unitary $U$ and its Hermitian conjugate into an imaginary time evolution tensor network. E.g., for a local unitary acting on two indices of the tensor network:

$$(4.18)$$

**Observation 49.** We can also insert a pair of any generalized local unitary $U$ and its Hermitian conjugate into the imaginary time evolution tensor network by performing a tensor-network move.

To this end we have to assume that a generalized local unitary as defined in Eq. (4.16) is not only an isometry when restricted to the locally supported subspace of the global ground state, but already when restricted to the locally supported subspace by a local patch of the tensor network representing the ground states. This assumption is very plausible for gapped systems: Due to the gap, if we contract the imaginary time evolution tensor network in time direction we filter out everything that is not in the ground state space exponentially quickly. Usually one has the same situation also when contracting the tensor network in the spatial direction (in fact this corresponds to what is known as an *area law* [42]). So it seems natural to assume that Eq. (4.16) holds already on a small patch of $\phi$ up to an error that is exponentially small in the diameter of this patch.

Now if we regard the insertion of $UU^\dagger$ as tensor-network move acting a patch of locality $l$ around the insertion, then the cut-out and pasted patch of the tensor network are indeed equal (up to an error that is exponentially small in $l$). E.g., for a nearest-neighbour Hamiltonian in $1+1$ we have

$$(4.19)$$

Here $x$ is short for $e^{-\Delta\beta H}$.

**Observation 50.** By applying a circuit tensor-network move we can insert pairs of $UU^\dagger$ into the imaginary time evolution tensor network at periodic time steps, where $U$ is a circuit generalized local unitary. E.g., in $1+1$ dimensions:

$$(4.20)$$

To this end we insert $U_{\text{loc}}U_{\text{loc}}^\dagger$ (where $U_{\text{loc}}$ is a generalized local unitary) by parallel tensor-network moves of the form Eq. (4.19) on non-overlapping patches. With this we can insert one pair $U_{\text{loc}}U_{\text{loc}}^\dagger$ per volume $V$ of the patch. So in order to insert the whole $UU^\dagger$ at all time steps we need $\mathcal{O}(Vn)$ layers of parallel tensor-network moves, where $n$ is the number of layers of $U$.

We still need to argue that we cannot do anything with circuit tensor-network moves that cannot be done with circuit generalized local unitaries.

**Remark 76.** Circuit tensor-network moves are formally more powerful than circuit generalized local unitaries, as they can transform an imaginary time evolution tensor network into a tensor network that does not directly represent an imaginary time evolution any more. In essence, the difference is that we can change the geometry of the tensor network arbitrarily, and that we can insert $SS^{-1}$ for arbitrary invertible $S$ instead of $UU^\dagger$ for unitary $U$. However, if we want to transform one imaginary time evolution tensor network into another via a circuit tensor-network move we cannot change the geometry, and the single tensors have to stay Hermitian, such that we are left with circuit tensor-network moves inserting $UU^\dagger$.

**Remark 77.** Usually different TL phases can be distinguished by invariants, which are just evaluations of the imaginary time evolution tensor network on different backgrounds. However, it is clear that tensor-network moves can by construction not change those invariants.

So we have seen that up to technical details the two definitions of phases are equivalent.

## 4.3 Extendibility

TLs of a type with a large/powerful set of basic moves are restricted by a large/powerful set of consistency conditions. For many such types the TLs modulo gauge tansformations form discrete sets (up to maybe a few continuous parameters). All we need to do then is to give labels to the elements of this discrete set and we are done with the classification. The only thing that is left to check is whether two elements of the discrete set represent the same phase or not. In the former case one can often find a simple circuit tensor-network move relating them. In the latter case one can often proof this by calculating invariants, i.e., the numbers obtained by evaluating the TL on different backgrounds. On the other hand, if a TL type does not have any basic moves then the formalism of TLs/CTLs is rather useless for practically classifying phases of matter.

Unfortunately, when we write down a physical model on a lattice we usually do not restrict it by any consistency conditions coming from basic moves. I.e., the corresponding TLs are exactly of this useless types. For models coming from condensed matter physics those lattice types are typically $n$-dimensional regular grids with translation invariance in different directions. Those do not even allow for any local changes of the lattice.

In order to be able to use the formalism of TLs for classifying phases of matter we introduce the concept of extension. A TL of one type $\mathcal{A}$ (with a small/no/non-powerful set of basic moves) is extendible if it can be interpreted as a TL of type $\mathcal{B}$ (with a large/powerful set of basic moves) via a TL mapping from $\mathcal{B}$ to $\mathcal{A}$. Even if classifying the TLs of type $\mathcal{A}$ was intractable, the classification of the TLs of type $\mathcal{B}$ might be tractable. In this case we can classify the (physical) TL phases of type $\mathcal{B}$ instead: If two TLs of type $\mathcal{B}$ are in the same phase then also their mappings (of type $\mathcal{A}$) are in the same phase.

We already defined extension for TLs. As for physical TL phases, in order for such an extension to be possible for a realistic model we have to take care of exponential tails and normalizations.

**Definition 164.** A **physical TL** is a generalization of a TL where the move axiom only holds approximately and up to normalization. More precisely for every basic move there is a sequence $P_l$ of patches of locality $l$ such that exchanging the tensor networks on $P_l$ before and after this basic move is a physical tensor-network move of error $\epsilon$, and $\epsilon$ vanishes exponentially with $l$.

**Definition 165.** A TL $A$ of some type $\mathcal{A}$ **physically extends** to another type $\mathcal{B}$, if there is a physical TL $B$ of type $\mathcal{B}$ together with a TL mapping from $\mathcal{B}$ to $\mathcal{A}$ that maps $B$ to $A$.

**Comment 12.** Saying that a TL $A$ physically extends to a TL type $\mathcal{B}$ is equivalent to saying that $A$ is in the same physical TL phase as some $A'$ such that $A'$ (exactly) extends to $\mathcal{B}$. Though physical extendibility has a much more direct picture: It says that one can directly generalize the definition of the model to a more flexible lattice type in a way that is consistent with the basic moves of that lattice type.

For extensions to be a useful tool, it has to be a rather generic feature of models (on translation-invariant grids) to be extendible to TLs with a powerful set of moves. Perhaps surprisingly this seems to be indeed the case: Models that can be physically extended in a simple way to lattices with more powerful moves, i.e., models that are effectively restricted by more consistency equations, seem to be more generic than models that cannot. I.e., those models seem to be more robust with respect to small perturbations of the tensors of the model.

**Remark 78.** It is very generic for models on translation-invariant lattices to be extendible in a simple manner to a CTL type with a "topological kind of lattice type". By the latter we roughly mean a lattice type with a very powerful set of basic moves that only preserve topological, homological or homotopical properties of some kind of background. Any model that has trivial, symmetry breaking or topological order can be physically extended to that kind of type. A bit less generic are models that cannot be extended to a lattice type with topological but only conformal invariance. Those correspond to critical systems at a phase transition.

Instead of trying to directly classify all translation-invariant models, we will instead associate those models to certain **classes of order**. Such a class of order is a TL type with a very powerful set of axioms such that a classification is tractable, together with a mapping to translation-invariant lattices. A model is in a class of order if it is extendible to the corresponding TL type with the corresponding mapping.

Let us consider some examples of how models can be physically extended to other lattice types and for which physical situations such an extension might be possible.

**Example 93.** $(1+1)$-dimensional translation-invariant classical partition functions or quantum imaginary time evolutions can usually (if they do not happen to be critical, or inhomogenous in some way) be physically extended to 2-dimensional simplicial complexes with Pachner moves. I.e., their definition can be extended from regular grids to irregular ones in a consistent way. The corresponding lattice mapping consists in simply interpreting the regular grid as simplicial complex (by, e.g., dividing each plaquette into two triangles). As a concrete example consider the TL on square grids that associates to each plaquette a delta tensor:

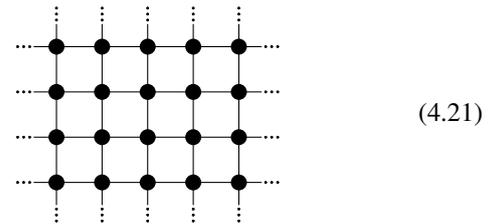

$$(4.21)$$

Its extension to arbitrary simplicial complexes is the delta TL.

**Example 94.** $(1+1)$-dimensional translation-invariant classical or quantum models with a (translation-invariant) physical boundary can usually be physically extended to 2-dimensional simplicial complexes with boundary, with Pachner moves in the interior and boundary attachment moves at the boundary. Physically this is the case, e.g., for a model with symmetry-breaking order with a boundary that projects onto one symmetry sector (or a direct sum of many such irreducible boundaries). After extending the model can be put onto lattices with a boundary that can be arbitrarily shaped, and go along any direction (i.e., in the quantum case also in the spatial direction).

**Example 95.** $(2+1)$-dimensional translation-invariant quantum models with points where the Hamiltonian differs from the rest of the bulk can usually be physically extended to 3/2CC-lattices with higher order bi-stellar flips, with one 3-region and one 1-region with a circle upper link. Physically such situations occur, e.g., for a system with topological order with anyons of one type. After extending the anyon world line can be shaped arbitrarily, instead of being a line in spacetime with constant space coordinates.

**Example 96.** Consider the tensor network that associates a real number $\alpha$ to every plaquette of a square lattice, which is a nor-malization TL as it does not have any contractions:

$$
\begin{array}{cccccc}
ⓐ & ⓐ & ⓐ & ⓐ & ⓐ & ⓐ \\
ⓐ & ⓐ & ⓐ & ⓐ & ⓐ & ⓐ \\
ⓐ & ⓐ & ⓐ & ⓐ & ⓐ & ⓐ \\
ⓐ & ⓐ & ⓐ & ⓐ & ⓐ & ⓐ
\end{array} \tag{4.22}
$$

This tensor network evaluates to $\alpha^V$ where $V$ is the volume of the square lattice measured in the number of plaquettes. It cannot be exactly extended to a topological lattice type with the standard mapping as in example (93), as the volume is not a topological invariant. However, as physical TLs allow dropping normalizations, it can be physically extended to a TL on topological lattices with this standard mapping. Also it is in the same physical phase as the mapping of the trivial TL on topological lattices.

Alternatively, it can be (exactly) extended to a TL on 0-dimensional cell complexes (i.e., sets of vertices), with the lattice mapping that takes one vertex for each plaquette of the square lattice. The extended TL associates the number $\alpha$ to each vertex of the 0-dimensional cell complex and has no basic moves.

**Example 97.** Consider a TL on square lattices which associates an identity matrix to every vertical edge, and the indices of nearby identity matrices are contracted. I.e.,

(4.23)

This TL cannot be (physically) extended to a topological lattice type with the standard mapping as in example (93), as it has a preferred direction: Information flows in the vertical direction, but none in the horizontal direction. In other words, if we evaluate the tensor network on a patch with boundary, then the degrees of freedom with the same horizontal component will be correlated with each other, but there will be no correlation between parts of the system that are horizontally separated.

Physically this tensor network corresponds to a classical 2-dimensional Ising-like model at zero temperature with coupling only in the $y$-direction, or a $1+1$-dimensional quantum qu-$d$-it system with Hamiltonian 0. It is not a gapped system as there is an exponential ground state degeneracy.

However, we can (physically) extend this TL to a topological lattice type with another mapping: Namely to 2/1CC-lattices with higher order bi-stellar flips, with one 2-region and one 1-region, in other words, lattices on 2-manifolds with codimension 1 defects. The corresponding lattice mapping interprets a square lattice as a 2/1CC-lattice where every vertical edge corresponds to the 1-region, i.e., with stacked vertical defect lines everywhere.

Alternatively this TL can be extended to 1-dimensional cell complex, with the corresponding lattice mapping given by taking one edge of the 1-dimensional cell complex for each vertical edge of the square lattice.

We can tilt this example such that the lines run in any direction. Also we can construct this stripe-tensor network in any other dimension.

**Example 98.** Similar to the previous examples, consider a TL on 3-dimensional cubic grids consisting of one independent 2-dimensional TL (say the delta tensor TL from example (93)) inside each horizontal layer of the grid. This TL cannot be extended to 3-dimensional cell complexes with the standard mapping, because of its apparent inhomogeneity. However, as in the example before, we can extend it to a TL on 3/1CC-lattices by a lattice mapping that puts defect membranes everywhere along the embedded slices, or alternatively directly to 2CC-lattices.

In general covering a translation-invariant tensor network in $n$ dimensions with $d$-dimensional slices (with $d < n$) each of which carries a $d$-dimensional TL in a non-trivial phase yields a TL in $n$ dimensions that cannot be extended to topological lattices with the standard mapping, but can However, be extended with a mapping involving defects. The above three examples were of that type with $d = 0, n = 2$, $d = 1, n = 1$ and $d = 2, n = 3$.

One can also construct models that have some underlying slicing, but where the slices interact with each other in a non-trivial way. This is what happens for fracton phases [43], which also clearly cannot be extended to topological lattices with the standard mapping, however, they can be extended to lattices with defects.

**Example 99.** Consider the following TL on 1-dimensional regular grids: On every edge there is a $2 \times 2$ Pauli-$x$ matrix, and neighbouring matrices are contracted:

$$
\cdots ⓧ\!\!-\!\!ⓧ\!\!-\!\!ⓧ\!\!-\!\!ⓧ\!\!-\!\cdots, \qquad -ⓧ- = \begin{pmatrix} 0 & 1 \\ 1 & 0 \end{pmatrix} \tag{4.24}
$$

Physically this corresponds to a 1-dimensional antiferromagnetic Ising model at zero temperature.

Evaluation of this TL on circle with $n$ edges does not only depend on the topology, but also on the parity of $n$, even in the thermodynamic limit: For $n$ even it yields 2, otherwise 0. That is why we cannot extend this TL to 1-dimensional cell complexes with Pachner moves using the standard mapping. However, we can extend it to 1-dimensional cell complexes with the following $3 \to 1$ move preserving the parity of $n$:

$$
\cdots\!\!-\!\!\bullet\!\!-\!\!\bullet\!\!-\!\!\bullet\!\!-\!\cdots \to \cdots\!\!-\!\!\bullet\!\!-\!\cdots \tag{4.25}
$$

This moved lattice type is also of a topological kind: It is in the same class as 1-dimensional cell complexes with a combinatorial 0-chain on them, i.e., with a $\mathbb{Z}_2$-element on every vertex and an additional move that flips the $\mathbb{Z}_2$-elements at the two vertices of an edge. The continuum picture for this are 1-manifolds with an element of their 0th $\mathbb{Z}_2$ homology group.

**Remark 79.** Every TL on a square lattice can be extended to a TL on 2/2CC-lattices with horizontal and vertical defect lines, and points where horizontal and vertical defect lines intersect. I.e., instead of square lattices those lattices look like a coarse square grid refined by an arbitrary irregular cellulation. However, such an extension is useless for classification purposes, as this TL type has a continuum of gauge-inequivalent TLs. Determining which of those TLs are in the same phase after the mapping to the original type is not easier than the original classification problem.

**Remark 80.** So the natural question is, are there any easy to check conditions under which TLs on regular grids can be approximately extended to topological lattice types? We do not know a precise answer to this question. However, it seems reasonable to conjecture the following: If a TL on $n$-dimensional regular grids is such that all mappings that coarse-grain by going to a new unit cell can be undone by a circuit tensor-network move, then it can be extended to a lattice type with $n$-manifolds as background. Those mappings involve all rotations and scalings, so at least this condition rules out all the above counter examples.

**Example 100.** The real-time evolution of a translation-invariant many-body local quantum Hamiltonian also defines a TL on regular grids that usually cannot be extended to topological lattices with the standard mapping. This is because the real-time evolution usually has some kind of light cone structure, which distinguishes the space and time directions. So 2-point expectation values between points that are separated by a space-time line whose velocity is larger than the Lieb-Robinson velocity are uncorrelated, whereas otherwise there can be non-trivial correlations. It is an interesting question though if real-time evolution tensor networks can also generically be extended to a CTL type with more powerful moves.

So the question on how the imaginary- and real-time evolution tensor networks relate to each other is an interesting perspective on thermalization. For example one could slowly Wick rotate from the imaginary time evolution tensor network to the real-time evolution tensor network and ask at which point a phase transition occurs. For more details see Section (6.2.2).

**Example 101.** For any $\lambda \in \mathbb{R}$, consider the TL on 2-dimensional cell complexes that associates $\lambda$ to each face, $\lambda^{-1}$ to each edge and $\lambda$ to each vertex. The evaluation of this tensor network for a lattice yields $\lambda$ to the power of the Euler characteristic of the background manifold. So each $\lambda$ is in a different phase. However, if we restrict the TL to square lattices, we can at each plaquette combine $\lambda$ from this plaquette, $\lambda^{4 \cdot (1/2)}$ from the 4 surrounding edges, and $\lambda^{4 \cdot (1/4)}$ from the 4 corners of the plaquette to 1. So we find a circuit tensor network move that transforms the TL to the trivial one. Thus, as TLs on regular grids they are all in the same phase. We cannot distinguish them by their invariants as with regular grids we can only form global tori which have 0 Euler characteristic.

## 4.4 Physical quantities in and situations modelled by (C)TLs

In this section we will give a few examples for what kinds of physical systems can be described by what kinds of TLs/CTLs and what physical interpretations the different objects of the CTL/TL have. Thereby we will concentrate on CTLs coming from TLs on lattice types of a topological kind with real/complex tensors.

**Physical situations modelled by (C)TLs**

**Remark 81.** Classical statistical models are described by TLs/CTLs with real tensors. However, not any TL with real tensors has to directly give rise to a valid classical statistical model: The TLs arising from statistical systems consist of tensors with only non-negative entries. There are TLs for which there exists no gauge transformation after which the TL tensors have this property. This does not directly imply that there cannot be any valid statistical (perhaps non-fixed point) model that is in the same physical TL phase as this TL. On the other hand, if the TL evaluates to a negative number on some background, then there cannot be any other non-negative TL in the same (exact) TL phase.

**Comment 13.** In this work, we do explicitly not pay too much attention to the *stability* (or *robustness*) of phases. A phase is robust if it does not change under arbitrarily small perturbations of the model, i.e., , the phase is an open set in the space of all models. Robustness is an important property of a phase, yet we find that one can talk about phases without caring for robustness. Examples for non-robust phases include any reducible phase, i.e., any model that is a direct sum of other models. If we add an arbitrarily small perturbation that prefers one (or a subset) of the direct sum components, we will immediately end up in the phase corresponding to this (subset of) component(s). Irreducible phases in $2 + 1$ dimensions (known as the standard topological order) have been shown to be robust in turn [44]. Another example for non-robust phases are critical models, as they occur at transitions between other phases. For the related but slightly more general concept of robustness of TLs, see Section (6.2.1). (Conventional robustness of phases is robustness of TLs on regular grids).

**Remark 82.** TLs on $n$-manifolds (and the corresponding CTLs) correspond to fixed point models of topological/symmetry breaking (gapped quantum) phases.

**Definition 166.** TLs on $n/1$-manifolds with the following central link (and the corresponding CTLs with the same boundary central link):

$$\bullet \qquad (4.26)$$

(i.e., $n$-manifolds with boundary) correspond to topological/symmetry breaking phases with topological/symmetry breaking (i.e., gapped) boundaries. We will also refer to them as **physical boundaries**.

**Definition 167.** TLs on $n/1$-manifolds with the following central upper link (and the corresponding CTLs):

$$\bullet \quad \bullet \qquad (4.27)$$

correspond to topological/symmetry breaking phases with topological/symmetry breaking (i.e., gapped) domain walls. We will also refer to them as **domain walls**.

**Definition 168.** TLs on framed $n/o$-manifolds with a central upper link consisting of a single $(o - 1)$-sphere, and with a framing of the central region with respect to the surrounding $n$-region (and the corresponding CTLs) correspond to **codimension $o$ defects**. Thus, we will also call them like that. An example are TLs on framed $3/2$-manifolds with a circle as central link

$$\bigcirc \qquad (4.28)$$

i.e., TLs living on 3-manifolds with embedded ribbons. Those correspond to **anyons**.

**Definition 169.** Codimension $n$ defects (for $n$-manifolds) correspond to **ground states** of models with topological/symmetry breaking order, as we will see in Observation (54). More generally, TLs on central $n/n$-manifold types are states in the ground state space of a system with defects, where the defect pattern is given by the central link of the $n/n$-manifold type.

**Example 102.** Consider a <span style="color:blue">TL</span> on a rather complicated 2/2-manifold type that allows for the following <span style="color:blue">background</span>:

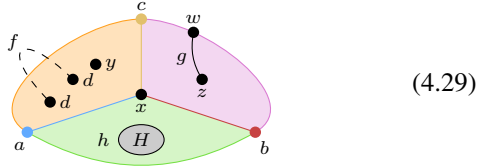

(4.29)

The picture shows a model with three different bulk phases coloured orange, yellow and olive that each have their own physical boundary. There are three domain walls labelled blue, purple and green that separate all different pairs of those bulk phases. There is one additional bulk phase coloured gray area labeled by $H$, that does not have a physical boundary, together with a domain wall to the orange bulk phase. Also there is a domain wall between the yellow bulk phase and itself, in other words a co-dimension 1 defect, coloured black and labelled by $g$. Then there is a 1-dimensional bulk phase labelled by $f$ that intersects with the olive phase at two points, which physically corresponds to having a pair of maximally entangled qu-$d$-its at those points. All those phases are constrained by consistency equations due to the topological moves, and the solutions to those form discrete sets. This is no longer true for the parts of the <span style="color:blue">TL</span> associated to 0-dimensional regions of the higher order manifold, like $a$, $b$, $c$, $d$, $x$, $y$, $w$, or $z$ above: The solutions to the consistency equations for such points form a vector space, and this vector space is exactly the ground state space of the model on an <span style="color:blue">index lattice</span> with a <span style="color:blue">background</span> that is the upper link of the 0-region. I.e., $y$ can be labelled by any ground state of the olive phase on a circle.

**Definition 170.** A **unitary domain wall** is a <span style="color:blue">TL</span> of a type that extends the <span style="color:blue">TL type</span> describing domain walls between to CTLs $A$ and $B$ by adding further moves. Those moves allow to change the embedded sub-manifolds representing the domain wall in certain topologically non-trivial ways. To this end define the $x$-**burger** with respect to region $A$ as the following cut-out piece of the corresponding $n/1$-manifolds: Take an $n$-ball coloured by $A$, and embed an $x$-ball such that the boundary of the $x$-ball is a sphere contained in the boundary of the $n$-ball. Cut out the tubular neighbourhood of that $x$-ball in the $n$-ball and fill in the same piece coloured by $B$, and let the two differently coloured $n$-regions be separated by the $(n-1)$-region representing the domain wall. Note that the $x$-burger with respect to $A$ has the same cutting boundary as the $(n-x)$-burger with respect to $B$. Now the **domain wall $x$-surgery move** is an operation that acts on the corresponding $n/1$-manifolds by cutting out the $x$-burger with respect to $A$ and pasting the $(n-x)$-burger with respect to $B$.

**Example 103.** Consider the following cut-out and pasted pieces corresponding to different domain wall surgery moves:

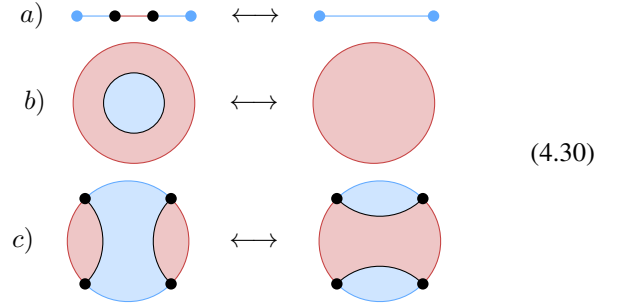

(4.30)

a) shows a domain wall 0-surgery move on 1/1-manifolds with respect to the purple region (being equal to the domain wall 1-surgery move with respect to the blue region). b) shows a domain wall 0-surgery move on 2/1-manifolds and c) shows a domain wall 1-surgery move for the latter.

**Comment 14.** Unitary domain walls are usually called *invertible domain walls* in the literature. This suggests that an equivalent definition should be "there exists another domain wall that, together with the considered domain wall, fuses to the trivial domain wall". This follows from the unitarity condition above, but not necessarily vice versa.

**Remark 83.** With the help of unitary domain walls we can formulate being in the same phase for <span style="color:blue">TLs</span>/CTLs in an alternative way: Two <span style="color:blue">TLs</span>/CTLs on $n$-manifolds/boundary $n$-manifolds are in the same phase if there exists a unitary domain wall between them.

**Remark 84.** When describing defects, domain walls, anyons or the like, we are usually interested in the phase relative to all non-trivial sub-types. This yields a finer classification than the overall phase. For example the rough and smooth boundary of the toric code are in the same phase globally, but are different phases relative to the bulk toric code: Going from one to the other involves applying a duality symmetry everywhere in the bulk and cannot be achieved by a transformation at the boundary alone. By the same argument the $e$ and $m$ anyon in the toric code are globally the same phase but different relative to the bulk toric code.

### Physical quantities in (C)TLs

A lot of the CTL tensors of CTLs like the ones described above have a very direct physical interpretation. Examples include

- ground states,
- isometric mapping from the "topological code space" to the physical ground state space,
- local Hamiltonian terms,
- RG-flows,
- data describing fusion and braiding of anyonic excitations,
- ribbon operators creating pairs of anyonic excitations,
- order parameters for classical symmetry-breaking order,
- modular matrices,
- ground state projectors,
- ground state degeneracies,

- zero-temperature density matrices,

- partition functions evaluated on different topological manifolds,

- TQFTs corresponding to the model,

- PEPS tensors representing the ground states and their virtual symmetries, and

- operators moving around domain walls.

**Observation 51.** Consider a TL on manifolds and the corresponding CTL on boundary manifolds with boundary surgery gluing. For a given space lattice $X$, we can construct the CTL background $X \times [0, 1]$ whose extended background is the extended background of $X$ times the interval, and whose index lattice is two copies of $X$. Consider the CTL tensor associated to this background and interpret is as a linear map $P$ from all indices associated to one copy of $X$ to all indices of the other copy. E.g., in 2 dimensions for a index lattice $X$ with circle extended background we get $X$ twice on the boundary of an annulus:

$$X = \bigcirc \qquad \rightarrow \qquad P_{ab} = a \, \text{⬭} \, b \qquad (4.31)$$

Due to the symmetry axiom, $P$ is a symmetric (self-adjoint in the complex case) operator. E.g.

$$a \, \text{⬭} \, b = b \, \text{⬭} \, a \quad \Rightarrow \quad P = P^T \quad (4.32)$$

Due to the gluing axiom, $P$ is a projector:

$$a \, \text{⬭} \, b = a \, \text{⬭} \, b \qquad (4.33)$$
$$\Rightarrow P^2 = P$$

In the context of quantum physics, $P$ is nothing but the *ground state projector*, and its support is the *ground state space*, and the tensors in the support the *ground states* on the index lattice $X$: The corresponding TL is the imaginary time evolution on $X \times [0, 1]$, and because we have a fixed point model we do not need to take the limit of infinite imaginary time.

For classical statistical models, $P$ is known as the *heat kernel*, and corresponds to the projector onto the eigenstates of the transfer matrix with the eigenvalues of the largest magnitude. For classical symmetry breaking models the support of $P$ is spanned by the different symmetry broken sectors.

**Observation 52.** Consider a TL on a $n/o$-manifold type and the corresponding CTL on the corresponding boundary $n/o$-manifold type with gluing at the boundary regions. For every index lattice $X$ we can consider the background $X \times [0, 1]$, whose index lattice consists of $X$ twice. Analogously to Observation (51) we find that the associated CTL tensor is again a symmetric (i.e., self-adjoint) projector. E.g., for a $1 + 1$-dimensional systems with a co-dimension 1 defect, and an index lattice $X$ with an extended background consisting of a circle with one defect, the corresponding operator would be:

$$X = \bullet \bigcirc \qquad \rightarrow \qquad P_{ab} = a \, \text{●—●} \, b \quad (4.34)$$

$P$ is again some kind of ground state projector, but this time it projects onto "ground states" that have a specific pattern of defects (given by $X$). It is also known as the *topological vector*

*space*. E.g., consider TL on a framed boundary $3/2$-manifold type with one 3-region and three embedded 1-regions, with a framing of each 1-region with respect to the 3-region, which physically could represent a $2+1$-dimensional phase with three types of anyons. Now consider the corresponding CTL and the index background $X$ consisting of a 2-ball with three embedded points representing the boundary 0-regions corresponding to the three bulk 1-regions:

 $\qquad (4.35)$

The corresponding supported space of the ground state projector for this $X$ is exactly the fusion space of the three anyon types.

Assume for a moment that the $n/o$-manifold type is such that there is a single boundary $(n - 1)$-region, and all other regions can be interpreted as sub-manifolds embedded into this boundary $(n - 1)$-region. Then we can for any index lattice $X$ consider the index lattice $\tilde{X}$ with all the embedded sub-manifolds removed. In principle one can embed the Hilbert space of $X$ into the Hilbert space of $\tilde{X}$. Then one can interpret the "ground states" on $X$ in terms of the Hamiltonian on $\tilde{X}$. Then one would find that those "ground states" on $X$ are "excitations" on $\tilde{X}$, localized at the embedded sub-manifolds. Though there is no guarantee and no need that those "excitations" are eigenstates of the Hamiltonian on $\tilde{X}$. Also the embedding of the one Hilbert space into the other is very unnatural and non-canonical and does not fit into the mindset of phases of matter. We will thus rather interpret anyons as 1-dimensional defect ribbons in a 3-dimensional space-time, which means that they are points in space where the Hamiltonian is different from the bulk Hamiltonian such that it locally projects onto a non-trivial topological sector.

**Observation 53.** In the previous observation we saw that the ground state projector on a index lattice $X$ is the CTL tensor associated to $X \times [0, 1]$. If we want to calculate the ground state degeneracy on $X$ we have to take the trace of the ground state projector. Due to the CTL axioms, this trace is equal to the CTL tensor associated to $X \times S_1$ (where $S_1$ is the circle). This background has an empty index lattice, so this CTL tensor is just a real number. E.g.

$$X = \bullet \bigcirc \qquad \rightarrow \qquad P_{aa} = \bigcirc \hspace{-1.2em} \bigcirc \qquad (4.36)$$

where the background on the right hand is a torus with an embedded non-contractible loop. In a quantum mechanical model, the trace of the ground state projector is nothing but the ground state degeneracy. We saw that it only depends on the topology (but not on the size) of $X$. This also implies that quantum mechanical models described by (C)TLs on manifolds (with boundary) are necessarily gapped.

Another way to see this is the following: One can find a "hour glass shaped" lattice representative of $X \times [0, 1]$ such that one only needs to cut a constant number of connections in the interior to separate the two copies of $X$ in the index lattice of $X \times [0, 1]$. Constant means that this number is independent of the size of $X$. This implies that the projector associated to $X \times [0, 1]$ has a constant rank when interpreted as linear map between the two copies of $X$.

**Observation 54.** Consider again a TL on (higher order) manifolds, and the corresponding CTL on (higher order) manifolds with boundary. For any background $\Phi_X$ with index lattice $X$, the corresponding CTL tensor $\phi_X$ is a ground state on $X$. This follows from the gluing axiom of the CTL by observing the following: Glue the whole index lattice of $\Phi_X$ to one of the two copies of $X$ that form the index lattice of $X \times [0, 1]$ from Observation (51). This yields the same background again. E.g.

$$ (4.37) $$

For a given $X$, every choice of $\Phi_X$ yields a ground state on $X$. There is However, no guarantee in general that the ground states obtained from different $\Phi_X$ themselves are different, or linear independent.

If we have another TL/CTL containing the original TL/CTL as a sub-type, then we can find new backgrounds of this new CTL that still have $X$ as their index lattice. Those new backgrounds correspond to new ground states. E.g., for the above $X$, if we introduce a new CTL that also allows for endpoints of the black bulk 1-regions inside the bulk 2-region, another background would be the following:

$$ (4.38) $$

Physically the purple points are ground states on a circle with one defect (which is the upper link of that 0-region), so the construction builds a ground state with two defects from a ground state with one defect on a circle. So in this case we already get continuously many ground states by the construction instead of single ones. In the most extreme case we can take a new TL with a 0-region whose upper link is just $X$. E.g.

$$ (4.39) $$

For CTLs of this type, the basic lattices have the extended background of $\Phi_X$ and the basic axioms are of the form Eq. (4.37) with this $\Phi_X$. So the TLs/CTLs of this type are exactly the ground states, and the construction yields exactly all ground states.

**Remark 85.** Consider a TL type on a central $n/n$-manifold type and the corresponding CTL type, representing ground states on the central link. Now add another basic gluing that on the level of backgrounds corresponds to gluing the central 0-region, and an index associated to every point of the 0-region that are contracted when glued. E.g., in the example above we have gluings like the following (in addition to gluing at the boundary regions):

$$ (4.40) $$

Note that this new CTL type does not correspond to a TL type. Consider a CTL lattice which has the central link of the original TL as extended background (Eq. (4.39) in our example). The index lattice consists of a part $X$ living on the boundary regions as well as a part living on the central 0-region. We can interpret the corresponding CTL tensor as a linear map $U$ from the index associated to the central point to all remaining indices distributed over the boundary regions. I.e., in our example we have schematically:

$$ U_{xa} = \quad (4.41) $$

Due to the gluing axiom and the local support convention, this linear map $U$ is an isometry:

$$ \Rightarrow \quad UU^\dagger = P $$
$$ \Rightarrow \quad U^\dagger U = \mathbb{1} $$
$$ (4.42) $$

where $P$ is the ground state projector on $X$. $U$ maps vectors in some abstract ground state space (known as *code space* in the topological codes community) to the actual ground states in the physical many-body Hilbert space associated to $X$. We will thus call it the **ground space embedding tensor** of $X$. $U$ is only defined up to an arbitrary orthogonal/unitary gauge transformation acting on the index associated to the central vertex. Note that, unlike the tensor ground state projector, the ground space embedding tensor is in general not symmetric under the index permutations corresponding to the symmetries of $X$. For lattices on (higher order) manifolds this is, e.g., the case for symmetries that have non-trivial homotopy (i.e., modular transformations on a torus), and yields a representation of the mapping class group on the code space.

In symmetry breaking order of classical statistical models, the ground space embedding tensor gives a full set of order parameters. As we have seen in Remark (47), expectation values of local operators correspond to embedded points where the tensor network differs from the rest. If we have an additional open index at this point, this corresponds to a collection of local observables. We can choose a basis such that each of those observables detects whether we are in one particular symmetry broken sector.

**Observation 55.** For some TL types on (higher order) manifolds and the associated CTL type, we find the following property: There exists a set of pairs of place types in the index lattice and small lattices called **local projector lattices**, such that:

• There is a way of gluing each local projector lattice to every place of the according type in a lattice, such that every background remains unchanged under such a gluing.

- The local projector lattice is symmetric under a reflection that exchanges the part that is glued to lattices with the other part.

- Imagine gluing the same local projector lattice twice to one place of the according type. Thereby the two local projector lattices are glued on top of each other. Now glue just the two local projector lattices on top of each other. We impose that this yields a lattice with the same background as the local projector lattice again.

- Imagine gluing local projector lattices to two places of some lattice of the according types that are near each other, such that they are partly glued on top of each other. Now glue just the two local projector lattices on top of each other in this way. We impose that the order of the gluing does not matter, i.e., both orders yield the same background.

- Imagine gluing the local projector lattices to all places of the associated types in a lattice. Thereby the local projector are glued in fact on top of each other, if the places of where they are glued overlap. Now take just the local projector lattices and glue them in this way. The resulting lattice depends only on the index lattice $X$ of the original lattice. We impose that its background equals $X \times [0, 1]$.

The above properties are not independent and usually are fulfilled all together.

Now consider the CTL-tensors associated to the local projector lattices. When glued to lattices, some of their indices are contracted with the indices of the lattice. Thus, we can interpret them as linear maps from those contracted indices to all other indices. Those linear maps have the following properties:

- According to the third property above, the linear maps are projectors.

- According to the second property above, the linear maps are symmetric (self-adjoint).

- According to the fourth property above, the linear maps all commute with each other, when applied to different places of an index lattice.

- According to the fifth property above, the product of all linear maps when applied to all places of some index lattice yields the ground state projector for this index lattice.

- According to the first property, the product of the ground state projector for some index lattice with such a linear map applied to any place yields the ground state projector again. So the linear maps leave every ground state on every index lattice invariant, when applied to any place.

If we interpret the TL as a quantum system, then we can use those local linear maps $P_p$ for different places $p$ on a index lattice $X$ to define a *Hamiltonian* whose ground states are exactly the ground states of the TL/CTL:

$$H_X := \sum_{p \in \text{places of } X} (1 - P_p), \quad (4.43)$$

where the offset of 1 normalizes the ground state energy to 0. This Hamiltonian is a sum of strictly local terms by construction. According to the properties above, it is a self-adjoint, frustration-free, commuting projector Hamiltonian. This implies that it is also gapped.

The way of representing that *ground state projector* as a product of local Hamiltonian projectors directly yields that the ground state projector is a TL operator.

**Example 104.** Consider a CTL type whose lattices are 3CCb-lattices, moves are bi-stellar flips in the interior, the index lattice is given by the boundary, gluing happens at the boundary faces, one index is associated to each boundary face, and when two faces are glued the corresponding indices are contracted. Such a CTL type has the properties described in Observation (55): For each vertex in the index lattice consider the faces around that vertex. For every possible configuration of faces, we can define a local projector lattice whose index lattice consists of two copies of the faces around the vertex. E.g.

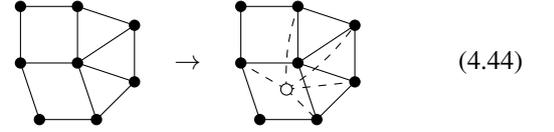

$$(4.44)$$

If we glue all faces to the back layer of the local projector lattice to the faces around a vertex in another lattice, the background of this lattice remains unchanged. E.g.

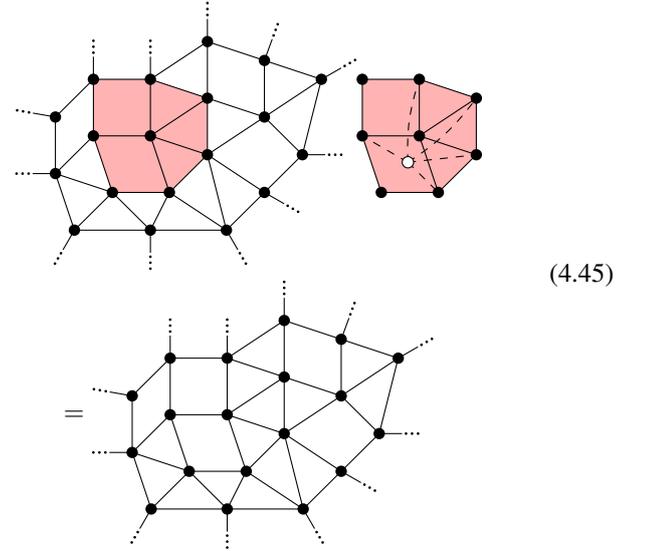

$$(4.45)$$

Now the CTL tensor associated to these projector lattices are linear maps that yield a local commuting-projector Hamiltonian.

**Remark 86.** The CTL-tensor on a manifold with empty boundary is a number. Physically this is what is known as the **partition function** of the model on the manifold.

**Observation 56.** Consider a TL on $n$-manifolds, together with a physical boundary for this TL. Now, as we have seen in Observation (54), every background with a given index lattice $X$ defines a ground state on $X$. Now, we can consider the background $X \times [0, 1]$ where $X \times 1$ is identified with $X$, and $X \times 0$ is identified with the physical boundary. E.g., for $n = 2$ on the level of extended backgrounds we have for the background of $X$ being a circle:

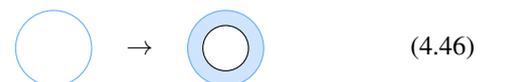

$$(4.46)$$

Or, for $n = 2$ and $X$ being a torus, we get a hollow solid torus, whose outer boundary corresponds to the boundary part 2-region whereas the inner boundary part is the bulk 2-region corresponding to the physical boundary:

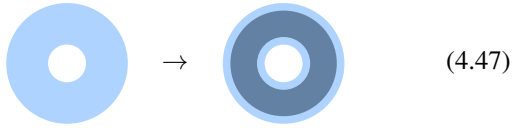

$$(4.47)$$

Now the TL on this CTL background that evaluates to the CTL tensor has open indices on the boundary 2-region, but none at the bulk 2-region (the physical boundary). As the thickness of the $[0, 1]$ direction can be kept small/constant, independent of the size of $X$, this TL has the same dimensionality as $X$. Such a tensor-network representation for a (ground) state is known as tensor-network state, or projected entangled pair state (PEPS) [45], or matrix product state [46] (MPS) in one dimension.

So in our framework, (topological, gapped) physical boundaries yield tensor-network representations of ground states. Chiral phases, which by definition do not admit a gapped boundary, also do not have this kind of tensor-network representation for their ground states. It is unclear, however, whether there are meaningful PEPS representations of some ground states that are not of this kind, and correspond to gapless boundaries with some restricted deformability (e.g., conformal). For more discussion see Section (6.1.5).

Non-chiral $2 + 1$-dimensional topological phases have been intensely studied via their tensor-network representations [3, 5, 4, 11]. In our framework we see that for a given model there is not only one such PEPS representation but a whole family of topologically different PEPS for (potentially) different ground states, corresponding to the different physical boundaries.

**Observation 57.** As in the previous observation, consider a TL on $n$-manifolds, together with a physical boundary, this time for $n = 3$. Now also consider defects within this physical boundary, i.e., lines embedded into this 2-dimensional surface. Now take the background from the previous observation, e.g., Eq. (4.47), together with defect lines on the inner, physical boundary. The TL on such a background looks like the 2-dimensional PEPS away from the boundary defects, and is different at the boundary defects. So the evaluation of the TL (i.e., the CTL tensor which is a ground state) does only depend on the background, i.e., it is invariant under moving around the boundary defects. Thus, the boundary defect lines along which the PEPS is modified can be moved around arbitrarily.

This phenomenon is known as *virtual symmetry* of the PEPS, and the consistency condition arising from local deformations of the boundary defect is called *pulling-through*.

**Observation 58.** Consider a TL on $n$-manifolds and the corresponding CTL on $n$-manifolds with boundary. With the help of the extended moves we can arbitrarily deform the index lattice. Those extended moves consist in gluing small lattices to the index lattice. E.g., consider the following patch of the boundary of a 3CCb-lattice with gluing at faces, which changes by gluing

a small lattice with ball extended background:

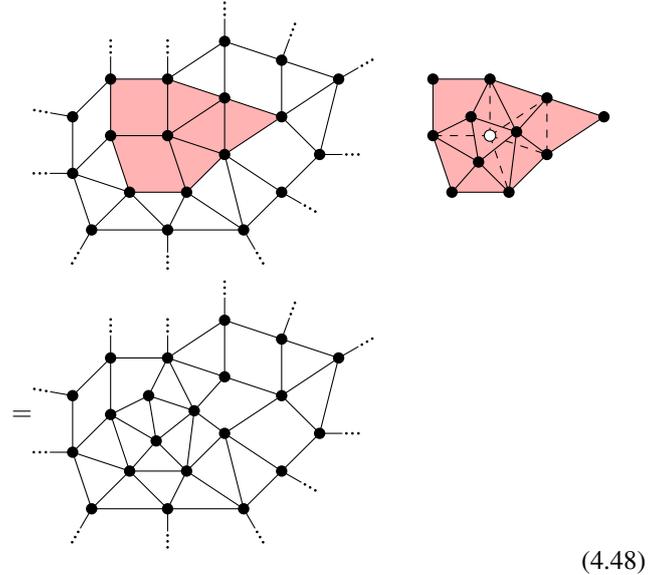

$$(4.48)$$

We can view the CTL tensors associated to the small lattices as linear maps from all the indices that are contracted during the gluing to all other indices. In the above example, if there is one index for each face, this would be a linear map from the 6 indices on the back to the 10 indices on the front layer. So the index lattice can be deformed locally by locally contracting with a tensor. With many extended moves we can fine or coarse grain the lattice, by applying a TL operator. Such a TL operator is known as *renormalization group flow* (*RG flow*). The invariance of models under such a RG flow is the origin of the name *fixed point models*.

For commuting-projector CTLs, we can directly invert every extended move by gluing the same small lattice a second time, but this time gluing the opposite part of the index lattice of the small lattice. For real/complex tensors, gluing the lattice in the opposite way corresponds applying to the transpose (adjoint) linear map. So we find that the product of the linear map and its transpose/adjoint leaves every ground state on that index lattice invariant. So the linear map is an isometry when restricted to the locally supported subspace of the ground states, in other words a generalized local unitary. So for commuting-projector CTLs the RG flow is a generalized local unitary circuit.

Note that this indicates that there might be non commuting-projector CTLs that do also not have an RG flow that can be decomposed into generalized local unitaries.

### Physical examples of (C)TL mapping/fusion

**Remark 87.** For every TL type on $n$-manifolds there is a TL mapping to a TL type describing co-dimension $o$ defects containing the original type. What this lattice mapping does is just using the bulk TL along the embedded sub-manifolds. Therefore we will call it the **trivial defect**. The corresponding lattice mapping takes a lattice of the resulting type (with defect sub-manifold) and maps it to a lattice of the original type (without defect sub-manifold) by replacing the lattice around the sub-manifold by a piece of bulk lattice. The tensor-network mapping consists of interpreting the bulk TL tensors as defect tensors.

**Remark 88.** Consider a TL type describing co-dimension 1 defects of TLs on $n$-manifolds. For every such TL type there is a fusion from twice this TL type to this TL type by taking the joint type over the two contained TL types on $n$-manifolds, and then applying a TL mapping. The effect of this fusion is what is commonly understood under the fusion of two co-dimension 1 defects, namely stacking them on top of each other and interpreting them as one single defect. The corresponding lattice mapping takes every lattice of the original type and maps it to a lattice of the joint product type by simply replacing the one defect $(n-1)$-sub-manifold in a lattice of the original type by a thin double-layer of the two defects of the joint product type. In which order we place the two defects next to each other within the double-layer can be determined by the framing of the one resulting defect.

**Remark 89.** The above fusion of co-dimension 1 defects can also be done for arbitrary co-dimension, such that the lattice mapping in the last step consists in replacing one sub-manifold by two sub-manifolds directly next to each other. Again how we put the two sub-manifolds next to each other depends on the framing. For example consider co-dimension 2 defects in 3-manifolds, i.e., anyons. We can fuse two anyons by replacing one ribbon by two ribbons, such that those ribbons are located on the two margins of this ribbon.

**Example 105.** Consider a CTL type on 3-manifolds with boundary with gluing at the boundary. There is the following CTL mapping from this type to another CTL type on 3-manifolds with embedded 1-manifolds and gluing at those 1-manifolds: The lattice mapping replaces the 1-region by a thin narrow tunnel (or *tube*) by cutting out a tubular neighbourhood, yielding a lattice on a 3-manifold with boundary. The lattice mapping just blocks the indices at one tube segment to yield the indices at one line segment. Those mappings are consistent with gluing and contraction. This CTL mapping is closely connected to what is known as the *Drinfel'd double* for fusion categories or weak Hopf algebras in the conventional algebraic language. The resulting CTL type is not a CTL coming from a TL. A concrete lattice implementation of such a CTL would yield algebraic structures related to braided fusion categories or quasi-triangular Hopf algebras.

**Example 106.** Start with a physical boundary of an TL on 3-manifolds. Consider the following TL mapping to a TL type on 2-manifolds: The lattice mapping consists in replacing the 2-manifold by a thin layer of 3-manifold everywhere that is sandwiched by a physical boundary on both sides. The tensor-network mapping consists in blocking the tensors along the direction perpendicular to the thin layer. As we have seen, a topological boundary yields a PEPS representation for a ground state of the model, so the result of the above mapping is the 2-dimensional tensor network obtained by taking the overlap of this PEPS representation with itself.

We can also get a physical boundary for this TL on 2-manifold by another TL mapping: The lattice mapping is as above for the interior of the 2-manifold, and at the boundary we just "fold over" the physical boundary of the 3-manifolds from one side of the thin layer to the other.

**Example 107.** Start again with a physical boundary for a TL type on 3-manifolds. Consider the following TL mapping to a TL type on 3-manifolds with embedded line-like defects (e.g., anyons): The lattice mapping consists in replacing the embed-ded 1-region by a thin tube by cutting out a tubular neighbourhood, like in example (105). Just that now the boundary of the tube consists of the physical boundary of the TL manifold, and not of the index boundary of a CTL manifold. The tensor-network mapping consists in blocking the tensors of the tube around each line segment. It does not yield open indices along the line, in contrast to example (105).

Physical boundaries correspond to holes in a topologically ordered model, which can be made arbitrarily large. The mapping above shrinks those holes to point-like punctures, i.e., anyons.

**Example 108.** Start with a physical boundary of a TL type $\mathcal{A}$ on $n$-manifolds together with a TL type $\mathcal{B}$ on $(n-1)$-manifolds. Consider the following fusion of $\mathcal{A}$ and $\mathcal{B}$ to $\mathcal{A}$: The lattice mapping consists in taking the lattice of $\mathcal{A}$ as it is (using the identity mapping) and mapping the lattice at the boundary of the $n$-manifold into a lattice of type $\mathcal{B}$ on the corresponding $(n-1)$-manifold. The tensor-network mapping is the identity mapping in the interior of the $n$-manifold and at the boundary blocks together the boundary tensors of $\mathcal{B}$ and the tensors of $\mathcal{A}$.

# Chapter 5

# Concrete (C)TL types

In this section we will study different concrete (C)TL types with real tensors on topological lattice types on different (framed) $n/o$-manifold types. For each of those types we will give a combinatorial definition, determine the basic tensors and basic axioms, give some families of solutions and/or concrete examples, discuss their properties and relation to other types, state their relation to established algebraic/categorial structures and give the physical context in which those models appear. Almost all CTL types described in this section come from TL types. We will use the language of CTLs though, as they provide us with the possibility to speak about tensors associated to different lattices in a more concrete way without having to specify which cutting boundary of a patch of TL we consider.

**Overview**

We will give CTLs of different types names like "X-CTL", where "X" is some string of letters describing the different ingredients of the type. "X" will start with "$n/o$" if the CTL is defined on a boundary $n/o$-manifold type. The following table summarizes the different types with their basic lattices, closely related algebraic/categorial structures, and their physical context.

| (C)TL type | basic lattices | algebraic structures | type of phases |
|---|---|---|---|
| 0cV-CTLs | single vertex | number 1 | trivial |
| 1V-(C)TLs | single edge | symmetric projector | 1D classical symmetry breaking (zero temperature), $0 + 1$D Hamiltonian with exact degeneracy, qu-$d$-it EPR pair |
| 1/1cVV-CTLs | stellar cone of some vertices | isometry | $0 + 1$D ground state collection, 1D classical order parameter collection |
| 2E-(C)TLs | triangle | *-algebra | classical/quantum 2- or $(1 + 1)$-dimensional symmetry breaking order (with/without time-reversal symmetry) |
| 2/1EE-(C)TLs | prism of stellar cone of some vertices | *-algebra representation | domain walls, line-like defects, boundaries (of multi-layer system) in 2D symmetry-breaking order |
| 2/2cEV-CTLs | 1-edge-boundary disk | isometry | $1 + 1$D ground state collection, 2D classical order parameter collection |
| 2/2cEEV-CTLs | stellar cone of $n$-gon | isometry | collection of $1 + 1$D quantum states with defects, collection of 2D classical order parameters |
| 3FE-(C)TLs | tetrahedron | unitary multi-fusion category | Topological or symmetry-breaking order in $2 + 1$ dimensions |
| 3E-(C)TLs | pillow, banana | weak Hopf *-algebra | Topological or symmetry-breaking order in $2 + 1$ dimensions |
| 3/1EE-(C)TLs | EE-pillow, EE-banana | ? | Domain walls, membrane-like defects, boundaries (of multi-layer system) in $2 + 1$ dimensions |
| 3/2EE-(C)TLs | snake, bellow | representation of Drinfel'd double | anyons |

## 5.1 0cV-CTLs

**Definition**

0**cV-CTLs** are a very trivial CTL type in 0 dimensions. "cV" stands for "central glued, vertex" as there is a basic gluing for the "central" vertices even if they are not part of a boundary, and indices are associated to vertices. Because of the gluing 0cV-CTLs are a CTL type that does not correspond to a TL type.

**Backgrounds**

The backgrounds are given by 0-manifolds, i.e., sets of points. There is one background gluing, namely surgery gluing. I.e., each pair of points of a 0-manifold can be glued and during the gluing they disappear.

**Lattices**

The lattices are 0CC-lattices, i.e., sets of vertices without any further structure. There are no basic moves (bi-stellar flips on 0CC-lattices are trivial). The index lattice of a lattice is the whole lattice. There is one basic gluing, namely cell gluing, i.e., two vertices can be glued and disappear.

**Tensors**

The tensors are real tensors (containing complex ones via realification) with one index type. The index prescription associates one index of this type to every vertex. When two vertices are glued the two associated indices are contracted.

### 5.1.1 Basic tensors and axioms

The lattice consisting of a single vertex forms a set of basic lattices. Every lattice can obtained by copies of this basic lattice by disjoint union alone already. Thus, the whole CTL is already determined by the associated basic tensor:

$$T1) \quad F_a := \bullet a \qquad (5.1)$$

The tensor associated to a lattice with $x$ vertices is just a tensor product of $x$ times this vector $F$.

All CTL axioms follow the single following basic axiom (see Remark (12), Remark (10) and Remark (8)):

$$A1) \quad \bullet a \ a \bullet = \\ \Rightarrow |F|^2 = F_a F_a = 1 \qquad (5.2)$$

Consider the lattice with two vertices. It can be glued to any other lattice with one of the vertices without changing that lattice. Thus, according to the local support convention we set the corresponding tensor to be the identity matrix:

$$A2) \quad a \bullet \quad \bullet b = a \text{———} b = \delta_{a,b} \qquad (5.3)$$

This convention can only be fulfilled if the basis is the trivial 1-element set and $F_a$ is the number 1. So the only 0cV-CTL is the trivial CTL.

### 5.1.2 Alternatives

Alternatively one can consider another CTL type called 0-**CTLs** that uses the same lattice type. In contrast to 0cV-CTLs the former do not have any basic gluing and no indices, such that all CTL tensors are numbers. The only constraints to those numbers arise from the disjoint union axiom. The single vertex is still a basic lattice. So the whole CTL is determined by the corresponding basic tensor, the number $F$. This $F$ can be arbitrary, and the number associated to a lattice consisting of $n$ vertices is $F^n$.

0-CTLs come from the TL type that associates one number to each vertex. Each single $F \in \mathbb{R}$ corresponds to a different gauge family and TL phase.

### 5.1.3 Physical interpretation

Physically 0cV-CTLs correspond to 0-dimensional classical statistical models or 0-dimensional quantum systems at finite temperature. In such systems the concepts of locality and thermodynamic limit become trivial, so there are no non-trivial phases. For 0V-CTLs different numbers correspond to different TL phases, which However, belong to a single physical TL phase, as they are all normalization TLs.

## 5.2 1V-(C)TLs

### 5.2.1 Definition

1**V-CTLs** are a rather trivial CTL type coming from a TL type for cell complexes on 1-manifolds. The "V" stands for "vertex" as this is where indices are associated to.

**Backgrounds**

The backgrounds of the TL type are 1-manifolds, thus the extended backgrounds of the CTL are boundary 1-manifolds. There is one background gluing, namely boundary surgery gluing.

**Lattices**

The lattices are given by 1CCb-lattices, i.e., collections of loops and intervals composed of line segments. The index lattice consists of the 0CC-lattice (i.e., the set of vertices) formed by the boundary of the 1CCb-lattices. The basic move consist of splitting an edge into two and vice versa:

$$\cdots\bullet\text{—}\bullet\text{—}\cdots \quad \longleftrightarrow \quad \cdots\text{—}\bullet\text{—}\bullet\text{—}\bullet\text{—}\cdots \qquad (5.4)$$

The basic gluing is boundary cell gluing, as shown in examples a) and b) of (83).

**Tensors**

The tensors are real tensors (containing complex ones via realification) with one index type. The index prescription associates one index of this type to each vertex of the index lattice, i.e., each boundary vertex. When two boundary vertices are glued together, the associated indices are contracted.

### 5.2.2 Basic tensors and axioms

A set of basic lattices is given by the single lattice consisting of a single edge:

$$\bullet\!\!-\!\!\!-\!\!\!-\!\!\bullet \tag{5.5}$$

Thus, the whole CTL is already determined by the corresponding basic tensor:

$$T1) \quad F_{ab} := a\,\bullet\!\!-\!\!\!-\!\!\bullet\,b \tag{5.6}$$

There is a set of basic axioms, one of which is the following axiom that is directly related to the basic move Eq. (5.4) of the (C)TL type:

$$A1) \quad a\,\bullet\!\!-\!\!\!-\!\!\bullet\,c = a\,\bullet\!\!-\!\!\bullet\!\!\!\!\color{red}\bullet\!\!\bullet\!\!\!-\!\!\bullet\,c$$
$$\Rightarrow F_{ab}F_{bc} = F_{ac} \tag{5.7}$$

In order to see this in the CTL language, we can consider the following history mapping: Replace every edge of a lattice by the basic lattice above, and replace every vertex in the interior by a basic gluing of the according two vertices of the basic lattices nearby. E.g.

$$\text{(5.8)}$$

For every lattice, this history mapping yields an axiom that we can view as a definition of the associated CTL tensor in terms of the basic tensor. Those defining axioms already imply the disjoint union axioms and gluing axioms of the CTL. Together with the axiom A1) they also imply the move axioms.

In order to also imply the symmetry axioms we have to add another basic axiom coming from the reflection symmetry of the basic tensor (which is already implicitly implied by the notation):

$$A2) \quad a\,\bullet\!\!-\!\!\!-\!\!\bullet\,b = b\,\bullet\!\!-\!\!\!-\!\!\bullet\,a$$
$$\Rightarrow F_{ab} = F_{ba} \tag{5.9}$$

### 5.2.3 Properties and classification

According to two basic axioms A1) and A2), the basic tensor $F_{ab}$ interpreted as a linear map from $a$ to $b$ is a symmetric (self-adjoint in the complex-real case) projector. According to the local support convention we can set $F$ to the identity matrix. The only free choice is the cardinality $d$ of the basis set. 1V-CTLs are classified by this dimension $d$.

### 5.2.4 Physical interpretation

In the world of classical statistical physics, 1V-CTLs describe 1-dimensional symmetry-breaking order at zero temperature. The corresponding fixed point model is equivalent to the 1-dimensional $d$-state Potts model at zero temperature (i.e., the Ising model for $d = 2$). Those models (for $d > 1$) have long-range correlations which are However, not robust under perturbations (such as adding temperature), even if they do not break the symmetry.

In the world of quantum physics, 1V-CTLs correspond to the $(0+1)$-dimensional models with an exact ground state degeneracy. The fixed point model is equivalent to the qu-$d$-it Hamiltonian 0. Arbitrary $(0+1)$-dimensional Hamiltonians with an exact $d$-fold ground state degeneracy are also in the same phase. When cooling down (i.e., performing imaginary time evolution) the system preserves its state within this degenerate subspace and thus transports this part of information through imaginary time. The CTL tensor on an interval is at the same time the ground state projector and the zero-temperature density matrix (up to normalization). Another interpretation of the CTL tensor on an interval when the endpoints are at equal time but separated in space is as a EPR pair (up to normalization).

In both worlds, non-trivial 1V-CTLs are non-robust (according to Com. (13)), as they are reducible. Moreover there are also unstable to essentially any permutations, even if they do not break the symmetry between the different direct sum components, namely all perturbations that lift the exact degeneracy between the largest eigenvalues of the TL matrix.

## 5.3  1/1cVV-CTLs

### 5.3.1 Definition

By 1/1cVV-CTLs we mean a collection of different CTL types that model point-like intersections of 1V-CTLs. "cVV" stands for "central glued, vertex, vertex", because there is an additional basic gluing for the central 0-region, and indices are associated to two kinds of vertices. Because of this additional gluing, 1/1cVV-CTLs do not come from a TL type.

**Backgrounds**

The extended backgrounds of different 1/1cVV-CTL types are boundary 1/1-manifolds of a central type with an arbitrary 0/0-manifold as boundary central link. E.g.

$$a)\;\; \color{blue}\bullet \qquad b)\;\; \color{red}\bullet\;\color{red}\bullet \qquad c)\;\; \color{blue}\bullet\;\color{blue}\bullet \qquad d)\;\; \color{red}\bullet\;{\color{orange}\bullet}\;\color{red}\bullet \tag{5.10}$$

1/1cVV-CTLs contain one 1V-CTLs as sub-type multiple times, once for each 0-region of the boundary central link. a) corresponds to 1V-CTLs with additional endpoints, b) corresponds to meeting points between two different 1V-CTLs, c) includes co-dimension 1 defects for a 1V-CTL, and d) incldes points where three different 1V-CTLs can meet.

There are two kinds of background gluings: First, there is surgery gluing for each of the boundary 0-regions, and second, surgery gluing for the central bulk 0-region (see examples a) and b) in (50), for the boundary central link d) in Eq. (5.10)).

**Lattices**

The lattices are 1/1CC-lattices living on the corresponding 1/1-manifolds. E.g., a lattice for the boundary central link d) in Eq. (5.10) can look like:

$$\text{(5.11)}$$

The index lattice is the 0/0CC-lattice corresponding to the boundary 0-regions and to the central bulk 0-region. For each 1-region there is one basic move, namely the higher order bi-stellar flip for this 1-region. Those moves are just the basic moves for the sub 1V-CTLs. For each boundary 0-region there is one basic gluing, namely cell gluing for this boundary region, just as for the corresponding sub 1V-CTL. There is one additional basic gluing, namely cell gluing for the central 0-region (see example a) in (73), for the boundary central link d)).

**Tensors**

The tensors are real tensors (containing complex ones via realification) with one index type for each boundary 0-region, and one index type for the central bulk 0-region. The index prescription associates one index of a given type to each vertex of the corresponding boundary or bulk 0-region. At all basic gluings of two vertices (of any boundary or the bulk 0-region), the two corresponding indices are contracted.

### 5.3.2 Basic tensors and axioms

A set of basic lattices is given by the basic lattices of all the sub 1V-CTLs together with the following lattice: There is one edge for each point of the boundary central link, coloured by the corresponding 0-region of that point. All those edges are joint together at one vertex of the central bulk 0-region. I.e., for the boundary central link d) in Eq. (5.10) we get the following basic lattices:

$$a) \quad \bullet \qquad b) \quad \bullet\!-\!\!-\!\bullet \qquad c) \quad \bullet\!-\!\!-\!\bullet \qquad d) \quad \bullet\!-\!\!-\!\bullet \qquad (5.12)$$

Here b), c) and d) are the basic lattices of the sub 1V-CTLs, and a) is the 1/1cVV basic lattice.

The 1/1cVV-CTL is fully determined by the corresponding basic tensors. For the boundary central link d) in Eq. (5.10) we have the following 1/1cVV basic tensor (in addition to the basic tensors of the sub 1V-CTLs.

$$T1) \quad G^x_{bcd} := \qquad (5.13)$$

A history for a given lattice can be obtained by the following defining mapping: For every edge take one copy of the 1V basic lattice for the corresponding 1-region, and for every vertex of the central bulk 0-region take one copy of the 1/1cVV basic lattice. Then glue all the basic lattices according to how they are located. This does not directly yield the original lattice as there are additional edges around every vertex of the central 0-region. To obtain the original lattice we have to add basic moves that fuse pairs of edges next to the vertices of the central 0-region into single edges.

In addition to the basic history moves for all the sub 1V-CTLs, there is the following additional basic history move: On the one hand there are two copies of the 1/1cVV basic lattices glued at the central vertices, and on the other hand there is the disjoint union of one of the 1V basic lattices for each point of the boundary central link. This 1/1cVV basic history move yields a basic axiom. E.g., for the boundary central link d) in Eq. (5.10) we get the following axiom:

$$A1) \qquad (5.14)$$

$$\Rightarrow G^x_{abc} G^x_{def} = \delta_{ad}\delta_{be}\delta_{cf}$$

This can be seen as follows: The defining history above is trivially consistent with disjoint unions and gluing at the 1V index vertices. Together with the basic history move A1) (and the 1V basic history move) it is also consistent with the gluing at the 1/1cVV vertices. As it is consistent with the defining histories for all the sub 1V-LTAs, it is also consistent with the 1V basic moves.

In addition to the basic axioms arising from the basic history moves we have the basic axioms arising from symmetries of the basic lattice. Those are only non-trivial when there are regions with more than one point in the boundary central link. E.g., for the boundary central upper link c) in Eq. (5.10) we have the following axiom (already implied by the notation):

$$A2) \quad a \bullet\!-\!\!\overset{x}{\bullet}\!\!-\!\bullet b \; = \; b \bullet\!-\!\!\overset{x}{\bullet}\!\!-\!\bullet a \qquad (5.15)$$

### 5.3.3 Basic properties and classification

Consider the lattice consisting of two 1/1cVV vertices of the central 0-region connected to one common edge for every point of the boundary central link. E.g.,

$$\bullet\!=\!\!=\!\bullet \qquad (5.16)$$

Gluing one 1cVV vertex of this lattice to any other lattice does not change the background of the latter, i.e., it can be undone by basic moves. So according to the local support convention we can set the associated tensor to the identity matrix. E.g.,

$$x \bullet\!-\!\!-\!\bullet y \; = \; x \!-\!\!-\! y = \delta_{x,y} \qquad (5.17)$$

This yields another axiom for the basic tensor T1). E.g.,

$$A3) \quad x \!-\!\!-\! y \; = \; x \bullet\!-\!\!-\!\bullet y = x \bullet\!\!\!\!\bullet y \qquad (5.18)$$
$$\Rightarrow G^x_{abc} G^y_{abc} = \delta_{x,y}$$

which together with the basic axiom A1) yields that $G^x_{abc}$ interpreted as a linear map from $a$, $b$ and $c$ to $x$ is an orthogonal map. By a gauge transformation on the index $x$ we can set it to the identity map. The basis set of $x$ is then the cartesian product of the basis sets of $a$, $b$ and $c$. So given all the sub 1V-CTLs, there is only one possible 1/1cVV-CTL, up to gauge transformations:

$$\qquad (5.19)$$

$$G^{(x,y,z)}_{abc} = \delta_{x,a}\delta_{y,b}\delta_{z,c}$$

### 5.3.4 Alternatives

Alternatively one can consider another CTL type called 1/1V-CTLs that uses the same lattice type, but without the basic gluing and without indices associated to the 2/1cVV vertices. The set of basic lattices is still the same, as the defining history didn't make use of the basic gluing at the 2/1cVV vertices.

The whole CTL is already determined by the corresponding basic tensor, together with the basic tensors of the sub 1V-CTLs. E.g.,

$$T1) \quad G^{bcd} := \qquad (5.20)$$

A set of basic axioms is given by the axioms of the sub 1V-CTL, together with the following additional basic axiom for each region of the boundary central link: On the left hand there is the 1V basic lattice of the according region glued together the 1/1V basic lattice, and on the right hand there is the 1/1V basic lattice alone. E.g.,

$$A1) \qquad \qquad (5.21)$$

$$G^{bcd}\delta_{d,a} = G^{acd}$$

Given that the 1V basic tensors are identity matrices according to the local support convention, this axiom is fulfilled for an arbitrary tensor $G$. So the 1/1V-CTLs for a given set of sub 1V-CTLs form a vector space. Also the set of gauge families is in general labeled by continuous parameters. E.g., for the boundary central link a) in Eq. (5.10) consisting of a single point, $G$ is an arbitrary vector (a tensor with one index) and the norm of this vector is constant under gauge transformations. Every different norm corresponds to one gauge family. When there are regions in the boundary central link represented by multiple points, we also need symmetry axioms, analogously to the axiom A2) for 1/1cVV-CTLs.

In contrast to 1/1cVV-CTLs, 1/1V-CTLs do come from a TL type on the 1/1-manifold type whose central link is the boundary central link of the CTL.

### 5.3.5 Physical interpretation

In the context of classical statistical physics, 1/1V-CTLs with boundary central link a) in Eq. (5.10) are topological boundaries of (zero-temperature) symmetry-breaking order in 1 dimension. Those boundaries correspond to an arbitrary superposition of symmetry-broken sectors. The corresponding fixed point model is the $d$-state zero-temperature Potts model with an arbitrary Boltzmann weight at the boundary. For the boundary central link b) they are co-dimension 1 defects within this Potts model, and in general they correspond to points where multiple different Potts models meet. Those are However, equivalent to multiple boundaries. 1/1cVV-CTLs represent the whole collection of such boundaries/defects/meeting points.

In the context of quantum mechanics, complex-real 1/1V-CTLs with boundary central link a) correspond to ground states of the exactly $d$-fold degenerate imaginary time evolution given by the sub 1V-CTLs. 1/1cVV-CTLs are isometries from some abstract ground state space into the many-body Hilbert space (which is trivial as the latter Hilbert space does not grow). In general they describe the ground states of the tensor product of one system for each point of the boundary central link.

## 5.4 2E-(C)TLs

### 5.4.1 Definition

2E-CTLs are a CTL type on 2-dimensional lattices coming from a TL type. The "E" stands for "edge" as this is where indices are associated to.

**Backgrounds**

The extended backgrounds of the CTL type correspond to boundary 2-manifolds, and the backgrounds of the TL correspond to 2-manifolds. There is one background gluing, namely boundary surgery gluing (see example c), d) and e) in (34)).

**Lattices**

The lattices are 2CCb-lattices with the following decoration: Each boundary edge has an orientation. The index lattice consists of the 1CC-lattice formed by the boundary (together with the edge orientations). The basic moves are bi-stellar flips in the interior (see example c) in (70)). The basic gluing is (higher level) boundary cell gluing (see examples c) and d) in (83) and a) in (84)), where we have to take care of the edge orientations: If two edges are glued we have to identify their lower links such that the orientations match. E.g., the following gluing yields a Möbius strip, whereas it would yield an annulus if we would reverse one of the edge orientations (see Remark (3)):

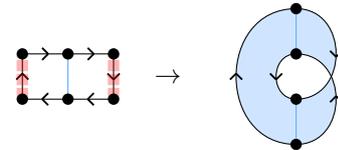

$$(5.22)$$

**Tensors**

The tensors are real tensors (containing complex ones via realification) with one index type. The index prescription associates one index to each boundary edge. When two edges are glued, the two associated indices are contracted.

### 5.4.2 Basic tensors and axioms

A set of basic lattices is given by the following single lattice:

$$(5.23)$$

We can get a defining history for a given lattice in the following way: Perform a barycentric subdivision of the underlying 2CCb-lattice (i.e., replace every cell by a vertex), except for the boundary edges. This yields a 2SCb-lattice. The orientation for all edges in the interior is canonically determined from the barycentric subdivision, and the orientation of the boundary edges is the one given by the original lattice. Now glue together all basic lattices according to how they are located. This yields a lattice that is connected to the original lattice by basic moves.

The whole 2E-CTL is determined by the associated basic tensor:

$$T1) \quad F_c^{ab} := \qquad (5.24)$$

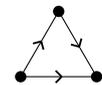

There is the following set of basic history moves (yielding basic axioms): 1) 3 basic axioms directly related to a 2-Pachner move, (or the according bi-stellar flip) corresponding to the three choices of opposite edges of the 3-simplex representing the pairs of glued edges:

$$\tag{5.25}$$

2) 4 basic axioms directly related to a 3-Pachner move (or the according bi-stellar flip) corresponding to the 4 vertices of the 3-simplex corresponding to the vertex in the middle on the right side:

$$\tag{5.26}$$

The defining history above is automatically consistent with disjoint union. Together with the basic history moves above it is also consistent with gluings and moves (as the basic history moves directly correspond to basic moves).

**Remark 90.** Consider the lattice consisting of a single 2-gon face with non-cyclic edge orientation. Gluing one of the edges of this lattice to other lattice does not change the background of the latter. By the symmetry and gluing axioms, the associated tensor interpreted as a linear map from one edge to the other is a symmetric (Hermitian in the complex-real case) projector:

$$\tag{5.27}$$

So by the local support convention we can set it to the identity matrix:

$$\tag{5.28}$$

**Remark 91.** We can reduce the number and complexity of the basic axioms by adding the 2-gon face with cyclic edge orientation to the set of basic lattices, such that we can use the associated basic tensor as some kind of auxiliary variable:

$$T2) \qquad \tag{5.29}$$

Now the following 5 axioms involving both basic tensors T1) and T2) are equivalent to the basic axioms above together with the local support convention in Eq. (5.28).

$$A1)$$
$$A2)$$
$$A3) \qquad \tag{5.30}$$
$$A4)$$
$$A5)$$

On the one hand the axioms A1) to A5) clearly follow from the CTL axioms and thus are implied by the basic axioms Eqs. (5.25, 5.26).

On the other hand A1) is equal to Eq. (5.25 for one particular edge orientation. Using A2) we can bring any of the triangles from the left side of A1) to the right in order to obtain Eq. (5.26) for one particular edge orientation. Then we can use A4) and A5) to invert the orientation of any of the non-glued edges on both sides of any axiom, whenever this does not lead to cyclic orientations of the triangles. Using A4) and A5) together with A3), we can also invert the orientation of the two glued edges on one side of any axiom. Using that we get all the axioms in Eqs. (5.25, 5.26) from the ones for one particular edge orientation.

So we have reduced the 7 rather complicated axioms Eqs. (5.25, 5.26) together with the local support convention Eq. (5.28) to 5 much simpler axioms.

### 5.4.3 Solutions

**Delta tensors**

For any finite basis set $B$, the **delta 2E-CTL** (see Observation (20)) is given by

$$F_c^{ab} := \delta_{a,b,c}, \text{ where } a, b, c \in B. \tag{5.31}$$

The tensor associated to an arbitrary lattice is just the tensor product of one delta tensor for each connected component, with as many indices as needed. E.g.,

$$\tag{5.32}$$

## Groups

Every (finite) group $G$, the **group** 2**E-CTL** has $G$ as basis set and the following basic tensor:

$$F_c^{ab} = \left(|G|^{-1/2}\right)\delta_{ab,c} = \begin{cases} |G|^{-1/2} \text{ if } ab = c \\ 0 \text{ otherwise} \end{cases}, \quad (5.33)$$

where $a$, $b$, $c$ label group elements and $\delta$ is the delta tensor on the set of group elements. So the basic tensor is just the normalized, linearized version of the group multiplication, i.e., the tensor representing the product of the group algebra.

The tensor associated to an arbitrary lattice with disk extended background has entry 1 for a configuration of group elements whose product going counter-clockwise around the edges of the index lattice yields the identity group element, and 0 otherwise. If the corresponding edge is oriented clockwise instead of counter-clockwise one has to take the inverse group element instead of the group element. Note that whether the product yields the group identity does not depend on what "counter-clockwise", so no orientation is actually needed. We also have to include a normalization factor of $|G|$ to the power of 1 minus half the number of edges. E.g.

$$c \begin{array}{c} b \\ \bullet \\ \bullet \\ d \end{array} a = \left(|G|^{1-4/2}\right)\delta_{a^{-1}bcd^{-1},\mathbf{1}} \quad (5.34)$$

where $\mathbf{1}$ denotes the identity group element. The gluing axiom of the CTL follows by the associativity of the group multiplication.

The following two tensors represent the linearized identity and inverse operation of the group:

$$\bullet\!\!\!\text{⟨}\,a = \delta_{a,\mathbf{1}} \qquad \begin{array}{c} b \\ \bullet\!\!\!<\!\!\!\bullet \\ a \end{array} = \delta_{a,b^{-1}} \quad (5.35)$$

## Double-line tensors

For any set $B$ the **double-line** 2**E-CTL** has as basis set $B \times B$ and the following basic tensor:

$$F_{(c\bar{c})}^{(a\bar{a})(b\bar{b})} = \left(|B|^{-1/2}\right) \begin{array}{c} \bar{b} \quad b \\ \bar{a} \quad a \\ \hline \bar{c} \quad c \end{array} \quad (5.36)$$

where $a, \bar{a}, b, \bar{b}, c, \bar{c} \in B$. The basic tensor $F$ (when read as operator from the two upper composite indices to the lower composite index) represents the (normalized version of the) product in the algebra of $|B| \times |B|$ matrices. The two non-overlined (overlined) components of each index represents the columns (rows) of the matrices:

$$(XY)_{\bar{c}c} = X_{\bar{c}i}Y_{ic} = X_{\bar{b}b}Y_{\bar{a}a}\delta_{\bar{b},\bar{c}}\delta_{b,\bar{a}}\delta_{ac} \quad (5.37)$$

The tensor associated to a general lattice with disk extended background has the same double-line structure as Eq. (5.36). For each edge the order of the overlined and non-overlined component of the associated index is determined from the orientation. The normalization factor is $|B|$ to the power of 1 minus half the number of edges. E.g.

$$(c\bar{c})\begin{array}{c}(b\bar{b}) \\ \bullet\!\bullet \\ \bullet\!\bullet \\ (d\bar{d})\end{array}(a\bar{a}) = \left(|B|^{1-4/2}\right) \begin{array}{c} b \quad \bar{b} \\ \bar{c} \\ c \end{array}\begin{array}{c} \bar{a} \\ a \\ d \quad \bar{d} \end{array} \quad (5.38)$$

The tensors associated to the following two lattices represent the transposition operation and the identity element of the matrix algebra:

$$\bullet\!\!\!\text{⟨}\,(a\bar{a}) = \left(|B|^{1/2}\right) \begin{array}{c} a \\ \bar{a} \end{array}$$

$$\begin{array}{c}(b\bar{b}) \\ \bullet\!\!<\!\!>\!\!\bullet \\ (a\bar{a})\end{array} = \begin{array}{c} \bar{b} \\ b \end{array}\!\!\!\text{———}\!\!\!\begin{array}{c} a \\ \bar{a} \end{array} \quad (5.39)$$

## Complex numbers

The **complex number** 2**E-CTL** (see Definition (88)) has as basis set $\{\mathbf{1}, \mathbf{i}\}$ and the following basic tensor:

$$F_c^{ab} := \mathbb{C}_{abc}^{\{in,in,out\}}, \text{ with } a, b, c \in \{\mathbf{1}, \mathbf{i}\} \quad (5.40)$$

The tensor associated to a general connected lattice is also the complex number tensor with as many indices as needed. The complex arrow orientations are determined by the (clockwise or counter-clockwise) orientation of each edge. This definition does not actually need an orientation on the lattice because of the arrow reversal property of the complex number tensors. E.g.,

$$c\begin{array}{c} b \\ \bullet\!\bullet \\ \bullet\!\bullet \\ d \end{array} a = c \begin{array}{c} b \\ <\!\!\hat{\mathbb{C}}\!\!< \\ \hat{} \\ d \end{array} a \quad (5.41)$$

The tensors associated to a single 1-gon or 2-gon face with cyclic edge orientations represent the identity and complex conjugation operation:

$$\bullet\!\!\!\text{⟨}\,a = \mathbb{C}_a \qquad \begin{array}{c} b \\ \bullet\!\!<\!\!>\!\!\bullet \\ a \end{array} = \mathbb{C}_{ab} \quad (5.42)$$

## Quaternions

The **quaternion** 2**E-CTL** has a 4-element basis set and the following basic tensor:

$$F_c^{ab} := \left(\frac{1}{2}\right) \begin{pmatrix} \mathbf{0}:1 & \mathbf{1}:1 & \mathbf{2}:1 & \mathbf{3}:1 \\ \mathbf{1}:1 & \mathbf{0}:-1 & \mathbf{3}:1 & \mathbf{2}:-1 \\ \mathbf{2}:1 & \mathbf{3}:-1 & \mathbf{0}:-1 & \mathbf{1}:1 \\ \mathbf{3}:1 & \mathbf{2}:1 & \mathbf{1}:-1 & \mathbf{0}:-1 \end{pmatrix} \quad (5.43)$$

where $a, b, c \in \{\mathbf{0}, \mathbf{1}, \mathbf{2}, \mathbf{3}\}$ where $a$ and $b$ correspond to the row and column, and $c$ is given in sparse notation (see Remark (15)). The basic tensor interpreted as a linear map from the indices $a$ and $b$ to $c$ are the (normalized version of the) product of the 4-dimensional real division algebra known as *quaternions*.

The tensor associated to the 2-gon face with non-cyclic edge orientation represents some sort of conjugation for the quaternion algebra:

$$
\bullet\!\!\overset{a}{\underset{b}{\rightrightarrows}}\!\!\bullet = \begin{pmatrix} 1 & 0 & 0 & 0 \\ 0 & -1 & 0 & 0 \\ 0 & 0 & -1 & 0 \\ 0 & 0 & 0 & -1 \end{pmatrix} \tag{5.44}
$$

**Direct sum of double-line tensors**

As some kind of combination of the double-line 2E-CTL and the delta 2E-CTL above one can consider a direct sum of many double-line 2E-CTLs with different basis sets $B_i$ for each element $i$ of a set $I$.

$$
F_{(\gamma c \bar{c})}^{(\alpha a \bar{a})(\beta b \bar{b})} := \tag{5.45}
$$

$$
w_\alpha = |B_\alpha|^{-1/2}
$$

where $\alpha, \beta, \gamma$ will be referred to as **irrep indices** and $a, b, c, a', b', c'$ as **block indices**. The basis set of the block indices $B_\alpha$ depends on the value of $\alpha = \beta = \gamma$ of the irrep indices (see Remark (14)). One whole composite index has dimension $\sum_\alpha |B_\alpha|^2$.

### 5.4.4 Properties and classification

**Remark 92.** As we are using real tensors, gauge transformation are orthogonal maps $G$ as described in Observation (4). E.g.

$$
\overset{b}{\underset{d}{\overset{\displaystyle\bullet}{c\!\rightarrow\!\bullet\!\rightarrow\!a}}} \quad \rightarrow \quad \overset{\bar{b}}{\underset{\bar{d}}{\overset{\displaystyle\bullet}{\bar{c}\!\rightarrow\!\bullet\!\rightarrow\!\bar{a}}}} \quad \begin{array}{l} \bar{d}\!-\!\boxed{G}\!-\!d \\ \bar{c}\!-\!\boxed{G}\!-\!c \\ \bar{b}\!-\!\boxed{G}\!-\!b \\ \bar{a}\!-\!\boxed{G}\!-\!a \end{array} \tag{5.46}
$$

2E-CTLs have a simple classification:

**Proposition 4.** Every 2E-CTL $A$ is equivalent to a direct sum of CTLs up to gauge transformations:

$$
A = \bigoplus_{i \in I} A_i \tag{5.47}
$$

where each $A_i$ is a stacking of a double-line 2E-CTL as in Eq. (5.36) and either 1) nothing/the trivial CTL, 2) the complex number 2E-CTL or 3) the quaternion 2E-CTL. We will refer to this gauge as **the standard form** of 2E-CTLs. In tensor network notation the standard form is denoted as

$$
F_{(\gamma\gamma' c\bar{c})}^{(\alpha\alpha' a\bar{a})(\beta\beta' b\bar{b})} := \tag{5.48}
$$

$$
w_\alpha = |B_\alpha|^{-1/2}
$$

where $x$ can be the basic tensor of either the trivial, the complex number or the quaternion 2E-CTL, depending on the value of the index $\alpha = \beta = \gamma$.

We do not give an explicit proof here, but the above statement is just a direct translation of the following fact to our language: Each finite-dimensional real C\*-algebra is isomorphic to a direct sum of matrix algebras over a) the reals, b) the complex numbers or c) the quaternions [47].

**Observation 59.** According to Prop. (4), the gauge families of 2E-CTLs form a discrete set labeled by three lists of increasing positive integers. The integers are the dimensions of the double-line 2E-CTLs in the direct sum and the three lists correspond to the parts with the trivial, complex number, and quaternion 2E-CTLs.

**Example 109.** Consider the group 2E-CTL for the group $\mathbb{Z}_3$. It is gauge equivalent to a direct sum of the trivial CTL and the complex number 2E-CTL (times the trivial double-line 2E-CTL). The gauge transformation that takes it from the basis spanned by group elements into its standard form is the following:

$$
G_i^{\alpha, \alpha'} \begin{pmatrix} \frac{1}{\sqrt{3}} & \frac{1}{\sqrt{3}} & \frac{1}{\sqrt{3}} \\ 0 & \frac{1}{\sqrt{2}} & \frac{-1}{\sqrt{2}} \\ \frac{2}{\sqrt{6}} & \frac{-1}{\sqrt{6}} & \frac{-1}{\sqrt{6}} \end{pmatrix} \tag{5.49}
$$

where $i \in \mathbb{Z}_3$ labels the columns. The rows correspond to the values $(0, \mathbf{1})$, $(1, \mathbf{1})$ and $(1, \mathbf{i})$ of $(\alpha, \alpha')$ in Eq. (5.48), whereas the indices $a$ and $a'$ are trivial (i.e., one dimensional).

For complex-real 2E-CTLs, the classification becomes even simpler:

**Proposition 5.** For complex-real 2E-CTLs the standard form does only contain the complex number 2E-CTL (instead of the trivial and quaternion 2E-CTL), i.e., it equals the complexification of a direct sum of double-line 2E-CTLs. In this case the gauge transformation that brings it into standard form is the realification of a complex-linear map, see Remark (29).

We do not give an explicit proof for this statement here, but it corresponds to the fact that each finite-dimensional complex C\*-algebra is isomorphic to a direct sum of full matrix algebras. The connection will become clear in Section (5.4.6).

**Observation 60.** According to Prop. (5) the gauge families of complex-real 2E-CTLs form a discrete set labeled by an increasingly ordered list of positive integers. Those integers are the dimensions of the double-line 2E-CTLs in the direct sum.

**Remark 93.** The complexifications of two 2E-CTLs can be gauge equivalent (i.e., have the same standard form) even if they are themselves not gauge equivalent. For example the complexification of the complex number 2E-CTL is gauge equivalent to the complexification of the direct sum of two trivial 2E-CTLs, or the complexification of the quaternion 2E-CTL is gauge equivalent to the complexification of the double-line 2E-CTL over a 2-element basis set.

**Comment 15.** Consider the complexification of a group 2E-CTL Eq. (5.33), i.e., the stacking with the complex number 2E-CTL. The gauge transformation that takes it into the standard form is the realification of a complex-linear map known as the *Fourier transform* of the group. Note that the standard form of the complexified group 2E-CTL can look quite different from the standard form of the group 2E-CTL: For example the standard form of the complexified $\mathbb{Z}_3$ group 2E-CTL is a direct sum of three times the complex number 2E-CTL in contrast to the direct sum of complex number 2E-CTL and trivial 2E-CTL in example (109).

**Observation 61.** Consider the real number that the delta 2E-CTL with basis set $B$ associates to a lattice $X$ with empty index lattice (or better, the corresponding background). If $X$ is connected, the corresponding history yields a connected tensor network of delta tensors. Such a tensor network evaluates to

$$T[X] = \sum_{b \in B} = |B| \tag{5.50}$$

If $X$ consists of $x$ connected pieces, we get the product of the single numbers:

$$T[X] = |B|^x \tag{5.51}$$

This shows that the delta 2E-CTLs for different $|B|$ are really in different TL phases.

**Observation 62.** Consider the number that the double-line 2E-CTL associates to a lattice $X$ with empty index lattice. Consider the tensor network corresponding to the defining history of a lattice $X$: For each basic lattice in the history we get a factor of $|B|^{-1/2}$, and for each vertex of the history where the corners of multiple basic lattices come together we get a closed loop, yielding $|B|$ when evaluated. So in total we get a weight of

$$T[X] = |B|^{\text{ #vertices of } X} |B|^{-1/2\,\text{ #triangles of } X} \tag{5.52}$$
$$= |B|^{\text{ #vertices of } X - \text{#edges of } X + \text{#faces of } X} = |B|^{\chi(X)}$$

where $\chi(X)$ is the Euler characteristic of the cell complex $X$. This shows that those CTLs for different $|B|$ are really in different TL phases.

**Observation 63.** Consider the number that a direct sum of double-line 2E-CTLs with basis sets $B_i$ associates to a lattice $X$ with empty index lattice. Let us start with a sphere background:

$$\tag{5.53}$$

Next let's consider a torus background:

$$\tag{5.54}$$

For a general connected surface as background we get the sum over one term of the form Eq. (5.52) for every $i \in I$:

$$T[X] = \sum_{i \in I} |B_i|^{\chi(X)} \tag{5.55}$$

And products thereof for multiple connected components.

**Observation 64.** Consider the number that the complex number 2E-CTL associates to a lattice with empty index lattice. Let us start with a connected, orientable surface as background.

The tensor network corresponding to a history for such a lattice is a connected network of complex number tensors, whose arrow directions are all consistent (due to the orientability, see Definition (88)). According to the fusion property of the complex number tensors, this tensor network evaluates to 2:

$$T[X] = 2 \tag{5.56}$$

For an orientable surface with $x$ connected pieces we thus get

$$T[X] = 2^x \tag{5.57}$$

On the other hand, if the background is a non-orientable surface, the tensor network corresponding to a history contains at least one cycle with non-matching complex arrow orientations. According to the arrow obstruction property of the complex number tensors this tensor network evaluates to 0 (see also Definition (88)). For example, for the cross-cap we find:

$$\tag{5.58}$$

Also every 2E-CTL that is the complexification of another one, and in general every complex-real 2E-CTL evaluates to 0 on non-orientable surfaces.

**Observation 65.** Consider the number that a group 2E-CTL associates to lattices with empty index lattice $X$, for some selected backgrounds. For $X$ having sphere background, we get:

$$T[X] = \quad = |G| \tag{5.59}$$

For $X$ having the background of a torus, we get:

$$T[X] = \quad = \sum_{g \in G} |C[g]| \tag{5.60}$$

Here $C[g]$ is the *centralizer* of $g$, i.e., the subset of all elements that commute with $g$ (which is also a subgroup).

As a third example consider a lattice $X$ with cross-cap background:

$$T[X] = \quad = \tag{5.61}$$
$$\text{Tr}([\cdot]^{-1}) = [\#g \in G : g^2 = 1]$$

**Observation 66.** Consider the tensor corresponding to an annulus whose boundary consists of two 1-gons:

$$\tag{5.62}$$

It is a symmetric (or self-adjoint in the complex-real case) projector when interpreted as map from one edge to the other:

$$\tag{5.63}$$

After this projector is applied to any index of a CTL tensor corresponding to a disk with arbitrary space, this index can be freely permuted with others. E.g.,

(5.64)

**Comment 16.** The projector in Eq. (5.62) is the projector onto the commutative part of the CTL, i.e., that subspace in which

$$F_c^{ab} = F_c^{ba} \qquad (5.65)$$

holds. In the conventional algebra language (see Section (5.4.6)) this subspace is known as the *centre* of a *-algebra.

### 5.4.5 Alternatives

**With normalizations**

**Definition 171.** Let us consider another CTL type using the same tensor type, lattice type and index prescription, but a different contraction prescription that uses normalizations. We will refer to CTLs of this type as **2En-CTLs**, where the "n" stands for "normalization". For a level 1 boundary cell gluing (i.e., gluing together two neighbouring edges such that the vertex in between disappears from the index lattice), we contract the associated indices via one of two normalization matrices $C^-$ or $C^+$. Which one of those two we use depends on whether the two edge orientations point towards or away from the vertex in between. E.g.,

(5.66)

The normalization matrices $C^-$ and $C^+$ have to be symmetric due to the symmetry axiom of CTLs.

**Remark 94.** Along the lines of Remark (35), 2En-CTLs can be transformed into a CTL type without normalizations by adding decorations to the index lattice. To this end add to each index vertex a flag 0 or 1 that indicates whether the normalization matrix $C$ has already been incorporated or not. The new CTL tensor equals the old CTL tensor where we contract the matrix $C$ to one neighbouring edge of each of the index vertices flagged by 1. The decoration restricts the gluing in the following way: 1) Two edges cannot be glued if the fused vertices are both flagged 1, otherwise the flags are added. 2) Two neighbouring edges can only be glued if the vertex separating them is flagged 1. The last rule ensures that we have included a normalization matrix every time two neighbouring edges are glued.

**Remark 95.** The following choice of CTL tensors and normal-

ization matrices defines a 2En-CTL:

(5.67)

$$v_\alpha = |B_\alpha|^{-1} w_\alpha^{-2}$$

where $x$ is the basic tensor of the complex number, the quaternion, or the trivial 2E-CTL, depending on the value of the index $\alpha = \beta = \gamma$. We will call this the **standard form** of 2En-CTLs.

**Proposition 6.** Every 2En-CTL can be brought into the standard form by gauge transformations. For complex-real 2En-CTLs, $x$ is always the complex number 2E-CTL, and the gauge transformation is the realification of a complex-linear map.

**Remark 96.** One particularly simple example of an 2En-CTL is the following one where all basis sets are trivial:

$$F = \alpha^{-1/2} \qquad C = \alpha \qquad (5.68)$$

corresponding to an 2En-CTL in its standard form with only one trivial block of dimension 1 and $w = \alpha^{-1/2}$.

The number associated to the background of a lattice $X$ with empty index lattice is:

$$T[X] = \alpha^{\#\text{vertices of } X} \alpha^{-1/2 \ \#\text{triangles of } X}$$
$$= \alpha^{\#\text{vertices of } X - \#\text{edges of } X + \#\text{faces of } X} = \alpha^{\chi(X)} \qquad (5.69)$$

where $\chi(X)$ is the Euler characteristic of the cell complex $X$.

**Remark 97.** The (C)TLs from the previous remark for different $\alpha$ are all in different TL phases, as the corresponding invariants are different. So we see that there is at least one continuous parameter necessary to label the different TL phases. On the other hand the tensor products of the CTLs with $\alpha = \alpha_1$ and $\alpha = \alpha_2$ will result in the CTL with $\alpha = \alpha_1 \alpha_2$. So for each CTL with $\alpha$ the tensor product with the CTL for $\alpha^{-1}$ yields the trivial phase. So those phases have inverses under stacking.

Also all those different (C)TLs are in the same physical TL phase, as they are just normalization TLs.

**Without orientations**

Let us see what happens if we drop the edge orientations of the index lattice but leave everything else the same. We still have a single basic tensor, namely the tensor associated to the triangle (without edge orientations). The triangle will become invariant under arbitrary flips and rotations, thus the corresponding basic tensor $F$ will be invariant under arbitrary index permutations. So if we look at the standard form of 2E-CTLs, double-line 2E-CTLs with a non-trivial basis set, as well as quaternion 2E-CTLs drop out, as they are non-commutative. Also the complex number 2E-CTLs drop out as we need the edge orientations to determine the complex arrow orientations. So the delta 2E-CTL is the only solution that survives.

### 5.4.6 Connection to *-algebras

**Definition 172.** A (finite-dimensional unital C)*-algebra over a complex vector space with basis set $B$ is given by three linear/anti-linear maps:

$$\text{unit } \eta : \mathbb{C}^1 \longrightarrow \mathbb{C}^B \quad [\eta(1) = \mathbf{1}]$$
$$\text{involution } t : \mathbb{C}^B \longrightarrow \mathbb{C}^B \quad [t(a) = a^*] \tag{5.70}$$
$$\text{product } \mu : \mathbb{C}^{B \times B} \longrightarrow \mathbb{C}^B \quad [\mu(a,b) = ab]$$

such that the axioms Eqs. (5.77, 5.79, 5.81, 5.83, 5.85) hold. $\eta$ and $\mu$ are linear whereas $t$ is anti-linear.

**Remark 98.** We will think of the three linear/anti-linear maps as the realifications of the tensors given by their coefficients in the canonical basis of $\mathbb{C}^B$, as described in Remark (41).

$$\eta_i = \begin{array}{c}\text{⊐}\end{array}\!\!— i \tag{5.71}$$

$$t^i_j = \; i \; —\!\!\triangleleft\!\!\triangleright\!\!— j \tag{5.72}$$

$$\mu^{ij}_k = \begin{array}{c} i \\ \text{⊐}\!\!\!\text{⊐}\!— k \\ j \end{array} \tag{5.73}$$

We will omit the complex arrow orientations in the future.

**Proposition 7.** For every complex-real 2E-CTL the complex tensors underlying the basic tensors define a *-algebra. The tensors $\eta$, $t$ and $\mu$ are given by

$$\eta_i = \bullet\!\!-\!\!\triangleright i \tag{5.74}$$

$$t^i_j = \begin{array}{c} i \\ \triangleleft\!\!\!\bullet\!\!\!\triangleright \\ j \end{array} \tag{5.75}$$

$$\mu^{ij}_k = \begin{array}{c} i \qquad j \\ \triangle \\ k \end{array} \tag{5.76}$$

We will show this by giving a list of all *-algebra axioms in the conventional form as well as for the corresponding tensors as described in Remark (41) in tensor-network notation. Then we show that the axiom holds for the CTL tensors Eq. (5.74) by giving the corresponding CTL axiom:

1. The identity is left invariant by the *-operation:

$$\mathbf{1}^* = \mathbf{1} \qquad t \circ \eta = \eta \qquad \text{⊐}\!\!\triangleright\!— = \text{⊐}\!— \tag{5.77}$$

$$\begin{array}{c}\bullet\!\!\!\bullet\!\!\!\triangleright a\end{array} = \bullet\!\!-\!\!\triangleright a \tag{5.78}$$

2. The *-operation is an *involution* by

$$(a^*)^* = a \qquad t \circ t = \mathbb{1} \qquad —\!\!\triangleright\!\!\triangleright\!— = —— \tag{5.79}$$

$$\begin{array}{c} a \\ \bullet\!\!\!\bullet\!\!\!\bullet \\ b \end{array} = \begin{array}{c} a \\ \bullet\quad\bullet \\ b \end{array} \tag{5.80}$$

which holds if we impose the local support convention Eq. (5.28).

3. The *identity property* is given by

$$a \cdot \mathbf{1} = a = \mathbf{1} \cdot a \qquad \mu \circ (\mathbb{1} \otimes \eta) = \mathbb{1} = \mu \circ (\eta \otimes \mathbb{1})$$

$$\begin{array}{c}\text{⊐}\!\!\!\text{⊐}\end{array} = ——— = \begin{array}{c}\text{⊐}\!\!\!\text{⊐}\end{array} \tag{5.81}$$

$$\begin{array}{c} a \\ \bullet\!\!\!\bullet\!\!\!\bullet \\ b \end{array} = \begin{array}{c} a \\ \bullet\quad\bullet \\ b \end{array} = \begin{array}{c} a \\ \bullet\!\!\!\bullet\!\!\!\bullet \\ b \end{array} \tag{5.82}$$

which again holds if we impose the local support convention.

4. The *-property* is given by

$$(ab)^* = b^* a^* \qquad t \circ \mu = \mu \circ (t \otimes t) \circ \text{Swap}$$

$$\begin{array}{c}\triangle\!\!\triangleright\!—\end{array} = \begin{array}{c}\triangleleft\!\!\!\triangle\end{array} \tag{5.83}$$

$$\begin{array}{c} a \;\; b \\ \triangle\!\!\!\triangle \\ c \end{array} = \begin{array}{c} a \;\; b \\ \triangle \\ c \end{array} = \begin{array}{c} b \;\; a \\ \triangle\!\!\!\triangle \\ c \end{array} \tag{5.84}$$

where $\text{Swap}$ is the operation that exchanges the two arguments.

5. *Associativity* is given by

$$(ab)c = a(bc) \qquad \mu \circ (\mu \otimes \mathbb{1}) = \mu \circ (\mathbb{1} \otimes \mu)$$

$$\begin{array}{c}\triangle\!\!\!\triangle\end{array} = \begin{array}{c}\triangle\!\!\!\triangle\end{array} \tag{5.85}$$

$$\begin{array}{c} b \;\; a \\ \triangle\!\!\!\triangle \\ c \quad d \end{array} = \begin{array}{c} b \;\; a \\ \triangle\!\!\!\triangle \\ c \quad d \end{array} = \begin{array}{c} b \;\; a \\ \triangle\!\!\!\triangle \\ c \quad d \end{array} \tag{5.86}$$

Now let us see to what extent the reverse of this statement is true.

**Remark 99.** A *-algebra cannot directly yield a 2E-CTL via the reverse identification Eq. (5.74). This is because *-algebras have a richer set of gauge transformations compared to 2E-CTLs: The indices of the *-algebra tensors are divided into input and output indices (which we indicated by an "arrow of time" from left to right), such that input indices are only contracted with output indices. So if we apply an arbitrary invertible linear map $G$ to all input indices and at the same time $G^{-1}$ to all output indices this will leave the *-algebra axioms invariant. E.g., the product changes as

$$\begin{array}{c}\triangle\end{array} \longrightarrow \begin{array}{c} —\boxed{G}\;\; \\ \qquad\quad\;\triangle\!—\boxed{G^{-1}}— \\ —\boxed{G}\;\; \end{array} \tag{5.87}$$

For the corresponding CTL tensor only orthogonal/unitary transformations $G$ would be allowed such that $G^{-1} = G^\dagger$.

**Remark 100.** As a consequence of Remark (99) there are some 2E basic axioms that are not invariant under gauge transformations for the *-algebra tensors: Whereas the basic axioms A1) and A3) are gauge invariant as equations for the *-algebra tensors, the other basic axioms A2), A4) and A5) are not, as input and output indices are contracted or equated:

$$
\begin{aligned}
A2) &\Rightarrow \quad \text{} \\
A4) &\Rightarrow \quad \text{} \\
A5) &\Rightarrow \quad \text{}
\end{aligned}
\tag{5.88}
$$

So whether those equations hold or not change when applying a gauge transformation to the *-algebra.

**Proposition 8.** Each *-algebra is equivalent to a complex-real 2E-CTL. I.e., there exists an invertible linear map $G$ such that after gauging with $G$ as in Eq. (5.87), Eq. (5.88) hold.

This follows from the well-known statement that every (finite-dimensional unital C)*-algebra is gauge equivalent to a direct sum of full matrix algebras, whose multiplication corresponds to the tensor Eq. (5.67) with $\omega_\alpha = 1 \forall \alpha$. We can further bring it into the standard form for complex 2E-CTLs with the following gauge matrix $G$:

$$
S^{(\alpha i j)}_{(\beta k l)} = \alpha \; \substack{i \\ j} \underbrace{\bullet}_{} \substack{k \\ l} \; \beta \; , \text{ with } v_\alpha = |B_\alpha|^{1/2}
\tag{5.89}
$$

**Remark 101.** We can construct the gauge transformation $S$ from Prop. (8) in the following way. Define the following matrix $X$:

$$
\text{}
\tag{5.90}
$$

$X$ is Hermitian as we can shift the involution from the $j$-leg to the $i$-leg with Eqs. (5.79, 5.83).

If we apply a gauge transformation $S$ to the *-algebra, $X$ changes as

$$
\text{}
\tag{5.91}
$$

If we want to use the *-algebra tensors as 2E-CTL tensors we have

$$
\text{}
\tag{5.92}
$$

So we have to find $S$ such that

$$
SXS^\dagger = \mathbb{1} \quad \Rightarrow \quad X = S^{-1}(S^{-1})^\dagger
\tag{5.93}
$$

For this to be possible, $X$ has to be a positive matrix. Then if $X = U\Lambda U^\dagger$ is the eigenvalue decomposition of $X$, $S$ is given by

$$
S = V(\Lambda)^{-1/2}U
\tag{5.94}
$$

for an arbitrary unitary $V$.

**Observation 67.** It is in particular the symmetry axioms of the CTL tensors that do not hold for the according *-algebra tensors if we do not fix the correct (non-orthogonal part of the) gauge. E.g., we have

$$
\text{}
\tag{5.95}
$$

but not

$$
a \longrightarrow b \;=\; b \longrightarrow a
\tag{5.96}
$$

and we have

$$
\text{}
\tag{5.97}
$$

but not

$$
\text{}
\tag{5.98}
$$

**Remark 102.** Similarly, every 2E-CTL (that is not necessarily complex-real) defines a real (unital finite-dimensional C)*-algebra, i.e., the equivalent set of tensors over a real vector space (where the involution is not complex anti-linear but just real linear), obeying the equivalent axioms. Those algebras are classified by a direct sum of matrix algebras over the reals, the complex numbers, or the quaternions, which are all examples of 2E-CTLs, so the correspondence goes in both directions.

### 5.4.7 Physical interpretation and connections to known models

2E-CTLs are essentially equivalent to the 2-dimensional lattice TFTs introduced in Ref. [24]. However, there are big differences in our way of presenting them and in various technical details. E.g., our framework makes it clear that instead of geometry pure combinatorics is all we need and that there is no orientation needed. Also the real (non-complex) case is included as well.

**Classical statistical physics**

The delta 2E-CTLs represent a classical 2-dimensional Ising model (for $|B| = 2$, otherwise $|B|$-state Potts model) at zero temperature and magnetic field. So they are fixed point models for symmetry breaking order. The tensor network that the TL associates to a patch of square lattice looks like:

$$
\text{}
\tag{5.99}
$$

The transfer matrix for this tensor network is directly a projector (here as a map from the bottom to the top indices):

$$
\text{}
\tag{5.100}
$$

Every tensor of the form

$$
v = \alpha_0 \left|\ldots, 0, 0, 0, \ldots\right\rangle + \alpha_1 \left|\ldots, 1, 1, 1, \ldots\right\rangle + \ldots
\tag{5.101}
$$

is a fixed point of this transfer operator. The solutions with $\alpha_i = 0$ for all $i$ apart from a single one correspond the symmetry broken sectors of the model. The symmetry-preserving fixed point with $\alpha_0 = \alpha_1 = \ldots$ is a classical analogue to the quantum *GHZ state* [48], and has long-range correlations.

The double-line 2E-CTLs are fixed point models for the trivial phase: The tensor network that the TL associates to a patch of square lattice consists of disconnected parts:

$$(5.102)$$

Here we omitted normalization factors. Due to the disconnectedness it is in the same TL phase as a normalization TL and therefore in the trivial physical TL phase.

The complex number 2E-CTL is an example of a CTL that does not directly describe a classical statistical model (see Remark (81)): There is no gauge transformation after which all entries of the basic tensor are non-negative. However, if we consider for example the direct sum of the trivial CTL with the complex number CTL, we do get one: This direct sum is gauge equivalent to the group CTL of $Z_3$ (as seen in example (109)) whose basic tensor has entries of $0$ and $1$ only. Similarly, all other group 2E-CTLs define valid classical statistical models that are gauge-equivalent to some non-trivial direct sum containing complex number or quaternion CTLs. It is unclear what physical meaning those new non-symmetry-breaking TL phases have, as they cannot be detected as long-range correlations of local observables (they can however, if we introduce defects).

Also the quaternion 2E-CTL has negative entries in any gauge. Additionally, they evaluate to negative numbers on some backgrounds, e.g., $-2$ for the real projective plane. So the whole TL phase given by the quaternions cannot correspond to a classical statistical model.

## Quantum many-body physics

In the world of quantum physics, complex-real 2E-CTLs describe $(1+1)$-dimensional fixed point models for quantum symmetry-breaking phases. There is a local commuting projector Hamiltonian: To each pair of neighbouring edges of the index lattice of a lattice we can glue a lattice consisting of a single 4-gon face without changing its background. E.g.,

$$(5.103)$$

Thus, each ground state is invariant under applying this CTL tensor interpreted as (the realification of) a complex-linear map from the two contracted indices to the two remaining indices:

$$(5.104)$$

This map is a symmetric (self-adjoint) projector due to the gluing and symmetry axiom. Now the overall Hamiltonian is given by

$$H = \sum_i (1 - P^{(i)}) \tag{5.105}$$

where $i$ runs over all vertices of the index lattice, and $P^{(i)}$ is the projector acting on the pair of edges around the vertex $i$.

For a complex-real 2E-CTL in its standard form, the global Hamiltonian projector looks like (omitting normalizations and complexification):

$$(5.106)$$

The symmetry-preserved ground state looks like (again omitting normalization and complexification):

$$(5.107)$$

which is also known as the *isometric form for (non-injective) MPS* [2]. It consists of a GHZ state and one EPR pair shared by each pair of neighbouring edges of the space. Apart from normalizations, we can get rid of those EPR pairs by a generalized local unitary circuit. So complex-real 2E-CTLs model quantum systems with long-range entanglement of GHZ type.

2E-CTLs also have an interpretation in the non complexreal case: They correspond to fixed point models of $(1+1)$-dimensional quantum phases protected by *time-reversal symmetry*. For quantum spin systems a time-reversal symmetry is a local anti-unitary operation that squares to identity (where local means that it is representable as a TL operator, or in other words a co-dimension 1 defect, which is unitary and non-framed). We can always find a local basis such that such an anti-unitary operator is just complex conjugation in this basis. So fixed point models with time-reversal symmetry are described by general (non complex-real) CTLs. The complexification of this CTL yields a complex-real CTL that is symmetric under complex conjugation.

Using our framework we find that there are three irreducible time-reversal symmetric phases in $1+1$ dimensions, corresponding to the trivial, the complex number and the quaternion 2E-CTLs.

In the group cohomology classification [49] there are two SPT phases protected by time-reversal symmetry. The trivial SPT phase corresponds to the trivial 2E-CTL. The non-trivial phase corresponds to the quaternion 2E-CTL. A fixed point model for the non-trivial SPT phase is known to be the following 1-dimensional cluster Hamiltonian [50]:

$$H = \sum_i -X_{i-1}Z_i X_{i+1}, T = K \bigotimes_i Z \tag{5.108}$$

Here $i$ labels the lattice sites, $X, Y, Z$ are the Pauli operators, $T$ is the anti-unitary operator representing the time-reversal symmetry, and $K$ is complex conjugation in the $Z$ basis. After a change of basis we get to a form where time-reversal is only complex conjugation, and the Hamiltonian is a real tensor:

$$H = \sum_i Y_{i-1}Z_i Y_{i+1} = -(XZ)_{i-1}Z_i(XZ)_{i+1}, T = K \tag{5.109}$$

The local ground state projector on three neighbouring qubits is given by

$$P = (1 - XZ \otimes Z \otimes XZ)/2 \tag{5.110}$$

If we block pairs of neighbouring sites into one double-qubit, the local ground state projector becomes the product of two ground state projectors above, acting on two double-qubits:

$$P_{\text{blocked}} = (1 - XZ \otimes Z \otimes XZ \otimes 1)$$
$$(1 - 1 \otimes XZ \otimes Z \otimes XZ)/4 \qquad (5.111)$$
$$= (1 - XZ \otimes Z \otimes XZ \otimes 1 - 1 \otimes XZ \otimes Z \otimes XZ$$
$$- XZ \otimes X \otimes X \otimes XZ)/4$$

On the other hand we can think of the 4-element basis set of the quaternion 2E-CTL as $2 \times 2$. One can easily see that

$$(5.112)$$

Here the index $a$ corresponds to the list entry, and $b$ and $c$ are the ingoing and outgoing indices on the double qubit Pauli products, respectively. Using Eq. (5.104) we get:

$$P = (1 - 1 \otimes XZ \otimes Z \otimes XZ - XZ \otimes Z \otimes XZ \otimes 1$$
$$- XZ \otimes X \otimes X \otimes XZ)/4 \qquad (5.113)$$

This is exactly equal to the projector $P_{\text{blocked}}$ above. So the quaternion 2E-CTL is exactly this time-reversal protected SPT phase.

The complex number 2E-CTL does not correspond to a SPT phase from the literature, as it is not in the trivial phase when we remove the symmetry: Removing the symmetry means taking the complexification of the complex number 2E-CTL without enforcing the complex conjugation symmetry. Now stacking the complex numbers with the complex numbers yields a direct sum of twice the complex numbers, i.e., the complexification of the (2-element) delta 2E-CTL. So without the symmetry the model is in the ordinary $\mathbb{Z}_2$ symmetry-breaking phase. We should thus consider this novel phase to be a SET rather than a SPT phase.

We can also perform the gauge transformation to obtain the complexified (2-element) delta 2E-CTL when we do enforce the symmetry. However, then we have to transform the symmetry representation in the same way: In the new basis the representation is given by $XK$ where $X$ is the flip between the symmetry-broken sectors and $K$ is complex conjugation. This is in a different phase than the same model with just $K$ as symmetry representation: They can be distinguished by invariants. For example we can put the models on a Klein bottle with a symmetry defect winding around the orientation-preserving loop: The model with only $K$ yields 4 as the parity flip around the Klein bottle and the time-reversal defect cancel out, whereas the model with $XK$ yields 0 as there is a closed loop with a single $X$ flip.

Also in the non-complexified formulation the complex number 2E-CTL and the (2-element) delta 2E-CTL can be distinguished by invariants: Whereas the complex number 2E-CTL yields 0 on all non-orientable manifolds, the delta 2E-CTL yields $2^C$ where $C$ is the number of connected components, even on non-orientable manifolds.

Note that the role of the symmetry in this new SET phase is quite interesting: Conventionally a symmetry can play two roles in the classification of phases: First, in symmetry-breaking phases imposing the symmetry makes the phase robust to perturbations (in 2 or more dimensions), but is not important for the definition of the phase: A symmetry-breaking model cannot be transformed to a trivial one even when breaking the symmetry. In SPT phases the symmetry is what makes the phase non-trivial, however, those phases do not need any symmetry for robustness. In the SET phase corresponding to the complex number 2E-CTL, the complex-conjugation symmetry takes both roles at the same time: It makes the phase different from the ordinary symmetry-breaking phase, and at the same time it makes it robust to perturbations (i.e., irreducible).

## 5.5 $2/1$EE-(C)TLs

### 5.5.1 Definition

By $2/1$**EE-CTLs** we actually refer to a collection of different CTL types coming from TL types. "EV" stands for "edge, vertex", as this is where indices are associated to.

**Backgrounds**

The extended backgrounds for the different CTL types are given by boundary $2/1$-manifolds with any $0/0$-manifold as boundary central link. The backgrounds of the corresponding TL types are $2/1$-manifolds with the same $0/0$-manifold as central link. E.g.,

$$a) \quad \bullet \quad b) \quad \bullet \quad \bullet \quad c) \quad \bullet \quad \bullet \quad d) \qquad (5.114)$$

Each $2/1$EE-CTL type contains 2E-CTLs multiple times as sub-type, namely once for each 0-region of the boundary central link. a) corresponds to a physical boundary of that sub 2E-CTL, b) to a domain wall between the two sub 2E-CTLs, c) to a defect line within the sub 2E-CTL and d) a line where the three sub 2E-CTLs meet.

If one region of the boundary central link has more than one point then we can also consider adding a framing of the central 1-region with respect to this region. Alternatively one can also consider a complete ordering of the points in the boundary central link.

**Example 110.** Consider the following examples for extended backgrounds (see Remark (3) for example d)).

$$(5.115)$$

a) to d) show different backgrounds for the boundary central links a) to d) in Eq. (5.114). If we had a framing or an ordering of the points within one region of the boundary central link, then the bulk 1-region in c) would have little arrows pointing perpendicular to that region.

The index background is the 1/1-manifold forming the boundary of the boundary 2/1-manifolds.

**Example 111.** Consider, e.g., the following extended backgrounds (left) together with their index background (right):

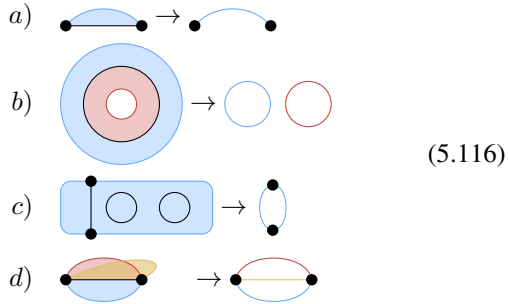

(5.116)

The background gluings are surgery gluings for all boundary regions of the boundary 2/1-manifolds. In the case a) and c) there are 2, in case b) there are 3, and in case d) there are 4 such regions that have a gluing: One boundary 1-region for each 0-region of the boundary central link and the central boundary 0-region.

**Example 112.** Consider, e.g., the following examples of 0-surgery gluings:

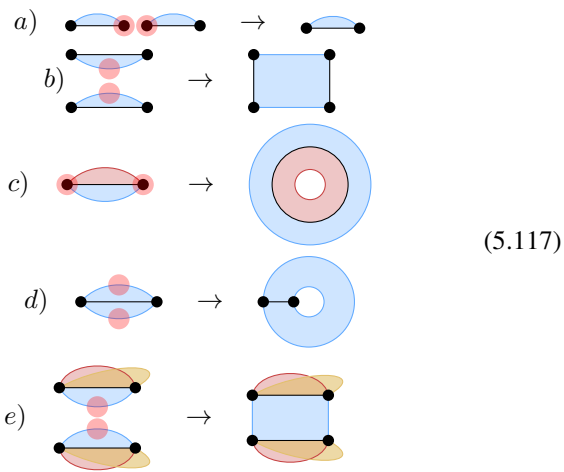

(5.117)

a) and b) correspond to the boundary central link a) in Eq. (5.114), whereas c), d) and e) correspond to b), c) and d), respectively.

### Lattices

The lattices are based on 2/1CCbt-lattices living on the extended backgrounds: The central boundary 0-region is 2-thickened. The edges of the boundary 1-regions carry orientations, just as the corresponding sub 2E-CTLs. The index lattice of a lattice is given by all the 2E edges and 2/1EE edges of the 1/1CC-lattice forming the boundary. The basic moves are given by higher order bi-stellar flips for all bulk regions.

In order to distinguish the 2/1EE edges from the other edges we will draw them fat.

**Example 113.** Consider the following examples for extended backgrounds (left) with lattices representing them (middle) as well as the corresponding index lattices (right):

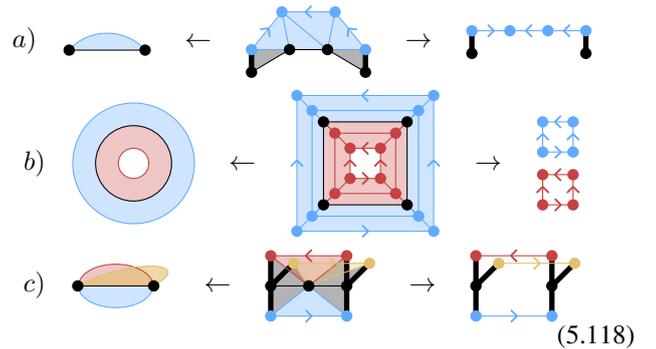

(5.118)

where a), b) and c) correspond to the central boundary links a), c) and d) in Eq. (5.114), respectively.

For boundary 1-region there is one basic gluing, namely higher order cell gluing at this region (respecting the edge orientation), just as for the sub 2E-CTLs. Additionally there is higher order cell gluing for the boundary 0-region, i.e., pairs of 2/1EE edges can be glued.

**Example 114.** Consider the following examples for gluings:

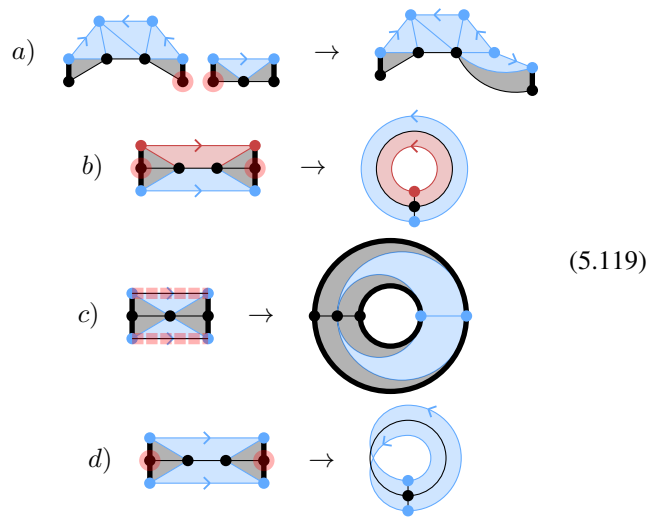

(5.119)

The gluings a), b) and c) represent the background gluings a), c), and d) in Eq. (5.117), respectively. Note that if the boundary central link has a region consisting of multiple vertices, there are different identifications of the adjacent 2/1EE edges that yield different gluings. E.g., in d) one can either obtain a Möbius strip as shown, but another gluing of the same 2/1EE edges would yield an annulus.

### Tensors

The tensors are real tensors with one index type for every region of the boundary central link, referred to as 2E indices, and one additional index type referred to as 2/1EE indices. The index prescription associates one 2E index of each type to each edge of the corresponding boundary 1-region. Additionally it associates one 2E index to every edge of the central boundary 0-region.

When two 2E edges or two 2/1EE edges are glued, the two associated indices are contracted.

### 5.5.2 Basic tensors and axioms

**Observation 68.** A set of basic lattices is given by the basic lattices of all sub 2E-CTLs together with the following lattice: The index lattice consists of two 2/1EE edges connected to 2E vertices which are pairwisely separated by one 2E edge each. The orientations for all those edges are aligned. The extended background is the 1-sphere completion of the boundary central link. Consider, e.g., the following basic lattices (middle), their extended backgrounds (left) and index lattices (right):

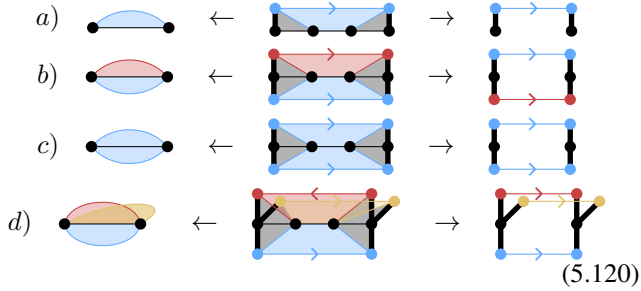

(5.120)

Here a), b), c) and d) correspond to the boundary central links a), b), c) and d) from Eq. (5.114), respectively.

A defining history for a given lattice can be obtained in the following way: Within every bulk 2-region, apply the defining history for the corresponding sub 2E-CTL (via the barycentric subdivision). For every edge of the central bulk 1-region take one copy of the basic lattice above. The edge orientations can either be chosen arbitrarily, or by performing a barycentric subdivision before. In the end glue all basic lattices according to how they are located.

**Observation 69.** According to Observation (68) the whole 2/1EE-CTL is already determined by the basic tensors of the sub 2E-CTLs and the basic tensor associated to the basic lattice above. E.g., for the boundary central link b) in Eq. (5.114) we get the following tensor (where we only draw the index lattice and assume the interior topology of the background, see Remark (1)):

$$T1) \quad G^{xy}_{ab} := x \begin{array}{c} a \\ \vdots \\ b \end{array} y \qquad (5.121)$$

**Proposition 9.** A set of basic history lattices is given by the basic history lattices of all sub 2E-CTLs and the additional basic history lattice consisting of 3 2/1EE basic lattices and one 2E basic lattice for each point of the boundary central link. E.g.,

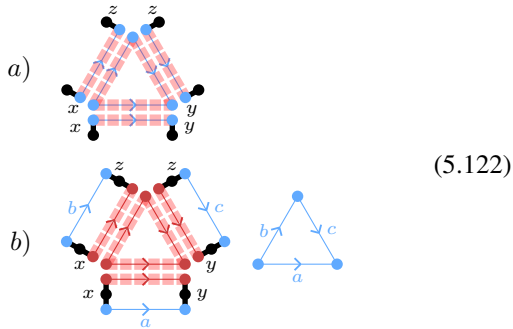

(5.122)

Here again a) and b) are for the boundary central links a) and b) from Eq. (5.114), respectively.

A defining history for a given history lattice is the following procedure: History lattices are in the same class as 3/1CCb-lattices with the same boundary central link as the 2/1EE-CTL itself. (The volumes (tetrahedra) of the bulk 3-regions correspond to basic moves of the corresponding sub 2E-CTL, and the faces (triangles) are the basic moves acting on the bulk 1-region of the original CTL lattices.) For every volume of each bulk 3-region take one 2E basic history lattice, and each face of the bulk 2-region by the 2/1EE basic history lattice. Then history glue all those basic history lattices according to how they are located. The remaining 2E basic lattices at the faces of the boundary 2-regions and 2/1EE basic lattices at the edges of the boundary 1-region form the desired history lattice.

**Observation 70.** According to Remark (24), the bi-partitions of the basic history lattices are a good candidate for basic history moves/axioms. All such bi-partitions have two of the 2/1EE basic lattices on one side and one on the other side. The 2E basic lattices can be arbitrarily on both sides. Those axioms look like e.g.:

$$
\begin{array}{c}
a) \\[30pt]
b) \\[30pt]
c)
\end{array}
\qquad (5.123)
$$

a) and b) correspond to the boundary central link a) in Eq. (5.114), whereas c) corresponds to the boundary central link b).

All those bi-partitions with all different edge orientations indeed define a set of basic history moves. The defining history above is already consistent with disjoint union and gluing. Together with the 2E basic history moves it is also consistent with the 2E moves. Together with the 2/1EE basic history lattices it is also consistent with the 2/1EE moves.

The basic history moves yield a full set of basic axioms which imply all the CTL axioms, except for the symmetry axiom. To this end we have to add axioms corresponding to all symmetries of the basic lattice (which are only non-trivial if there are regions in the central boundary link with more than one point).

**Remark 103.** The number and complexity of axioms can be significantly reduced by using the additional basic tensor T1) together with the axioms A1) to A5) for all the sub 2E-CTLs.

Then we get the following equivalent set of axioms:

$$(5.124)$$

A1) is the basic axiom from Eq. (70) where all the 2E basic lattices are on the side with only one 2/1EE basic lattice, for one particular edge orientation. A2) is the axiom coming from the fact that gluing the 2E basic tensor T2) to all 2E edges of the 2/1EE basic lattice reverses all the edge orientations. Both are exemplarily shown for the boundary central link b) in Eq. (5.114). A3) is the basic axiom coming from the symmetries of the 2/1EE basic lattice (which is actually already implied by our notation). It is exemplarily shown for the boundary central link c), as for a), b) or d) it would be trivial.

Together with the 2E axiom A2) we can bring any of the 2E basic lattices in A1) above to the right side. Together with the 2E axioms A4) or A5) and the axiom A2) above we can invert the orientations of any non-glued edge in A1) above. With the 2E axioms A4) or A5), and A3) we can invert the orientation of any glued edge in the 2/1EE axiom above. So A1) to A3) together with the 2E axioms A1) to A5) for all sub 2E-CTLs imply all the basic axioms from Observation (70).

**Remark 104.** Alternatively one can use a larger set of simpler basic lattices: For each region of the boundary central link take the lattice consisting of one 2E edge (of the corresponding region) and two 2/1EE edges. The vertices to which the 2/1EE edges are connected are the same except for the ones connected to the 2E edge. E.g., for the boundary central link b) in Eq. (5.114) we get the two basic lattices:

$$(5.125)$$

For the associated basic tensors we get axioms similar to A1) and A2) above: e.g., for the boundary central link b) in Eq. (5.114) we have the following axioms A1) and A2):

$$(5.126)$$

and the same for the purple region instead of the blue region. The symmetry axiom A3) above is not needed any more. Ad-ditionally we get consistency axioms for the basic tensors corresponding to the different regions. E.g.,

$$(5.127)$$

### 5.5.3 Solutions

**Standard physical boundary**

For every 2E-CTL there is a 2/1EE-CTL with a single point as boundary central link, which has this 2E-CTL as the corresponding sub CTL. The 2/1EE indices have the same basis set as the 2E indices. The 2/1EE basic tensor is given by

$$(5.128)$$

With this choice the 2/1EE axioms reduce to the 2E axioms.

**Trivial defect line**

For every 2E-CTL there is a 2/1EE-CTL with boundary central link c) in Eq. (5.114). The basis set for the vertex indices is the same as that of the edge 2E indices. The 2/1EE basic tensor is defined as

$$(5.129)$$

This is just a realization of a trivial defect.

**Group homomorphism**

Consider two group 2E-CTLs for groups $G$ and $H$, and a group homomorphism

$$\alpha : G \to H \qquad (5.130)$$

giving rise to a tensor

$$(5.131)$$

which can be used to define a domain wall between the two 2E-CTLs. In order to have a basic tensor fulfilling all axioms we have to supplement it with a trivial defect of $G$ (let the blue 2E-CTL be the one of $H$ and purple the one of $G$):

$$(5.132)$$

**Projective group representations**

**Definition 173.** A **projective representation** of a group $G$ over some vector space with basis set $B$ is given by two

complex-real tensors

$$\tag{5.133}$$

with $g, h \in G$, $\quad a, b \in B$

Actually, it suffices if only the indices $a$ and $b$ are realified. The following axioms must hold:

$$\tag{5.134}$$

i.e., $\quad \tilde{\rho}(gh) = H(g,h)\tilde{\rho}(g)\tilde{\rho}(h)$

$$\tag{5.135}$$

i.e., $\quad |H(g,h)| = 1 \quad \forall g, h \in G$

For each (complexified) double-line 2E-CTL as in Eq. (5.36) with basis set $B$ and each finite group and projective representation over $B$ there is a 2/1EE-CTL whose basis set for the 2/1EE indices is given by the set of group elements $G$ times $B \times B$. The basic tensor is given by

$$\tag{5.136}$$

If we restrict the representation to a fixed group element $g$ we get a matrix $\tilde{\rho}(g)$. This matrix for any $g$ (just like any orthogonal/unitary matrix) also defines a 2/1EE-CTL by

$$\tag{5.137}$$

However, such a 2/1EE-CTL is directly gauge equivalent to the trivial defect by a gauge transformation on the 2/1EE indices. So also the 2/1EE-CTL in Eq. (5.136) is gauge equivalent to a direct sum of $|G|$ trivial defects.

**Comment 17.** This solution reminds one to the setup for SPT phases in $1 + 1$ dimensions in Ref. [3]: The tensors Eq. (5.36) are nothing but the canonical isometric form of injective MPS where the physical symmetry action equals the virtual one. However, we found that those symmetries are topologically trivial as co-dimension 1 defects. So, unlike for some SET phases, we cannot classify such SPT phases as TL phases of invertible defects. In Section (6.1.3) we will outline how those SPT phases also fit into our framework.

**Double-line physical boundary**

For each set $B$ there is a 2/1EE-CTL with a single point as boundary central link whose sub 2E-CTL is the double-line 2E-CTL for $B$. The basis set for the 2/1EE indices is $B$ (in contrast to $B \times B$ for the 2E indices). The 2/1EE basic tensor is given by

$$\tag{5.138}$$

**Complex numbers**

For an arbitrary boundary central link and a set of one arrow direction for each of its vertices the complex number tensors define a 2/1EE-CTL whose sub 2E-CTLs are the complex number 2E-CTL. The 2/1EE basic tensor is the complex number tensor, whose arrow directions for the 2E indices are the chosen arrow directions for the vertices of the boundary upper link: e.g., for a boundary central link with three points corresponding to three different regions, ingoing arrow directions for the purple point, and outgoing orientations for the blue and purple points, we get:

$$\tag{5.139}$$

**Quaternions and complex numbers**

Consider boundary central link and choose for each of its regions either the trivial, the complex number or the quaternion sub 2E-CTL. Then for all points that have the complex number 2E-CTL choose one arrow direction. For this setup, a 2/1EE-CTL can be constructed in the following way: We have to distinguish two cases: either there is an even or an odd number of points of the boundary central link that have the quaternion sub 2E-CTL.

Let us start with the case when the number of such points is even. Pair up all points of the boundary central link that have a quaternion sub 2E-CTL. Now for each pair of points take the trivial defect between the corresponding quaternion 2E-CTLs. For the remaining quaternion 2E-CTL (if there is one) take the standard boundary. For all points that have a complex number sub 2E-CTL, take the complex number 2/1EE-CTL for the boundary central link restricted to those points, and the chosen arrow directions. For all points that have a trivial sub 2E-CTL, take the standard (trivial) boundary. Now fuse all those 2/1EE-CTL for parts of the boundary central link together to obtain one 2/1EE-CTL for the total boundary central link.

Now consider an odd number of points in the boundary central link that have the quaternion sub 2E-CTL. Again pair up the points with the quaternion sub 2E-CTL, and take the trivial defects between all those pairs. This time there is one remaining point. For this point and all the points that have the complex number sub 2E-CTL choose the following 2/1EE-CTL with a 4-element basis set for the 2/1EE indices: The basic tensors in the form of Eq. (5.125) are given by the following: For the point with the quaternion sub 2E-CTL it is given by the standard physical boundary for the quaternion 2E-CTL:

$$x \blacktriangledown_{\overset{a}{\longrightarrow}} y \;=\; \overset{a}{\underset{x \to \boxplus \to y}{\downarrow}} \tag{5.140}$$

For points with the complex sub 2E-CTL it is given by e.g.:

$$x \blacktriangledown_{\underset{a}{\longrightarrow}} y \;=\; \begin{pmatrix} \mathbf{1}:1 & & & \mathbf{i}:-1 \\ & \mathbf{1}:1 & \mathbf{i}:1 & \\ & \mathbf{i}:-1 & \mathbf{1}:1 & \\ \mathbf{i}:1 & & & \mathbf{1}:1 \end{pmatrix} \tag{5.141}$$

where $x$ and $y$ label the row and column of the matrix, and $a$ is presented in sparse form, see Remark (15). If the arrow direction is outgoing instead of ingoing, we have to add a complex conjugation to the $a$ index.

Now fuse together all the trivial 2/1EE-CTLs for the points with trivial sub 2E-CTL, the trivial defects for all pairs of points with quaternion sub 2E-CTL and the 2/1EE-CTL described above for the selected remaining quaternion point and all complex number points. This yields a 2/1EE-CTL for the boundary central link with the given sub 2E-CTLs that we will refer to as **complex-quaternion** 2/1**EE-CTL**.

**Delta projection**

For every set $B$ with an element $b \in B$ there is a 2/1EE-CTL that is a physical boundary for the delta 2E-CTL for $B$. The basis set for the 2/1EE indices is trivial (a one-element set), and the basic tensor is given by

$$x \underset{\bullet}{\overset{a}{\longrightarrow}} y \;=\; \overset{a}{\underset{(b)}{\mid}} \tag{5.142}$$

and $x$ and $y$ are trivial. The vector $b$ has entry 1 for $a = b$ and entry 0 otherwise, i.e., $b_a = \delta_{b,a}$. Consider an arbitrary boundary central link and for each region one delta sub 2E-CTL, together with an element of the according basis set $B$ for each point in the region. We can consider the 2/1EE-CTL for this setup by fusing all the physical boundaries like above for all the points in the boundary central link with the corresponding set elements. To such a 2/1EE-CTL we will refer to as a delta projection 2/1EE-CTL.

### 5.5.4 Properties and classification

**Remark 105.** Consider the background with only two 2/1EE edges and no 2E edges. E.g.,

$$a) \quad \bigwedge \qquad b) \quad \bigblack \qquad c) \quad \bowtie \tag{5.143}$$

where a), b) and c) correspond to the boundary central links a), b) and d) in Eq. (5.114).

This lattice is symmetric under reflection at the vertical axis. Gluing two copies at one 2/1EE edge each yields another copy, and gluing it to any 2/1EE edge of any lattice does not change its background. So the associated CTL tensor is a symmetric (self-adjoint) projector when interpreted as linear map from left to right. So we can set it to identity by the local support convention. E.g.,

$$x \blacklozenge y \;:=\; x \;\text{——}\; y \tag{5.144}$$

**Observation 71.** The gauge transformations for 2/1EE-CTLs consist of applying one independent orthogonal/unitary matrix for all index types, i.e., one for every region of the boundary central link, and one for the 2/1EE indices.

**Proposition 10.** Consider a 2/1EE-CTL such that all sub 2E-CTLs are either the trivial, the complex number, or the quaternion 2E-CTL. Every such 2/1EE-CTL can be turned into a complex-quaternion 2/1EE-CTL for some complex arrow orientations by a gauge transformation acting only on the 2/1EE indices. All such complex-quaternion 2/1EE-CTLs with the same complex arrow orientations (but different pairings of the quaternion points) are gauge equivalent. Also simultaneously flipping the complex arrow orientations for all points can be achieved by a gauge transformation on the 2/1EE indices. If we also allow for gauge transformations on the 2E indices of a region, we can also simultaneously flip the complex arrow orientations of all points of that region.

**Proposition 11.** For every 2/1EE-CTL $A$ we can bring all the sub 2E-CTLs in the standard form by gauge transformations on the 2E indices. In this form the 2E-CTLs are a direct sum whose components are the tensor product of a double-line 2E-CTL and the trivial, complex number, or quaternion 2E-CTL. Now by further gauge transformations on the 2/1EE indices we can bring it in direct sum form $A = \bigoplus_{i \in I} A_i$, where each $A_i$ is the tensor product of a delta projection 2/1EE-CTL, a double-line 2/1EE-CTL, and complex or a complex-quaternion 2/1EE-CTL. We will refer to the latter CTLs as **in the standard form**. E.g., for the boundary central link a) in Eq. (5.114) the 2/1EE-CTL basic tensor can be denoted in generalized tensor-network notation as

$$(u,v,w) \overset{(a,b,c,d)}{\underset{\bullet\quad\quad\bullet}{\longrightarrow}} (x,y,z) \;:=\; \begin{array}{c} d \;\; c \\ \vert \;\; \vert \\ w \;\overset{a\vert b}{\underset{(\mu)}{\bigcirc}}\; z \\ u \;\text{——}\;\otimes\;\; x \\ v \;\text{————}\; y \end{array} \tag{5.145}$$

where the basis sets of the blue fat index type, the black index type and the black dotted index type all depend on the index value of the black fat index. The tensor $\mu$ is the linearized version of a function that associates to every element of the basis set of the black fat index type one element of the corresponding basis set of the fat blue index type. The $x$-tensor is either the trivial tensor, the complex number tensor or quaternion tensor, depending on the value of the black fat index.

As another example consider the boundary central link b) in

Eq. (5.114):

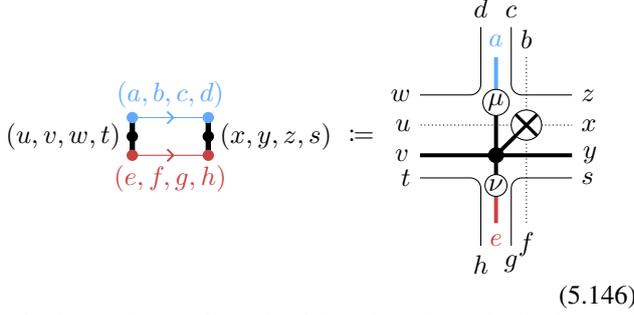

$$(u,v,w,t) \xrightarrow[(e,f,g,h)]{(a,b,c,d)} (x,y,z,s) :=$$

(5.146)

Now both $\mu$ and $\nu$ are linearized functions from the basis set of the black fat index type to the blue and purple fat index types. The $x$-tensor depends on whether $a$ and $e$ are trivial, complex number, or quaternionic blocks: If one of them is trivial, then it is the trivial, complex number or quaternion tensor with one index for the non-trivial side, as in Eq. (5.145). If both are complex number or quaternion blocks, then it is the complex number or quaternion tensor, with one index for both sides. For the quaternions, the arrow directions have to point both outwards, whereas for the complex numbers, they can point either both inwards (both outwards) or one inwards and one outwards. The last case is when one block is complex numbers and the other block quaternions. In this case we choose a product of the quaternion standard boundary Eq. (5.140) and the 4-dimensional physical boundary of the complex numbers Eq. (5.141). The order of the product is irrelevant as the two representations commute.

**Remark 106.** For complex-real 2/1EE-CTLs, all sub 2E-CTLs are complex-real, so their standard forms are a direct sum of tensor products of the double-line and complex number CTLs. So if we put the 2/1EE-CTL into the standard form all of the complex-quaternion 2/1EE-CTLs will be just complex number 2/1EE-CTLs.

### 5.5.5 Mappings and fusions

**Observation 72.** Every 2E-CTL can be mapped to the trivial co-dimension 1 defect for that 2E-CTL, as already shown in Eq. (5.129).

**Observation 73.** 2/1EE-CTLs with the empty manifold as boundary central link are directly equivalent to 1V-CTLs.

**Observation 74.** We can fuse a pair of 2/1EE-CTL and 1V-CTL to another 2/1EE-CTL by putting the 1-manifold on which the 1V-CTL lives along the defect line of the 2/1EE-CTL. Concretely we can get the basic tensor of the new 2/1EE-CTL by the tensor product of the old 2/1EE-CTL and 1V basic tensors, where we take as the new 2/1EE indices composites of the old 2/1EE indices and the 1V indices.

**Observation 75.** Consider a 2E-CTL together with two physical boundaries (which are 2/1EE-CTLs of type b) in Eq. (5.114)). Then we can restrict to backgrounds that consist of thin stripes of 2E-CTL framed by those physical boundaries. This yields a CTL effectively living on 1-manifolds with boundary, which are all equivalent to 1V-CTLs. I.e., we can fuse any pair of physical boundaries of a 2E-CTL to a 1V-CTL.

**Observation 76.** To generalize the previous observation consider a 1st order 1-manifold $M$ whose 1-regions are labeled by 2E-CTLs and whose 0-regions are labeled by 2/1EE-CTLs with their upper link as boundary central link. Now we can restrict to backgrounds that are $M$ times some 1-manifold with boundary and think of them as effectively 1-dimensional tube-like structures. This again yields a CTL on 1-manifolds with boundary.

**Observation 77.** For every set of 2E-CTLs there is a fusion that yields a 2/1EE-CTL for the separate 2E-CTLs and the stacking fusion of all of them. To this end replace the 2-region corresponding to the stacked 2E-CTL by a stack of all CTLs, replace the central 1-region by a line where the 2-regions of the separate 2E-CTLs join together to a stack.

**Observation 78.** Every 2/1EE-CTL can be fused to a physical boundary for the stack of all the sub 2E-CTLs. Conversely, from the previous observation we see that every set of 2/1E-CTLs together with a physical boundary of the stacked CTL can be fused to a 2/1EE-CTL with the 2E-CTLs as sub CTLs. So instead of searching for 2/1EE-CTLs with complicated boundary central link we can search for physical boundaries of the stacked CTLs formed by one CTL for each of the points in the boundary central link. Note that however, the division into gauge families and phases is different for both cases. On the other hand again, the gauge families and phases relative to all the sub 2E-CTLs are equal for both cases.

### 5.5.6 Connection to *-algebra representations

In this section we will discuss the relation between 2/1EE-CTLs and known algebraic structures. According to Observation (78) we focus on 2/1EE-CTLs with a single point as boundary central link, i.e., physical boundaries.

**Definition 174.** A **representation** of a (finite-dimensional unital C)*-algebra with basis set $B$ is given by some vector space with basis set $R$ and a linear map

$$\rho : \mathbb{C}^{B \times R} \longrightarrow \mathbb{C}^R \tag{5.147}$$

such that the axioms Eqs. (5.149, 5.150) hold. We will think of $\rho$ as the realification of a complex tensor:

$$\rho_y^{ax} = \begin{array}{c} a \\ \downarrow \\ x \to \boxed{\rho} \to y \end{array}, \quad a \in B, \quad x, y \in R \tag{5.148}$$

We will omit the complex arrow orientations in the future.

Now the axioms are given by

- 

$$\rho(\mathbf{a})\rho(\mathbf{b}) = \rho(\mathbf{ab}) \qquad \rho_y^{ax}\rho_z^{by} = \mu_c^{ab}\rho_z^{cx}$$

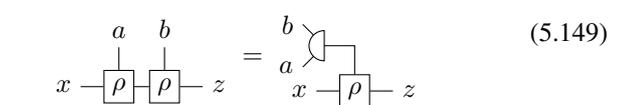

(5.149)

- 

$$\rho(\mathbf{1}) = \mathbb{1} \qquad \rho_y^{ax}\eta_a = \delta_y^x$$

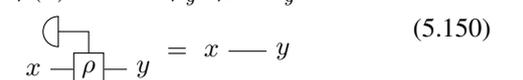

(5.150)

**Proposition 12.** Every complex-real 2/1EE-CTL (with a single point as boundary central link) defines a representation of the *-algebra corresponding to the sub 2E-CTL via Section (5.4.6). The corresponding tensor $\rho$ is given by

$$
x \underset{\phantom{x}}{\overset{a}{\boxed{\rho}}} y \;:=\; x \bullet \!\!\overset{a}{\longrightarrow}\!\! \bullet y
\tag{5.151}
$$

To demonstrate the validity of the proposition we will now write down the axioms for $\rho$ as a representation and the corresponding CTL axiom.

- Eq. (5.149) corresponds exactly to the 2/1EE basic axiom A1).

- Eq. (5.150) holds due to the following CTL axiom and the local support convention in Remark (105).

$$
x \bullet \!\!\!\begin{array}{c}\ \end{array}\!\!\! \bullet y \;=\; x \blacktriangle y \;=\; x \text{——} y
\tag{5.152}
$$

**Remark 107.** A *-algebra representation cannot directly yield a 2/1EE-CTL via the reverse identification Eq. (5.151). This is because *-algebra representations have a larger set of gauge transformations compared to 2/1EE-CTLs: The *-algebra representation tensor has one ingoing algebra index, one ingoing representation index, and one outgoing representation index. The *-algebra representation axioms only involve contractions of outgoing with ingoing indices. So if we apply an arbitrary invertible linear map $H$ to all representation input indices and at the same time $H^{-1}$ to all representation output indices this will leave the *-algebra representation axioms invariant. Additionally we can apply a gauge transformation $G$ for the *-algebra as in Remark (99), which then also has to act on the algebra index of the representation.

$$
x \underset{\phantom{x}}{\overset{a}{\boxed{\rho}}} y \;\rightarrow\; x \underset{\boxed{H}}{\overset{a\;\boxed{G}}{\boxed{\rho}}} \boxed{H^{-1}} y
\tag{5.153}
$$

For the corresponding CTL tensor only orthogonal/unitary transformations $G$ and $H$ would be allowed.

**Proposition 13.** For every *-algebra representation there are gauge transformations $G$ and $H$ as in Remark (107) such that the representation tensor (together with the *-algebra tensors) form a 2/1EE-CTL.

This follows from the known fact that every representation of a *-algebra is gauge equivalent to a direct sum of irreducible representations, each with some multiplicity. So we can first bring the *-algebra into its block diagonal form via a gauge transformation $G$ on the algebra indices, as in Remark (99). Then we can apply another gauge transformation $H$ on the representation indices to bring the representation into the following form:

$$
\begin{array}{c}
\overset{a\alpha b}{\underset{u}{\underset{v}{\underset{w}{\boxed{H}}}}}\;\boxed{\mathcal{G}}\;\boxed{\rho}\;\boxed{H^{-1}}\;\overset{x}{\underset{y}{\underset{z}{}}}
\end{array}
=
\begin{array}{c}
\overset{a\ \ b}{\underset{u}{\underset{v}{\underset{w}{}}}}\!\!\overset{\alpha}{\bullet}\!\!\overset{x}{\underset{y}{\underset{z}{}}}
\end{array}
\tag{5.154}
$$

for some invertible matrix $H$. The wiggly line is called **multiplicity index**, its dimension can depend on the value of the irrep index and is known as the **multiplicity** of the respective irrep. We can write the right hand side in the following alternative form, where the wiggly and the fat index types are blocked together and the realification is explicitly spelled out:

$$
\begin{array}{c}
\overset{d\ \ c}{\underset{w}{\underset{u}{\underset{v}{}}}}\!\!\!\overset{a}{\underset{\mu}{\downarrow}}\!\!\overset{b}{\underset{\downarrow}{}}\;\underset{x}{\underset{y}{\boxed{C}}}\;z
\end{array}
\tag{5.155}
$$

which is nothing but the standard form for a complex-real 2/1EE-CTL.

### 5.5.7 Alternatives

Similar to 2En-CTLs, one can introduce normalizations, such that if two glued 2/1EE edges are connected to a common 2E vertex, then we have to use a normalization matrix $C$ for the contraction. This way we can replace the double-line 2E-CTLs with their double-line physical boundaries by the continuous family of euler-characteristic 2En-CTLs with boundary.

### 5.5.8 Physical interpretation

#### Classical statistical physics

In the context of classical statistical physics, 2E-CTLs correspond to 2-dimensional models with 1-dimensional boundaries, or lines where multiple 2-dimensional classical models intersect. The delta projection 2/1EE-CTLs are physical boundaries that fix one of the symmetry broken sectors for the symmetry-breaking model that the sub 2E-CTL describes. The double-line 2/1EE-CTL is in the trivial physical TL phase: Just as the double-line sub 2E-CTL the corresponding TL is equivalent to an euler-characteristic normalization. The physical boundaries for the complex number or quaternion CTLs do not directly represent classical statistical systems (see Remark (81)), just as the corresponding sub 2E-CTLs. However, e.g., the standard physical boundary of a group 2E-CTL has non-negative tensor entries only and is gauge equivalent to a direct sum of the trivial, complex number, and quaternion CTLs and their physical boundaries.

#### Quantum physics

In the quantum world complex-real 2/1EE-CTLs describe $1+1$-dimensional quantum spin systems with a $0+1$ (or $1+0$)-dimensional physical boundary. General 2/1EE-CTLs correspond to quantum spin systems with a local anti-unitary (i.e., time-reversal) symmetry.

According to Observation (56) 2/1EE-CTLs (with a single point as boundary central link) also yield tensor network representations of ground states of the sub 2E-CTLs. The corresponding MPS tensor is simply the 2/1EE-CTL basic tensor:

$$
\begin{array}{c}
\cdots\!\!\overset{a_1\ \ a_2\ \ a_3}{\underset{\phantom{.}}{\circ\text{—}\circ\text{—}\circ}}\!\!\cdots
\end{array}
=
\begin{array}{c}
\overset{a_1\ \ a_2\ \ a_3}{\underset{\phantom{.}}{\text{♦ ♦ ♦ ♦}}}
\end{array}
\tag{5.156}
$$

For the standard physical boundary 2/1EE-CTL, this tensor network representation also corresponds to the tensor network arising from the stellar tiling of an index lattice:

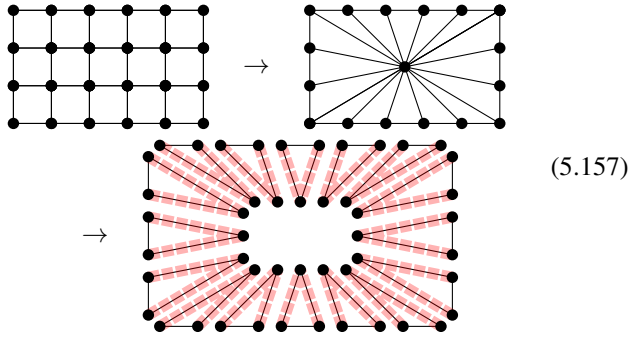

$$(5.157)$$

Note, however, that there is no actual lattice allowing for those tilings for arbitrarily big index lattices, as the central vertex then obviously has to allow an infinite amount of possible local features of some type.

For a complex-real 2E-CTL in its standard form this tensor network looks like:

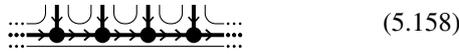

$$(5.158)$$

which is the same as the isometric form of MPS in Eq. (5.107) apart from the delta-line at the bottom which yields a loop giving a normalization when we have periodic boundary conditions.

## 5.6  2/2cEV-CTLs

### 5.6.1  Definition

By 2/2**cEV-CTLs** we refer to a collection of different CTL types on different types of boundary higher order manifolds, that contain 2E-CTLs as a sub-type. "c" stands for "central glued" as there is an additional gluing at the central bulk 0-region. "EV" stands for "edge, vertex" as this is where indices are associated to.

#### Backgrounds

The extended backgrounds are boundary 2/2-manifolds, with an arbitrary 1-manifold as boundary central link, i.e., an arbitrary number of loops. E.g.

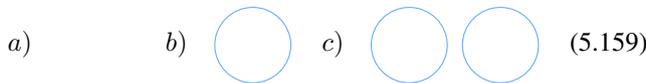

$$(5.159)$$

However, the case a) is trivial and all cases with more than one loop like c) can be obtained from the single loop b) by fusion as we will see later. So we can restrict to the case b) for now.

**Example 115.** Consider the following examples of backgrounds:

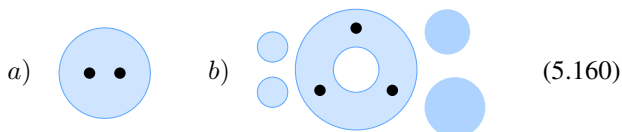

$$(5.160)$$

a) shows a disk with two embedded points. b) shows a disjoint union of an annulus with three embedded points, two disks, one sphere and one torus.

2E-CTLs are a sub-type by restricting to backgrounds without embedded points.

There are two kinds of background gluings: 1) Surgery gluing at the boundary 1-region as for the sub 2E-CTL, and 2) surgery gluing at the central bulk 0-region (i.e., the embedded points).

**Example 116.** Consider the following examples for surgery gluing at the central bulk 0-region (see Remark (3)):

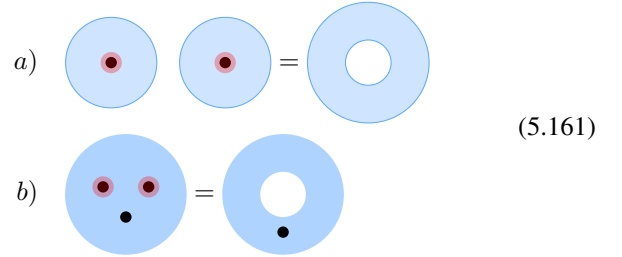

$$(5.161)$$

a) shows gluing a background consisting of two disks with one embedded point each, yielding an annulus. b) shows gluing two out of three points embedded into a sphere, yielding a torus with one embedded point.

#### Lattices

The lattices are based on 2/2CC-lattices living on the backgrounds. Thereby we will restrict the upper link of the vertices of the central bulk 0-region (referred to as 2/2cEV vertices) to a 1-gon (which can be extended to arbitrary $n$-gons, we do not gain anything essential from this). There are two kinds of decorations: 1) The edges of the boundary 1-region (referred to as 2E edges) have orientations as decoration, just as for the sub 2E-CTL. 2) The 2/2cEV vertices have a chirality, which we will indicate by lines pointing clockwise or counterclockwise. Note that there is no underlying orientation, so after a flip the chirality changes but we still have the same lattice. Combinatorially this chirality can be modelled by choosing a favorite vertex-edge-relation-cell of the 1-gon upper link of the 2/2cEV vertex.

**Example 117.** Consider the following example for a patch of a lattice:

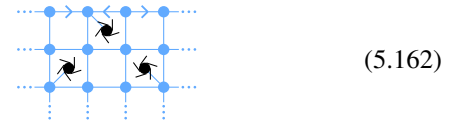

$$(5.162)$$

The index lattice consists in taking the boundary 1CC-lattice including the edge orientations (just as for the sub 2E-CTL) plus taking one disconnected vertex for each 2/2cEV vertex.

One basic gluing is cell gluing at the 2E edges, just as for the sub 2E-CTL. The other basic gluing is cell gluing at the 2/2cEV vertices. Thereby the chiralities have to match. Consider the following example for gluing at the 2/2cEV vertices:

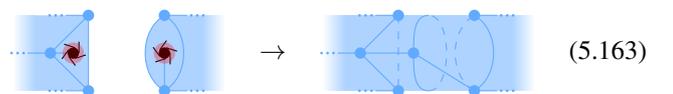

$$(5.163)$$

#### Tensors

The tensors are real tensors with two index types, that we will refer to as 2E indices and 2/2cEV indices. The index prescription associates one 2E index to every 2E edge, just as for the sub 2E-CTL. Additionally it associates one $2/2$cEV index to every $2/2$cEV vertex.

If two 2E edges or two $2/2$cEV vertices are glued, the associated indices are contracted.

### 5.6.2 Basic tensors and axioms

**Observation 79.** A set of basic lattices is given by the basic lattice of the sub 2E-CTL and the following additional basic lattice (with the corresponding index lattice on the right):

 (5.164)

Its index lattice consists of a 1-gon and a single $2/2$cEV vertex. Its extended background is a disk with an embedded point.

This is easily seen by giving an according history mapping: For a given lattice replace every $2/2$cEV vertex by the basic lattice above, and then perform the barycentric subdivision as for the sub 2E-CTL. Gluing all those basic lattices according to how they are related and performing a circuit move yields the desired lattice.

**Observation 80.** The whole CTL is already determined by the 2E-CTL basic tensor and the following additional basic tensor:

$$T1) \quad G_a^x := \begin{array}{c}\bullet\end{array} \!\!\!\!\!\!\!\!\!\!\!\!\!\!\!\! {}^x \!\!\!\!\rightarrow a \quad (5.165)$$

**Proposition 14.** A set of basic history lattices consists of the basic history lattice of the sub 2E-CTL, together with the following history lattice:

 (5.166)

This can be seen in the following way: History lattices are in the same class as $3/2$CCb-lattices with the same boundary central link as the CTL lattices: A history lattice can be obtained from a $3/2$CCb-lattice by replacing every face of the boundary 2-region by a 2E basic lattice, every vertex of the boundary 0-region by a $2/2$cEV basic lattice, every volume of the bulk 3-region by a 2E basic move, and every edge of the bulk 1-region by a move involving the $2/2$cEV vertices.

Replacing every volume of the bulk 3-region by a 2E basic history lattice and every edge of the bulk 1-region by a $2/2$cEV basic history lattice, and gluing all basic lattices according to how they are located, yields a history mapping for the history lattices.

**Remark 108.** Consider the following lattice (with the extended background of a sphere with two embedded points):

 (5.167)

It can be glued to any $2/2$cEV vertex of any lattice without changing the background of the latter. So according to the local support convention we can set the associated tensor to the identity matrix:

$$x \bullet\!\!\!-\!\!\!-\!\!\!- \bullet y \;=\; x \text{------} y \quad (5.168)$$

**Proposition 15.** According to Remark (24) the bi-partitions of the basic history lattice are good candidates for a set of basic axioms. There are two kinds of such bi-partitions. First there is a bi-partition with the two $2/2$cEV basic lattices on one side and the two 2E lattice on the other side, yielding the following axiom:

 (5.169)

Second there is a bi-partition with one of the $2/2$cEV basic lattices on one side and the rest on the other side, yielding the following axiom:

 (5.170)

Third there is a bi-partition with one $2/2$cEV basic lattice and one 2E basic lattice on each side, yielding the following axiom:

 (5.171)

These axioms do indeed define a set of basic axioms: The defining history mapping from Observation (79) yields a defining axiom for every CTL tensor. Then the disjoint union, symmetry and gluing axioms of the CTL follow already from those defining axioms. Together with the $2/2$cEV basic axioms above and the 2E basic axioms, the defining axioms also imply the move axioms for the basic moves involving the $2/2$cEV vertices and the 2E basic moves.

**Remark 109.** The number and complexity of basic axioms can be strongly reduced. The following set of axioms (together with the 2E basic axioms A1) to A5) is equivalent to the basic axioms above together with the local support convention:

$$A1) \quad  \quad (5.172)$$

$$A2) \quad $$

The axiom A2) is precisely the basic axiom in Eq. (5.169). We can glue one $2/2$cEV basic lattice to both sides of A2) and use A1) to obtain the basic axiom Eq. (5.170). On the other hand, we can glue a 2E basic lattice to the index $b$ and a $2/2$cEV basic lattice to the index $a$ on both sides of A2). After applying A1), the left hand side of A2) becomes the left hand side of Eq. (5.171). On the right side we can apply the 2E basic axioms A1) to A5), and Eq. (5.170) to obtain the right hand side of Eq. (5.171). Thus, the three basic axioms Eqs. (5.169, 5.170, 5.171) follow from the axioms A1) and A2).

### 5.6.3 Solutions

**Double-line CTL**

For each double-line 2E-CTL there is the following 2/2cEV-CTL: The basis set for the 2/2cEV indices is the trivial one-element set. The 2/2cEV basic tensor is given by

 (5.173)

**Complex number CTL**

The complex number 2/2cEV-CTL with the complex number sub 2E-CTL is defined by the following 2/2cEV basic tensor:

 (5.174)

**Quaternion CTL**

The following 2/2cEV-CTL has the quaternion sub 2E-CTL: The basis set of the 2/2cEV indices is the trivial one-element set, and the basic tensor is given by

 (5.175)

**Delta CTL**

The delta 2/2cEV-CTL with the delta sub 2E-CTL is defined by the following basic tensor:

 (5.176)

### 5.6.4 Classification

**Observation 81.** According to Observation (66) the right hand side of the axiom A2) is a symmetric projector $P$. If we read the basic tensor $G_a^x$ as a linear map from $x$ to $a$ then the axioms A1) and A2) read

$$G^T G = \mathbb{1} \qquad GG^T = P \qquad (5.177)$$

Thus, $G$ is an isometry, i.e., an orthogonal map when restricted to the support of $P$ on the index $a$. Such an isometry is unique up to a gauge transformation $G \to GO$ for an orthogonal map $O$.

So for a fixed sub 2E-CTL there is only one 2/2cEV-CTL up to gauge transformations.

**Remark 110.** For a 2/2cEV-CTL we can bring the sub 2E-CTL into its standard form via a gauge transformation on the 2E indices. Then the 2/2cEV-CTL is fixed up to a gauge transformation on the 2/2cEV indices. So by such a gauge transformation we can bring the 2/2cEV basic tensor into the following form:

 (5.178)

Here we used generalized tensor-network notation: The basis set $B_\alpha$ of the $a$ index depends on the value of the $\alpha$ index. $\mu_\alpha = |B_\alpha|^{-1/2}$ and the $x$-tensor equals the 2/2cEV basic tensor of the trivial, complex number or quaternion 2/2cEV-CTL. Note that both in the trivial and quaternion case the basis of the $y$ index is trivial.

**Comment 18.** According to Com. (16), the support of the projector $P$ is exactly the centre in the *-algebra language. The *-algebras of the delta and complex number 2E-CTLs are commutative, so the centre is the full space (thus the 2/2cEV basis equals the 2E basis). The *-algebras corresponding to the double-line and quaternion 2E-CTLs on the other hand are non-commutative and have a trivial centre consisting of only the identity element of the *-algebra (thus the 2/2cEV basis is trivial).

### 5.6.5 Subtypes, mappings and fusions

**Observation 82.** There is a mapping from 2E-CTLs to 2/2cEV-CTLs with a single loop as boundary central link: The lattice mapping cuts out a small disk around each point of the central bulk 0-region, adding a small circle of boundary 1-region. Then tensor mapping blocks the 2E indices associated to each such small circle and uses this composite index as 2/2cEV index of the corresponding point of the central bulk 0-region. The 2/2cEV-CTL obtained in this way is the (up to gauge transformations) unique 2/2cEV-CTL with the original 2E-CTL as sub CTL.

**Observation 83.** For every 2/2cEV-CTL type $\mathcal{A}$ with $x$ loops as boundary central link there is a CTL mapping from the 2/2cEV-CTL type $\mathcal{B}$ with a single loop as boundary central link to $\mathcal{A}$: The lattice mapping replaces every point of the central 0-region of $\mathcal{A}$ (with $x$ loops as upper link) by $x$ points of the central 0-region of $\mathcal{B}$ (with a single loop as upper link). The tensor mapping takes the $x$ indices associated to the $x$ points and blocks them together to one composite index associated to original single point. The lattice mapping at the boundary 1-region is trivial. Gluing two such points of $\mathcal{A}$ is equivalent to pairwisely gluing the corresponding groups of $x$ points of $\mathcal{B}$ after the lattice mapping.

**Observation 84.** Conversely, there is the following mapping from $\mathcal{B}$ to $\mathcal{A}$: The lattice mapping replaces the bulk 2-region and boundary 1-region of $\mathcal{B}$ by $x$ layers of the 2-region and boundary 1-region of $\mathcal{A}$. The points of the central bulk 0-region of $\mathcal{B}$ are replaced by one point of the central bulk 0-region of $\mathcal{A}$ where all $x$ layers are joined together. The tensor mapping consists in blocking the $x$ indices on the $x$ different layers at each place of the boundary 1-region of $\mathcal{B}$ and block them together to one single index associated to this place.

**Observation 85.** Consider the 2/2cEV-CTL type with the empty 1-manifold as boundary central link. Its extended backgrounds are disjoint unions of boundary 2-manifolds and a set of isolated vertices. I.e., the trivial 1cgV-CTLs and 2E-CTLs are both sub-types of this type, or better, this type is the product of the latter two types. Accordingly there are mappings from both sub-types to this type and vice versa.

**Observation 86.** We can restrict ourselves to lattices with empty boundary 1-region. E.g., the extended background of such a lattice could look like:

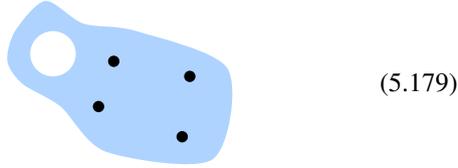

(5.179)

representing a torus with 4 embedded points.

This defines a mapping (which is even a sub-type) to a CTL type on 2-manifolds with embedded points, and gluing at those points. Such an CTL type is closely related to what is called an *axiomatic TQFT*, as outlined in Section (6.1.7).

### 5.6.6 Physical interpretation

**Classical statistical physics**

2/2cEV-CTLs do not come from a TL type, thus they do not directly yield tensor networks for the partition function. Consider the boundary central link being a single circle. Those CTLs yield a tensor network in the form of Observation (46), where at the embedded points we do not only have one single local observable, but a collection of local observables labeled by open 2/2cEV indices. For the delta 2E-CTL as sub CTL this collection is spanned by one observable for each symmetry broken sector which projects onto this sector. So the 2/2cEV-CTL corresponds to a complete collection of local order parameters. The CTL tensor for a lattice with empty boundary 1-region and $x$ points of the central bulk 0-region (like Eq. (5.179)) represents $x$-point correlators for those order parameters.

In the case for multiple loops as boundary central link we get the same thing for multiple stacked copies of the system.

**Quantum physics**

Consider a complex-real 2E-CTL and the corresponding quantum mechanical model. Now for a given 2E index lattice $X$ consider a 2/2cEV-CTL with the background of $X$ as boundary central link. Now consider the 2/2cEV lattice that has $X$ as index lattice and whose extended background is the stellar cone of $X$. The tensor associated to this lattice can be interpreted as a linear map from the 2/2cEV index at the single point of the central bulk 0-region to the 2E-indices on $X$. This linear map takes vectors in some abstract ground state space to the actual ground states in the physical Hilbert space.

For a single circle as boundary central link, the CTL tensors for lattices with empty boundary 1-region can also be interpreted as imaginary time $n$-point correlation functions for a collection of local observables.

## 5.7 2/2cEEV-CTLs

### 5.7.1 Definition

By 2/2**cEEV-CTLs** refer to a collection of different CTL types. They contain 2/1EE-CTLs as well as 2E-CTLs as subtypes, potentially multiple times. "c" stands for "central glued" and "EEV" for "edge, edge, vertex", as this is where the indices are associated to.

**Backgrounds**

The extended backgrounds are boundary 2/2-manifolds with a boundary central link consisting of one loop that is divided into $x$ segments where each segment forms its own 1-region, separated by $x$ points where each point forms its own 0-region. E.g.

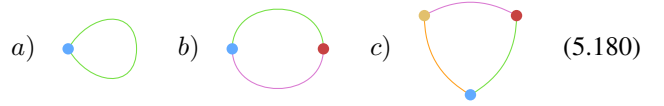

(5.180)

There is a sub 2E-CTL for each 1-region and a sub 2/1EE-CTL for each 0-region of this boundary central link. The boundary central link of the sub 2/1EE-CTL is the upper link of the corresponding 0-region.

There is a one-to-one correspondence between 2/2cEEV-CTLs of arbitrary boundary central link with $x$ sub 2/1EE-CTLs and the boundary central link a) with one sub 2/1EE-CTL which is the stacking fusion of the $x$ former 2/1EE-CTLs. Formally this one-to-one correspondence comes from a CTL mapping, see Observation (90). Thus, for finding CTLs we could restrict to the type a). However, the gauge families and also the phases for those two types are different. Only the phases relative to all the sub 2/1EE-CTLs are in one-to-one correspondence.

**Example 118.** Here are some examples for extended backgrounds of the different types:

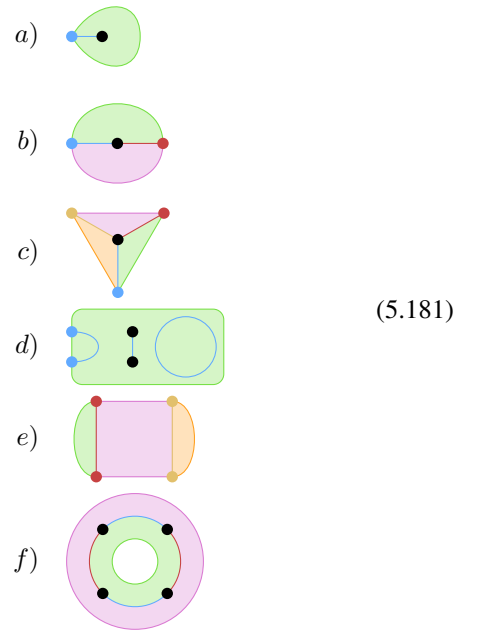

(5.181)

Examples a), b), c), d), e) and f) show backgrounds of type a), b), c), a), c), b) from Eq. (5.180), respectively. a), b) and c) are the stellar cone of the respective boundary central link.

There are multiple background gluings: There is surgery gluing for every boundary 1-region and every boundary 0-region, just as for the sub 2/1EE-CTLs. Additionally, there is surgery gluing at the central bulk 0-region.

**Example 119.** Consider the following examples for back-ground gluings involving the central bulk 0-region:

$$(5.182)$$

where the examples a) and b) correspond to the boundary central links a) and c) in Eq. (5.180), respectively. a) shows a disk with an embedded interval and gluing the two endpoints of this interval, yielding a disk with a handle and a loop winding around one of its non-contractible loops.

The sub 2E-CTL for a 1-region of the boundary central link is obtained by restricting to lattices where only the according bulk 2-region and boundary 1-region of the lattices are non-empty. The sub 2/1EE-CTL for a 0-region of the boundary central link is obtained by restricting to lattices where only the according bulk 1-region and boundary 0-region of the lattices, and all regions in their upper links, are non-empty.

**Lattices**

The lattices are 2/2CCbt-lattices on the backgrounds. Just as for the sub 2/1EE-CTLs, the boundary 0-regions are thickened. Just as the sub 2E-CTLs, the edges of the boundary 1-regions carry orientations. Additionally the vertices of the central bulk 0-region carry a chirality, which can be omitted if the boundary central link consists of more than two line segments.

One could without loss of generality restrict to a single allowed upper link for the vertices of the central bulk 0-region, namely the boundary central link interpreted as a 1/1CC-lattice.

**Example 120.** Consider the following examples for extended backgrounds and lattices representing those backgrounds:

$$(5.183)$$

a) and b) correspond to the central boundary links c) and b) in Eq. (5.180). For a) we do not need to indicate the chirality of the vertices of the central bulk 0-region, as there is no mirror symmetry.

The index lattice consists of the 1/1CCt-lattice forming the boundary (including the edge orientations) as well as the isolated vertices of the central bulk 0-region (without the chirality). From now on we will mostly draw the index lattice only and assume that the rest of the background is clear from the context.

**Example 121.** Consider the following examples for extended backgrounds (left), the according index backgrounds (middle) and index lattices with this index background (right):

$$(5.184)$$

a), b) and c) correspond to the boundary central links a), c) and a), respectively.

The basic gluings include cell gluing at the edges and vertices of the boundary 1-regions and 0-regions, just as for the sub 2E-CTLs and sub 2/1EE-CTLs. Additionally one basic gluing is given by cell gluing of the vertices of the central bulk 0-region.

**Tensors**

The tensors are real tensors, with one index type for each 0-region and each 1-region of the boundary central link and one additional index type. There is one index of the according type associated to the 2E edges and 2/1EE edges of each boundary 0-region and boundary 1-region, just as for the corresponding sub-types. There is one index of the additional type associated to every vertex of the central bulk 0-region.

When two 2E edges or two 2/1EE edges are glued, the associated indices are contracted, just as for the sub 2E-CTLs and sub 2/1EE-CTLs. Also when two vertices of the central bulk 0-region are glued, the associated indices are contracted.

### 5.7.2 Basic tensors and axioms

**Proposition 16.** A set of basic extended backgrounds is given by the 2E and 2/1EE basic extended backgrounds, together with the stellar cone of the boundary central link. Furthermore a set of basic lattices is given by the 2E and 2/1EE basic lattices together with the following basic lattice representing the basic extended background: Each boundary 0-region is represented by a single face, the boundary 1-regions consist of a single vertex, each bulk 2-region has a 4-gon face, and the bulk 1-regions are single edges connected to one vertex representing the central bulk 0-region.

e.g., consider the following basic lattices (middle), together with their basic extended backgrounds (left), and index lattices

(right):

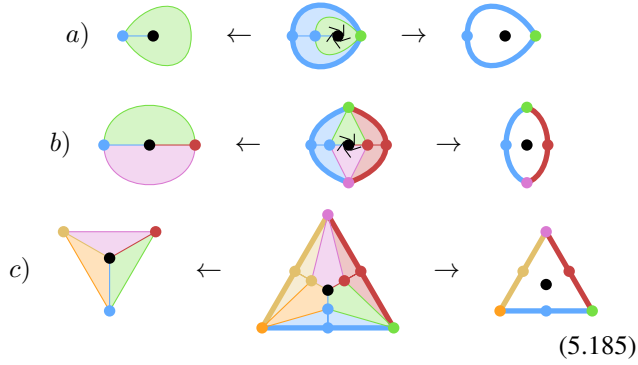

$$(5.185)$$

where a), b) and c) correspond to the boundary central links a), b) and c) in Eq. (5.180), respectively.

A corresponding history mapping goes as follows: First apply a lattice mapping to obtain a lattice where all faces have triangle lower link and all vertices of the bulk 0-region have the standard upper link. Replace every face of each bulk 2-region by the corresponding 2E basic lattice, every edge of each bulk 1-region by the corresponding 2/1EE basic lattice and every vertex of the bulk 0-region by the 2/2cEEV basic lattice above. Gluing all basic lattices according to how they are located yields the desired lattice.

**Observation 87.** Each 2/2cEEV-CTL is fully determined by the 2E and 2/1EE basic tensors together with the tensor corresponding to the 2/2cEEV basic lattice Eq. (5.185). E.g.

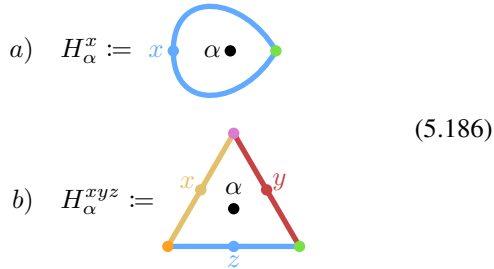

$$(5.186)$$

where a) and b) are for the boundary central links a) and c) in Eq. (5.180), respectively.

**Proposition 17.** A set of basic history lattices is given by the 2E and 2/1EE history lattices together with the 2/2cEEV basic history lattice consisting of two 2/2cEEV basic lattices, and one 2/1EE basic lattice for each 0-region of the boundary central link. Consider, e.g., the following 2/2cEEV basic history lattices:

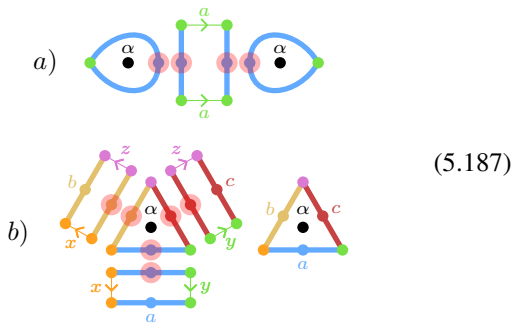

$$(5.187)$$

where a) and b) correspond to the boundary central links a) and c) in Eq. (5.180). This can be seen as follows: History lattices are in the same class as 3/1CCb-lattices with the same boundary central link as the lattices. We can replace every volume of the bulk 3-regions by a 2E basic history lattice, replace every face of the bulk 2-regions by a 2/1EE-CTL basic history lattice, and replace every edge of the bulk 1-region by the 2/2cEEV basic history lattice above.

**Observation 88.** According to Remark (24) a good candidate for basic axioms are those arising from bi-partitions of the basic history lattice Eq. (5.187). First, there is one decomposition with the two 2/2cEEV basic lattices on the left hand and all the 2/1EE basic lattices on the right. This yields e.g., the following axioms:

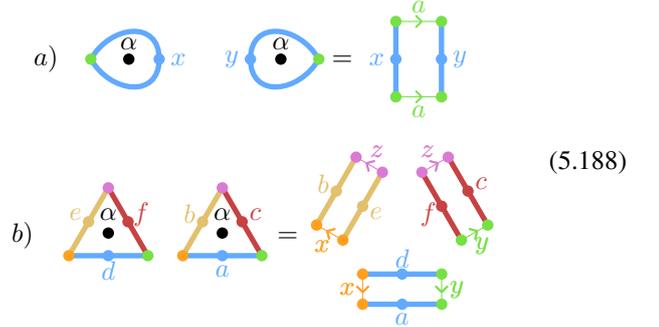

$$(5.188)$$

where a) and b) correspond to the boundary central links a) and c) in Eq. (5.180).

Second, there are bi-partitions with one of the 2/2cEEV basic lattices on every side, and all the 2/1EE basic lattices arbitrarily on the left or right side. This yields e.g., the following axioms:

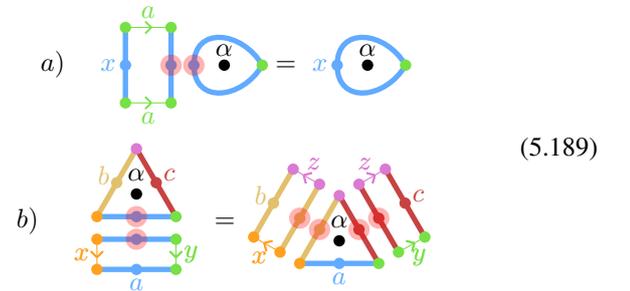

$$(5.189)$$

Those axioms are a full set of basic axioms: The history mapping above yields a defining axiom for every lattice. The disjoint union, symmetry and all 2E and 2/1EE gluing axioms directly follow from those defining axioms. The 2/2cEEV gluing axioms and the 2/2cEEV move axioms follow from the defining axioms together with the basic axioms above.

**Remark 111.** Consider the 1-ball completion of the boundary central link and some lattice with this extended background. This lattice can be glued to every 2/2cEEV vertex of any lattice without changing the background of this lattice. Thus, according to the local support convention we can set the associated tensor to the identity matrix. E.g.,

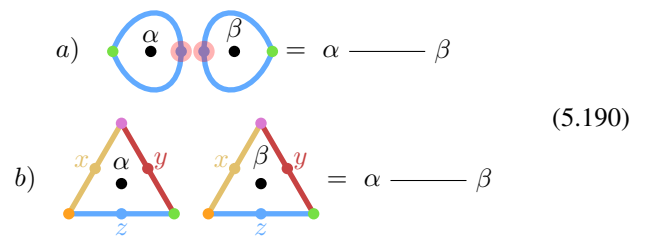

$$(5.190)$$

**Remark 112.** All the axioms in Eq. (5.189) already follow from the axiom Eq. (5.188) together with the local support convention Eq. (5.190). So we have e.g.:

$$A1) \qquad = \quad \alpha \rule{2cm}{0.4pt} \beta$$

$$A2) \qquad x \quad y \quad = \quad x \quad y \qquad (5.191)$$

for the boundary central link a) in Eq. (5.180).

All the axioms in Eq. (5.189) are implied in the following way: Bring one of the 2/2cEEV basic lattices in Eq. (5.188) to the right side by gluing a 2/2cEEV basic lattice to both sides and applying Eq. (5.190) to the left side. Then glue the desired selection of 2/1EE basic lattices to both sides. Use the 2/1EE-CTL basic axioms to invert this selection on the right side. Then use Eq. (5.189) and Eq. (5.190) on the right side to obtain the desired axiom.

### 5.7.3 Solutions

**Double-line CTL**

Consider a collection of double-line 2E-CTLs and tensor products of two double-line 2/1EE-CTLs between them. For this configuration of sub CTLs there exists the following double-line 2/2cEEV-CTL. The basic tensor is e.g., for the boundary central link c) in Eq. (5.180):

$$(5.192)$$

where $B_1$, $B_2$ and $B_3$ are the basis sets of the three sub 2E-CTLs.

**Complex number CTL**

Consider a collection of complex number 2E-CTLs with complex 2/1EE-CTLs between pairs of them. Each of the 2/1EE-CTL can either preserve or invert the complex arrow orientations. We can consider two cases: 1) There is an odd number of 2/1EE-CTLs where the complex arrow orientation is inverted. In this case there is no 2/2cEEV-CTL with this configuration of sub CTLs, or more precisely there is only the zero CTL, where the 2/2cEEV indices have the empty basis set. 2) There is an even number of 2/1EE-CTLs where the complex arrow orientation is inverted. In this case there is the complex number 2/2cEEV-CTL with the basic tensor given by the complex number tensor with only outgoing arrow directions. E.g.,

$$(5.193)$$

**Delta projection CTL**

Consider a collection of delta 2E-CTLs and tensor products of delta projection CTLs as 2/1EE-CTLs between them. In this case all the 2/1EE indices have trivial basis. So choosing the 2/2cEEV indices to also have trivial basis and the 2/2cEEV basic tensors to be trivial defines a 2/2cEEV-CTL with the desired sub CTLs.

### 5.7.4 Classification

**Observation 89.** The right hand side of the axiom A2) is a symmetric projector $P$. If we read the basic tensor $H^x_{(a_1,\ldots)}$ as a linear map from $x$ to $(a_1,\ldots)$ then the axioms A1) and A2) read

$$H^T H = \mathbb{1} \qquad\qquad HH^T = P \qquad (5.194)$$

Thus, $H$ is an isometry, i.e., an orthogonal map when restricted to the support of $P$ on the index $(a_1,\ldots)$. Such an isometry is unique up to a gauge transformation $H \to HO$ for an orthogonal map $O$.

So for fixed sub 2E-CTLs and sub 2/1EE-CTLs there is only one 2/2cEEV-CTL up to gauge transformations.

### 5.7.5 Subtypes, mappings and fusions

**Remark 113.** There is a mapping from 2/2cEEV-CTLs by restricting to lattices whose boundary 1-regions and boundary 0-regions are all empty. The CTL type that this mapping targets only has one background gluing, namely surgery gluing of the central bulk 0-region. CTLs of this type are very similar to what is known as a *TQFT with defects*, see also Sec. (6.1.7) (However, we do not strictly have only defects but possibly also domain walls between different phases). More precisely such a CTL corresponds to a defective TQFTs restricted to only a finite set of boundary components. So every 2/2cEEV-CTL gives rise to such a defective TQFT. There is no fundamental topological argument why the converse should be true, however, it seems to be the case that this actually happens to be true.

**Observation 90.** For every 1-region of a 2/2cEEV-CTL type with a boundary central link with $n$ segments, there is a mapping to a 2/2cEEV-CTL with a boundary central link consisting of only one segment. The sub 2E-CTL of the new 2/2cEEV-CTL is the old sub 2E-CTL associated to the selected 1-region. The sub 2/1EE-CTL of the new 2/2cEEV-CTL stacking fusion of sub 2/1EE-CTLs of the original 2/2cEEV-CTL. The lattice mapping is like the stacking fusion, just that we also replace every point of the bulk 0-region where the single bulk 1-region ends by a point where the $n$ layers of 1-regions meet. For example consider the lattice mapping from a 2/2cEEV-CTL of type c) in Eq. (5.180) applied to some extended background:

$$(5.195)$$

The tensor mapping uses the old 2/2cEEV index as the new 2/2cEEV index. The mapping of the 2E and 2/1EE indices is the same as for the stacking fusion.

**Observation 91.** There is a CTL mapping from 1/1cVV-CTLs to 2/2cEEV-CTLs. The boundary central link of the 1/1cVV-CTL consists of only the 0-regions of the boundary central link of the 2/2cEEV-CTL. The according lattice mapping takes only the boundary 0-regions, bulk 1-regions and the central bulk 0-region and forgets about the rest. The tensor mapping associates the trivial basis to all 2E indices, and takes the 1V indices as 2/1EE indices and the 1/1cVV indices as 2/2cEEV indices.

### 5.7.6 Alternatives

Similar to 2/2EV-CTLs we can consider 2/2EVV-CTLs which are 2/2cEEV-CTLs except that there are no indices and gluing at the central bulk 0-region. 2/2EVV-CTLs do come from a TL in contrast to 2/2cEEV-CTLs. The 2/2EVV basic lattices are the same as the 2/2cEEV basic lattices, just that the associated basic tensors do not have the 2/2cEEV index. There is one basic axiom, namely the one in Eq. (5.189) with all 2/1EE basic lattices on one side and none on the other side (just that there are no 2/2cEEV indices). Every vector in the image of the 2/2cEEV basic tensor interpreted as an isometry yields the basic tensor of a 2/2EVV-CTL with the same sub 2E and 2/1EE-CTLs. So the different 2/2EVV-CTLs with a fixed configuration of sub CTLs form a vector space. Those different 2/2EVV-CTLs are also in different TL phases relative to all sub CTLs.

### 5.7.7 Physical interpretation

#### Classical statistical physics

In the context of classical statistical models, 2/2cEEV-CTLs are collections of local observables located at meeting points of defects/domain walls. Thereby local observables that are only microscopically different but globally equivalent are represented by the same observable in the collection.

#### Quantum physics

Consider a complex-real 2/2cEEV-CTL and the joint product of all its sub 2E-CTL and sub 2/1EE-CTLs. Now consider a 2/2cEEV lattice whose extended background is the stellar cone of the boundary central link with index lattice $X$. The tensor associated to this lattice can be interpreted as a linear map from the 2/2cEEV index at the single point of the central bulk 0-region to the 2E indices and 2/1EE indices on $X$. $X$ is also an index lattice of the joint product of sub CTLs. So the linear map takes vectors in some abstract ground state space to the actual ground states in the physical Hilbert space of the joint product of sub CTLs. 2/2EVV-CTLs are individual ground states.

2/2cEEV-CTL tensors for backgrounds whose index lattice consists of 2/2cEEV vertices only can also be interpreted as imaginary time $n$-point correlation functions of collections of observables at the meeting points of different defects/domain walls.

## 5.8 3FE-(C)TLs

### 5.8.1 Definition

3FE-CTLs are a CTL type on 3-dimensional lattices coming from a TL type. The "FE" stands for "face, edge", as this is where indices are associated to.

#### Backgrounds

The extended backgrounds of 3FE-CTLs correspond to boundary 3-manifolds, and the backgrounds of the TLs are 3-manifolds. There is one background gluing, namely boundary surgery gluing (see example f) and g) in example (34)).

#### Lattices

The lattices are based on 3CCb-lattices. The index lattice is given by the 2CC-lattice forming the boundary of the 3CCb-lattice. There are two kinds of decorations: First, every index edge carries an orientation. Second, for every index face we specify a favorite connected index edge. We will indicate the orientation by an arrow and the favorite edge by placing a small half circle at the respective edge at the side of the respective face. E.g., consider the following patch of an index lattice:

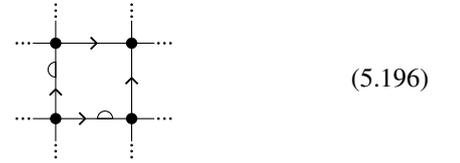

$$(5.196)$$

The lower edge is the favorite edge of the middle face. We will only work on the level of backgrounds and according to Remark (1) draw only the index lattice and describe the rest in words if unclear. If the face shape (together with the edge orientations) has no permutation or reflection symmetries, then the favorite edge can be fixed for this particular face shape, such that we can omit drawing it.

**Remark 114.** One can without loss of generality restrict to 3CCb-lattices whose volumes have a tetrahedron as lower link. Further one can restrict to index edge orientations that are non-cyclic (such orientations are known as *branching structures*). As the non-cyclic triangle does not have any symmetries, the favorite index edge can be fixed to be, e.g., the one that is oriented oppositely to the other two edges.

The basic gluings are given by (higher order) cell gluing, see d) and e) in example (73) or b) in example (74). Thereby the index edge orientations and the favorite index edges have to match when identifying the two glued index faces.

#### Tensors

The tensors are of a mixed type: They are real tensors with two index types one of which is equipped with a preferred basis. Also the tensors are generalized tensors in the sense of Remark (32).

There is one **edge index** with basis set $B^E$ associated to every index edge and one **face index** with basis set $B^F$ associated to each index face. The edge indices are equipped with a preferred basis. $B^F$ depends on 1) the upper link of the face together with the edge orientations and favorite edge and 2) the values of the edge indices in the preferred basis. I.e., the total basis set $B$ for the tensor associated to a complete index lattice

$X$ is given by a disjoint union:

$$B[X] = \bigoplus_{\substack{i_a, i_b, \ldots \in B^E \\ \text{for } a, b, \ldots \in E[X]}} \bigotimes_{f \in F[X]} B^F_{i_{f1}, i_{f2}, \ldots} \qquad (5.197)$$

where $E[X]$ is the set of edges of $X$, $F[X]$ is the set of faces of $X$, and $f1, \ldots$ are the boundary edges of $f$.

When a cell gluing of two index faces is performed, the indices are contracted in the following way: The edge indices around the glued faces are contracted pairwise with the 3-index delta tensor, such that we have one remaining index for each pair of indices. This makes sense as there is one remaining edge for each pair of edges around the glued faces, to which we can associate this remaining index. Also the two face indices are contracted, which is well defined according to Remark (32) as all the indices on which their basis set depends are also contracted. E.g., gluing of two triangle faces is given by

$$(5.198)$$

There are normalizations that we have to include into the contraction prescription for higher level cell gluing: Those are a tensor $C_i$ with one edge index $i$, and a number $D$. First consider gluing two index faces that share a common edge, which disappears during the gluing. In this case we have to contract the corresponding edge index with the tensor $C_i$ such that it disappears. E.g.

$$(5.199)$$

Second consider gluing two faces that share two common edges connected to the same vertex such that this vertex disappears. In this case, we need to multiply with the number $D$ (in addition to using the previous rule twice for the two disappearing edges).

### 5.8.2 Basic tensors and axioms

**Definition 175.** For any polygon $P$ with oriented edges and favorite edge, define the **pyramid** with respect to $P$ as the background whose index lattice has one **bottom face** that looks like $P$, and one **tip vertex** that is connected via one edge to each vertex of the bottom face. All of those edges are oriented towards the tip and yield triangles with non-cyclic edge orientation between them. E.g., consider the following pyramids for a 1-gon and a 4-gon:

$$(5.200)$$

The pyramid with respect to the triangle c) with non-cyclic edge orientations is called the **basic tetrahedron**.

**Observation 92.** A set of basic lattices is given by the set of pyramids with respect to all $P$ that are allowed as index lattices (together with the basic tetrahedron if not already included).

A defining history for a given lattice can be obtained by the following procedure: Performing a barycentric subdivision in the interior yields a simplicial complex, with pyramid-like volumes at the boundary. Replace every simplex in the interior with the basic tetrahedron, and every pyramid-like volume at the boundary with the according pyramid. Then glue all basic lattices according to how they are located.

**Definition 176.** For every polygon $P$ with oriented edges and favorite edge, define the **face-pillow** with respect to $P$ as a lattice whose index lattice has two faces with the shape of $P$ (one of which reflected). E.g.

$$(5.201)$$

**Remark 115.** Gluing one face of the face-pillow with respect to some $P$ to a face of a lattice with shape $P$ does not change its background. The associated tensor can be interpreted as a linear map from the bottom to the top face index that is parametrized by the edge indices. We can impose the local support convention by setting it to the identity matrix for all edge index values. E.g., for $P$ being a triangle we get:

$$(5.202)$$

**Observation 93.** The tensor associated to the pyramid with respect to $P$ can be interpreted as a linear map $U_I$ from the bottom face index to the edge and face indices on the top part (parametrized by the edge indices $I$ of the boundary edges of the bottom face). E.g.

$$(5.203)$$

If we take two copies of the pyramid and glue together all their faces on the top part we get the face-pillow with respect to $P$. Together with the local support convention this implies that the linear map $U$ is an isometry (parametrized by the edge indices of the bottom face):

$$(5.204)$$

$$U_I^\dagger U_I = \mathbb{1} \quad \forall I$$

**Definition 177.** For any 2E index lattice $P$ with oriented edges, define the **double-pyramid** with respect to $P$ as the background whose index lattice has two **tip vertices**, one **belt edge** for every edge of $P$, and one **belt vertex** that connects two belt edges and two edges from this vertex to each of the two tip vertices, for every vertex of $P$. All edges are oriented towards the tips. E.g., consider the double-pyramid for the following 4-gon:

$$(5.205)$$

Note that the double-pyramid for $P$ does not depend on a choice of favorite edge of $P$, whereas the pyramid for $P$ does. For 2E index lattices $P$ that are the disjoint union of multiple components of ball-like background, we can define the double-pyramid with respect to $P$ as the disjoint union of the double-pyramids of all components.

**Proposition 18.** The tensor associated to the pyramid is already determined by the tensor associated to the double-pyramid (with respect to the same $P$), up to a gauge transformation on the index at the bottom face of the pyramid.

This is easy to see: 1) Gluing the bottom and top part of two copies of the double-pyramid yields a double-pyramid again, and the double-pyramid is symmetric under reflections at the equator plain. Thus, the associated tensor is a symmetric projector $P_I$ when interpreted as a linear map from the indices at the bottom half to the indices a the top half, parametrically depending on the irrep values $I$ of the belt edges. (Actually we also have to deal with the normalizations correctly.)

2) The double-pyramid can be obtained from gluing together two copies of the pyramid at their bottom faces. So the isometry $U_I$ from Observation (93) fulfills $P_I = U_I U_I^\dagger$, or more

precisely:

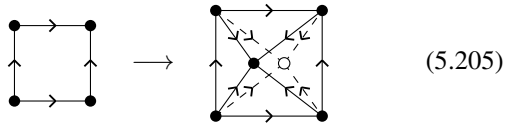

$$(5.206)$$

Such an isometry is unique up to a orthogonal gauge transformation $U_I \to U_I O_I$ on the index at the bottom face of pyramid.

**Remark 116.** Each double-pyramid with respect to a polygon $P$ can be glued together from copies of the basic tetrahedron: A simplicial complex representing the double-pyramid can be obtained by adding one edge from the bottom to the top tip of the double-pyramid, and for every "equator" edge adding a tetrahedron spanned by this edge and the added edge between the tips. Replacing every such tetrahedron by a copy of the basic tetrahedron and gluing them accordingly yields the double-pyramid.

$$(5.207)$$

Thus, the whole 3FE-CTL is already determined by the tensor associated to the basic tetrahedron:

$$T1) \quad F_{d\ f\gamma\delta}^{abce\alpha\beta} :=$$

$$(5.208)$$

**Remark 117.** History lattices are in the same class as 4CCb-lattices (or more specifically 4SCb-lattices): The 4-cells (4-simplices) of the bulk represent bi-stellar flips (or more specifically Pachner moves), the boundary volumes (tetrahedrons) are copies of the basic tetrahedron and the boundary faces represent basic gluings. A set of basic history lattices is given by the following single history lattice:

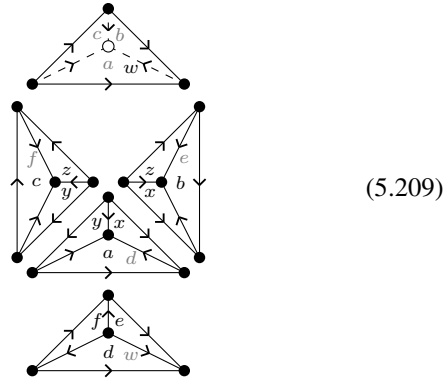

(5.209)

This is just a decomposition of a 4-simplex into 5 tetrahedra. A defining history for a given history lattice can be obtained by the following procedure: Replace every 4-simplex of the 4SCb-lattice by one copy of the basic history lattice, then history glue all pairs of basic tetrahedra that are shared by neighbouring 4-simplices according to how they are located. The tetrahedra at the boundary 3SC-lattice remain unglued and form the desired history lattice.

**Remark 118.** According to Remark (24) the bi-partitions of the basic history lattice Eq. (5.209) are good candidates for basic history moves (yielding basic axioms). There are two kinds of such bi-partitions: First, there are 10 bi-partitions with 2 basic tetrahedra on the left and 3 on the right side, corresponding to 10 different choices of the edge that is shared by all 3 tetrahedra the one right side. E.g., for one particular choice of edge orientations we get:

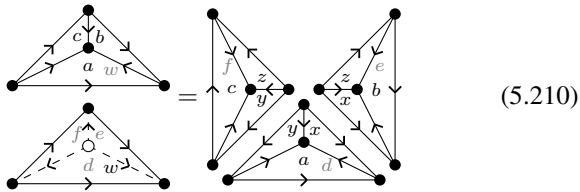

(5.210)

Second, there are 5 different bi-partitions into one basic tetrahedron on the left and 3 on the right, corresponding to the 5 different choices of the tetrahedron on the left. E.g., for one particular edge orientation we get:

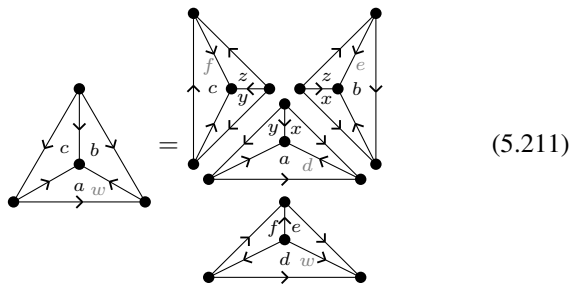

(5.211)

Those 15 history moves define indeed a set of basic history moves: The defining history above is already consistent with the disjoint union and basic gluings. Together with the 15 basic history moves above it is also consistent with basic moves (note that the two kinds of axioms are directly related to the $2 \to 3$ and $1 \to 4$ Pachner move).

**Remark 119.** The above 15 basic axioms are quite numerous and complex. We can drastically reduce the complexity and

number of axioms by introducing the following auxiliary tensors:

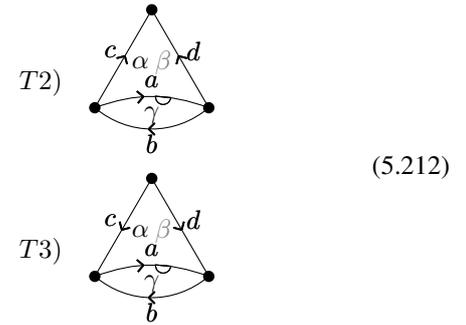

(5.212)

Then the following axioms are equivalent to the 15 basic axioms above plus the local support convention:

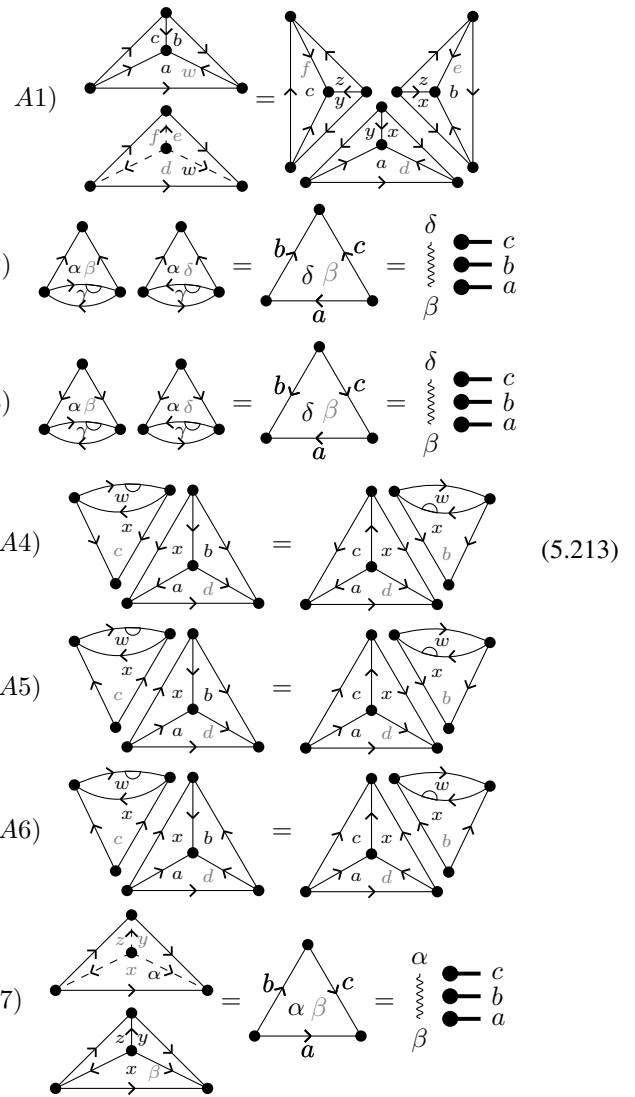

(5.213)

So instead of 15 axioms all involving 5 of the complex basic tensors T1) we have a single axiom involving two copies of T1) and 6 additional axioms involving at most two copies of T1) and at most two copies of the simpler auxiliary tensor T2).

This can be seen in the following way: A1) is one of the axioms Eq. (5.210) for one particular edge orientation. The axioms A2), A3), A4), A5) and A6) can be used to change the edge orientations: The basic tensor T2) can be glued to any face of a basic tetrahedron (yielding an additional 2-gon face). With A2) or A3) we can add or remove two copies of T2) to one single tetrahedron face. With A4), A5) or A6) we can slide one of the copies of T2) over to another neighbouring face of the tetrahedron, thereby inverting the orientation of the edge in between. If we want to invert the orientation of one of the edges that occur at both sides of the axioms Eq. (5.210) we can add the same two copies of T2) on both sides to the two neighbouring faces. Then we slide one copy over to the other face and remove the two copies of T2) at one face, thereby inverting the edge orientation. This way we get the 9 other axioms in Eq. (5.210). We also get all variants of A7) for different edge orientations. Finally we can use those variants of A7) to bring one of the two basic tetrahedra on the left side of Eq. (5.210) over to the right side of Eq. (5.210), which yields all 5 axioms in Eq. (5.211).

### 5.8.3 Solutions

**Delta tensors**

For every finite set $B$ there is the **delta 3FE-CTL**: The basis set for the edge indices is trivial and $B^F = B$. The basic tensor and normalizations are given by

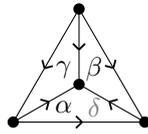

$$C = 1 \qquad D = 1 \tag{5.214}$$

Alternatively one could have encoded the delta tensor into the edge indices: To this end we take the basis set of the face indices to be 1 if all the edge index values at the adjacent edges are equal, and 0 otherwise. So the basic tensor equals the number 1 if all edge index values are equal, and does not have any entry otherwise.

**Complex numbers**

The **complex number 3FE-CTL** is defined as follows: The basis set for the edge indices is the trivial one-element set. The basis set for the face indices is $\{1, \mathbf{i}\}$. The basic tensor is the complex number tensor:

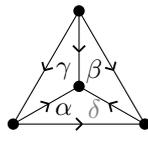

$$C = 1 \qquad D = 1 \tag{5.215}$$

**Double-line tensors**

For every set $B$ there is the following **double-line 3FE-CTL**: The basis set for the edge indices is the trivial one-element set, and $B^F = B \times B \times B$. The basic tensor and normalizations are given by

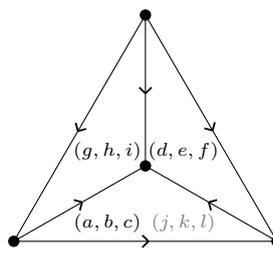

$$C = |B|^{-1} \qquad D = 1 \tag{5.216}$$

**Groups**

For each finite group there is the following **group 3FE-CTL**. The basis set of the edge indices is given by the set of group elements. The basis set for the face indices is 1 if the product of group elements around the face equals the identity element of the group. In the product, one has to replace each group element by its inverse, depending on weather the corresponding edge orientation is clockwise or counter-clockwise. E.g.:

$$|B^F( \ d \ \ b \ )| = \delta_{ab^{-1}cd,1} \tag{5.217}$$

So the basic tensor is given by

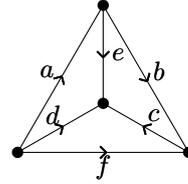

$$C_i = \bullet\!\!-\!\!\bullet \ i \qquad D = |G|^{-1} \tag{5.218}$$

**Group cocycles**

For every finite group $G$ with $U(1)$ group 3-cocycle $\omega$ there it the **twisted group 3FE-CTL**. The basis sets of the edge and face indices is the same as for group 3FE-CTLs. The basic tensor is determined by the cocycle $\omega$:

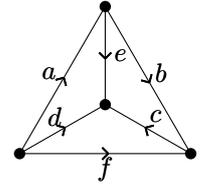

$$C_i = \bullet\!\!-\!\!\bullet \ i \ , \qquad D = |G|^{-1}. \tag{5.219}$$

Now the [basic axiom](#) A1) reduces to the cocycle equation for $\omega$:

$$\omega(a, b, c)\omega(a, bc, d)\omega(b, c, d) = \omega(ab, c, d)\omega(a, b, cd). \tag{5.220}$$

**Groupoids**

**Definition 178.** A (finite) groupoid over a source/target set $X$ with one set $B_{xy}$ for $x, y \in X$ is given by a unit $1_x \in B_{xx}$, an inverse $(\cdot)_{(xy)}^{-1} : B_{xy} \to B_{yx}$ and a product $\mu_{xyz} : B_{xy} \times B_{yz} \to B_{xz}$, such that those maps obey axioms similar to the group axioms.

Every groupoid defines a 3FE-CTL in a very similar way that groups do: The basis set for the edge indices is the set of groupoid elements, and the basis set of the face indices is 1 if the groupoid elements on the surrounding edges multiply to 1. Note that to this end for groupoids also the sources and targets of adjacent edges have to match. E.g.

$$|B^F( \ d \ \ b \ )|$$

$$= \delta_{T[a],T[b]} \quad \delta_{S[b],S[c]} \quad \delta_{T[c],S[d]} \quad \delta_{T[d],S[a]} \quad \delta_{ab^{-1}cd,1} \tag{5.221}$$

Here $S$ and $T$ are the source and target of the groupoid elements, respectively.

### 5.8.4 Basic properties

**Observation 94.** Let us have a look at the possible gauge transformations for 3FE-CTLs. First, the edge indices are equipped with a preferred basis, so gauge transformations on them are not general orthogonal matrices but only permutation matrices. Second, the basis of the face indices depends on the shape of the face and on the values of the indices at the adjacent edges. Therefore, the gauge transformations on those indices are orthogonal matrices $G_I$ depending on the shape and on the values $I$ of the surrounding edge indices. As for every contraction of a pair of face indices also the adjacent edge indices are contracted, such a gauge transformation leaves the CTL axioms invariant. E.g.,

$$ (5.222) $$

**Observation 95.** In addition to the gauge transformations above there is another ambiguity involving both the CTL tensors and the normalizations. We can rescale

$$
\begin{aligned}
F &\longrightarrow \alpha F, \\
C_i &\longrightarrow \alpha^{-1} C_i, \\
D &\longrightarrow \alpha D,
\end{aligned}
\tag{5.223}
$$

by an arbitrary $\alpha \in \mathbb{R}$. This leaves the basic axioms invariant: The additional factors $\alpha$, $\alpha^{-1}$ and $\alpha$ associated to the $F$-tensors, internal edges and internal vertices on both sides of the $2 \longrightarrow 3$ and $4 \longrightarrow 1$ moves cancel out exactly. CTLs related in this way are in the same phase, as the factors $\alpha$ alone as an 3FE-CTL with trivial indices is in the trivial phase (they calculate the euler characteristic of the lattice which is trivial in 3 dimension). If we reformulate the 3FE-CTLs without normalizations (as is vaguely described in Remark (35)), this ambiguity disappears automatically.

**Remark 120.** The local support convention assures that the indices around one face are fully supported. We can go one step further and impose that the indices around two adjacent faces are fully supported. To this end we take a lattice that can be glued to a pair of faces of any other lattice without changing it. Then we interpret the associated tensor as a linear map from the indices on the glued side to the remaining indices and set this linear map to the identity matrix (apart from some normalization). E.g.,

$$ (5.224) $$

$$ = \mathbb{1} \otimes \mathbb{1} \otimes \mathbb{1} \quad \forall a, b, c, d $$

Note that unlike the local support convention we might loose something by imposing this condition. Indeed the delta 3FE-CTL violates it. However, we can also implement the delta 3FE-CTL via the edge indices, and it seems that we can still get models for the same phases after imposing the condition above. For complex-real 3FE-CTLs we can impose an adapted condition by taking the complexification of the right hand side in Eq. (5.224).

**Remark 121.** Take the basic tetrahedron and associate two of its faces to a back layer and the other two to a front layer. Now regard the corresponding basic tensor as a linear map from the back to the front layer. Gluing the back layers of two copies of the basic tetrahedron yields one of the lattices that were set to the identity matrix by the condition in Remark (120). E.g.,

$$ (5.225) $$

Thus, the linear map corresponding to the basic tetrahedron is orthogonal (unitary in the complex-real case), up to normalization.

**Observation 96.** $|B^F|$ does not depend on the favorite edge of the face $F$. To see this consider the lattice whose index lattice consists of two copies of $F$, with different favorite edges. E.g., the following lattices for a 2-gon and a 4-gon:

$$ (5.226) $$

Gluing two copies of such a lattice together at the same face yields a face-pillow (whose favorite edge depends on which faces we glue). According to the local support convention the tensor associated to any face-pillow is the identity matrix, which implies that the tensor associated to one of the lattices above is orthogonal. This implies (again together with the local support convention) that the dimension of the face indices cannot depend on the favorite edge of the face.

**Observation 97.** Assume that we impose the condition in Remark (120). $|B^F_{a,b,\ldots}|$ can be interpreted as an (non-negative) integer valued tensor with indices $a, b, \ldots$ (for different face shapes $F$). Consider two adjacent faces $F_1$ and $F_2$ sharing one edge and the face $F_{1,2}$ obtained from removing the edge in between the two faces. The tensors for the different $F$ fulfill:

$$
|B^{F_{1,2}}_{a,b,\ldots,x}| = |B^{F_1}_{a,\ldots,x}||B^{F_2}_{b,\ldots,x}|
\tag{5.227}
$$

Here $x$ labels the index of the edge in between $F_1$ and $F_2$ and is summed over (see Remark (10)).

This can be seen in the following way: Consider the lattice whose index part consists of $F_1$ and $F_2$ on the front layer and $F_{1,2}$ on the back layer. E.g., for $F_1$ and $F_2$ being triangles and $F_{1,2}$ being a 4-gon we get:

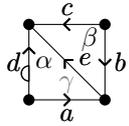

$$(5.228)$$

We can interpret the associated tensor as a linear map from the back layer to the front layer. If we glue two copies of this lattice at the front layer we get the face-pillow for $F_{1,2}$ to which the associated tensor is the identity linear map according to the local support convention. If we glue two copies of this lattice at the back layer we get a lattice to which the associated tensor is the identity linear map according to the condition in Remark (120). Thus, the linear map above is orthogonal, so the overall dimensions on the front and back layer have to be equal. For the overall dimension of the front layer we need to sum over the index on the edge between $F_1$ and $F_2$, yielding Eq. (5.227).

Consider a 2-gon face with non-cyclic edge orientations. If we put this face as $F_2$ adjacent to some face $F_1$, the resulting face $F_{1,2}$ equals $F_1$ again. So according to Eq. (5.227) we must have

$$|B^F\left(\begin{array}{c} b \\ \rightleftarrows \\ a \end{array}\right)| = \delta_{a,b} \qquad (5.229)$$

Consider the 2-gon face with non-cyclic edge orientations. If we take this face for both $F_1$ and $F_2$, then $F_{1,2}$ is the 2-gon face with non-cyclic edge orientations above. Thus, according to Eq. (5.227) the corresponding integer valued tensor is an involution when interpreted as a map from one edge to the other. Integer valued involutions are nothing but bijective maps $* : I \to I$ with $* \circ * = \mathrm{Id}$, i.e., for each edge index basis element $a$ there is one "inverse" element $a^*$ such that $a = a^{**}$. So we have

$$|B^F\left(\begin{array}{c} b \\ \rightleftarrows \\ a \end{array}\right)| = \delta_{a,b^*}. \qquad (5.230)$$

If we work with complex-real 3FE-CTLs with the adapted condition in Rem(120) we still get Eq. (5.227), for $|B^F|$ being half of the dimension of the actual realified tensors (what would be conventionally called the dimension of an index of a complex tensor).

**Comment 19.** Consider the tensor a) in Eq. (5.226). According to Observation (97) it consists of one number for every pair of $a$ and $a^*$ (or a complex phase in the complex-real case). Because of the cyclic symmetry of the face the bases of the two face indices are the same if $a = a^*$, so for this case this number is invariant under gauge transformations. In the language of unitary fusion categories, this complex number is known as the *Frobenius-Schur indicator*. The numbers associated to lattices that shift the favourite edge of a cyclic $n$-gon by one, for all edge indices having identical value $a$, are known as *$n$-th Frobenius-Schur indicator*.

### 5.8.5 Subtypes, mappings and fusions

**Observation 98.** Taking the double-pyramid with respect to a 2E index lattice defines a lattice mapping from 2E index lattices to the 3FE index lattices that can be extended to a mapping of the two glued, moved lattice types. For every edge of $P$ there are two triangular faces of the double-pyramid: Gluing two edges of a 2E lattice corresponds to gluing the two corresponding pairs of faces. E.g.,

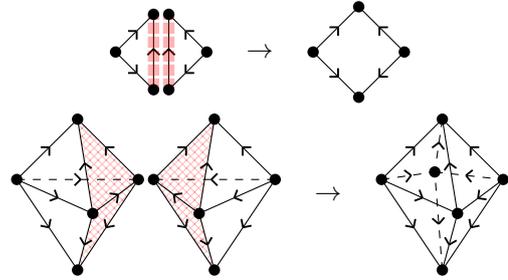

$$(5.231)$$

Copying each of the indices of the edges connected to the tips, and then block together all 7 indices around the pair of faces defines a tensor mapping. This yields a CTL mapping from 3FE-CTLs to 2E-CTLs.

**Definition 179.** For every 2E lattice $P$, define the **face-banana** for $P$ as the lattice whose index lattice has two vertices, one edge connecting them for every index vertex of $P$ and two 2-gon face between those edges for every index edge of $P$. The edge orientations are chosen such that all edges are oriented towards the same vertex and the favourite edge of each face is chosen to be the edge that corresponds to the vertex towards which the associated index edge of $P$ is oriented. E.g., consider the face-banana for the following 4-gon:

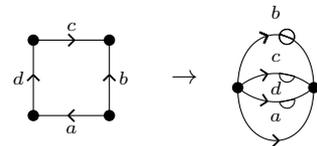

$$(5.232)$$

The face-banana for a 2E lattice whose index lattice consists of multiple disconnected components is the disjoint union of the face-bananas for the connected pieces.

**Remark 122.** Taking the face-banana can be formalized as a lattice mapping. Gluing two edges of a 2E lattice corresponds to gluing the associated faces of the face-banana. We can copy each of the edge indices of a face-banana and block every face index with one copy of the adjacent edge indices. Associating this blocked index with the corresponding edge of the 2E lattice defines a tensor mapping. Together with the lattice mapping this defines a CTL mapping from 3FE-CTLs to 2E-CTLs (actually 2En-CTLs, as we have to take care of the normalizations).

**Comment 20.** Take the tensors associated to the face-bananas and fix the same basis element $i$ for all edge indices. The TL mapping from Remark (122) restricted to this fixed element $i$ also yields a 2E-CTL, whose index is not the 3FE face index alone. E.g., for the delta 3FE-CTL and the only basis element $i$ this yields the delta 2E-CTL. According to Observation (97), the condition in Eq. (120) forces this 2E-CTL to be trivial (the number one), or the complex numbers in the complex-real case.

In the language of *-algebras and unitary fusion categories, this [2E-CTL](#) is known as *automorphism algebra* of the simple object $i$. For ordinary topological order in $2+1$ dimensions (i.e., without symmetry breaking order) this automorphism algebra is indeed trivial. But it is existing if we have GHZ-like entanglement in our system. (Also if we use fermionic tensors with spin structure lattices it can be non-trivial, see Ref. [51].)

### 5.8.6 Alternatives

By **3F-CTLs** we refer to a [CTL type](#) that is very similar to 3FE-CTLs. The lattices, [index lattices](#), moves and gluings are the same. However, the tensors are just plain real tensors with one [index type](#). There is one index associated to every index face. As for 3FE-CTLs the basis set of the index depends on the shape, edge orientations and favourite edge of the face. There is no dependency on values of edge indices however. When two faces are glued, the associated indices are contracted. If two faces next to each other are glued such that the edge connecting them disappears, we have to contract with some normalization matrix instead of the trace. This matrix depends on the shape of the glued face and on the location of the disappearing edge within this face.

3FE-CTLs and 3F-CTLs are in the same class, i.e., there are invertible [CTL mappings](#) in both directions. The [lattice mappings](#) are trivial. The tensor mapping in one direction simply takes the face indices of the 3F-CTL as face indices of the 3FE-CTL, together with trivial edge indices. The tensor mapping in the other direction copies each of the edge indices. Then for each face block the face index with one copy of all the edge indices surrounding the face. E.g., for some local patch of the [index lattice](#):

$$(5.233)$$

where $\xi = (\alpha, a_1, b_1, c_2, d_2)$, $\quad \epsilon = (\beta, b_2, e_1, f_2)$.

The advantage of 3F-CTLs is that the index distribution and contraction is simpler. In particular, the new face indices are all independent, and we do not have a direct-sum structure as for 3FE-CTLs. On the other hand, we now need a larger index dimension of the new face indices compared to 3FE-CTLs. 3FE-CTLs are a bit more explicit: We need much less numbers to write down the basic tensor (associated to the tetrahedron), as the dependency of the face index basis on the edge indices already poses a lot of constraints.

### 5.8.7 Connection to unitary multi-fusion categories

**Definition 180.** A **unitary multi-fusion category** consists of some discrete data, namely: A finite set $I$ whose elements are called (isomorphism classes of) **simple objects**, sets of non-negative integers $N_c^{ab} \forall a, b, c \in I$ called **fusion dimensions**, $N^{ab} \forall a, b \in I$ called the **inversion**, and $N_a \forall a \in I$ called the **unit**. Those have to obey a few equations,

$$\sum_c N_c^{ab} N_e^{cd} = \sum_c N_c^{bd} N_e^{ac}, \qquad (5.234)$$

$$\sum_x E_x N_b^{ax} = \delta_{a,b}, \qquad (5.235)$$

$$N_x^{aa^*} = E_x, \qquad (5.236)$$

$$\sum_b N^{ab} N^{bc} = \delta_{a,c}. \qquad (5.237)$$

The last equation says that $N^{ab}$ associates to each $a$ a unique $b$. The corresponding involution is usually denoted by $b = a^*$. The actual algebraic data is given by a complex tensor $[\tilde{F}_d^{abc}]_{f\gamma\delta}^{e\alpha\beta}$ (where we put the tilde to differentiate it from the 3FE basic tensor) with $\alpha \in N_e^{bc}$, $\beta \in N_d^{ae}$, $\gamma \in N_f^{ab}$ and $\delta \in N_d^{fc}$, satisfying the following set of equations

$$\sum_\delta [\tilde{F}_e^{fcd}]_{g\delta\gamma}^{l\delta\nu} [\tilde{F}_e^{abl}]_{f\alpha\delta}^{k\lambda\mu} = \sum_{h\sigma\psi\rho} [\tilde{F}_g^{abc}]_{f\alpha\beta}^{h\sigma\psi} [\tilde{F}_e^{ahd}]_{g\sigma\gamma}^{k\lambda\rho} [\tilde{F}_k^{bcd}]_{h\psi\rho}^{l\mu\nu}, \qquad (5.238)$$

$$\sum_{f\gamma\delta} [\tilde{F}_d^{abc}]_{f\gamma\delta}^{e\alpha\beta} \; \overline{[\tilde{F}_d^{abc}]_{f\gamma\delta}^{g\mu\nu}} = \delta_{e,g} \delta_{\alpha,\mu} \delta_{\beta,\nu}. \qquad (5.239)$$

**Proposition 19.** Every complex-real 3FE-CTL defines a unitary multi-fusion category when we impose the [local support convention](#) and the complex-real version of the condition in Remark (121). This multi-fusion category is given by

$$I := B^E,$$

$$N_c^{ab} := |B^F \left( \begin{smallmatrix} a \\ & \cdot \\ & \cdot \end{smallmatrix} b \right)|,$$

$$N^{ab} := |B^F \left( \begin{smallmatrix} a \\ \cdot \\ b \end{smallmatrix} \right)|, \qquad (5.240)$$

$$N_a := |B^F \left( \bigcirc a \right)|,$$

$$[\tilde{F}_d^{abc}]_{f\gamma\delta}^{e\alpha\beta} := F_d^{abc}{}_{f\gamma\delta}^{e\alpha\beta} \; \sqrt{C_e} \sqrt{C_f}. \qquad (5.241)$$

Note that there is no summation over $e$ and $f$ in the last equation.

This is easily verified: The axioms Eq. (5.234) follow immediately from Observation (97). Eq. (5.238) with Eq. (5.241) is immediately equivalent to the axiom Eq. (5.210), where the edge orientations are such that the edge $1 - 4$ of the overall 4-simplex is the one in the middle on the right side (due to the branching structure the vertices of the 4-simplex can be labeled by 0 to 4 such that the edge orientations point towards the vertex with higher number). Eq. (5.239) follows directly from the complex-real case of Remark (121), where the back (or front) layer is given by the faces $0 - 1 - 2$ and $0 - 2 - 3$.

**Comment 21.** In the context of fusion categories, the $C_i$ are usually denoted $d_i$ and are called the *quantum dimensions*.

**Remark 123.** Consider the following condition that we conjecture to hold for all multi-fusion categories: There exist real non-zero $C_i$ (for $i \in I$) such that also the following equations hold:

$$\sum_{c\alpha\delta} [\tilde{F}_d^{abc}]_{f\gamma\delta}^{e\alpha\beta} \quad \overline{[\tilde{F}_d^{gbc}]_{f\nu\delta}^{e\alpha\mu}} \quad \frac{C_c}{\sqrt{C_e C_f}} = \delta_{a,g}\delta_{\beta,\mu}\delta_{\gamma,\nu},$$

$$\sum_{b\alpha\gamma} [\tilde{F}_d^{abc}]_{f\gamma\delta}^{e\alpha\beta} \quad \overline{[\tilde{F}_g^{abc}]_{f\gamma\nu}^{e\alpha\mu}} \quad \frac{C_b}{\sqrt{C_e C_f}} = \delta_{d,g}\delta_{\beta,\mu}\delta_{\delta,\nu}.$$

(5.242)

Any multi-fusion category fulfilling this condition yields a 3FE-CTL.

This can be seen in the following way: The conditions above are equivalent to Remark (121) for any choice of back and front layer. On the other hand the basic axiom Eq. (5.210) for one particular choice of edge orientations follows from Eq. (5.238). Starting from there we can use Remark (121) to bring basic tetrahedra from the left to the right side and vice versa. This way we obtain Eq. (5.238) for all other edge orientations and also Eq. (5.239) for all edge orientations.

So if the conjecture above is true then there is a one-to-one mapping between unitary multi-fusion categories and complex-real 3FE-CTLs fulfilling the condition in Remark (120

**Comment 22.** The condition Eq. (5.242) is immediately true if we impose that the basic tensor $F$ that we would obtain from $\tilde{F}$ according to Eq. (5.241) has the following property: The orientation of every edge can be changed by replacing the simple object label by its inverse. This results in an invariance of the tensor $\tilde{F}$ under the full tetrahedral symmetry group, generated by

$$[\tilde{F}_d^{abc}]_{f\gamma\delta}^{e\alpha\beta} = \overline{[\tilde{F}_{c}^{b^*a^*d}]_{f^*\gamma\delta}^{e\beta\alpha}} = \overline{[\tilde{F}_d^{fb^*e}]_{a\gamma\beta}^{c\alpha\delta}} \quad \sqrt{\frac{d_e d_f}{d_c d_a}}$$

$$= \overline{[\tilde{F}_e^{a^*fc}]_{b\gamma\alpha}^{d\delta\beta}} \quad \sqrt{\frac{d_e d_f}{d_d d_b}}.$$

(5.243)

Those are the kind of $\tilde{F}$-tensors used in Ref. [17].

However, $F$-tensors with this form of tetrahedral symmetry are only a genuine subset of unitary fusion categories as well as of 3FE-CTLs. For example the twisted group 3FE-CTLs (and the corresponding unitary fusion categories) are in general not of this form.

### 5.8.8 Physical interpretation and connection to existing models

#### Classical statistical systems

In the context of classical statistical physics, the delta 3FE-CTLs correspond to fixed point models for symmetry-breaking order in 3 dimensions. Also other 3FE-CTLs with non symmetry-breaking correlations correspond to classical statistical models (not necessarily all of them however). It is unclear to us in which sense those models are in a different phases in the conventional language.

#### Quantum physics

In the context of quantum physics, complex-real 3FE-CTLs correspond to fixed point models for non-chiral topological order in $2 + 1$ dimensions. They also include symmetry-breaking order and mixes of symmetry-breaking and topological order.

**Observation 99.** Let us construct a Hamiltonian for 3FE-CTLs following Observation (55). The minimal possible covering of a given index lattice by local projector lattices is the following: For every vertex consider all the surrounding faces. Take the lattice whose index lattice has a back and front layer that both look like all those surrounding faces. The CTL tensor associated to such a local projector lattice is a projector (up to normalizations). E.g., for a vertex that is surrounded by 5 triangles we get the following local projector lattice:

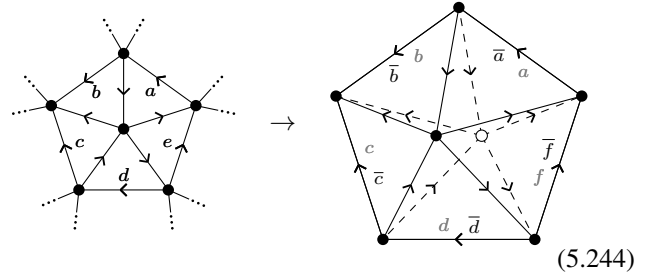

(5.244)

Gluing together all those local projector lattices for all vertices of an index lattice $X$ yields a lattice with background $X \times [0, 1]$. Note that if there was one single vertex for which not at least all surrounding faces are included in some local projector lattice, this would not work anymore, as this vertex would be part of both $X \times 0$ and $X \times 1$. In order to make the associated tensor a projector we have to normalize by $D^{-1}$ once and by $C_i^{-1/2}$ for every of the indices at the edges within the back or front layer.

**Remark 124.** Technically, the vector spaces of the CTL tensors are not even many-body quantum state spaces, as they do not have a tensor product structure, but a more direct sum structure. There is a simple work-around for this that is usually assumed in first place in the literature: For any shape of face $F$ determine the maximal basis set $B^F$ for the face index depending on the values of the edge indices. Then, if the basis set of face indices is smaller for particular edge index values, embed the vector space spanned by the smaller set into the one spanned by the maximal set. Of course, such an embedding cannot be chosen in a canonical way. After embedding, the basis set of face indices does not depend any more on the values of the edge indices and the total vector space is a simple tensor product again.

With the connection to fusion categories in Section (5.8.7), the physical models or manifold invariants defined by 3FE-CTLs are essentially equivalent to the Turaev-Viro state-sum construction originally introduced in Ref. [12] for the quantum subgroups of $sl_2(\mathbb{C})$ and later generalized to arbitrary spherical fusion categories by Barrett and Westbury [13]. A more recent introduction can be found in Ref. [52]. The Turaev-Viro state-sum on a triangulation equals the evaluation of the tensor network that the TL behind a 3FE-CTL associates to a lattice.

Turaev-Viro invariants were later put in the context of topological order with a Hamiltonian formulation under the name "string-nets" by Levin and Wen [17, 53]. The string-net ground states on a trivalent grid are given by the ground states of the corresponding 3FE-CTL on the index lattice which is the dual of this grid (which is a triangular lattice then). The string-net Hamiltonian corresponds to our local Hamiltonian projector from Observation (99).

All those known models are only roughly equivalent but differ in various technical details of the algebraic structures that are put into the constructions. In our construction the algebraic structures emerge directly from topological invariance, so the technical assumptions of those algebraic structures are by construction those that are necessary and sufficient for topological quantum models.

For example, in Ref. [17] they imposed an explicit tetrahedral symmetry of the $F$-symbols (i.e., that the edge orientations can be inverted by simply inverting the corresponding simple object labels), which yields a canonical local anti-unitary (i.e., time-reversal) symmetry. However, such a tetrahedral symmetry is not necessary for topological invariance or unitarity of the resulting physical models. 3FE-CTLs do not have it and thus also cover the *twisted quantum double models* from Ref. [54] or the non-tetrahedral symmetric generalizations of abelian string-nets in Ref. [55]. We believe that they also cover examples of string-net models with non-abelian fusion that lack tetrahedral symmetry.

Our models are very directly connected to the Turaev-Viro models and make the connection between this state-sum construction and the Hamiltonian (string-net) formulation very apparent. A very similar approach can be found in Ref. [23].

Our models also include GHZ-like entanglement (symmetry breaking order) as the delta 3FE-CTLs show. They also allow for non-trivial mixtures between GHZ-like order and genuine $2+1$-dimensional topological order, as the groupoid 3FE-CTLs show. We believe that this amounts to using multi-fusion categories instead of fusion categories in the usual Turaev-Viro construction.

Our models come with a very natural way of extending the models to arbitrary cell complexes (not just triangular ones). While this does of course not yield any new models or even phases, it provides us with a nice interpretation of different algebraic constructions in terms of CTL tensors for non-standard lattices, as for example the (higher order) Frobenius-Schur indicators or the endomorphism algebra of simple objects.

Non complex-real 3FE-CTLs correspond to quantum models protected by a local anti-unitary (i.e., time-reversal) symmetry.

### 5.8.9 Concrete examples

**Example 122.** Consider the following twisted group 3FE-CTL for the group $Z_2$. The normalizations are given by $C_i = 1 \forall i$ and $D = 2$. The basic tensor has the full tetrahedral symmetry and all $Z_2$ elements are self-inverse, thus the edge orientations can be ignored. There are only 3 edge index configurations to specify the tensor entries for, which we will represent by drawing the edges with the trivial group element black, and the edges with non-trivial group element red:

$$
\begin{array}{ccc}
\triangle = 1 & \triangle = -1 & \triangle = 1
\end{array}
\tag{5.245}
$$

This CTL is equivalent to the simplest non-trivial twisted quantum double model [54], also known as *double-semion model*. The example also can be found in Ref. [17].

**Example 123.** As an example with multiple fusion channels (i.e., $N_{abx} \neq 0$ for more than a single $x$ for some $a, b$), consider an 3-FE-CTL with 2 irreps, 1 and $\tau$. $N_1 = N_{11} = N_{\tau\tau} = N_{111} = N_{\tau\tau1} = N_{\tau\tau\tau} = 1$, and all fusion dimensions except for cyclic permutations of those are 0. The quantum dimensions are given by $v_1 = 1$,

$$
v_\tau = \sqrt{\phi} = \sqrt{\frac{1 + \sqrt{5}}{2}},
\tag{5.246}
$$

$1 + \phi = \phi^2$. As in the $Z_2$ twisted example, the irreducible representations are self-inverse and the tetrahedron tensor with entries independent of the edge orientations. There remain five configurations to specify

$$
\begin{array}{ccc}
\triangle = 1 & \triangle = v_\tau^{-1} & \triangle = v_\tau^{-2} \\[1em]
\triangle = v_\tau^{-2} & \triangle = -v_\tau^{-4} &
\end{array}
\tag{5.247}
$$

## 5.9  3E-(C)TLs

### 5.9.1  Definition

3E-CTLs are a CTL type on 3-dimensional lattices coming from a TL type. The "E" stands for "edge" as this is where indices are associated to.

**Backgrounds**

Just as for 3FE-CTLs, the extended backgrounds are given by 3-manifolds with boundary. There is one background gluing, namely boundary surgery gluing.

**Lattices**

The lattices are based on 3CCb-lattices. The index lattice is given by the 2CC-lattice forming the boundary of the 3CCb-lattice. There are the following decorations: Every index edge carries an orientation and a dual orientation, i.e., for every edge we specify one favorite adjacent vertex and one favorite adjacent face. The index lattices have a very easy implementation of Poincaré duality: When exchanging vertices with faces we also interchange orientations with dual orientations. The basic moves are bi-stellar flips in the interior, i.e., the backgrounds are determined by the index lattice plus an abstract label for the topology of the interior. We will mostly work on the level of backgrounds by drawing the index lattice and describing the interior topology in words if ambiguous.

We will indicate the orientations and dual orientations of the index edges by putting a half arrow pointing along the orientation on the side where the dual orientation points to. I.e., the following edge:

$$
\cdots\!\!\bullet\!\!-\!\!\bullet\!\!\cdots
\tag{5.248}
$$

has the face above and the vertex to the right as its favourite face and vertex.

There is one basic gluing. The gluing locations are pairs of index edges. Intuitively one can imagine gluing of two edges as bringing the edges together such that their orientations and dual orientations match, until they touch each other and then "popping up" the line where they touch, i.e., removing the edges and fusing the two newly neighbouring faces at each side of this line together. E.g.,

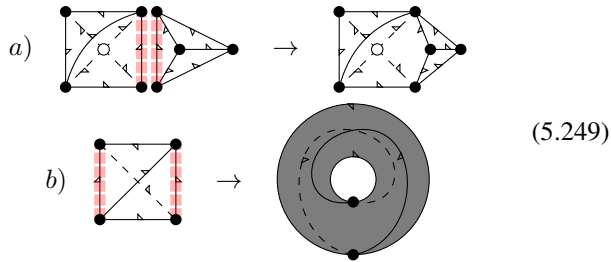

$$(5.249)$$

a) Shows a lattice with the extended background of two 3-balls glued to a single 3-ball. b) shows gluing two edges of a lattice with the extended background of a single 3-ball yielding a lattice with the extended background of a solid torus.

### Tensors

The tensors are real tensors with one index type. One index is associated to every index edge.

When two index edges that are well separated from each other are glued together, the associated indices are contracted. In rather degenerate cases when they are near each other we have to slightly modify the contraction using the following three different normalizations: A tensor $C$ with two indices, and two numbers $D$ and $E$. First, when we glue two index edges that are connected to a common face and a common vertex, we have to contract the associated indices using the matrix $C$:

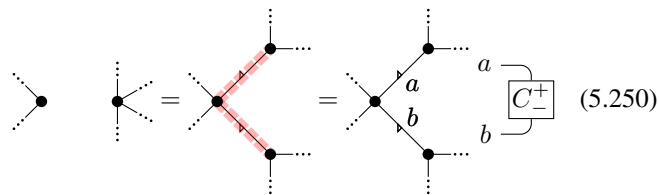

$$(5.250)$$

More precisely we need 4 different matrices $C_-^-$, $C_+^-$, $C_-^+$ and $C_+^+$. The superscript indicates whether the orientation of the glued edges point towards the common vertex or not and the subscript whether the dual orientations of the glued edges point towards the common face or not. Often all matrices are the same in which case we omit the sub- and superscripts. All those matrices have to be symmetric as by the symmetry axiom there is no way to determine which index to connect to which edge.

Second, when we glue two edges that share a connected face and both connected vertices such that this face in between disappears, we additionally have to multiply by $D$:

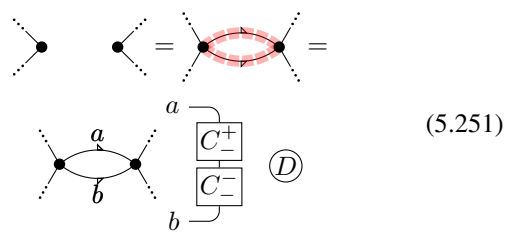

$$(5.251)$$

Third, when we glue two edges that share a connected vertex and both connected faces such that this vertex in between dis-

appears, we additionally have to multiply by $E$:

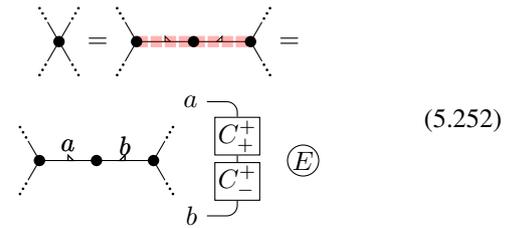

$$(5.252)$$

### 5.9.2 Basic tensors and axioms

**Basic lattices and tensors**

**Definition 181.** For every 2E background $X$, define the **pillow** with respect to $X$ as the following 3E background $Y$: The extended background of $Y$ equals the extended background of $X$ times the interval, where $X$ times each of the endpoints is contracted to a point each. If $X$ has multiple disconnected pieces then $Y$ also has (according to the definition before they would all touch at two points, but the precise combinatorial formulation of the lattices will not be sensitive to that). For every connected component of $X$, the index lattice of $Y$ consists of 2 faces that both have the shape of the connected component. I.e., for every edge of $X$ there is one edge of $Y$. One might visualize this as taking $X$, fixing its index lattice, and inflating the interior part into 3 dimensions. The dual orientations are chosen such that they all point to the same face. E.g., consider the following 2E backgrounds (left) and their corresponding pillows (right):

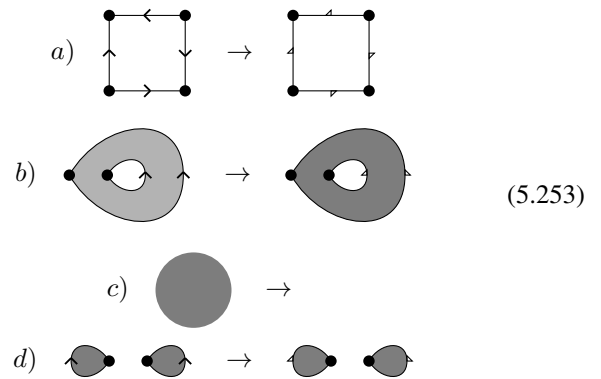

$$(5.253)$$

a) shows the standard case with disk extended background of $X$ yielding a 3-ball extended background of $Y$. b) shows an annulus extended background of $X$ yielding a solid torus extended background of $Y$. c) shows a sphere extended background of $X$ yielding a 3-sphere extended background of $Y$ (which has an empty 2-skeleton, so unfortunately the corresponding drawing is empty). d) shows an extended background of $X$ consisting of two disks, yielding an extended background of $Y$ consisting of two 3-balls.

**Definition 182.** For every 2E background $X$, define the **banana** with respect to $X$ as the following 3E background $Y$: The extended background is the same as for the pillows, but now the index lattice of $Y$ has 2 vertices (for every connected component) and one edge that connects them for every edge of $X$ with 2-gon faces in between. The dual orientations of those edges match the orientations of the edges of $X$. The orientations are chosen such that all point towards the same vertex.

E.g., consider the following 2E background (left) and the corresponding banana (right):

 (5.254)

**Observation 100.** The pillow and the banana with respect to the same $P$ are Poincaré duals to each other.

**Observation 101.** Taking the pillow of a 2E background commutes with symmetries, disjoint union and gluing. E.g., consider the following gluing of 2E backgrounds (above) and the corresponding gluing of the associated pillows (below):

 (5.255)

Taking the pillow can be formalized as a mapping from 2E lattices to 3E lattices. The same statement holds for the bananas by Poincaré duality.

**Observation 102.** Every background can be glued together from pillows and bananas.

We can obtain a history for a given lattice in the following way: Choose one favourite adjacent volume for every face and one favourite adjacent vertex for every edge in the interior of the 3CCb-lattice. Replace each face of this decorated lattice by the pillow with respect to the lower link of this face. Dually, replace each edge in the interior of the 3CCb-lattice by the banana with respect to the upper link of this edge. For edges in the index lattice the upper link is a (decorated) 1CCb-lattice with an interval background. We can close this 1CCb-lattice to a 1CC-lattice by one edge whose orientation is given by that of the original index edge. We replace every edge in the index lattice by the banana of this closed upper link.

So in the interior, each banana at each edge of the 3CCb-lattice has one edge for each connected face, and each face in the interior has one edge for each connected edge. So at each pair of connected face and edge in the interior there is a pair of edges of the respective pillow and banana that can be glued together. The bananas corresponding to the edges at the boundary have one edge that is not glued and that yields the edge of the index lattice lattice in the end. So we found a history for the desired lattice.

**Observation 103.** Due to the pillow (banana) mapping from Observation (101) we see that every pillow (banana) can be glued together from copies of the pillow (banana) with respect to the 2E basic lattice. Combined with Observation (102) we see that the pillow and the banana with respect to the 2E basic lattice form a set of basic lattices. So the whole 3E-CTL is

already determined by the associated basic tensors:

$$T1) \quad F_c^{ab} := \text{}$$

$$T2) \quad G_c^{ab} := \text{}$$ (5.256)

We also need the normalizations $C$ and $D$ or $E$.

**Basic history moves and axioms**

**Definition 183.** With the pillow mapping we can transform every 2E history $X$ into a 3E history that we call the **pillow history** of $X$. E.g.,

 (5.257)

Analogously we can define the **pillow history lattice** of a 2E history lattice. All pillow history lattices can be history glued from the pillow history lattice of the 2E basic history lattice:

 (5.258)

Dually we can define the **banana history** of a 2E history and the **banana history lattice** of a 2E history lattice, and every banana history lattice can be history glued from the banana history lattice of the 2E basic history lattice.

**Definition 184.** For every pair of 2E lattices $P$ and $Q$, define the **pillow-banana history lattice** of $P$ and $Q$ as the history lattice that consists of a copy of the banana with respect to $Q$ for each edge of $P$ and a copy of the pillow with respect to $P$ for each edge of $Q$. So for each pair of edges of $P$ and $Q$ there is one edge of a banana and one edge of a pillow. Each such pair of edges is glued together. E.g., for $P$ and $Q$ both being the basic triangle:

 (5.259)

**Proposition 20.** Every history lattice can be history glued from pillow history lattices, banana history lattices and pillow-banana history lattices.

A history for a given history lattice $X$ can be obtained in the following way: History lattices are in the same class as 4CCb-lattices. Replace each volume of the 4CCb-lattice by the pillow history lattice of its lower link. Also replace each edge by the banana history lattice of its upper link. Also replace every face by the pillow-banana history lattice of its upper and lower link. For the edges/faces at the boundary of the 4CCb-lattice we complete the upper link by a single face/edge. Now in the interior there is one pair of pillows for each pair of connected volume and face, and one pair of bananas for each pair of edge and connected face. We can history glue at all those pairs of pillows and bananas. At the boundary of the 4CCb-lattice we get unglued pillows and bananas yielding the desired $X$.

**Observation 104.** For some 2E lattice $P$ let $P'$ be the lattice obtained after gluing two edges of $P$. Then for any $Q$ the pillow-banana history lattice with respect to $Q$ and $P'$ can be obtained from the pillow-banana history lattice with respect to $Q$ and $P$ by history gluing the two bananas corresponding to the two glued edges of $P$. So each pillow-banana history lattice with respect to some $Q$ and $P$ can be history glued from pillow-banana history lattices with respect to $Q$ and the (triangle) 2E basic lattice.

Dually, each pillow-banana history lattice with respect to some $Q$ and $P$ can be history glued from pillow-banana history lattices with respect to the 2E basic lattice and $P$. So in total all pillow-banana history lattices can be history glued from the pillow-banana history lattice with respect to twice the 2E basic lattice shown in Eq. (5.259).

**Observation 105.** According to Remark (24) the bi-partitions of the basic history lattices are good candidates for basic history moves (yielding basic axioms). There are three kinds of bi-partitions of the pillow-banana history lattice Eq. (5.259).

First there are 9 bi-partitions into one banana and one pillow on the right and two bananas and two pillows on the left, corresponding to the 3 different choices for the banana and 3 different choices of the pillow on the right. E.g.,

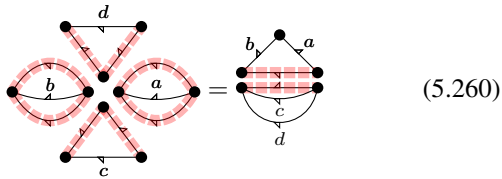

$$(5.260)$$

Second, there are 3 bi-partitions into one single pillow on the right and the rest on the left, corresponding to the three different choices of the pillow on the right. E.g.

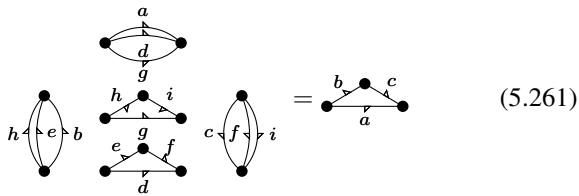

$$(5.261)$$

Third, there are the 3 dual bi-partitions to the second ones: One banana on the right and the rest on the left.

There are two kinds of bi-partitions of the pillow gluing lattice corresponding to the two kinds of bi-partitions of the 2E basic gluing lattice. Dually there are two kinds of bi-partitions of the banana gluing lattice.

**Remark 125.** The axioms from the observation before indeed form a complete set of basic history lattices (yielding basic axioms): The defining history given in Observation (102) is already consistent with disjoint union and basic gluing. We still need to show that the defining history of a lattice after a basic move can be obtained by applying basic history moves to the defining history before the move.

To demonstrate this we restrict to lattices whose underlying 3CCb-lattices look like 3SCb-lattices in the interior, and show that the Pachner moves can be accomplished by a sequence of basic history moves. We keep in mind that history lattices for 3SC-lattices (without gluing) are 4SC-lattices, where each 4-simplex represents a Pachner move. Such a 4SC-lattice can be history glued from one pillow history lattice for each volume, one banana history lattice for each edge and one pillow-banana gluing lattice for each face of the 4SC-lattice as described in Prop. (20). So a Pachner move corresponding to a 4-simplex can be accomplished by applying the basic history moves arising from bi-partitions of the basic history lattices associated to the volumes, edges and faces that are part of the 4-simplex. We just need to find an ordering of the basic history lattices such that each history gluing represents a valid history move. Such orderings exist for both Pachner moves, e.g., we can replace the following $1 \to 4$ Pachner move replacing the tetrahedron (0123) with four other tetrahedra with sequence the pillow, banana and pillow-banana gluing moves corresponding to the volumes, edges and faces below:

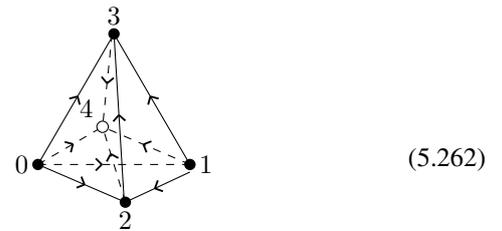

$$(5.262)$$

$$(012), (0124), (123), (12), (124), (1234),$$
$$(234), (24), (024), (0234), (02), (23), (023)$$

Here e.g., (012) corresponds to applying the basic history move arising from a bi-partition from the pillow-banana history lattice associated to the face (012).

**Simplification of the axioms**

The procedure above yields a large number of rather complex axioms. In the following we show how to obtain a simpler set of equivalent axioms.

**Observation 106.** The pillow mapping can be made a CTL mapping from 3E-CTLs to 2En-CTLs. The corresponding tensor mapping takes the 3E edge index as the 2E index at the corresponding edge. E.g., for the 2En basic lattice:

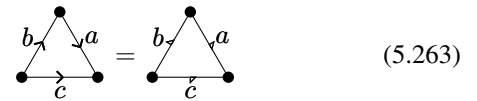

$$(5.263)$$

During a gluing of edges, the indices of 2En-CTLs are contracted, just as the indices of the corresponding pillows. Just that for the pillows we have to include normalizations when we glue neighbouring edges. Thus, the 3E-CTL restricted to the pillows yields a 2En-CTL, with the normalizations given by

$$a - \boxed{C_{2\mathrm{En}}^{\pm}} - b \; = \; \overset{\textcircled{E}}{\underset{a - \boxed{C_{3\mathrm{E}-}^{\pm}} - \boxed{C_{3\mathrm{E}+}^{\pm}} - b}{}} \qquad (5.264)$$

The dual statement holds for the bananas. The only difference for the normalizations is that we have to use the number $D$ instead of $E$ and combine different superscripts of $C$ with a fixed subscript.

**Remark 126.** Consider the pillow for the 2-gon with non-cyclic edge orientations. Gluing one of its edges to any edge of any lattice does not change the background of the latter. So according to the local support convention we can set the associated tensor to the identity matrix:

$$a \bullet\!\!\!\!\diamond\!\!\!\!\bullet b = a \;\text{———}\; b \tag{5.265}$$

**Observation 107.** According to Observation (106) we can take the simplified set of tensors T1) and T2) and axioms A1) to A5) for 2E-CTLs from Section (5.4.2) and apply the pillow mapping to get a simplified set of 3E basic tensors and axioms, which are equivalent to the axioms arising from bi-partitions of the basic pillow history lattice together with the local support convention. Thereby we get the following auxiliary basic tensor:

$$T3) \qquad \bullet\!\!\!\!\diamond\!\!\!\!\bullet^{\,a}_{\,b} \tag{5.266}$$

With this auxiliary tensor we have the following axioms:

$$
\begin{aligned}
A1) \quad & \cdots = \cdots \\
A2) \quad & \cdots = \cdots \\
A3) \quad & \cdots = \cdots \\
A4) \quad & \cdots = \cdots \\
A5) \quad & \cdots = \cdots
\end{aligned}
\tag{5.267}
$$

Dually we can use the banana mapping to get a simplified set of tensors and axioms that is equivalent to the axioms arising from bi-partitions of the basic banana history lattice:

$$T4) \qquad \bullet\!\!\!\!\diamond\!\!\!\!\bullet^{\,a}_{\,b} \tag{5.268}$$

$$
\begin{aligned}
A6) \quad & \cdots = \cdots \\
A7) \quad & \cdots = \cdots \\
A8) \quad & \cdots = \cdots \\
A9) \quad & \cdots = \cdots \\
A10) \quad & \cdots = \cdots
\end{aligned}
\tag{5.269}
$$

Further, we have the following consistency axioms between the pillows and bananas,

$$
\begin{aligned}
A11) \quad & \cdots = \cdots \\
A12) \quad & \cdots = \cdots
\end{aligned}
\tag{5.270}
$$

With those axioms we only need one of the axioms arising from the decompositions of the pillow-banana history lattices:

$$A13) \quad \cdots = \cdots \tag{5.271}$$

For any equation between bananas and pillows we can contract the tensor T3) to both sides and use A4) or A5) to invert the orientation of any non-glued edge of any pillow on each side. Analogously we can contract the tensor T4) to both sides and use A9) or A10) to invert the dual orientation of all non-glued edges of every banana. We can use A8) together with A9) or A10) and A11) to invert the dual orientations of all edges of any pillow together with the dual orientations of the edge of any banana that is glued to this pillow. Analogously we can use A3) together with A4) or A5) and A12) to invert the orientations of all edges of any banana together with the orientations of the edge of any pillow that is glued to this banana.

Applying this to A13) we get all the 9 basic axioms in Eq. (5.260). Also we can glue a pillow to both sides of A13) and use A2) on the right side to obtain one of the 3 axioms in Eq. (5.261). We can again use the axioms as in the previous paragraph to obtain the other 2 axioms in Eq. (5.261). Analogously we can glue a pillow to both sides of A13) and use A7) on the right side to obtain one of the axioms dual to Eq. (5.261), and from this get also the other 2 axioms.

Thus, all the basic axioms together with the local support convention follow from the axioms A1) to A13) above.

### 5.9.3   Solutions

**Delta tensors**

For every (finite) set $B$, the **delta 3E-CTL** (see Observation (20)) with the basis set of edge indices given by $B$ are a solution to the axioms. The basic tensors and normalizations are given by

$$
\begin{aligned}
F_k^{ij} &:= \delta_{i,j,k}, \\
G_k^{ij} &:= \delta_{i,j,k}, \\
C &:= \mathbb{1}, \\
D &:= 1, \\
E &:= 1.
\end{aligned}
\tag{5.272}
$$

**Observation 108.** The 2En-CTL formed by the bananas and the one formed by the pillows is the delta 2E-CTL.

The tensor $T[X]$ associated to an arbitrary lattice is determined by the following property: For a given configuration of set elements at the index edges, the tensor entry is given by

- 1, if for all set elements in each connected component of the lattice are the same.

- 0, otherwise.

**Groups**

For every (finite) group $G$, the **group 3E-CTL** is a 3E-CTL with the basis set of edge indices given by the set of group elements. The basic tensors and normalizations are given by

$$
\begin{aligned}
F_k^{ij} &:= \delta_{ij,k}, \\
G_k^{ij} &:= \delta_{i,j,k}, \\
C &:= \mathbb{1}, \\
D &:= 1, \\
E &:= |G|^{-1}.
\end{aligned}
\tag{5.273}
$$

**Observation 109.** The 2En-CTL formed by the bananas is the delta 2E-CTL. The 2En-CTL formed by the pillows is the group 2E-CTL, apart from the missing normalization factor.

The tensor $T[X]$ associated to an arbitrary lattice (up to some normalization factor) is determined by the following property: For a given configuration of group elements at the edges, the tensor entry is given by

- 1 if for all faces of $X$, the product of group elements associated to the edges around the face in cyclic order is the group identity. In the product we have to take the inverse of the group element instead of the group element if the edge orientation points against the chosen cyclic order. For a lattice with non-ball extended background, also the products of group elements along any bounding path of index edges (i.e., one that is the intersection of some membrane in the interior with the boundary) have to yield the group identity.

- 0 if the configuration violates one of the constraints above.

**Complex numbers**

The **complex number 3E-CTL** (see Definition (88)) is a 3E-CTL with the basis set of edge indices equal to {real, imag}. The basic tensors and normalizations are given by

$$
\begin{aligned}
F_k^{ij} &:= 2^{-1/2} \mathbb{C}_k^{ij}, \\
G_k^{ij} &:= 2^{-1/2} \mathbb{C}_k^{ij}, \\
C &:= \mathbb{1}, \\
D &:= 1, \\
E &:= 1.
\end{aligned}
\tag{5.274}
$$

For a general lattice $X$ the associated tensor is just a tensor product of one complex number tensor for each connected component of $X$, with one index for each index edge. Because of the permutation property of the complex number tensors, it does not matter which index we associate to which edge. The complex arrow orientation depends on whether the dual orientation of the edge points to the left or to the right when looking along the orientation of the edge. Note that there is no orientation, so what is "left" or "right" can only be non-canonically chosen for each connected component. This is still consistent because of the arrow reversal property of the complex number tensors.

**Quadruple-line tensors**

For each (finite) set $B$ the **quadruple-line 3E-CTL** is a 3E-CTL with the edge index basis set equal to $B^4$. The basic tensors and normalizations are given by

$$
\begin{aligned}
C &:= \mathbb{1}|B|^{-1}, \\
D &:= 1, \\
E &:= 1,
\end{aligned}
\tag{5.275}
$$

where $a_1, \ldots, c_4 \in B$.

For a general lattice $X$ the associated tensor (up to an overall normalization) can be constructed as follows: For each pair of connected index vertex and index face take one identity matrix and associate one of its two indices to each of the two edges that are connected to both the vertex and face. Each edge is connected to two neighbouring vertices and faces that form four connected vertex-face-pairs in total, so each edge index gets four $B$-indices that way. We can order those via the orientation and dual orientation of the edge. The CTL tensor does not depend on the interior topology of the background apart from normalization.

**Observation 110.** Both the 2En-CTLs of bananas and pillows are given by a tensor product of twice the double-line 2E-CTL. Just that the indices are grouped in a different way (namely $(1,2),(3,4)$ vs. $(1,3),(2,4)$, and that we have a normalization $C_{2En} = |B|^{-2}$.

**Delta-double-line tensors**

For each (finite) set $B$ the **delta-double-line 3E-CTL** is the 3E-CTL with the edge index basis set equal to $B \times B$. The basic tensors and normalizations are given by

$$C := \mathbb{1},$$
$$D := 1,$$
$$E := |B|^{-1},$$

where $a, \ldots, \overline{c} \in B$.

The tensor corresponding to an arbitrary lattice $X$ can be constructed in the following way (up to a potential normalization): Put a delta tensor on every index face of $X$ with one index for each edge adjacent to that face. Then associate to each edge the composite of the two indices coming from the two faces connected to it. The tensor does not depend on the interior topology of a background, except for a potential normalization.

**Observation 111.** The 2En-CTL of bananas is the double-line 2E-CTL (where the normalization is put to $C$), and the 2En-CTL of pillows is given by two copies of delta 2E-CTL.

**Groupoids**

For every (finite) groupoid $G$ there is the **groupoid 3E-CTL** with the basis set of edge indices given by the disjoint union of all $B_{xy}$. So we can write each index formally as a triple $(xyi), x, y \in X, i \in B_{xy}$ where the dimension/type of the $i$-index depends on the value of the $x$ and $y$ indices as described in Remark (32). The basic tensors and normalizations are given by

$$F_{(yzk)}^{(uvi)(wxj)} := \delta_{v,w}\delta_{u,y}\delta_{x,z}\delta_{ij,k}$$

$$D := 1,$$
$$E := 1,$$
$$v_y = (\sum_{z \in X} |B_{yz}|)^{1/4},$$
$$w_y = (\sum_{z \in X} |B_{yz}|)^{-1/2}.$$

**Remark 127.** It turns out that in general one cannot normalize the $F$ and $G$ tensors such that the normalization matrices become both trivial. Groupoid 3E-CTLs are the first of the solutions so far for which this is not possible.

For a general lattice, the corresponding tensor (up to normalization) can be constructed as follows: For any configuration of triple-indices, the tensor entry is 1 if

- For each vertex, the sources/targets of the surrounding edges are all the same. We have to take the source or target depending on whether the corresponding edge points inwards or outwards.

- For every face, the groupoid elements/inverse groupoid elements of the surrounding edges multiply to an identity of the groupoid. Note that this product is always defined when the previous point holds.

- The item before also holds for bounding cycles of edges

and 0 otherwise.

**Observation 112.** Groupoid 3E-CTLs can be seen as a generalization of both of delta-double-line 3E-CTLs and group 3E-CTLs: A group is just a groupoid with $X$ being the trivial 1-element set. The delta-double-line tensors are a groupoid with $B_{ij}$ being the trivial one-element set for all $i,j$ and the product of two elements being the only element with the correct source and target. Also the delta 3E-CTL corresponds to a groupoid CTL with $X$ being the delta tensor basis set and $B_{ij}$ being the empty set if $i \neq j$ and the trivial one-element set if $i = j$, i.e., the groupoid with only identity elements.

### 5.9.4 Basic properties

**Observation 113.** In addition to gauge transformations we have the following ambiguity: We can rescale the CTL tensors and normalizations by

$$
\begin{aligned}
F &\longrightarrow \alpha^{-1}F, \\
G &\longrightarrow \beta^{-1}G, \\
C &\longrightarrow \alpha\beta C, \\
D &\longrightarrow \alpha^{-2}D, \\
E &\longrightarrow \beta^{-2}E,
\end{aligned}
\tag{5.278}
$$

for arbitrary $\alpha, \beta \in \mathbb{R}$ without changing the CTL axioms. Those could be used to e.g., set $D$ and $E$ to 1. General pillows and bananas are rescaled by

$$
\begin{aligned}
T[\text{pillow}] &\longrightarrow \alpha^{2-\#\text{edges}}T[\text{pillow}], \\
T[\text{banana}] &\longrightarrow \beta^{2-\#\text{edges}}T[\text{banana}].
\end{aligned}
\tag{5.279}
$$

**Observation 114.** Let us look at where the normalizations occur in a the tensor network corresponding to a history. Such a history associates pillows and bananas to the faces and edges of a 3CCb-lattice as described in Observation (102). At every vertex in the interior we get one factor of $D$, no matter in which order we glue together the basic lattices. Dually we get a factor $E$ at every volume. At every $0, 3$-relation-cell (corner) of the 3CCb-lattice there is a loop of contractions. At one index of this loop we have to insert a normalization matrix $C$, the exact position does not matter.

Now take the trivial 3E-CTL and rescale it as in Eq. (5.278) with $\alpha = 1$ and some $\beta$. Then the number associated to a gluing lattice coming from a 3CC-lattice $X$ is:

$$
\begin{aligned}
T[X] &= \beta^{2\#1-\#1/3}\beta^{-2\#0}\beta^{\#0/3} = \\
&\beta^{(\#2/3-\#1/3+\#0/3)+(-\#2/3+2\#1-2\#0)} = \\
&\beta^{(\sum_{x\in 3\text{-cells}}\#2_x-\#1_x+\#0_x)+(-2\#2+2\#1-2\#0)} = \\
&\beta^{2\#3-2\#2+2\#1-2\#0} = \beta^{2\chi(X)} = 1
\end{aligned}
\tag{5.280}
$$

where $\#i$ is the number of $i$-cells, $\#i/j$ is the number of $i, j$-relation-cells and $\#i_x$ is the number of $i$-cells in the lower link of $x$. We used that the lower link of every 3-cell has ball-like background and therefore its (2-dimensional) Euler characteristic is 2. $\chi(X)$ is the Euler characteristic of $X$ which is always 0 in odd dimensions due to Poincaré duality.

This indicates that the rescaled trivial CTL is still in the trivial phase and rescaling does not change the phase of the CTL. The dual statements hold for rescalings with $\beta = 1$ and some $\alpha$.

### 5.9.5 Subtypes, mappings and fusions

#### Doughnut CTL

**Definition 185.** For a given 2E background $X$ define the **doughnut** with respect to $X$ as the following 3E background $Y$: The extended background of $X$ is the extended background of $Y$, where we identify the extended background of $X$ with one special point of this 1-sphere. The index lattice of $X$ is obtained in the following way: Replace every 2E edge with a 3E edge. The orientations of the 3E edges match the orientations of the 2E edges, and the dual orientations of the 3E-CTL edges all point in the same direction. For every 2E edge there is an additional 3E edge starting and ending at the vertex where the orientation of the 2E edge points to and winding once around this vertex times the 1-sphere. The orientations of those 3E edges are chosen to match the dual orientations of the other kind of 3E edges. The dual orientations are chosen to match the orientations of the corresponding 2E edges. In the end, each edge of the 2E background corresponds to a pair of edges of the corresponding doughnut. E.g., consider the following 2E backgrounds and the doughnuts with respect to them:

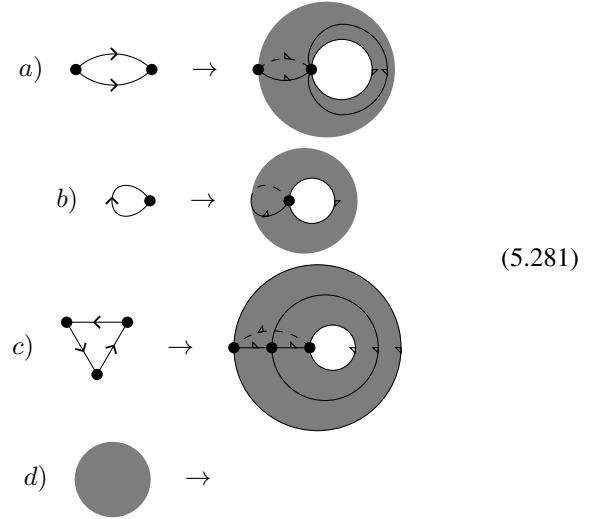

$$
\tag{5.281}
$$

In the cases a), b) and c) the 2E backgrounds are meant to have a disk extended background, thus the corresponding doughnuts have a solid torus as extended background (hence the name). d) shows a 2-sphere that is mapped to a 3-torus, both having an empty index lattice.

**Observation 115.** Gluing two edges of a 2E-CTL background commutes with gluing the corresponding pair of edges in the corresponding doughnut. E.g., the following shows one 2E gluing and the corresponding 3E gluing below:

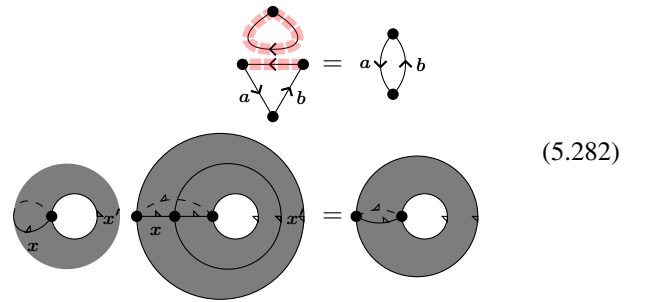

$$
\tag{5.282}
$$

In fact, taking the doughnut can be extended to a lattice mapping from 2E lattices to 3E lattices. On the other hand one can define a tensor mapping from the 3E tensors to the 2E tensors by taking the composite of the two indices at the two 3E edges corresponding to a 2E edge as the 2E index. Contracting two 2E indices corresponds to contracting both pairs of 3E indices in the composite, just as gluing two 2E edges corresponds to gluing both associated 3E edges. So combining both mappings we get a CTL mapping from 3E-CTLs to 2E-CTLs.

E.g., the 2E-CTL tensor associated to a triangle by the mapped 2E-CTL is obtained in the following way from the 3E-

CTL tensor associated to the corresponding doughnut:

$$(5.283)$$

This implies that all doughnuts can be glued from the doughnut associated to the 2E basic lattice.

**Equivalence to 3FE-CTLs**

In this section we will demonstrate that 3FE-CTLs and 3E-CTLs are equivalent.

First we will quickly demonstrate that if we forget about the normalizations, then the two CTL types are in the same class. To this end we will give a CTL mapping $\mathcal{C}$ from 3FE-CTLs to a 3E-CTL a CTL mapping $\mathcal{D}$ from 3E-CTLs to 3FE-CTLs, such that both $\mathcal{C} \circ \mathcal{D}$ and $\mathcal{D} \circ \mathcal{C}$ are exact.

The lattice mapping of $\mathcal{C}$ is the following. For a given 3E lattice $X$ the corresponding 3FE lattice $Y$ looks the same in the interior. The index lattice of $Y$ is obtained from the 2CC-lattice by adding a vertex to the centre of every face of the index lattice of $X$, and adding edges between this central vertex and all other vertices of the face. The orientations of those inserted edges are chosen such that they all point towards the central vertex. The favourite edge for each triangle is chosen to be the old edge of the 3E lattice. The following picture shows a patch of a 3E index lattice and its mapping to a 3FE index lattice:

$$(5.284)$$

Now let's consider the tensor mapping of $\mathcal{C}$. We first take two copies of each of the indices at the newly introduced edges. For each 2E index edge there are two 3FE index faces. So we can associate to each 3E index edge the composite of the following 7 indices: 1) The index at the old edge in the middle of the two faces, 2) one of the copies of the indices at the four new edges adjacent to the two faces, and 3) the two indices at the two faces. The ordering of the single indices within the composite is determined by the orientation and dual orientation of the 3E index edge, i.e., when one reverses the dual orientation, the indices corresponding to the left and right triangle of $E$ are interchanged. Of course, the actual order of indices is a pure matter of convention, the ordering just has to be done consistent with the orientations everywhere.

This indeed defines a CTL mapping: Gluing two edges is consistent with gluing the two pairs of corresponding faces, and contraction of the composite index is consistent with the copying and contraction of the individual indices.

The lattice mapping of $\mathcal{D}$ is given by replacing every edge by two edges connected to the same vertices and separated by a 2-gon face. The orientations of the 3E index edges are chosen to match those of the 3FE index edge. The dual orientations are taken to point towards the 2-gon face in between. E.g.,

$$(5.285)$$

The tensor mapping of $\mathcal{D}$ associates the trivial basis set to the 3FE edge indices. To every 3FE index face the composite of all the 3E edge indices is adjacent to that face. The ordering within the composite is determined by the orientations of the 3FE index edges and the favourite edge of the 3FE index face.

This indeed defines a CTL mapping: Gluing two faces of a 3FE lattice is consistent with gluing all the associated edges of the corresponding 3E lattice and contracting two 3FE face indices is consistent with contracting all the individual indices in the composite.

The above mapping is obscured by the normalizations. Because of them we cannot give formal CTL mappings between 3E-CTLs and 3FE-CTLs. However, one can still construct for each 3E-CTL a 3FE-CTL that is in the same phase and vice versa. Let us start with the construction of a 3E-CTL from a 3FE-CTL.

We can use the same lattice mapping as above, but we have to include normalization factors into the tensor mapping: We have to multiply the 3FE indices at the old edges (those in between the pairs of triangles) by the square root of the 3FE normalization $C$. The following picture shows how the tensor mapping includes normalizations and how the 3E normalizations can be obtained from the 3FE normalizations as

$$D_{3E} = D_{3FE},$$
$$E_{3E} = D_{3FE}.$$
$$(5.286)$$

We have only shown one of the 3E $C$ normalizations exemplary. For the other 3E $C$ normalizations the 3FE normalization $C$ has to be added to different indices: $c$ for $C_+^+$, $e$ for $C_-^+$, $b$ for $C_+^-$ and $d$ for $C_-^-$.

Constructing a 3FE-CTL from a given 3E-CTL is a bit more complicated. Here we do not take the lattice mapping above, but simply transform a 3FE lattice into a 3E lattice by taking the same underlying 3CCb-lattice, with the orientations of the 3E index edges matching the orientations of the 3FE index edges and the dual orientations chosen arbitrarily.

For each shape of a 3E index face with surrounding edges we can consider the following lattice: It has one vertex for every vertex of the face, two faces with the shape of the face, and for every edge of the face two edges with a 2-gon face in between. E.g., for a 4-gon face:

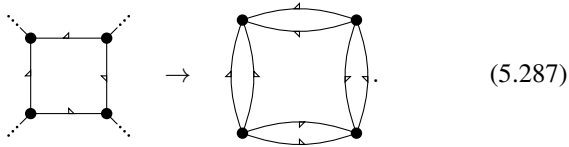

(5.287)

Gluing together all the lower edges of one copy of this lattice together with all upper edges of another we get the same lattice again. Also the lattice is symmetric under reflections interchanging the lower and upper half. So if we include one normalization factor $C$ on each upper edge the associated tensor interpreted as linear map from the lower to the upper edges is a symmetric projector $P$. The lattice can be glued from one pillow and one banana for each edge. E.g.,

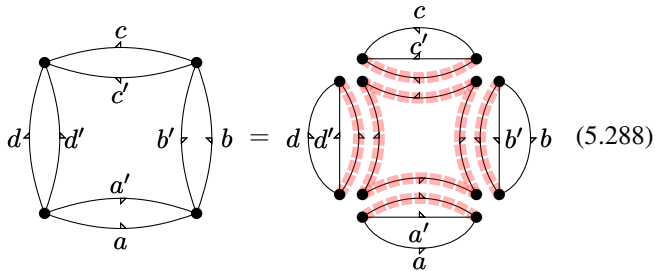

(5.288)

If we perform a gauge transformation that brings the 2En-CTL of bananas into the standard form of 2En-CTLs then this equation reads:

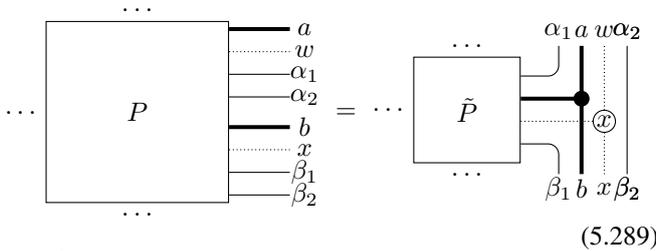

(5.289)

Here $\tilde{P}$ is the CTL tensor associated to the corresponding pillow, apart from the following normalizations: On every edge there is one normalization matrix $C$ from the original projector $P$ and one factor $w_a$ (where $a$ is the value of the corresponding edge index), coming from the pillow tensor in its standard form.

Note that in general $x$ is the algebra of reals, complex numbers, or quaternions depending on the irrep label. For complex-real CTLs $x$ can only be the complex numbers. Different dual orientations of the edges of the face $F$ correspond to different orderings of the composite indices. In our example inverting the dual orientation will interchange the labels $\alpha_1 \leftrightarrow \alpha_2$ and $\beta \leftrightarrow \beta_2$.

The projector $P$ acts trivially on one of the block indices for each edge (in our case, $\alpha_2 \leftrightarrow \beta_2$, for other dual orientations $\alpha_1 \leftrightarrow \beta_1$). It does not change the value of the irrep label ($a = b$) and acts on the real/complex/quaternionic index just by real/complex/quaternionic multiplication. It only acts arbitrarily on the second block index ($\gamma$ in our case). One can thus regard the tensor $\tilde{P}$ for a fixed set of irrep labels as a mixed real/complex/quaternionic self-adjoint projector acting on the block and real/complex/quaternionic indices (for complex-real CTLs it is just the realification of a complex projector). For every irrep label configuration we can find an isometry $U$ that maps the block and real/complex/quaternionic indices to indices another type (with potentially smaller basis), such that $\tilde{P} = U U^\dagger$. The type of the second index of $U$ will be the type of the 3FE face indices:

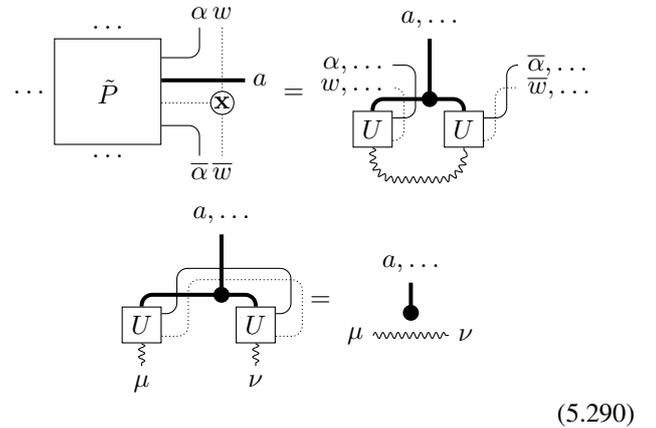

(5.290)

We get one such orthogonal/unitary root $U$ for each (type of) face. If the face will have a rotation or reflection symmetry, the projector must also have the corresponding symmetry. That is, it is invariant under the tensor product $d \otimes d$ of one symmetry $d$ on each side of the projector. $d$ is the representation of the full reflection/rotation symmetry group $D$ of the oriented face that acts by the corresponding index permutation. $D$ is a subgroup of the full dihedral symmetry group $D_n$ of a face with $n$ non-oriented edges. So we have $d_g U_I U_I^\dagger d_g^\dagger = U_I U_I^\dagger$. But we cannot always choose $U_I$ such that $d_g U_I = U_I$, we can only ensure that $d_g U_I c_g = U_I$ for some representation $c$ of $D$ acting on the fusion index. As $U$ does not have rotation/reflection symmetry it must also depend on a choice of a favourite edge of the face.

With this preparation we are able to construct the TL operator that maps the 3E-CTL tensors to the 3FE-CTL tensors. We start by applying a gauge transformation such that the 2En-CTL of bananas is in the standard form. Now every edge index becomes a composite of one irrep, one real/complex/quaternionic and two block indices. We then make three copies of the irrep index via the delta tensor and include irrep-dependent factors $w$ and $v$ from Eq. (5.67). We also duplicate the real/complex/quaternionic index by using the corresponding real/complex/quaternionic algebra tensor:

$$\tag{5.291}$$

We can now associate the indices $a_1, x_1, \alpha_1$ to the face on one side of the edge and $a_2, x_2, \alpha_2$ to the face on the other side. Which of the two $a$, $x$ and $\alpha$ indices we associate to which side can be decided based on the dual orientation of the edge (actually it does not matter for the $a$ indices, and also not for the $x$ labels if $\mathbf{x}$ is not the quaternions which are non-commutative).

Finally for each face of $X$ we take all indices associated to the face via its edges and feed them into the isometry $U$ for this shape of face. The remaining index of $U$ becomes the fusion index of the 3FE-CTL and the third copy if the irrep index becomes the edge index of the 3FE-CTL. At this step we also need the favourite edge of the face.

$$C_a = |B_a| v_a^2 D = w_a^{-2}$$
$$D_{3\mathrm{FE}} = E_{3\mathrm{E}}$$
$$\tag{5.292}$$

Imagine gluing two faces of a 3FE lattice. The same operation for the corresponding 3E lattice can be achieved by connecting each pair of the adjacent edges via a 3-edge banana. Each banana contains a factor of $w$, so we get one factor of $v$ at every edge coming from the 2En normalization $C$ (see Eq. (5.67)), and in the end we get once the 3E normalization $D$. On the other hand consider the contraction of indices of the 3FE tensor Eq. (5.292) when gluing two 3FE faces: The isometries $U$ cancel each other and we get twice the factor of $\sqrt{D}$. Moreover two 3FE edges fuse into a single one, so we end up with remaining factors $w$ and $v$ from Eq. (5.291) at every edge. So we see normalization factors we get from the two contractions are consistent.

### 5.9.6 Connection to involutive weak Hopf *-algebras

**Definition 186.** A **finite-dimensional unital associative co-algebra** over a complex vector space with basis set $B$ is given

by 2 linear maps:

co-unit
$\epsilon : \mathbb{C}^B \longrightarrow \mathbb{C}^1$

co-multiplication
$\Delta : \mathbb{C}^B \longrightarrow \mathbb{C}^{B \times B}$ $\quad [\Delta(a) = a_{(1)} \otimes a_{(2)}]$

$$\tag{5.293}$$

that satisfy the axioms Eqs. (5.296, 5.297). We will think of the 2 linear/anti-linear maps as tensors via the realification of their coefficients in the canonical basis of $\mathbb{C}^B$ as described in Remark (41).

$$\epsilon^i = \quad i \; \blacktriangleright \tag{5.294}$$

$$\Delta^k_{ij} = \quad k \; \blacktriangleright\!\!\!\begin{smallmatrix} i \\ \\ j \end{smallmatrix} \tag{5.295}$$

Then the axioms can be written as

- **Co-identity property**:

$$\epsilon(a_{(1)}) \otimes a_{(2)} = a_{(1)} \otimes \epsilon(a_{(2)}) = a$$
$$(\mathbb{1} \otimes \epsilon) \circ \Delta = (\epsilon \otimes \mathbb{1}) \circ \Delta = \mathbb{1}$$
$$\tag{5.296}$$

- **Co-associativity**:

$$(a_{(1)})_{(1)} \otimes (a_{(1)})_{(2)} \otimes a_{(2)} = a_{(1)} \otimes (a_{(2)})_{(1)} \otimes (a_{(2)})_{(2)}$$
$$(\Delta \otimes \mathbb{1}) \circ \Delta = (\mathbb{1} \otimes \Delta) \circ \Delta$$
$$\tag{5.297}$$

**Observation 116.** If we make the identification

$$\tag{5.298}$$

with the tensors in Eq. (5.71), the axioms Eqs. (5.296, 5.297) and Eqs. (5.81, 5.85) become identical. So in the language of tensors, the notions of (unital associative finite-dimensional) algebra and co-algebra are identical.

**Definition 187.** An **involutive weak Hopf (finite-dimensional C)*-algebra** over a complex vector space with basis set $B$ is given by 6 linear/anti-linear maps:

unit
$\eta : \mathbb{C}^1 \longrightarrow \mathbb{C}^B$ $\quad [\epsilon(1) = \mathbb{1}]$

multiplication
$\mu : \mathbb{C}^{B \times B} \longrightarrow \mathbb{C}^B$ $\quad [\mu(a, b) = ab]$

co-unit
$\epsilon : \mathbb{C}^B \longrightarrow \mathbb{C}^1$

co-multiplication
$\Delta : \mathbb{C}^B \longrightarrow \mathbb{C}^{B \times B}$ $\quad [\Delta(a) = a_{(1)} \otimes a_{(2)}]$

antipode
$S : \mathbb{C}^B \longrightarrow \mathbb{C}^B$

involution
$t : \mathbb{C}^B \longrightarrow \mathbb{C}^B$ $\quad [t(a) = a^*]$

$$\tag{5.299}$$

such that the axioms Eqs. (5.302, 5.304, 5.307, 5.309, 5.311, 5.313, 5.315) hold. $\eta$, $\mu$, $\epsilon$, $\Delta$ and $S$ are linear whereas $t$ is anti-linear.

**Remark 128.** We will think of the three linear/anti-linear maps as tensors via the realifications of their coefficients in the canonical basis of $\mathbb{C}^B$ as described in Remark (41).

(5.300)

**Proposition 21.** Every complex-real 3E-CTL defines a involutive weak Hopf *-algebra, which has as underlying basis set just the basis set for the edge indices of the CTL (without the complex index), and whose 6 defining tensors are equal to the CTL tensors in the following way:

(5.301)

To demonstrate the validity of the proposition we give a list of all involutive weak Hopf *-algebra axioms in the conventional form as well as for the corresponding tensors as described in Remark (41) in tensor-network notation (see Remark (11). Then we show that the axiom holds for the CTL tensors Eq. (5.301) by giving the corresponding gluing axiom. Note that those two equations are not exactly equal in some cases, but equal after adding the same normalizations $C$ to both sides.

1. The unit, involution and multiplication form a (finite-dimensional unital C)*-algebra, see Section (5.4.6).

This holds for the tensors given in Eq. (5.301): The pillows form a 2En-CTL as we have seen in Observation (101). The CTL tensors corresponding to $\eta$, $\mu$ and $t$ are just the pillows associated to the 2En-CTL tensors corresponding to $\eta$, $\mu$ and $t$ in Eq. (5.74). As we have seen in Section (5.4.6), those tensors define a (finite-dimensional unital C)*-algebra for every 2En-CTL.

2. The co-unit and co-multiplication form an co-associative co-unital algebra, see Definition (186). This holds for the tensors given in Eq. (5.301): The bananas form a 2En-CTL as we have seen in Observation (101). The CTL tensors corresponding to $\epsilon$ and $\Delta$ are just the bananas associated to the S-2-CTL tensors corresponding to $\eta$ and $\mu$ in Eq. (5.74). As we have seen in Section (5.4.6) those tensors define a unital associative algebra for every 2E-CTL. So, according to Observation (116), $\epsilon$ and $\Delta$ satisfy the equivalent axioms of a co-algebra.

3. Multiplicativity of the co-multiplication says that the co-multiplication is an algebra morphism (from the algebra to two copies of the algebra) and is also the main axiom that holds for bi-algebras

$$(\mu \otimes \mu) \circ (\mathbb{1} \otimes \text{swap} \otimes \mathbb{1}) \circ (\Delta \otimes \Delta) = \Delta \circ \mu$$

(5.302)

(5.303)

4. Weak multiplicativity of the co-unit:

$$\epsilon \circ \mu \circ (\mu \otimes \mathbb{1}) = (\epsilon \otimes \epsilon) \circ (\mu \otimes \mu) \circ (\mathbb{1} \otimes \Delta \otimes \mathbb{1})$$
$$= (\epsilon \otimes \epsilon) \circ (\mu \otimes \mu) \circ (\mathbb{1} \otimes (\tau \circ \Delta) \otimes \mathbb{1})$$

(5.304)

(5.305)

where $\tau$ is the operator exchanging two components of a tensor product.

(5.306)

5. Weak co-multiplicativity of the unit:

$$(\Delta \otimes \mathbb{1}) \circ \Delta \circ \eta = (\mathbb{1} \otimes \mu \circ \mathbb{1}) \circ (\Delta \otimes \Delta) \circ (\eta \otimes \eta)$$
$$= (\mathbb{1} \otimes (\mu \circ \tau) \otimes \mathbb{1}) \circ (\Delta \otimes \Delta) \circ (\eta \otimes \eta)$$

(5.307)

$$(5.308)$$

6. The standard Hopf algebra axiom is weakened to:

$$\mu \circ (\mathbb{1} \otimes S) \circ \Delta =$$
$$(\epsilon \otimes \mathbb{1}) \circ (\mu \otimes \mathbb{1}) \circ (\mathbb{1} \otimes \tau) \circ (\Delta \otimes \mathbb{1}) \circ (\eta \otimes \mathbb{1}) \quad (5.309)$$

$$(5.310)$$

7. Another axiom involving the antipode:

$$S = \mu \circ (\mu \otimes \mathbb{1}) \circ (S \otimes \mathbb{1} \otimes S)$$
$$\circ (\Delta \otimes \mathbb{1}) \circ \Delta \quad (5.311)$$

$$(5.312)$$

8. The compatibility condition between the star operation $t$ and the co-multiplication:

$$\Delta \circ t = (t \otimes t) \circ \Delta \quad (5.313)$$

$$(5.314)$$

9. The involutivity of the antipode:

$$S \circ S = \mathbb{1} \quad (5.315)$$

$$(5.316)$$

which is true with the local support convention.

Let us see to what extent the reverse direction of the above proposition holds.

**Remark 129.** Involutive weak Hopf *-algebras have a different set of gauge transformations than 3E-CTLs: The tensors have a distinction between input and output indices such that input indices are only contracted with output indices. Gauge transformations consist of applying an invertible linear map $S$ to all input indices and $S^{-1}$ to all output indices, in contrast to an orthogonal/unitary linear map applied to all indices for 3E-CTLs.

Therefore we cannot simply invert the identification Eq. (5.301) to get CTL tensors from weak Hopf *-algebra tensors and hope to get a 3E-CTL. The question has to be: For a given involutive weak Hopf *-algebra, does there exist a gauge transformation $S$ and normalization matrices $C$, such that after gauging with $S$ the CTL tensors obtained via Eq. (5.301) yield a valid 3E-CTL?

**Proposition 22.** It is known that for any (finite-dimensional) weak Hopf *-algebra the co-unit and co-multiplication also form a (co-)*-algebra with the involution given by

$$(5.317)$$

See for example [18].

**Remark 130.** The 3E basic axioms A11), A12) and A13) directly follow from the weak Hopf *-algebra axioms Eq. (5.313), the dual equation to this that holds according to the proposition above, and Eq. (5.302). The other 3E basic axioms do not directly follow from the weak Hopf *-algebra axioms, as they partly depend on the correct choice of gauge. However, we know from Section (5.4.6) that there is a gauge transformation for the *-algebra part of the Hopf algebra such that the 3E basic axioms A1) to A5) hold. Dually we now that there is a gauge transformation for the co-*-algebra part in which the 3E basic axioms A6) to A10) hold. However, it is not clear whether one can find a single gauge transformation for both the *-algebra and the co-*-algebra such that A1) to A5) and A6) to A10) hold simultaneously. If this is possible then every involutive weak Hopf *-algebra yields a 3E-CTL. Is seems that this actually might be the case, though we do not have a proof.

**Remark 131.** The involutivity of the antipode is important: There exist weak Hopf *-algebras that are not involutive. As the involutivity is a gauge-independent property, those non-involutive weak Hopf *-algebras cannot yield 3E-CTLs.

## 5.9.7 Physical interpretation and connection to existing models

In classical statistical physics 3E-CTL correspond to 3-dimensional classical models in a topological phase. We do not know however, whether there are existing classical systems that can describe such a phase.

In quantum physics complex-real 3E-CTLs are quantum fixed point models for topological order in $2 + 1$ dimensions. They are essentially equivalent to the well-known *quantum double models* that were originally introduced as generalizations of the toric code [8] to arbitrary groups (see also Ref. [56]). As anticipated by the original paper, they were later generalized to Hopf C*-algebras in Ref. [18] and then to weak

Hopf algebras in Ref. [25]. 3E-CTLs are very similar to the latter generalizations as becomes apparent in Section (5.9.6). Note, however, that they differ in many technical details, and that our basic axioms A1) to A13) are directly derived from the topological invariance at a physical level, and should thus be sufficient and necessary to yield a meaningful microscopic model for topological order.

It is known that the weak Hopf algebra models yield models equivalent to string-net models [17]. This equivalence was also demonstrated in Ref. [25]. Via the equivalence of 3FE-CTLs and 3E-CTLs we give a very simple graphical proof of this equivalence in our setting.

**Remark 132.** One can construct a frustration-free local commuting-projector Hamiltonian for 3E-CTLs, following Observation (55). A sufficient set of local projector lattices is given by taking for each face and each vertex the lattice that can be glued to all the edges adjacent to this face or vertex without changing the background. E.g., consider the following local projector lattices for a face and a vertex with 4 adjacent edges each:

 (5.318)

For the associated tensors to become actual projectors we have to add one normalization $C$ to each edge, and one overall normalization $D$ for faces or $E$ for vertices.

One can explicitly show that a neighbouring face and vertex projector, overlapping at two edges, commute by the CTL axioms:

 (5.319)

## 5.10   3/1**EE-(C)TLs**

### 5.10.1   Definition

By 3/1**EE-CTLs** we refer to a collection of different CTL coming from TL types on 3-dimensional lattices with 2-dimensional defects. The "EE" stands for "edge, edge" as this is where indices are associated to.

**Backgrounds**

Their extended backgrounds are given by framed boundary 3/1-manifolds with an arbitrary higher order 0-manifold as boundary central link. E.g.

 (5.320)

a) corresponds to a physical boundary of a 3E-CTL, b) to a domain wall between different 3E-CTLs, c) a co-dimension 1 defect within one 3E-CTL and d) a membrane where three 3E-CTLs meet. The bulk 2-region has a framing with respect to each of the bulk 3-regions. For the boundary central links above, this framing is only non-trivial for c). In this case the framing means that the co-dimension 1 defect has an orientation within the surrounding bulk.

The backgrounds of the corresponding TL type are 3/1-manifolds with the same higher order 0-manifold as central link.

3/1EE-CTLs contain 3E-CTL as a sub-type once for each 0-region of the boundary central link, by restricting to backgrounds where only the corresponding bulk 3-region and boundary 2-region are non-empty. We will therefore refer to lattice or tensor parts associated to those regions or sub-type by the prefix 3E, and refer to things associated to the bulk 2-region or boundary 0-region by the prefix 3/1EE.

**Example 124.** Consider the following examples for extended backgrounds of 3/1EE-CTLs (where we only draw the 1-regions and 2-regions, see Remark (3)):

(5.321)

where a), b), c) and d) correspond to the boundary central links a), b), a) and b), respectively. a) shows a 3-ball (the bulk 3-region) whose boundary is divided by a 1-sphere (the boundary 1-region) into two 2-balls (the bulk 2-region and the boundary 2-region). b) shows a 3-ball that is split into two 3-balls (the bulk 3-regions) by a sphere (the bulk 2-region). Also its boundary 2-sphere is divided into two 2-balls (the two boundary 2-regions) by a 1-sphere (the boundary 1-region). c) shows a 3-ball (the bulk 3-region) whose boundary 2-sphere is split into three 2-balls (the bulk 2-region) and the remaining surface (the boundary 2-region) by three 1-spheres (the boundary 1-region). d) shows a solid torus that is split into two parts (the bulk 3-regions) along the torus by an annulus (the bulk 2-region). Also the boundary 2-torus is split into two annuli (the boundary 2-regions) by two 1-spheres (the boundary 1-region).

The background gluings are surgery gluing for all boundary regions (i.e., for the different boundary 2-regions and the boundary 1-region).

**Lattices**

The lattices are based on 3/1CCb-lattices on the extended backgrounds. The index lattice is given by restricting to the 2/1CC-lattice forming the boundary of the 3/1CCb-lattice. There are the following decorations: The 3E index edges carry orientations and dual orientations, just as the 3E sub-type. Also the 3/1EE edges carry an orientation, but no dual orientation. To distinguish them better from the 3E edges we will draw them with a thicker line style.

**Example 125.** Consider the following examples for index lattices (right) and the corresponding index backgrounds (left):

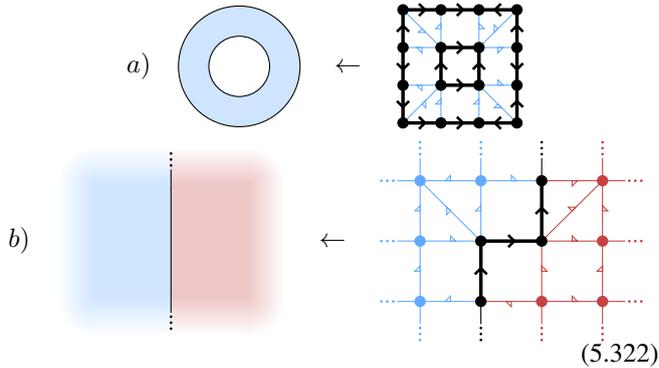

(5.322)

a) and b) correspond to the boundary central links a) and b) in Eq. (5.320), respectively. a) shows an annulus with physical boundary. b) shows a patch of a domain wall in a bigger lattice.

The basic moves are region bi-stellar flips for the bulk 3-regions and the bulk 2-region.

The basic gluing is gluing of two 3E index edges (in the same region), just as for the sub 3E-CTLs. Additionally the 3/1EE edges can be glued. Gluing two 3/1EE edges is the same as gluing two 3E edges. The only difference is that the former do not have dual orientations that would have to match.

**Tensors**

The tensors are real tensors with one 3E index type for each region of the boundary central link, and one additional 3/1EE index type. One 3E index is associated to each 3E index edge, which are contracted when the edges are glued, just as for the sub 3E-CTL. One 3/1EE index is associated to every 3/1EE index edge. Also those indices are contracted when the according edges are glued.

We should also introduce normalizations: When two 3/1EE index edges that share one vertex are glued, such that this vertex disappears, then we have to contract via the 3E normalization matrix $C$ and an additional normalization matrix $B$. However, we won't be entirely strict about forgetting this normalization matrix in this section.

## 5.10.2 Basic tensors and axioms

### Basic lattices and tensors

**Definition 188.** For any 2E background $X$, define the **EE-pillow** with respect to $X$ as the 3/1EE background whose index lattice consists of an upper and a lower face whose lower link is given by the index lattice of $X$. Its extended background is given by the extended background of $X$ times the interval. So it can be pictured as $X$ inflated to a pillow. E.g.

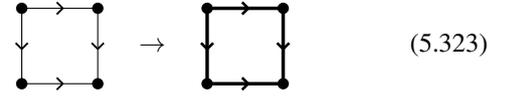

(5.323)

**Definition 189.** For any 2/1EE background $X$, define the **EE-banana** with respect to $X$ as the following 3/1EE background: Its index lattice consists of two 3/1EE vertices that are connected by 3E edges and 3/1EE edges. The upper link of the 3/1EE vertices is equal to the index lattice of $X$. The orientations of the 3E edges and 3/1EE edges all point towards the same 3/1EE vertex. The dual orientations of the 3E edges are equal to the orientations of the 2E edges of the index lattice of $X$. E.g.,

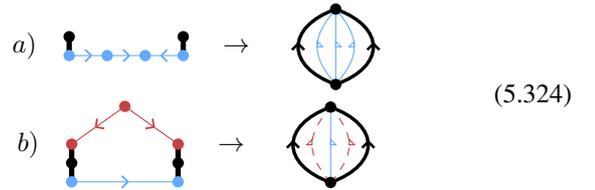

(5.324)

where a) and b) correspond to the boundary central links a) and b) in Eq. (5.320). The extended background the EE-banana with respect to $X$ equals the background of $X$ times the interval. If $X$ consists of multiple connected components, then the EE-banana with respect to $X$ is the disjoint union of the EE-bananas with respect to each component.

**Observation 117.** Every 3/1EE background can be glued together from copies of the 3E basic lattices, EE-bananas and EE-pillows.

We can get a defining history for a given lattice in the following way: Consider the underlying 3/1CCb-lattice. Replace every 3E face by a pillow and every 3E edge by a banana, just in the history mapping for 3E-CTLs from Observation (102). Replace face of the bulk 2-region by the EE-pillow with respect to the lower link of that face. Replace every edge of the bulk 2-region by the EE-banana with respect to the upper link of the edge. Finally, we glue together all pillows, bananas, EE-pillows and EE-bananas according to how they are located within the 3CCb-lattice.

**Observation 118.** Taking the EE-pillow with respect to a 2E background commutes with symmetries, disjoint union and gluing. E.g.

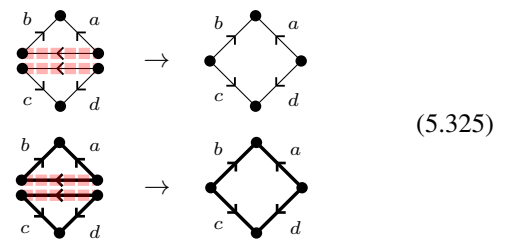

(5.325)

It can be extended to a lattice mapping from 2E lattices to 3/1EE lattices. Using the 3/1EE indices as 2E indices defines a tensor mapping from the 3/1EE tensors to the 2E tensors, yielding a CTL mapping from 3/1EE-CTLs to 2E-CTLs. I.e., the EE-pillows form a 2E-CTL.

This implies that every EE-pillow can be glued from copies of the EE-pillow with respect to the 2E basic lattice Eq. (5.24).

**Observation 119.** Taking the EE-banana with respect to a 2/1EE background commutes with symmetries, disjoint union and gluing. E.g.

(5.326)

It can be extended to a lattice mapping from 2/1EE lattices to 3/1EE lattices. Using the 3/1EE indices as 2/1EE indices and the 3E indices as 2E indices defines a tensor mapping from the 3/1EE tensors to the 2/1EE tensors, yielding a CTL mapping from 3/1EE-CTLs to 2/1EE-CTLs, i.e., the EE-bananas form a 2/1EE-CTL.

This implies that all EE-bananas can be glued from copies of the EE-banana with respect to the 2/1EE basic lattice Eq. (5.121).

**Observation 120.** According to Observation (117, 118, 119) a set of basic lattices is given by 3E basic lattices, the EE-pillow with respect to the 2E basic lattice and the EE-pillow with respect to the 2/1EE basic lattice.

So the whole CTL is already determined by corresponding basic tensors

$$T1) \quad H^{abc} := \begin{array}{c} a \\ \text{\includegraphics} \\ c \end{array} b \ ,$$

$$a) \quad I_x^{ab} = a \langle x \rangle b \ ,$$

$$T2) \quad b) \quad I_{xy}^{ab} = a \langle x | y \rangle b \ ,$$

$$c) \quad I_{xyz}^{ab} = b \langle x | y | z \rangle a \ ,$$

(5.327)

where the examples a), b) and c) correspond to the boundary central links a), b) and d) in Eq. (5.320). Note that there are no faces between the 3E-CTL edges of the different regions: Each edge lives in its own boundary 2-region. For b) we tried to indicate this by moving one of the edges to the back layer, which does However, not work for c) any more.

### Basic history moves and axioms

**Definition 190.** We can use the EE-pillow mapping to transform every 2E history lattice $X$ into a 3/1EE history by replacing every 2E-CTL basic lattice by the corresponding EE-pillow, yielding what we call the **EE-pillow history lattice** with respect to $X$. E.g.,

(5.328)

**Definition 191.** For every 2/1EE-CTL history lattice $X$ (with the same boundary central link), define the **EE-banana history lattice** with respect to $X$ as the history lattice where each each 2E basic lattice of $X$ is replaced by the corresponding banana and each 2/1EE basic lattice of $X$ is replaced by the corresponding EE-banana. E.g.,

(5.329)

where a) and b) correspond to the types a) and b) in Eq. (5.320).

**Definition 192.** For every pair of a 2E background $X$ and a 2/1EE background $Y$ (with the same boundary central link) define the **EE-pillow-banana history lattice** with respect to $X$ and $Y$ as the following 3/1EE history lattice: For every edge of $X$ take one copy of the EE-banana with respect to $Y$. For every 2E edge of $Y$ take one copy of the pillow with respect to $X$, and for every 2/1EE edge of $Y$ take one copy of the EE-pillow with respect to $X$. Then for every pair of edges $x$ and $y$ of $X$ and $Y$, we glue the edge $x$ of the copy of the (EE-)pillow corresponding to $y$ to the edge $y$ of the copy of the EE-banana corresponding to $x$. E.g.

(5.330)

**Observation 121.** Every history lattice can be history glued from copies of pillow history lattices, banana history lattices, pillow-banana history lattices, EE-pillow history lattices, EE-banana history lattices and EE-pillow-banana history lattices.

History lattices are based on 4/1CCb-lattices (with the same boundary central link): The basic lattices and gluings are living on the 3/1CC-lattice forming its boundary as described in Prop. (117). The cells of the bulk regions correspond to basic moves. A history for a given history lattice can be obtained in the following way: Within the bulk 4-regions, replace every edge by a banana history lattice, every volume by a banana-pillow history lattice and every face by a banana-pillow history lattice, just as for the sub 3E-CTL. Within the bulk 3-region, replace every volume by the EE-pillow basic lattice with respect to its lower link, every edge by the EE-banana history lattice with respect to its upper link and every face by the EE-banana-pillow history lattice with respect to its lower link and its upper link. Then glue the history lattices at the basic lattices according to how they are located in the $4/1CCb$-lattice. The basic lattices remaining unglued at the boundary form the desired history lattice.

**Observation 122.** Using the EE-pillow mapping, every EE-pillow history lattice can be history glued from the EE-pillow history lattice with respect to the 2E basic history lattice:

$$\tag{5.331}$$

**Observation 123.** Every EE-banana history lattice can be history glued from the EE-banana history lattice with respect the $2/1EE$ basic lattice. E.g.

$$\tag{5.332}$$

**Observation 124.** Every EE-pillow-banana history lattice can be history glued from the EE-pillow-banana history lattice with respect to the 2E basic lattice and the $2/1EE$ basic lattice. E.g.

$$\tag{5.333}$$

For the boundary central link a) in Eq. (5.320).

**Observation 125.** According to Observation (121, 122, 123, 124), a set of basic history lattices is given by the EE-pillow history lattice, the EE-banana history lattice and the EE-pillow-banana history lattice Eqs. (5.331, 5.332, 5.333).

**Observation 126.** According to Remark (24) the bi-partitions of the basic history lattice are a good guess for a set of basic axioms.

The bi-partition of the EE-pillow basic history lattice and the EE-banana basic history lattice are completely analogue to the bi-partitions of the basic 2E history lattice and of the basic $2/1EE$ history lattice (Remark (70)).

There are 4 kinds of bi-partitions of the basic EE-banana-pillow history lattice. First there are 6 are bi-partitions with one EE-pillow and one EE-banana on the right side, and the rest on the left side. The 6 choices correspond to which of the 2 EE-pillows and which of the 3 EE-bananas are on the right side. E.g., for the boundary central link a) in Eq. (5.320) one such bi-partition is:

$$\tag{5.334}$$

Second there are bi-partitions with one EE-banana and one pillow on the right side, and the rest the left. There is one such bi-partition for each 0-region of the boundary central link, determine which of the pillows is on the right. E.g., for the boundary central link a) in Eq. (5.320) there is only one such bi-partition:

$$\tag{5.335}$$

Third there are 2 bi-partitions with one single EE-pillow on the right side, and the rest on the left side. The 2 choices correspond to which of the two EE-pillows is on the right side. E.g., for the boundary central link a) in Eq. (5.320) one such bi-partition is the following:

$$\tag{5.336}$$

Fourth there are 3 bi-partitions with one single EE-banana on the right side and the rest on the left side. The 3 choices correspond to the 3 choices of the EE-banana on the right. E.g., for the boundary central link a) in Eq. (5.320) one such bi-partition is the following:

$$\tag{5.337}$$

The above bi-partitions do indeed form a full set of basic axioms.

**Simplified set of axioms**

**Remark 133.** We can impose the local support convention for the $3/1EE$ indices, by setting the EE-pillow with respect to the non-cyclic 2-gon to the identity matrix:

$$a \;\blacklozenge\; b \;=\; a \;\text{———}\; b \tag{5.338}$$

**Observation 127.** Analogously to 2E-CTLs we can add the following auxiliary tensor:

$$T3) \qquad \overset{a}{\underset{b}{\diamond}} \qquad (5.339)$$

Then the 7 axioms arising from bi-partitions of the EE-pillow basic history lattice together with the local support convention are equivalent to the following axioms: and the following additional axioms:

$$(5.340)$$

**Observation 128.** Analogously to 2/1EE-CTLs all the axioms arising from bi-partitions of the EE-banana basic history lattice together with the local support convention are equivalent to the following set of axioms (shown here for the boundary central link a) in Eq. (5.320):

$$(5.341)$$

**Observation 129.** Instead of Eqs. (5.334, 5.335, 5.336) together with the local support convention we can take the following equivalent axioms (here exemplary for the boundary central link a) in Eq. (5.320)):

$$(5.342)$$

Using A9), A2), A3), A4) and the 3E axioms A2), A3) and A4) we can invert any of the orientations of 3E edges and 3/1EE edges of any axiom involving pillows, EE-pillows and EE-bananas. Using A6) and the 3E axioms A8) and A11) we can also invert any of the dual orientations of the 3E edges. So from A10) and A11) we get all the other axioms in Eqs. (5.334, 5.335). Using the axiom A8) we can move the EE-banana on the right of A10) to the left, and obtain one of the axioms Eq. (5.336), and invert orientations to obtain all the other axioms. Using the 3E axiom A2) we can move the pillow on the right side of A11) to the left and obtain on of the axioms Eq. (5.337), and invert orientations to obtain all the other axioms.

### 5.10.3 Solutions

**Standard physical boundary**

There is a lattice mapping from 3E-CTLs to physical boundaries for 3E-CTLs, i.e., 3/1EE-CTLs with a single vertex as boundary central link. The sub 3E-CTL of the latter is the original 3E-CTL. Well refer to this 3/1EE-CTL as the **standard physical boundary** 3/1**EE-CTL**. The basic tensors are given by

$$(5.343)$$

With this choice of tensors, the 3/1EE basic axioms directly follow from the 3E basic axioms.

**Dual standard physical boundary**

There is a second CTL mapping from 3E-CTLs to a physical boundary for 3E-CTL that is dual to the standard physical boundary. Again the sub 3E-CTL is the original 3E-CTL. Well refer to this 3/1EE-CTL as the **dual standard physical** **boundary** $3/1$**EE-CTL**. The basic tensors are given by

$$(5.344)$$

Again, the $3/1$EE basic axioms reduce to the 3E axioms.

Note that $3/1$EE-CTLs violate Poincaré duality, as the EE-bananas have 3E edges, but not the EE-pillows. One could construct the dual type of $3/1$EE-CTLs where the EE-pillows have 3E edges but not the EE-bananas. Then we can look at the standard boundary of a 3E-CTL as this dual $3/1$EE-CTL. The above $3/1$EE-CTL is just this standard dual $3/1$EE-CTL physical boundary expressed as a non-dual $3/1$EE-CTL.

**Trivial defect**

There is a mapping from 3E-CTLs to co-dimension 1 defects for 3E-CTLs, i.e., a $3/1$EE-CTLs with the boundary central link c) in Eq. (5.320). The sub 3E-CTL is the original 3E-CTL. The $3/1$EE basic tensors are given by

$$(5.345)$$

**Group permutation action**

**Definition 193.** A **permutation action** $P$ of a group $G$ on a set $B$ is a map

$$P : G \times B \to B \qquad (5.346)$$

such that the following axioms axioms Eqs. (5.348, 5.349, 5.350) below hold. We will think of $P$ as a tensor

$$(5.347)$$

Now the axioms for $P$ in both conventional and tensor network language are:

1. $P$ is a function, i.e., as linear map it is a permutations in the preferred basis that $B$ defines.

$$(5.348)$$

where the blue and black dot are the delta tensors for the sets $G$ and $B$, respectively.

2. $P$ is bijective, i.e.,

$$\forall (g, b) \in G \times B \quad \exists a \in B : P(g, a) = b \qquad (5.349)$$

3. $P$ is a representation of the group $G$, i.e.,

$$\forall (g, h, a) \in G \times G \times B : P(g, P(h, a)) = P(gh, a)$$

$$(5.350)$$

For each group $G$ and permutation action of $G$ on a set $B$, define the **group permutation** $3/1$**EE-CTL** as the following physical boundary of a 3E-CTL, i.e., $3/1$EE-CTL with a single vertex as boundary central link: The sub 3E-CTL is the dual to group 3E-CTL for $G$. The $3/1$EE basic tensors are the following:

$$(5.351)$$

The $3/1$EE basic axioms A10) and A11) follow from the axiom Eq. (5.348). The $3/1$EE basic axiom A7) follows from the axiom Eq. (5.350). The $3/1$EE axioms A1) to A5) follow from the fusion property of the delta tensors.

For the regular representation of the group as permutation action, the group permutation $3/1$EE-CTL is equivalent to the standard physical boundary $3/1$EE-CTL for the dual group 3E-CTL. For the trivial permutation on the trivial one-element set, the group permutation $3/1$EE-CTL is equivalent to the dual standard physical boundary $3/1$EE-CTL for the dual group 3E-CTL.

**Duality switching domain wall**

There is a mapping from 3E-CTLs to domain walls between 3E-CTLs, i.e., $3/1$EE-CTLs with the boundary central link b) in Eq. (5.320). The two sub 3E-CTLs are the original 3E-CTL and its dual 3E-CTL. The $3/1$EE-CTL basic tensors are given by

$$(5.352)$$

**Complex conjugation defect for complex numbers**

For the complex number 3E-CTL there are two irreducible co-dimension 1 defects: The trivial defect above, and the complex conjugation defect. The 3/1EE-CTL basic tensors of the latter are given by

$$(5.353)$$

We see that when going from one side of the defect to the other (i.e., from the index labeled $x$ to $y$ in the second equation above), the complex arrow orientations get reversed. So the defect corresponds to complex conjugation along all indices that cut the defect membrane in the corresponding TL.

### 5.10.4 Properties

**Remark 134.** The rescaling Eq. (5.278) does not change the phase of the 3E-CTL. For a 3/1EE-CTL with a single point as boundary central link, we can rescale the EE-pillows and EE-bananas in the same way as the pillows and bananas, and rescale $B \to \alpha\beta^{-1}B$. This does not change the CTL axioms.

Let us work out where the normalizations appear in the corresponding TL on 3/1CC-lattices (on 3-manifolds with boundary): At each boundary vertex we get an additional factor of $\alpha\beta^{-1}$, at each boundary face we get a factor of $\alpha^{-1}$, at each boundary edge we get a factor of $\beta^{-1}$, and in the interior we get factors as described in Observation (114). Evaluating the rescaled trivial TL on a lattice $X$ on some manifold with boundary yields $\alpha^{2\chi(X)} = \alpha^{\chi(\partial X)}$.

So the rescaling with $\alpha$ does change the phase of the 3/1EE-CTL, whereas there are no indications that rescaling by $\beta$ does (and in fact it does not). The rescaling of the 3E-CTL corresponds to fusing the 2En-CTL Eq. (5.68) with $\alpha_{2En} = \alpha_{3E}^{-1}$ to the standard physical boundary.

**Observation 130.** 3/1EE-CTLs with a single point as boundary central link yield tensor-network representations of ground states of their sub 3E-CTLs.

$$(5.354)$$

e.g., for the standard boundary 3/1EE-CTL the tensor network looks like:

$$(5.355)$$

### 5.10.5 Physical interpretation and connection to existing models

In the world of quantum physics, complex-real 3/1EE-CTLs are fixed point models 2+1-dimensional topological order, with co-dimension 1 defects, domain walls, or physical boundary.

Our models are in the same class as the gapped boundaries presented in Ref. [26], though they have a different microscopic structure. The latter models could be written as a CTLs with 3FE-CTLs as sub-type: The boundary 1-region is thickened. Two new index types are associated to the faces and edges of the boundary 1-region. Again the edge indices have a preferred basis, and the face indices have a basis set depending on the value of the edge indices. Gluing happens at the faces. Face indices are contracted normally and edge indices are contracted via the delta tensor.

Apart from this we are not aware of any general construction for gapped boundaries that resembles 3/1EE-CTLs.

### 5.10.6 Concrete examples

**Duality defect of the toric code**

The toric code 3E-CTL is self-dual in the sense that it is equal to its dual CTL up to a gauge transformation. This gauge transformation is just the Fourier transform $\mathcal{F}$ for the group $\mathbb{Z}_2$, also known as Hadamard transformation:

$$x - \boxed{\mathcal{F}} - y = 2^{-1/2}\begin{pmatrix} 1 & 1 \\ 1 & -1 \end{pmatrix} \qquad (5.356)$$

From this fact, we can construct a co-dimension 1 defect, i.e., a 3/1EE-CTL that has as boundary central link a single 0-region consisting of two vertices. This defect is just the composition of the duality switching domain wall and the gauge transformation given by the Hadamard above:

$$(5.357)$$

Such a duality defect exists for all 3E-CTLs whose dual 3E-CTL is gauge equivalent to the original one. In particular this is the case for all (complexified) 3E-CTLs arising from abelian groups, containing all cyclic groups $\mathbb{Z}_n$.

## 5.11 $3/2$**EE-(C)TLs**

### 5.11.1 Definition

By $3/2$**EE-CTLs** we refer to a CTL type coming from a TL type on 3-dimensional lattices with 1-dimensional embedded structures. They contain 3E-CTLs as a sub-type. The "EC" stands for "edge, corner" as this is where indices are associated to.

**Backgrounds**

The extended backgrounds of the CTL are framed $3/2$-manifolds with boundary with a circle as boundary central link:

$$(5.358)$$

The bulk 1-region has a framing with respect to the surrounding bulk 3-region. I.e., we should picture this bulk 1-region as a thin strip/ribbon instead of a line. The sub 3E-CTL is obtained by restricting to extended backgrounds with empty bulk 1-region and boundary 0-region.

**Example 126.** Consider the following examples for CTL backgrounds, without the framing:

$$(5.359)$$

a) shows a ball with two points on the boundary connected by a line through the bulk. b) shows a solid torus with a line running through the bulk along the non-contractible loop. c) shows a ball with two braided loops inside. d) shows a thickened sphere

(ball with a ball removed from the middle) with one point on each of the two boundary components, connected by one line through the bulk.

In order to get an actual extended background we have to replace every line by a ribbon. Then e.g., in c) each of the two loops can be arbitrarily twisted.

Just as for the sub 3E-CTLs, one background gluing is surgery gluing at the boundary 2-region. Additionally, another background gluing is at the boundary 0-region.

**Example 127.** Consider the following examples of background gluing at the boundary 0-region:

$$(5.360)$$

a) shows gluing two balls with a line through them yielding a single ball with a line through. b) shows gluing the two points where a line pierces out of a ball, yielding a solid torus with a line winding around the non-contractible loop. c) shows gluing two of the four points where two lines embedded into a ball pierce out of the ball, yielding a solid torus with a line piercing through.

**Lattices**

The lattices are based on $3/2$CCbt-lattices living on the extended backgrounds, where the bulk 1-region and boundary 0-region are 1-thickened. Thereby the allowed backgrounds of the upper and lower links are the following: The faces of the bulk 1-region need an interval lower link and a single point as upper link. The edges of the boundary 0-region need an interval upper link and a single point as lower link. So the thickened 1-region can be imagined as a ribbon of faces within the cell complex, where one margin of the ribbon consists of edges of the bulk 3-region and the other margin is left open. So the 1-thickening already models the framing. The decorations inherited from the sub 3E-CTL namely an orientation and a dual orientation for each 3E index edge. Additionally each $3/2$EE bulk edge is decorated a chirality, i.e., one of the two relation-cells between the connected vertex and face is special.

**Comment 23.** If the lattice was oriented, the chirality decoration could be used to determine whether a given $3/2$EE edge is "right handed" or "left handed". Now there is no orientation, but we can still determine whether two $3/2$EE edges on the same connected component of lattice have equal or opposite "handedness".

We will work on the level of backgrounds and only draw the index lattice. When the interior topology is not clear from the context we will describe it in words. We will draw the index lattices in the same way as the 3E index lattices, apart from that there are additional edges corresponding to the boundary 0-region. We will draw the "centre" of those edges as a vertex, and indicate their chirality by lines spiraling left or right. E.g., consider the following patch of a index lattice (right) and its extended background (left):

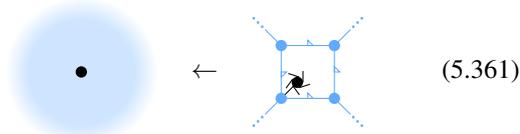

$$(5.361)$$

There are two basic gluings: First, 3E index edges can be glued, just as for the sub 3E-CTLs. Additionally two of the 3/2EE index edges can be glued. Thereby the chiralities have to match. Just as for the 3E index edges this can be pictured as cutting along each of the edges and then gluing the two pieces of boundary obtained in this way. Each such edge is in some corner, i.e., it is connected to exactly one vertex and one face. During the gluing, those faces and vertices are fused together. E.g., consider the following gluing of two 3/2EE index edges:

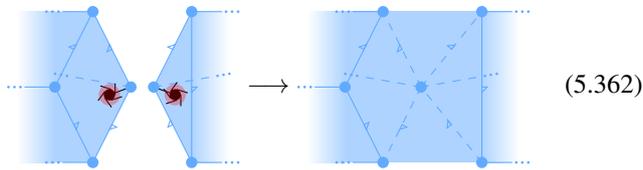

$$(5.362)$$

**Tensors**

The tensors are real tensors with two index types. There is one 3E index associated to every 3E index edge, just as for the sub 3E-CTLs. Additionally there is one 3/2EE index associated to every 3/2EE corner. When two corners or edges are glued the associated indices are contracted.

### 5.11.2 Basic tensors and axioms

**Basic lattices and tensors**

3/2EE-CTLs inherit many properties from 2/1EE-CTLs. Whenever we speak of 2/1EE-CTLs in this section we mean the type with boundary central link consisting of a single point.

**Definition 194.** For every 2/1EE extended background $X$, define the **snake** as the following 3/2EE extended background: Taking $X$ times the 1-sphere, and shrink $x$ times the 1-sphere to a single point for all $x$ of the bulk 1-region and of the boundary 0-region. So the 2/1EE bulk 1-region becomes the 3/2EE bulk 1-region, the 2/1EE bulk 2-region becomes the 3/2EE bulk 3-region, the 2/1EE boundary 0-region becomes the 3/2EE boundary 0-region, and the 2/1EE boundary 1-region becomes the 3/2EE boundary 2-region. Consider the following examples for 2/1EE extended backgrounds and their snakes:

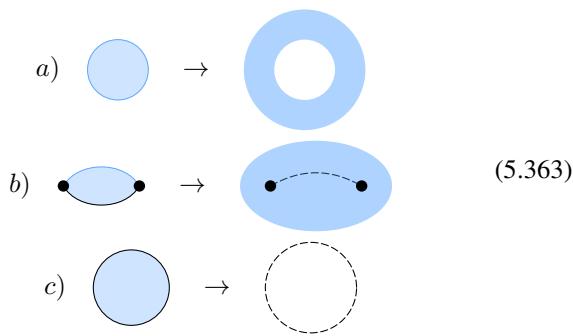

$$(5.363)$$

a) demonstrates that a disk whose boundary is formed by the boundary 1-region is mapped to a solid 2-torus. b) shows a disk whose boundary consists half of the boundary 1-region and half of the bulk 1-region, yielding a 3-ball with a line running from pole to pole through the bulk of the ball. c) shows a disk whose boundary is formed by the bulk 1-region, yielding a 3-sphere with an embedded loop.

**Definition 195.** For each 2/1EE background $X$ define the **snake** for $X$ as the following 3/2EE background. Its extended background is the snake of the extended background of $X$. The index lattice is obtained by replacing every 3E edge by a 3E edge and replacing every 2/1EE edge by a 3/2EE edge. The orientations of the 3/2EE edges match the orientations of the 2E edges and the dual orientations of the 3E edges and chiralities of the 3/2EE edges are all chosen to point in the same direction.

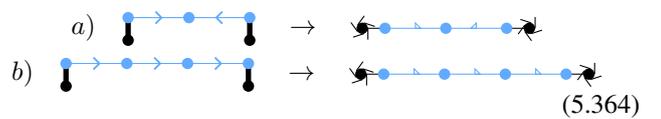

$$(5.364)$$

Note that taking the snake for a 2/1EE background is a generalization of taking the pillow for a 2E background.

**Definition 196.** For each 2/1EE background $X$ define the **bellows** with respect to $X$ as the following 3/2EE background: Its extended background is the snake of the extended background of $X$. Its index lattice consists of one 3E edge for every 2E edge of $X$. For each 2E vertex adjacent to two 2E edges there is one 2-gon face that has the two corresponding 3E edges as boundary. The dual orientations of the 3E edges are determined by the orientations of the 2E edges. The orientations of the 3E edges and chiralities of the 3/2EE edges are chosen to all point in the same direction. Then every 2/1E edge is replaced by a 3/2EE edge. If $X$ is connected then the index lattice of its bellows consists of only one vertex. E.g., consider the bellows for the following backgrounds:

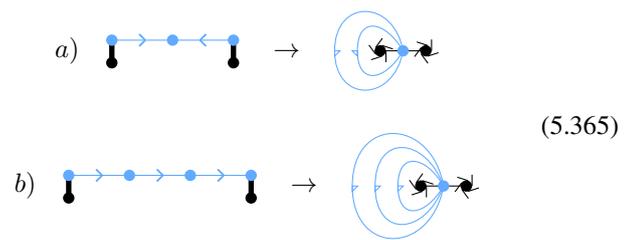

$$(5.365)$$

Note that taking the bellows for a 2/1EE background is a generalization of taking the banana for a 2E background.

**Observation 131.** Every lattice can be glued together from snakes and bellows.

We can construct a history for a given lattice in the following way: Choose orientations for all edges of the bulk 3-region and dual orientations for all faces. Just as for the sub 3E-CTL, replace every face of the bulk 3-region by the pillow for its lower link and every edge of the bulk 3-region by the banana for its lower link. Now replace every face of the bulk 1-region by the snake for its lower link and every edge of the bulk 1-region by the bellows for its upper link.

**Remark 135.** Taking the snake can be formalized to a lattice mapping from 2/1E lattices to 3/2EE lattices, such that gluing two 2E or 2/1EE edges or the corresponding 3E or 3/2EE edges commutes with taking the snake.

This implies that all snakes can be glued together from the snakes associated to the 2/1E basic lattices:

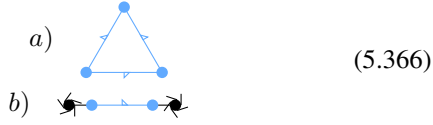

(5.366)

**Remark 136.** Taking the bellows can be formalized to a lattice mapping from 2/1EE lattices to 3/2EE lattices, such that gluing two 2E or 2/1EE edges or the corresponding 3E or 3/2EE edges commutes with taking the bellows.

This implies that all bellows can be glued together from the bellows associated to the 2/1E basic lattices:

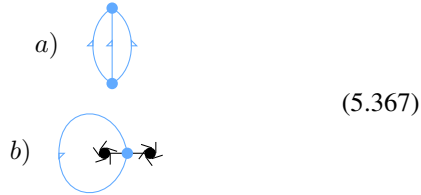

(5.367)

**Observation 132.** Combining Observation (131) with Remark (135, 136) we find that a set of basic lattices is given by the 3E basic lattices together with the snake and the bellows for the 3/2EV basic lattice. So the whole CTL is already determined by the associated tensors:

$$T1) \quad H_a^{xy} = x \overset{a}{\bullet\!\!-\!\!\bullet} y$$

$$T2) \quad I_a^{xy} = a \overset{}{\bullet} \; x \overset{}{\bullet\!\!-\!\!\bullet} y$$

(5.368)

**Basic history moves and axioms**

**Definition 197.** We can use the snake mapping to transform every 2/1EE history lattice $X$ into a 3/2EE history lattice by replacing every 2/1EE basic lattice by the corresponding snake. We will call those history lattices the **snake history lattice** for $X$. Consider, e.g., the following snake history lattice:

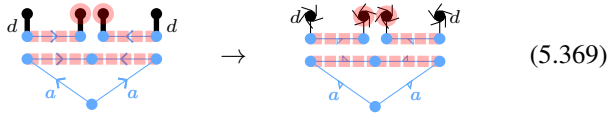

(5.369)

**Definition 198.** We can use the bellows mapping to transform every 2/1EE history lattice $X$ into a 3/2EE history lattice by replacing every 2/1EE basic lattice by the corresponding bellows. We will call those history lattices the **bellows history lattice** for $X$. Consider, e.g., the following bellows history lattice:

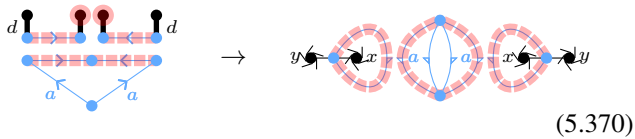

(5.370)

**Definition 199.** For every pair of 2/1EE index lattices $X$ and $Y$, define the **snake-bellows history lattice** as the following 3/2EE history lattice. First consider a 3/2EE face $x$ with upper link $X$ that is connected to a volume. The lower link of the face corresponding to $x$ in the lower link of the volume is a 2E index lattice that we will denote $D[X]$: $D[X]$ has 2 2E edges for every 2E edge of $X$, and one additional 2E edge for every 2/1EE edge of $X$. Now take for each 2E edge of $X$ the pillow for $D[Y]$, and for each 2/1EE edge of $X$ the snake for $Y$. Dually take for each 2E edge of $Y$ the banana for $D[X]$ and for each 3/1EV edge of $Y$ the bellows for $X$. For each pair of (2E or 2/1EE) edges of $X$ and $Y$, there is one (pair of) (2E or 2/1EE) edge of one pillow or snake, and one (pair of) (2E or 2/1EE) edge of one banana or bellows of the history lattice. The edges of the pillows or snakes are glued together with the corresponding edges of the bananas or bellows. E.g., consider the following snake-bellows history lattice:

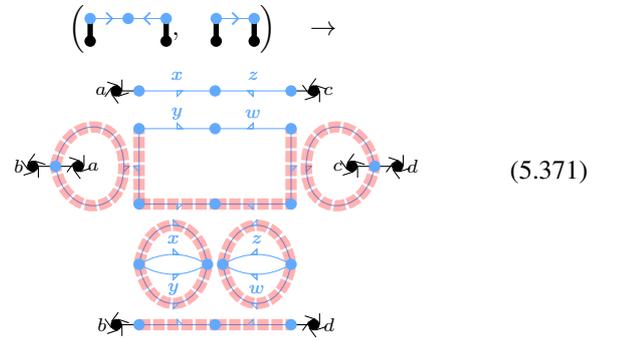

(5.371)

**Observation 133.** Every 2/1EE history lattice can be history glued from snake history mappings, bellows history mappings, and snake-bellows history mappings (together with the 3E basic history lattices).

History lattices are based on 4/2CCb-lattices with the same central lower link as the lattices. The cells of the bulk 4- and 2-regions correspond to basic moves, whereas the cells of the boundary 3- and 1-regions correspond to basic lattices and basic gluings. A history for a given history lattice can be constructed in the following way: Replace

- each volume of the 4/2CCb-lattice by the snake history lattice for its lower link (for volumes of the bulk 4-regions those reduce to pillow history lattices).

- each edge of the 4/2CCb-lattice by the bellows history lattice for its upper link (for edges of the bulk 4-regions those reduce to banana history lattices).

- each face of the 4/2CCb-lattice by the snake-bellows history lattice for its upper and lower link (for edges of the bulk 4-region this reduces to the pillow).

By this we get two bellows for each edge-face-relation-cell (which are actually a banana of not both the edge and face are part of the bulk 2-region), and two snakes for each volume-face-relation-cell (which are actually pillows if not both the volume and the face are part of the bulk 2-region). Gluing all those pairs of basic lattices yields remaining basic lattices at the 3/2CC-lattice forming the boundary of the 4/2CCb-lattice that form the desired history lattice.

**Observation 134.** Consider the snake-bellows history lattice for two 2/1EE index lattices $X$ and $Y$. Gluing two 2E (or 2/1EE) edges of $Y$ is consistent with history gluing the corresponding two pillows (or snakes) of the snake-bellows history lattice. Dually gluing two 2E (or 2/1EE) edges of $X$ is consistent with history gluing the corresponding two bananas (or bellows) of the snake-bellows history lattice. Also disjoint unions in the $X$ or $Y$ component are consistent with disjoint unions of the snake-bellows history lattice. Thus, all snake-bellows history lattices can be history glued from the one with $X$ and $Y$ being the 2/1EE basic lattice (where on the right hand side we decomposed the 4-gon pillow/banana into two 3E basic pillows/bananas, respectively):

$$(5.372)$$

Analogously to 2/1EE-CTLs, also all snake history lattices and bellows history lattices can be history glued from the one for the 2/1EE basic history lattices.

**Observation 135.** Combining Observation (133) and Observation (134) we see that a set of basic history lattices is given by the snake history lattice and bellows history lattice for the 2/1EE basic lattice, together with Eq. (5.372).

**Observation 136.** According to Remark (24), the bi-partitions of the basic history lattices are good candidates for basic history moves/axioms. For the basic snake history lattice, there are two kinds of bi-partitions, yielding 6 different history moves in general, analogous to 2/1EE-CTLs. The same is true for the basic bellows history lattice.

Let us consider the possible bi-partitions of Eq. (5.372): There are four kinds of such bi-partitions: First, there can be one bellow on the right and the rest on the left side. Second, there are the dual bi-partitions with one snake on the right side. Third, there are bi-partitions with one snake and one bellows on the right side, and the rest on the left side. Fourth, there are bi-partitions with one snake, one bellows, one pillow and one banana on both sides.

**Simplification of the axioms**

**Remark 137.** Consider the background whose index lattice consists of only two 3/2EE edges separated by a 3E vertex.

Gluing one 3/2EE edge of this lattice to any 3/2EE edge of any other lattice does not change the background of the latter.

Thus, we can implement the local support convention by setting the associated tensor to the identity matrix:

$$(5.373)$$

**Observation 137.** We can drastically reduce the number and complexity of axioms. First, analogously to 2/1EE-CTLs, the following set of axioms is equivalent to those arising from bi-partitions of the snake basic history lattice:

$$(5.374)$$

Second, analogously to 2/1EE-CTLs, the following set of axioms is equivalent to those arising from bi-partitions of the bellows basic history lattice:

$$(5.375)$$

Third, we can introduce the following auxiliary tensor:

$$T3) \qquad x \, \text{—} \, y \qquad (5.376)$$

With this tensor, the following set of axioms is equivalent those arising from all bi-partitions of the basic snake-bellows history lattice together with the local support convention:

$$(5.377)$$

Let us see how the axioms arising from bi-partitions of the basic snake-bellows history lattice follow from the axioms above: A10) is one of the axioms with one snake and one bellows on the right side. We can use A8) to bring the snake from the right to the left side to obtain one of the axioms with only a bellow on the right side. Dually we can bring the bellows from the right to the left side using A9).

Now we can use A2) together with the 3E axiom A3) to invert the orientation of any 3E edge of any snake in any axiom. Dually we can use A4) together with the 3E axiom A8) to invert the dual orientation of any 3E edge of any bellows. Also we can use A5) and A6) to invert the dual orientations of the 3E edge and chirality of the 3/2EE edges of any snake. Dually we can use A5) and A7) to invert the orientation of the 3E edge and the chirality of the 3/2EE edges of any bellows. So we see that we can arbitrarily change the orientations, dual orientations and chiralities in any axiom.

Using this all axioms with one snake, one bellows, or one snake and one bellows on the right side follow from one single axiom of that kind, respectively. Also the axiom with one snake, one bellow, one pillow and one banana on each side becomes trivial after changing the orientations, dual orientations, or chiralities on one side.

### 5.11.3 Solutions

**Cyclic groups**

For every positive integer $n$ and $0 \leq x, y < n$, the **cyclic group** 3/2**EE-CTL** is the following 3/2EE-CTL: The sub 3E-CTL is given by the group 3E-CTL for the cyclic group or order $n$. The 3/2EE indices have trivial (one-element) basis set, and the basic snake and bellows are given by

$$(5.378)$$

Here the elements of the cyclic group were identified with the numbers from 0 to $n - 1$.

The snake 2/1EE-CTL corresponds to an irreducible representation of the group, whereas the bellows 2/1EE-CTL corresponds to an irreducible representation diagonal *-algebra over the set of group elements, i.e., projections onto different group elements.

### 5.11.4 Subtypes, mappings and fusions

**Trivial defect**

There is a CTL mapping from 3E-CTLs to 3/2EE-CTL such that the original CTL is the sub 3E-CTL of the resulting CTL. We will refer to the 3/2EE-CTL as the **trivial defect** for the 3E-CTL. The lattice mapping replaces every 3/2EE edges by a 3E edge and a 3E vertex, and the tensor mapping takes the 3E index as 3/2EE index. E.g., the 3/2EE basic tensors are given by

$$(5.379)$$

**Anti-particle**

There is a mapping from 3/2EE-CTLs to 3/2EE-CTLs such that the sub 3E-CTLs of the original and the resulting 3/2EE-CTLs are the same. We will refer to the resulting 3/2-CTL as the **anti-particle** of the original 3/2EE-CTL. The lattice mapping consists in inverting the chiralities of all 3/2EE edges, and the tensor mapping is trivial (as the lattice mapping does not change the index distribution). E.g.,

$$(5.380)$$

Note that the anti-particle and its original 3/2EE-CTL are in the same phase (relative to the sub 3E-CTL), though for complex-real 3/2EE-CTLs the connecting TL operator is anti-unitary.

**Dual CTL**

There is a mapping from 3/2EE-CTLs to 3/2EE-CTLs such that the mapping restricted to the sub 3E-CTL is the dual 3E-CTL. We will refer to the resulting 3/2EE-CTL as the **dual** 3/2EE-CTL. The lattice mapping interchanges 3E vertices and 3E faces of the index lattice. The 3/2EE edges connected to a pair of 3E face and 3E vertex become 3/2EE edges connected to the associated 3E vertex and 3E face, its chirality is determined by the same vertex-face relation-cell. The tensor mapping is trivial, as the lattice mapping does not change the index distribution. E.g.

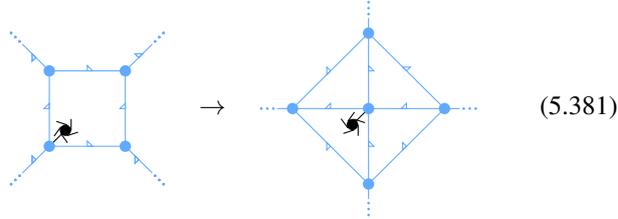

(5.381)

The 3/2EE-CTL and its dual are in the same phase, though they cannot be in the same phase relative to the sub 3E-CTL, as the two sub 3E-CTLs are different (namely dual to each other).

**Anyon fusion**

Consider two 3/2EE-CTLs with the same sub 3E-CTL. We can take the joint type relative to this sub 3E-CTL, and then apply the following mapping: The lattice mapping consists in replacing every 3/2EE edge by two 3/2EE edges of the two different input 3/2EE-CTLs. The two 3/2EE edges are connected to the same vertex and face as the original one, and their order and chiralities are determined by the chirality or the original one. The tensor mapping consists in taking the composite of the indices at each pair of 3/2EE edges as the new 3/2EE index. E.g.,

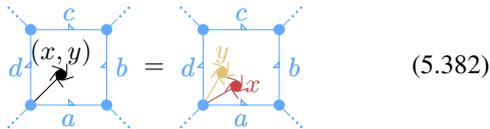

(5.382)

**Caterpillar CTL**

**Definition 200.** Consider 2/1EE-CTLs with a single point as boundary central link. Define the **caterpillar** with respect to a 2/1EE background $X$ as the following 3/2EE background $Y$: The extended background of $Y$ is the snake of the extended background of $X$. The index lattice of $Y$ is obtained by replacing every 2E edge by two kinds of 3E edges, and every 2/1EE edge by a 3/2EE edge. The 3E edges of the one kind go along the snake (along $X \times \{0\}$), whereas the 3E edges of the other kind wind around the snake (along $\{x\} \times S_1$). They start and end at the same vertex that the 3E edge of the first kind is oriented towards. The orientations of the former 3E edges and the dual orientations of the latter 3E edges are chosen to match the orientations of the corresponding 2E edges. The dual orientations of the former 3E edges, the orientations of the latter 3E edges, and the chiralities of the 3/2EE edges are chosen to point all in the same direction.

Consider the following examples for 2/1EE backgrounds (left) and their corresponding caterpillars (right):

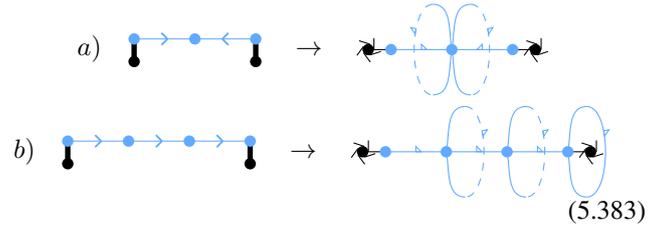

(5.383)

Here the extended background of the 2/1EE backgrounds a) and b) is a disk whose boundary is divided into the boundary 1-region and the bulk 1-region. The corresponding extended backgrounds of the caterpillars are a 3-ball with a line going from one pole to the other through the bulk.

Note that the doughnut for a 2E background is a special case of the caterpillar of a 2/1EE background.

**Observation 138.** Just as for the doughnuts in Observation (115), gluing two edges of a 2/1EE background commutes with gluing the two associated pairs of edges of the corresponding caterpillars. Taking the caterpillar can be extended to a lattice mapping from 2/1EE lattices to 3/2EE lattices. One can define a tensor mapping by taking as 2/1EE index the corresponding 3/2EE index and as 2E index the composite of the two 3E indices of the associated two kinds of 3E edges (just as for the doughnut mapping). Combining both mappings yields a CTL mapping from 3/2EE-CTLs to 2/1EE-CTLs.

e.g., the 2/1EE basic tensor is determined by the 3/1EC-CTL in the following way:

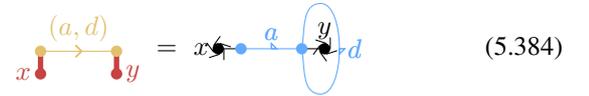

(5.384)

This implies that the every caterpillar can be glued together from copies of the caterpillar with respect to the 2/1EE basic lattice shown above.

**Observation 139.** A set of basic history lattices is given by the following history lattice, together with the 3E basic history lattices:

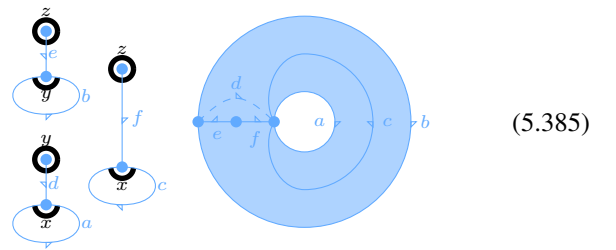

(5.385)

This can be seen as follows: History lattices are based on 4/2b-lattices. The cells of the boundary 1- and 3-region correspond to basic lattices, and the cells of the bulk 2- and 4-region correspond to basic moves. We can without loss of generality restrict to triangular faces of the bulk 2-region. We can get a history for a given history lattice in the following way: Every triangle of the bulk 2-region is replaced by the basic history lattice above. Edges, faces and volumes of the bulk 4-region are replaced by EE-pillow history lattices, EE-banana history lattices and EE-pillow-banana-history lattices as for 3E-CTLs.

The basic doughnut on the right side can be glued from 3E basic lattices in e.g., the following way:

$$(5.386)$$

### 5.11.5 Connection to the representations of the Drinfel'd quantum double

**Definition 201.** For any involutive weak Hopf C*-algebra $H = (\eta, \mu, \epsilon, \Delta, S, t)$, its **dual opposite** $(H^*)^{\mathrm{cop}} = (\tilde{\eta}, \tilde{\mu}, \tilde{\epsilon}, \tilde{\Delta}, \tilde{S}, \tilde{t})$ is the involutive weak Hopf C*-algebra given by

$$
\begin{aligned}
\tilde{\eta} &= \epsilon^\dagger, \\
\tilde{\mu} &= \tau \circ \Delta^\dagger, \\
\tilde{\epsilon} &= \eta^\dagger, \\
\tilde{\Delta} &= \mu^\dagger, \\
\tilde{S} &= S^\dagger, \\
\tilde{t} &= S \circ t^\dagger,
\end{aligned}
\tag{5.387}
$$

where $\tau$ is the swap operator exchanging two components of a tensor product. In terms of tensors (the tensors of $(H^*)^{\mathrm{cop}}$ are in blue), one gets

$$(5.388)$$

**Definition 202.** The **Drinfel'd double** [27] $D(H)$ of an involutive weak Hopf C*-algebra $H$ (over a vector space with basis set $B$) is an involutive weak Hopf C*-algebra with basis set $B \times B$. In the conventional mathematics language one thinks of the second component of the vector space $\mathbb{C}^{B \times B} = \mathbb{C}^B \otimes \mathbb{C}^B$ as the dual vector space of the first component. That is, one writes its elements as linear functions such that $d(a)$ for $d$ in the second and $a$ in the first component yields a number. Then one equips the first component with the (co-)multiplication of $H$ and the second component with the (co-)multiplication of $(H^*)^{\mathrm{cop}}$. In this context the multiplication of $D(H)$ is denoted

by [57]:

$$(a, d)(b, e) = (b_{(1)}(S(d_{(1)}))b_{(3)}(d_{(3)})ab_{(2)}, d_{(2)}e) \tag{5.389}$$

Translated to our tensor-based language, expressions like $b(a)$ represent the trace over the two elements. So we get:

$$(5.390)$$

**Observation 140.** The *-algebra associated to the doughnut 2E-CTL (via Eq. (5.74)) equals the *-algebra of the quantum double of the weak Hopf *-algebra associated to the 3E-CTL via Eq. (5.301):

$$(5.391)$$

In the third step we used a decomposition of the doughnut corresponding to the 2E basic lattice into 3E basic lattices.

**Observation 141.** Every 3/2EE-CTL defines a representation of the *-algebra of the quantum double of the weak Hopf *-algebra associated to the sub 3E-CTL. The representation tensor is given by the caterpillar for the 2/1EE basic lattice:

$$(5.392)$$

This can be seen as follows: According to the previous observation, the *-algebra of the quantum double of the weak Hopf *-algebra associated to the sub 3E-CTL is the *-algebra associated to the doughnut 2E-CTL. On the other hand caterpillar 2/1EE-CTL has the doughnut 2E-CTL as sub CTL. Thus, the caterpillar 2/1EE-CTL defines a representation of the *-algebra associated to the doughnut 2E-CTL.

**Remark 138.** Conversely, as for 3E-CTLs and weak Hopf *-algebras, it is not immediately clear that each representation of the quantum double yields a 3/2EE-CTL. On a rough level discarding technical details, this is true: An alternative 3/2EE basic lattice is given by the caterpillar for the 2/1EE basic lattice. Then the main basic axiom is the following:

 (5.393)

This axiom is equivalent to the basic caterpillar being a representation of the quantum double. Though in order to get a complete set of basic axioms we have to add multiple auxiliary tensors and axioms. Some of those axioms are not invariant under invertible basis changes which are the gauge transformations for *-algebra representations (in contrast to orthogonal maps for 3/2EE-CTLs). Thus, the technical part consists in finding a gauge in which all the 3/2EE basic axioms hold for the *-algebra representation tensors.

### 5.11.6 Physical interpretation and connection to existing models

Complex-real 3/2EE-CTLs correspond to anyons in the topologically ordered models corresponding to the sub 3E-CTLs, such as an $e$ or $m$ anyon in the toric code. The ground states on a given index lattice are states with anyons of one particular type at all 3/2EE-CTL edges.

**Remark 139.** Note that as mentioned in Observation (52) we do not interpret anyons as excitations (i.e., eigenstates of a Hamiltonian), but rather as points in space (or lines in space-time) where the Hamiltonian terms (or tensors of the imaginary time evolution) differ in a topologically non-trivial way (which might be called a co-dimension 2 defect). In this viewpoint it is very natural that the presence of anyons at a point leads to additional degrees of freedom (i.e., a different Hilbert space) at this point. In the concrete case of 3/2EE-CTLs we indeed have one additional 3/2EE index for each occurrence of an anyon.

One could add new lattice sites around the excitation, leading to new bulk degrees of freedom, in which we can non-canonically embed the additional anyonic degrees of freedom. Then the states with and without anyon are in the same Hilbert space. We will of course generically find that the states with anyons interpreted as states in the model without defects are not ground states anymore, but violate the local Hamiltonian terms. As there is a lot of arbitrariness of how to embed the anyon Hilbert space locally into the bulk Hilbert space, it seems very plausible that one can find a way such that ground states with anyons are exactly (higher) eigenstates of the local Hamiltonian terms of the bulk system. So in some sense the two viewpoints are equivalent. However, our viewpoint is strictly more general and as we find much more natural, mathematically beautiful and well-defined. In particular, the notion of anyons can be formulated as a simple information-theoretic concept and can be defined without complicated things like "energy" or "eigenstates".

**Remark 140.** Strictly speaking, 3/2EE-CTLs do not only cover anyons, but also 1) co-dimension 2 defects projecting onto one particular symmetry broken sector if the sub 3E-CTL is reducible (i.e., contains symmetry-breaking character). 2) They cover arbitrary direct sums of anyons/symmetry-breaking defects, such as a composite $e \oplus m$-particle in the toric code.

**Remark 141.** We can also write down a local commuting-projector Hamiltonian for those physical models with anyons: Away from the defect points in the index lattice, the Hamiltonian is that of the sub 3E-CTL with of vertex and face terms, see Section (5.9.7). At vertices or faces of the index lattice that are connected to a 3/2EE edge we have to modify the Hamiltonian terms accordingly. E.g., for a vertex/face with a 4-gon upper/lower link and connected to a 3/2EE edge, we get:

 (5.394)

Where the tensors on the right hand side yield the local ground state projector when interpreted as linear map from the indices $a, b, c, d, x$ to the indices $\bar{a}, \bar{b}, \bar{c}, \bar{d}, \bar{x}$, and when the normalizations are included. A Hamiltonian is given by $\sum_P (1 - P)$ for all such local ground state projectors $P$.

**Remark 142.** In topologically ordered systems there are operators that move the anyons around that have support in a 1-dimensional region. Those operators are known as *string or ribbon operators* [8, 17]. For the group quantum double models the explicit form of those ribbon operators was derived in Ref. [56]. In our language, those ribbon operators are simply CTL tensors associated to special lattices.

A *ribbon* is a path of corners (i.e., pairs of connected vertex and face), where in each step either the vertex or the face moves to an adjacent vertex or face. Equivalently, it is a sequence of edges, where at each edge either the connected vertex or the connected face changes. A ribbon operator creates or annihilates two anyons at the two end corners, or moves an anyon from one to the other end corner.

For a given ribbon, the **double-ribbon** is the following background: Its extended background is the 1-ball completion of the stellar cone of the boundary central link (a 3-ball with two points at the poles connected by a line through the bulk). Its index lattice has two 3/2EE edges at the start- and end corner. For each 3E edge of the ribbon the double-ribbon has two 3E edges, that are connected to the same pair of vertices and to the same pair of faces: For edges of the ribbon where the vertex changes, both 3E edges of the double ribbon go "along the snake" and are separated by a 2-gon face. For edges of the ribbon where the face changes, the two edges of the double-ribbon both go half "round the snake" in different directions and meet at one vertex there. We can take one out of each pair of edges of the double-ribbon to divide its index lattice into a back and a front layer. Those layers are both copies of the 1-dimensional patch of the ribbon. So gluing the back layer to the 1-dimensional patch of the ribbon yields the ribbon the same lattice again, just that there are additional 3/2EE edges at the start- and end corner of the ribbon. If they were already 3/2EE edges present at the start- or end corner, we can also glue those with the 3/2EE edges of the background. In this case existing 3/2EE edges are removed by the gluing. So we see that if we interpret the tensor associated to the double-ribbon as a linear map from the back to the front layer, then it is just the ribbon operator that creates/annihilates/moves anyons. E.g., the following shows how a ribbon operator that moves an anyon from one corner to another along the ribbon formed by the edges labeled by $a, h, o, e, m, b$:

We see that ribbon operators are TL operators on the 1-dimensional ribbon patches as lattices. In other words they are what is known as a matrix product operator (MPO) in the tensor-network community.

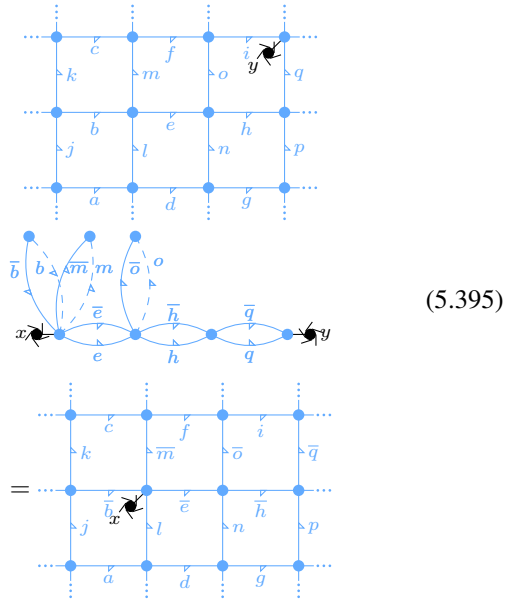

$$(5.395)$$

The double-ribbon can be glued together from snakes, bellows, bananas and pillows, e.g., for the example above:

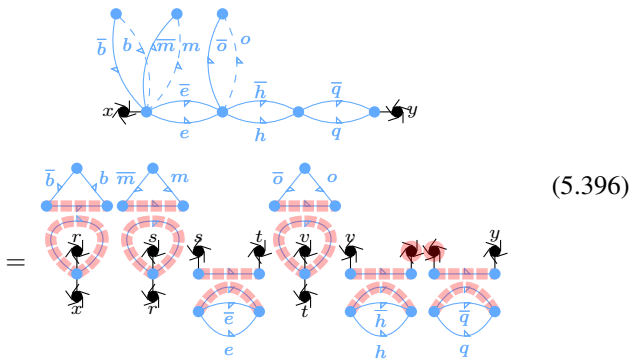

$$(5.396)$$

# Chapter 6

# Outlook

## 6.1 Other CTL types

### 6.1.1 Fermionic phases

In this section we will describe how we can use the framework of CTLs/TLs to get fixed point models for fermionic topological phases. We will have to use both a different lattice and a different tensor type: Per definition working with fermionic models means using fermionic tensors everywhere. This definition is the most general and simple definition of what *fermionic* means. Now it is perfectly fine to search for "topological fermionic" fixed point models on simplicial complexes with Pachner moves, however, we will make the following two findings: 1) Solutions to the axioms are rather ugly, and there are hardly any non-trivial ones. 2) Models defined on regular grids cannot be generically extended to simplicial complexes with Pachner moves. Those two shortcomings are resolved by using another "topological" type of lattice: Lattices with a combinatorial representation of a spin structure. This suggests that we use this lattice type instead and that we should better use the term "topological fermionic order" for models defined on this lattice type.

We do not have a good understanding why a spin structure is needed to resolve the two shortcomings above, but the reason why there are no non-trivial fixed point models without taking a spin structure into account might be extracted from Ref. [58].

#### Backgrounds

First, the TL backgrounds are of a different type. Instead of bare manifolds up to homeomorphism they will be spin manifolds up to spin homeomorphism. The CTL extended backgrounds are spin manifolds with a boundary. There are continuum descriptions for spin structures. However, we find those hard to deal with on both an intuitive pictorial and concrete computational level, so we will only give the combinatorial formulation below. In our reasoning the combinatorial formulation together with moves will replace the continuum formulation anyways.

#### Lattices

For CTLs describing fermionic phases, we can use the same lattices as for the bosonic case, just that we have to add a combinatorial, discrete version of a spin structure. Below we will describe how such a lattice type can look like in any dimension. We will not give the complete definition though. We start with $n$SCb-lattices for the corresponding space-time dimension $n$.

On such a simplicial complex we can define a discrete version of homology.

**Definition 203.** A ($\mathbb{Z}_2$) $m$-**chain** on a $n$SCb-lattice is a map that associates an element of $\mathbb{Z}_2$ to every $m$-simplex of the lattice. The **trivial** $m$-**chain** associates the trivial group element to all simplices.

**Definition 204.** The **boundary** $\partial(c)$ of a $m$-chain $c$ is the following $(m-1)$-chain: The $\mathbb{Z}_2$ element of $\partial(c)$ associated to a $(m-1)$-simplex is the $\mathbb{Z}_2$ product of the $\mathbb{Z}_2$ elements of $c$ associated to all $m$-simplices corresponding to the vertices of the link of the $(m-1)$-simplex.

**Definition 205.** A $m$-chain whose boundary is the trivial $(m-1)$-chain is called a $m$-**cycle**. A $m$-chain that is the boundary of a $(m+1)$-chain is called an **exact** $m$-**cycle**. As the name suggests, and exact $m$-cycle is automatically a $m$-cycle.

The central ingredient for constructing a lattice type describing spin manifolds is a local prescription (for all $n$) which for a given $n$SCb-lattice $X$ yields an $(n-2)$-chain $\omega(X)$ on $X$, such that $\partial\omega(X) = \omega(\partial X)$. $\omega(X)$ will be called the **obstruction cycle** of $X$. $\omega$ (or better the homology class of $\omega$) will be a discrete combinatorial analogue of what is called *Stiefel-Whitney obstruction class* of the manifold described by $X$.

A concrete local prescription defining the obstruction cycle $\omega(X)$ for $n = 2$ and $n = 3$ is given e.g., in Ref. [59]. A general one is sketched in Ref. [58].

**Definition 206.** A **combinatorial spin structure** on a $n$SCb-lattice $X$ is an $(n-1)$-chain $\eta_X$ such that $\partial\eta_X = \omega(X) \cup \eta_{\partial X}$ where $\eta_{\partial X}$ is an $(n-2)$-chain on $\partial X$ such that $\partial\eta_{\partial X} = \omega(\partial X)$.

Now the lattice type that we need are pairs of a $n$SCb-lattice $X$ and a combinatorial spin structure on $X$. The basic moves consist of 1) the basic moves for $n$SCb-lattices (i.e., Pachner moves) and 2) for each $n$-simplex of $X$, flipping $\eta_X$ on all $(n-1)$-simplices around the $n$-simplex. The latter moves can be used to let the combinatorial spin structure fluctuate without changing its homology. Pachner moves change $X$ and with it $\omega(X)$ and with it the combinatorial spin structure. So they also have to change the combinatorial spin structure locally. However, the way in which the spin structure changes does not matter because different ways are connected by the spin structure moves anyways.

The combinatorial spin structure also affects the basic gluings: When two boundary $(n-1)$-simplices of a $n$SCb-lattice are glued, $\eta_{\partial X}$ has to take the same values on the $(n-2)$-simplices around them. I.e., $\eta_{\partial X}$ is also part of the index lattice.

**Tensors**

Also the tensor type that we associate to lattices is different: We have to use fermionic tensors instead of ordinary real/complex tensors. The following definitions are equivalent to using Grassmann variables as e.g., in Ref. [60], however, our language is boiled down to the essential part: An ordering bookkeeping.

**Definition 207.** A $\mathbb{Z}_2$-**graded tensor** is a (complex-)real tensor where the basis set $B$ is the disjoint union of an **even** $B_0$ and an **odd** part $B_1$, and the entries of the tensor are only non-zero when the total parity is even. For each index and $b \in B$ we will write $|b| = i$ if $b \in B_i$. So we have

$$T_{i_1, i_2, \ldots} = 0 \quad \text{if} \quad |i_1| + |i_2| + \ldots = 1 \qquad (6.1)$$

where the sum is in $\mathbb{Z}_2$.

**Definition 208.** An **ordered tensor** is a tensor together with an ordering of its indices. Define the **fermionic transposition move** acting on two indices $i_x$ and $i_y$ of an ordered $\mathbb{Z}_2$-graded tensor that are consecutive in the ordering as the following operation: 1) Reverse the ordering of the indices $i_x$ and $i_y$. 2) Multiply the tensor by

$$T_{i_1, i_2, \ldots} \longrightarrow (-1)^{|i_x||i_y|} T_{i_1, i_2, \ldots} \qquad (6.2)$$

A **fermionic tensor** is an equivalence class of ordered $\mathbb{Z}_2$-graded tensors under fermionic transposition moves.

**Definition 209.** In order to compute the **tensor product** of two fermionic tensors, pick one representative (i.e., one particular order) of each of the fermionic tensors. Then a representative of the tensor product can be obtained by taking the ordinary tensor product of the two $\mathbb{Z}_2$-graded tensors and concatenating the two orderings. I.e., all indices of one tensor come before all indices of the other tensor, but the ordering within each tensor stays the same. Which tensor comes first does not matter because of the $\mathbb{Z}_2$-grading of the tensors.

**Definition 210.** We can **contract** two indices of a fermionic tensor in the following way: Choose a representative of the fermionic tensor such that the two contracted indices are consecutive in the order. Then a representative of the contraction is given by the ordinary contraction of the two indices of the $\mathbb{Z}_2$-graded tensor with the same ordering just that the two contracted indices are erased.

This is all we need to describe fermionic phases of matter: Take simplicial complexes with spin structure instead of just simplicial complexes, and fermionic tensors instead of ordinary real tensors. Then using our formalism one will obtain fixed point models for fermionic phases.

In $1 + 1$ dimensions our framework yields a CTL type that is closely related to *-super-algebras. Similar to (non-fermionic) 2E-CTLs, those are classified by direct sums of irreducible blocks. In the fermionic complex-real case there are two kinds of such irreducible blocks known as even and odd type [61] (also known as $m$-type and $q$-type), in contrast to ordinary complex-real 2E-CTLs. The blocks can have different sizes, however, blocks of the same type with different sizes are in the same phase (up to Euler-characteristic normalization). The simplest $q$-type block alone corresponds to a physical model known as the Kitaev chain (Ref. [62]). The simplest $m$-type block corresponds to a trivial (complex-real) phase.

In $2 + 1$ dimensions our framework yields a CTL type that describes all the models given in Refs. [53, 10], as well as the more general models given in Refs. [51, 11]. If we use the fermionic analogue to 3FE-CTLs (with combinatorial spin structure and where the face indices are fermionic), the (fermionic analogue of) 2E-CTL of face-bananad for a fixed edge index value is exactly the non-trivial *automorphism algebra* of the (isomorphism class of the) simple object corresponding to this edge index value. Those can be of $m$ and $q$ type as mentioned in the section before.

### 6.1.2 Free fermion phases

For gapped fermionic models which are free (i.e., quadratic/non-interacting), a rather complete classification of phases is known for a long time [28] (for a more recent treatment, see, Ref. [63]). However, it is not entirely clear so far how this classification is connected to a genuine many-body classification via local unitaries as in Ref. [53]. This is partly due to the fact that the nature of efficiently solvable models that are used to classify phases is very different: On the one hand we have commuting-projector models, and on the other hand free-fermion models, which are almost disjoint sets (except for the Kitaev chain). Also, all the models that have been classified (in e.g., Ref. [51]) by commuting-projector models do have a gapped boundary, whereas quadratic models (in more than 1 spatial dimension) generically do not allow for a gapped boundary (which is related to them being called *chiral*). Thus, it might seem that there is no way to connect the two worlds.

We believe that such a connection can be established on a very direct physical level using our framework. First of all, using a TL type with a more complicated way of associating tensor networks to lattices we can write down exactly solvable models without having a commuting-projector Hamiltonian or a gapped boundary, see Section (6.1.5). This way we can circumvent known no-go theorems [16] to write chiral models as commuting-projector models.

Second, restricting to quadratic systems corresponds to using another tensor type that we will refer to as *Gaussian tensors*. Gaussian tensors are essentially equivalent to matchgate tensors [64, 65], just that we will formalize it to fit our needs: The basis sets are finite sets just as for real tensors, but now the set elements will correspond to different *fermionic modes* combined into a single degree of freedom, instead of local configurations of one degree of freedom. A Gaussian tensor with $n$ indices is an anti-hermitian complex $X \times X$ matrix where $X = B \times n = B \oplus B \oplus B \ldots$. In the case of only one mode per index, $B$ is a one-element set and the tensor is represented by a $n \times n$ matrix, known as *covariance matrix*. One might also introduce different index types with different basis sets. $X$ is the direct sum of all those different basis sets: $X = B_1 \oplus B_2 \oplus B_3 \ldots$. The tensor product of two Gaussian tensors is given by the direct sum of the two matrices. The contraction of two indices then corresponds to an operation that changes the representing matrix and thereby removes two rows and the equivalent columns.

The relation between free-fermion phases and genuine many-body phases can be formalized as a tensor mapping from Gaussian tensors to fermionic tensors: The basis $B'$ of the fermionic tensor is given by bit strings over the basis set $B$ of the Gaussian tensor: $B' = \{0, 1\}^B$, where 0 is even and 1 is odd. The entry of the fermionic tensor for a given ordering and index configuration can be obtained in the following way: 1) Write the Gaussian tensor as a matrix with the ordering of rows and columns given by the ordering of the fermionic tensor. 2) The index configuration corresponds to a bit string by concatenating the bit strings of all individual indices. 3) The bit string defines a sub-matrix by taking only the rows and columns with entry 1. 4) The desired tensor entry is the Pfaffian of this sub-matrix.

One can also define a further tensor type that we will refer to as *particle-number preserving Gaussian tensors*. This time there are ingoing and outgoing indices, and tensors are $X \times Y$ matrices where $X$ ($Y$) is the direct sum of ingoing (outgoing) basis sets. The tensor product is the direct sum of matrices, and there is another operation for contracting two indices. Again there is a tensor mapping to fermionic tensors, this time by simply taking the determinant of the corresponding sub-matrix.

Using (particle-number conserving) Gaussian tensors has the convenient feature that one can efficiently work with them in the number of indices: In contrast to real tensors, their storage, tensor product and contraction is of polynomial complexity in the number of indices. For fixed point models the number of indices is not a quantity that scales anyways, however, in practice complicated types of fixed point models might yield equation with lots of indices, such that solving those equations can be quite cumbersome. For Gaussian tensors however, it will never be a problem.

### 6.1.3 SET/SPT phases

In quantum physics one often imposes symmetries coming from fundamental laws of nature or arising from effective descriptions. Including symmetries means restricting to models that are invariant under a representation of a symmetry group. This changes the classification of phases in two ways: On the one hand, some phases might disappear as they are not compatible with the kind of symmetry, on the other hand deforming one model into another will be restricted by the symmetries, so some phases will split up into multiple subsets.

Perhaps the most important kind of symmetry are on-site symmetry representations, i.e., the global symmetry representation is a tensor product of local representations on non-overlapping patches. Topological phases with such on-site symmetries are known as *symmetry protected topological phases* (SPT phases) if the system would be in the trivial phase after discarding the symmetry [1, 49, 3], or *symmetry enriched topological phases* (SET phases) otherwise. Such models with on-site symmetries can be easily described with our framework by using another tensor type, that we will refer to as *symmetric tensors*, see first part in this section below.

One can also imagine symmetries that are not on-site, but still local in the sense that they are representable by a tensor-network operator: In topologically ordered systems such symmetries are formally the same as co-dimension 1 defects which are unitary (or better, groups thereof under fusion). We can then classify the phase of the symmetry as a topological defect. If we apply this kind of classification to phases with an on-site symmetries we will end up with a much coarser classification, as in many SPT phases the symmetry itself as a topological defect is trivial. In particular, a trivial phase can never have non-trivial (irreducible) topological defects. However, in a SET phase, or in a SPT phase with an anti-unitary symmetry the symmetries can be in a non-trivial phase. An example would be the time-reversal or duality defect of the toric code. The classification of such symmetries can be done with our framework by using another lattice type, that we will refer to as *homology lattices* (see the second part of this section below).

Space-group symmetries such as parity symmetry are a third kind of symmetry. On regular grids they are simply implemented by choosing a lattice type with the desired symmetries. In topologically ordered systems the situation is slightly different: Due to the topological deformability of the lattice type there are local TL operators that perform all possible space-group transformations. Now one can take an explicit space-group symmetry and compose it with the TL operator coming from the topological deformability that reverses the space-group transformation. This yields a local TL operator that leaves the system invariant. So for topological systems, explicit space-group symmetries are just ordinary local symmetries implemented by TL operators and we can apply the classification described in the previous paragraph. Note that by working with complex tensors via their realifications, space-group symmetries involving a parity flip yield anti-unitary TL operator symmetries. Such anti-unitary TL operator symmetries are often called *time-reversal symmetry*.

#### Symmetric tensors

To model systems with an on-site symmetry representation we can use symmetric tensors instead of ordinary real tensors.

**Definition 211.** A **symmetric tensor** with respect to a group $G$ is a real (or complex) tensor that 1) comes equipped with a representation $G$ for every index type, acting on the corresponding vector space and 2) is in the trivial sector of the representation of $G$ formed by the tensor product of the representations for every index. In other words, the tensor is invariant under the simultaneous action of all the local representations.

**Remark 143.** When applying a gauge transformation to a symmetric tensor we have to both apply the orthogonal map to the index and conjugate the local representation with it.

**Observation 142.** Ordinary real/complex tensors are a special case of symmetric tensors where all the local representations are trivial. Formally this is captured by a tensor mapping from real/complex tensors to symmetric tensors.

**Remark 144.** If we want to assert whether two TLs with symmetric tensors are in the same phase, all the tensors involved in the tensor-network moves have to be symmetric tensors as well. Also the TL operators that one has to find in order to show that two CTLs are in the same phase have to consist of symmetric tensors. It might at first sight seem paradoxical that one gets a more refined classification by using a more general tensor type (as real tensors are also symmetric tensors). However, also the model itself has this additional information on a choice of symmetry representation. In our viewpoint we can in principle apply arbitrary transformations to the model, but we have to transform the symmetry representations in the same way. So classifying symmetric phases means classifying pairs of model and symmetry representation.

Often people think about SPT/SET phases as having a fixed representation and classifying models with this representation. If we choose a "time" direction and a fix symmetry representation of all indices (depending on whether they are ingoing or outgoing in the time direction), all the TL operators have to be symmetric under this fixed representation. So our formalism contains the viewpoint with a fixed representation as a special case. However, our setting is much more natural as one can easily compare models with different local Hilbert spaces, and arbitrary changes of the microscopic degrees of freedom are the essence of phases of matter.

It is a common misunderstanding that from writing down a Hamiltonian alone it makes sense to ask in which SPT phase a model is. In general one has to explicitly state what the symmetry representation is (though a particular representation is often implicitly assumed in condensed matter models). For SPT phases where the symmetry itself as a topological defect is in a trivial phase one can write down one single model that can be in every single irreducible phase, depending on which symmetry representation we choose. Also the question whether a particular model has a symmetry or not only makes sense if the symmetry itself has to be in a non-trivial phase, otherwise the answer is always yes.

#### Homology lattices

In order to classify the topological sector of the symmetry itself as a topological defect we can use a CTL type with another lattice type, which we will call *homology lattices*. The extended backgrounds are given by $n$-manifolds with boundary together with an $(n-1)$th $G$-homology class on the manifold (that restricts to an $(n-2)$th $G$-homology class on the boundary). Let us give one concrete example of a lattice type that describes this kind of backgrounds:

**Definition 212.** A $n$**HSCb-lattice** ("HSC" for "homology simplicial complex") is a $n$SCb-lattice together with a dual orientation of the $(n-1)$-simplices and an element $g \in G$ associated to each $(n-1)$-simplex, such that: For every $(n-2)$-simplex in the interior, the cyclic product of the group elements of the $(n-1)$-simplices in its link is equal to the identity. More precisely we take the group element or its inverse depending on whether the dual orientation matches the cyclic order in which we take the product or not. Note that if that product is the identity, then it does not matter where we start and which direction we go. The $(n-2)$-simplices of the boundary also carry a dual orientation within the boundary, and we can associate to them the product of the group elements on the $(n-1)$-simplices in their link (which is not identity, but this time the start and direction of the product can be canonically chosen by the dual orientation of the boundary $(n-2)$-simplices).

The Pachner moves are still basic moves, just that now we also have to transform the group labels accordingly. Additionally, for every $g \in G$ there is another basic move called **group-label move** that multiplies all group elements associated to the $(n-1)$-simplices forming the boundary of a $n$-simplex with $g$. Whether we use the left- or right-regular action for each multiplication depends again on the dual orientation of the $(n-1)$-simplex relative to the $n$-simplex. The group labels of the boundary $(n-2)$-simplices remain unchanged under local moves, whereas all other labels can change.

$n$HSCb-lattices can be glued at their boundary $(n-1)$-simplices. For the gluing to be possible, the dual orientations of the glued boundary $(n-1)$-simplices and their connected $(n-2)$-simplices have to match. Then during the gluing the group element labels for the $(n-1)$-simplices and $(n-2)$-simplices are multiplied as they are fused together.

$n$HSCb-lattice without boundary also define a lattice type and are referred to as $n$HSC-lattices.

The lattices for a TL type describing SET/SPT phases in $n$ dimensions are given by $n$HSC-lattices. The lattices for the corresponding CTL type are $n$HSCb-lattices, with the index lattice given by the $(n-1)$HSC-lattice forming the boundary of the $n$HSCb-lattice. A background consists of the index lattice together with the topology plus the homology class of the $G$-labeling in the interior.

Let us quickly sketch a possible CTL type on $n$HSCb-lattices: As tensors we can take real tensors with one index type, and associate one such index to every boundary $(n-1)$-simplex. When two boundary $(n-1)$-simplices are glued the associated indices are contracted.

**Definition 213.** Define the $g$-**labeled** $n$**-dimensional simplicial pillow** as the background whose index lattice is given by two $(n-1)$-simplices that are connected to the same $(n-2)$-simplices. The dual orientations of those $(n-2)$-simplices all point towards the same $(n-1)$-simplex and they all carry the same $G$-label. The extended background is a ball with the trivial homology class.

**Observation 143.** The CTL associates to each $g$-labelled $n$-dimensional simplicial pillow a 2-index tensor, i.e., a matrix. Gluing the front face of the $g$-labelled simplicial pillow with the back face of the $h$-labelled simplicial pillow yields the $gh$-labelled simplicial pillow. At the same time the two associated matrices are multiplied. Thus, the collection of those matrices for all $g \in G$ forms a representation of the group $G$. So we see that this particular CTL type models an on-site symmetry representation. One can also obtain more general symmetries by introducing additional indices at the boundary $(n-2)$-simplices.

**Observation 144.** Consider a $n$HSCb-lattice where all the group element labels are trivial. Then glue the $g$-labelled $n$-dimensional simplicial pillow to every boundary $(n-1)$-simplex, for a fixed $g$. This will not change the background of the lattice, as the original lattice can be restored by applying the group-label move for $g$ to all $n$-simplices of the lattice and performing some Pachner moves near the boundary. On the other hand we saw that gluing with the $g$-labelled simplicial pillow corresponds to acting with a representation of $G$ for the group element $g$ on the associated CTL tensor. Thus, all CTL tensors for lattices with trivial group element labels are invariant under the representation of $G$. So the CTL restricted to those lattices yields a CTL with symmetric tensors as in the paragraph before. However it can be the case that the CTL with homology lattices is in the trivial phase even when the CTL with symmetric tensors is not.

**Remark 145.** The $g$-labelling in the lattices already looks a lot like a gauge field. Still, in our description the $g$-labelling is explicitly part of the combinatorial structure on which the physical model is defined on, so it is a static background rather than a dynamical field. However, in our framework the process of gauging has a very simple picture: Gauging means promoting the gauge field from a static background to dynamic degrees of freedom, i.e., also summing over the different $g$-label configurations in the state sum. To this end insert into the TL at each $(n-1)$-simplex a tensor implementing the representation of $G$ applied between the two neighbouring $n$-simplices. Use a delta tensor to obtain one copy of the group-element label of the representation tensor, for each $(n-2)$-simplex adjacent to the $(n-1)$-simplex. Additionally we insert at each $(n-2)$-simplex a group multiplication tensor, connected to the indices coming from its adjacent $(n-1)$-simplices. This yields a TL for the gauged phase.

**Remark 146.** Homology lattices are also the kind of lattice to which a lot of models on regular grids can be extended. E.g., this is the case when the ground state of a quantum system breaks translation invariance, or informally speaking, the unit cell of the ground state is larger than the unit cell of the model. A consequence of this is that the ground state degeneracy depends on the number of unit cells of the model. One example would be an anti-ferromagnet, i.e., a model where neighbouring qubits in a chain are forced to have odd parity. A more non-trivial and non-fixed point example for this is the Kitaev honeycomb model [66] in the Abelian gapped phase (which is referred to as weak breaking of translational symmetry).

### 6.1.4 Non-Hermitian/dissipative phases

In unitary quantum mechanics, the imaginary time evolution tensor network consists of Hermitian operators. In our realified language this implies that the tensor-network is invariant under flipping the time direction. This implies that if we have some sort of topological invariance then this invariance does not depend on an orientation.

Though fundamental models are usually assumed to be unitary, in open quantum systems one often has effective models with a dissipative time evolution. If we trotterize such a time evolution the local tensors are not Hermitian any more. This does not automatically imply that those models cannot be in a Hermitian (non-oriented) phase, but it at least leaves the possibility open.

Also classical statistical systems do not have an in-built reflection symmetry. I.e., it is possible that there are classical statistical systems on regular grids that cannot be extended to a topological lattice type but only to an oriented topological lattice type. However, at the moment we do not have an example for this, and e.g., all symmtery-breaking systems also do not depend on an orientation.

Such non-hermitian/dissipative/classical models can be described in our language by a TL type with oriented manifolds as background. To this end we can use a lattice type with a combinatorial structure similar to the nSCb-lattice with combinatorial spin structure for fermionic CTLs. The major difference is that obstruction cycle $\omega$ is now a $(n-1)$-cycle in a nSCb-lattice, i.e., a subset of $(n-1)$-simplices with empty boundary.

The local prescription that defines $\omega$ on a given lattice is rather straight forward: If we imagine the nSCb-lattice embedded in an oriented space we can use the branching structure to assign to each $n$-simplex whether it is left- or right-handed. As we do not have such an embedding there is no way to combina-torially determine the handedness of a single $n$-simplex. However, for two neighbouring $(n-1)$-simplices one can combinatorially decide whether they have equal or different handedness. If the handedness changes then the $n-1$-simplex is part of the obstruction cycle $\omega$.

Now a combinatorial representation of an orientation is given by a $n$-chain that has $\omega$ as its boundary. Opposed to combinatorial spin structures the local moves that allow to deform this $n$-chain are trivial, so for each homology class of orientations there is only one representative. For every connected component there are always two choices of this $n$-chain which are related by swapping all $\mathbb{Z}_2$ elements.

The algebraic structures related to CTLs on oriented manifolds are similar to the non-oriented case, just that they loose their notion of "unitarity", or "*-property". E.g., in 2 dimensions we would get something like special Frobenius algebras instead of *-algebras, or in 3 dimensions we would get something like non-unitary fusion categories instead of unitary ones.

Consider a complex-real CTL arising from a TL on topological lattices in $n$ dimensions. By adding a combinatorial orientation to the lattice type we can transform such a complex-real CTL into a complex one by choosing the complex arrow orientations depending on how some decoration is handed relative to the combinatorial orientation. Now imagine flipping the lattice: The complex-real CTL lattice does not change and thus the complex-real tensor is invariant. However, the orientation of the complex CTL lattice is flipped, and with it our way of interpreting the complex-real tensors as complex tensors: Flipping all the complex arrow orientations amounts to a complex conjugation of the complex tensor. So for complex CTLs on oriented lattices coming from a complex-real CTL on lattices without orientation, a flip of the orientation corresponds to a complex conjugation. So tensors associated to lattices that have a reflection symmetry (without the orientation) are Hermitian matrices when interpreted as a linear map between the two sides that are exchanged by the reflection. For CTLs allowing for a commuting-projector Hamiltonian this implies that this Hamiltonian is Hermitian. So we see that in quantum mechanics with unitary real and thus Hermitian imaginary time evolution, we never need an orientation when we use the realified language.

### 6.1.5 Chiral phases

All known (zero correlation-length) fixed point models of topological order, such as the Levin-Wen model or the Kitaev quantum double allow for a topological (i.e., gapped) boundary. There is however, a huge class of phases (namely integer or fractional quantum Hall systems) that do not have this property. Such phases are usually referred to as *chiral* because the braided fusion categories describing their anyon statistics have non-zero chiral central charge.

**Comment 24.** The word "chiral" is used in many different contexts with different meanings which often leads to confusion. E.g., in Ref. [17] it is used for models with a time-reversal (i.e., local anti-unitary) and parity symmetry (as we saw in Section (6.1.4) the two are the same in Hermitian topological models). The models given in Ref. [17] have the unnecessary constraint of a tetrahedral symmetry of the $F$-tensor that exactly enforces the existence of a time-reversal/parity symmetry. In Refs. [54, 55, 67] those restrictions were partly removed to also encompass phases without time-reversal symmetries. Consequently those models were referred to as "chiral". However, all those models are contained in 3FE-CTLs and thus do allow for a topological (i.e., gapped) standard boundary, so they are not chiral in the latter sense.

We believe that our framework is able to describe general chiral phases of matter. To this end one has to use a more general (C)TL type for the same topological lattice type. I.e., we need a TL type where the prescription of how we associate a tensor network to lattices locally depends on a larger region surrounding the place where the tensor is associated to, or a CTL type where the index lattice has more decorations that are invariant under basic moves in the interior. In particular we could take a universal (C)TL type, i.e., a type to which any other (C)TL type on the same lattice type can be mapped. We will present such a (C)TL type in a future publication.

All of the CTL types we discussed in Section (5) were not universal, i.e., the dependence of the tensor networks on the lattices was too simple. So those CTL types have properties that are specific for them, and do not exist for general CTL types on the same lattices.

1. For the (C)TL type in Section (5) one can define a standard topological (i.e., gapped) physical boundary.

2. Topological physical boundaries directly correspond to tensor-network representations of the ground states, in the sense of *MPS* or *PEPS* (see Observation (56)). So for the (C)TL types in Section (5) we automatically have such tensor-network representations.

3. For the (C)TL types in Section (5) one can define a local commuting-projector Hamiltonian whose imaginary time evolution is in the same phase.

4. For every general (C)TL on $n$-dimensional topological lattices consider a background with $n$-ball extended background, and a bi-partition of its index lattice into two $(n-1)$-ball patches separated by an $(n-2)$-sphere region. Now it is always possible to find a lattice representative for the background and bi-partition of this lattice that yields the desired bi-partition of the index lattice, such that: The number of indices that one has to cut when bi-parting the tensor network on the lattice is $\mathcal{O}(L)$ where $L$ is the size of the $(n-2)$-sphere region that biparts the index lattice, measured in the number of cells in that region. So the logarithm of the rank of the CTL tensor interpreted as a linear map between the two sides of the bi-partition of the index lattice is bounded by the volume of the region that separates the two parts. Such a behaviour of physical systems is known as an *area law*.

5. As have seen in Observation (53), all CTLs on topological lattices correspond to gapped models.

Our conjecture that universal (C)TL types also describe chiral phases is motivated by two arguments:

The first argument is the following: Chiral phases fulfil the properties 4) and 5) above that hold for every possible CTL type, but violate exactly 1), 2) and 3) which are only guaranteed to hold for the special CTL types from Section (5): 1) Chiral phases do not have a gapped boundary by (our) definition. 2) It is an open question whether chiral phases admit tensor-network representations (i.e. MPS or PEPS) of their ground states. Though there seem to be no practical problem to work with such tensor-network representations numerically [68, 69], there are indications that there are problems with those representations on a more pedantic and fundamental level. Our equivalence of PEPS representations and topological boundaries shows that at least those well-behaved PEPS representations arising from topological boundaries do not exist for chiral phases. It seems questionable to us that a conformal boundary of a topological system can also yield a meaningful PEPS representation. 3) It is known that there do not exist any local commuting projector Hamiltonians for chiral phases [16]. 4) Chiral phase do fulfil the area law, as all topological phases. 5) Chiral phases are gapped, as all topological systems.

On the other hand it seems highly plausible that there do exist CTLs of a universal type that violate one of the properties 1), 2) or 3). If this is the case then those have to correspond to chiral phases.

**Comment 25.** For every general CTL type on topological lattices, every background has a lattice representative with a hyperbolic tiling in the interior. The tensor network corresponding to this tiling has a structure that looks like a tree with additional connections between neighbouring branches. Tensor networks of such a structure can be used to describe critical/gapless systems [65], and are formally similar to the *multi-scale entanglement renormalization ansatz* (*MERA*) [70] (though the latter obey a light cone structure and are therefore not topological).

The second argument is more direct: First, the correlation length of any topological model can be decreased arbitrarily by blocking. On the other hand, every (chiral or non-chiral) topological model with zero correlation length is in the same phase as a CTL of a universal type. So every topological model is in the same phase as a CTL of a universal type, unless there are phases whose models can have correlation lengths that are arbitrarily close to 0 but never exactly reach 0. The latter seems rather implausible to us.

**Remark 147.** One might ask what such CTLs are good for if they do not directly correspond to a quantum model (i.e., a local Hamiltonian). We would like to note four things concerning this question: 1) tensor networks themselves can be viewed as a more general and simpler approach to quantum mechanics, alternative to a Hamiltonian formulation with a continuous time. Measurement statistics (i.e., their probability distributions) are simply given by evaluations of tensor networks and there is no need to talk about states, Hamiltonians, etc. at any point. 2) Pedantically speaking, the fixed point model and the Hamiltonian are different models as well: The Hamiltonian formulation always has a non-zero correlation length in time direction whereas the CTL does not. The CTL would correspond to a Hamiltonian with an infinite energy penalty for states violating the ground state space projector. 3) The practical use of fixed point models is to have a classification of phases, and one blue print for each phase, in which all topological invariants can be computed in an efficient and certified way. CTLs yield all those things: They give a finite set of equations that serves as classification and an exactly solvable construction to calculate invariants. Also they can serve as a comparison for phase determination algorithms. 4) There might be a continuous extension (see Sec. (6.2.3)) that actually yields a Hamiltonian: E.g. one can add continuous variables to the lattice type that model angles of corners in the cell complex. We can then consider a sequence of lattices with a fixed time direction such that the angles perpendicular to the time direction get smaller and smaller. This sequence approximates a Trotterized continuous imaginary time evolution. We can now consider the local projector lattices from the local Hamiltonian construction (in the non-chiral case) but now with small angles perpendicular to the time direction. For universal CTL types the associated tensor will not be a projector in general. Instead we can take as Hamiltonian the derivative of the associated tensors with respect to the angle, evaluated at angle 0. Note that the Hamiltonian obtained in this way is still local but not commuting and not consisting of projectors any more. A very similar Hamiltonian construction is known for critical models fulfilling a Yang-Baxter equation [71].

### 6.1.6 Critical/conformal phases

As we have also seen in Observation (53), (C)TLs on topological lattice types can only classify gapped phases. Gapless models seem to be less generic, as they usually occur at a phase transition between different gapped phases. Or, maybe exactly because of this, such critical models are very interesting and important.

We believe that our formalism also has the potential to describe gapless/critical models. The latter are usually classified by conformal field theories, i.e., field theories that are invariant under conformal transformations. At least in two dimensions those conformal transformations are very powerful, though they are still only a subset of arbitrary homeomorphism. This suggests that if we want to describe critical models with TLs we have to search for a lattice type with moves that are almost but not quite as powerful as the Pachner moves in simplicial complexes, and that are a discrete lattice analogue of continuum transformations. The Yang-Baxter equation [71] in integrable critical models look a lot like consistency equations for the tensors for such a moved lattice type. There exist some constructions in these directions, based on circle packings or isoradial graphs [72], but we haven't found a purely combinatorial formulation yet.

Unfortunately the connection between the continuum and discrete (if there exists any) formulation of conformal models is not that clear: In the topological case the backgrounds of the lattice type were in one-to-one correspondence with topological manifolds up to homeomorphism. Such a one-to-one correspondence is impossible in the conformal case, as the equivalence classes of conformal manifolds under conformal transformations do not form a discrete set.

### 6.1.7 Axiomatic TQFTs

The mathematical literature about topological phases of matter is governed by *axiomatic TQFTs*. Those are not local physical theories but only describe the $n$-point functions potentially arising from such a physical theory. In this section we will sketch that those can be formalized as CTLs.

Let us start by defining axiomatic TQFTs [22] in a tensor-network language.

**Definition 214.** A **unitary $n$-dimensional topological quantum field theory** (short unitary TQFT) is a map $T$ that associates to each oriented $n$-manifold with boundary a complex tensor with one index for each connected component of the boundary. The type of the index depends only on the topology of the corresponding connected $(n-1)$-manifold. $T$ has to obey the following axioms:

- Symmetries of the manifold with boundary $M$ (that permute boundary components) correspond to symmetries of the tensor $T[M]$ (by permuting the indices accordingly). Symmetries that involve changing the orientation come with an additional complex conjugation (in some predetermined basis).

- The tensor associated to the disjoint union of two manifolds with boundary is the tensor product of the tensors corresponding to the two manifolds.

$$
\begin{array}{ccc}
A, B & \xrightarrow{\quad T \quad} & T[A], T[B] \\
\otimes \downarrow & & \downarrow \otimes \\
A \otimes B & \xrightarrow{\quad T \quad} & T[A \otimes B] = T[A] \otimes T[B]
\end{array}
\tag{6.3}
$$

- The tensor corresponding to a manifold where two components of the boundary that are equal (up to homeomorphism) are glued is the tensor of the original manifold where the two corresponding indices have been contracted.

$$
\begin{array}{ccc}
A & \xrightarrow{\quad T \quad} & T[A] \\
\text{gluing} \downarrow & & \downarrow \text{contraction} \\
A' & \xrightarrow{\quad T \quad} & T[A'] = T[A]'
\end{array}
\tag{6.4}
$$

**Remark 148.** The TQFT axioms look very similar to the axioms for a CTL on manifolds with boundary, with the following important difference: The indices are not locally distributed over the boundary (i.e., "index lattice") of the lattice but there is only one index for a whole connected component. Also not only two small lattice elements of the boundary are glued, but two connected components as a whole.

Unitary TQFTs can be formalized as a CTL. More precisely, if we select a set of $(n-1)$-manifolds, there is a CTL type that corresponds to a unitary TQFT restricted to those $(n-1)$-manifolds as boundary components.

The CTL extended backgrounds are $n/n$-manifolds consisting of one $n$-region into which different 0-regions are embedded. Every one of the selected $(n-1)$-manifolds is the upper link for one of the 0-regions. The background gluings are given by surgery gluing for each of the 0-regions. In other words we have $n$-manifolds with embedded points with different kinds of singularities, and points with the same kind of singularity can be glued.

The CTL lattices are $n/n$CC-lattices on the backgrounds. The $0/0$CC-lattice obtained by restricting to the 0-regions forms the index lattice. There have to be additional decoration that allow to associate a chirality to each point of the 0-regions.

The tensors are real tensors with one index type for every 0-region, and for every vertex of each 0-region there is one corresponding index. When two index vertices are glued the associated indices are contracted.

A CTL of this type corresponds to a unitary TQFT in the following way: For a given CTL background, cutting out a small neighbourhood around every embedded point of every 0-region yields the TQFT $n$-manifold with boundary, up to the orientation. In the other direction forgetting the orientation and gluing to each boundary component of the TQFT manifold the stellar cone of the corresponding $(n-1)$-manifold yields the TQFT extended background. The CTL tensor is the realification of the TQFT tensor. Thereby the complex arrow orientations are determined according to whether the chirality decorations are right- or lefthanded relative to a chosen orientation of the TQFT manifold. Because of the unitarity property of the TQFT the CTL tensor does not depend on which orientation we choose (see also Remark (40)).

It is quite obvious that the gluing axioms of the TQFT are equivalent to those of the CTL using this translation.

### 6.1.8 Extended TQFTs

We conjecture that the CTL framework is also capable of describing *extended TQFTs* [73]. Ordinary axiomatic TQFTs associate numbers to $n$-manifolds and vector spaces to $(n-1)$-manifolds. A $d$-extended TQFT also associates (higher) categories to $(n-2)$-manifolds, $(n-3)$-manifolds,..., $(n-d)$-manifolds. 0-extended $n$-dimensional TQFTs correspond to CTLs on $n$-manifolds with trivial index lattice and no gluings. 1-extended $n$-dimensional TQFTs are just ordinary TQFTs from the previous sections, and correspond to CTLs on $n$-manifolds with the index lattice given by embedded points.

This suggests that in general there is the following equivalence: Every $d$-extended TQFT can be formalized as a CTL when we restrict to a finite amount of possible $(n-x)$-manifolds for $0 < x \leq d$. The CTL extended backgrounds are higher order manifolds with one $n$-region and one $(x-1)$-region for each selected $(n-x)$-manifold of the TQFT. The upper link of each $(x-1)$-region is exactly the corresponding $(n-x)$-manifold. The index lattice is given by the lattice restricted to the $x$-regions for $x < n$. The gluing is surgery gluing for all those regions. The CTL axioms will deliver some algebraic structures that correspond to the categorical structures associated to the selected $(n-x)$-manifolds.

For example 3-dimensional 2-extended TQFTs would correspond to CTLs on 3-manifolds with embedded points and a net of lines connecting these points (one might also take ribbons instead of lines and small disks instead of points). Let us give a rough sketch of a combinatorial formulation for such a CTL: The lattices are 3/3CC-lattices living on those 3/3-manifolds, the index lattice given by the cells of the 0- and 1-regions. Gluing can be done at the edges of the 1-regions and the vertices of the 0-regions. Indices are also associated to those edges and vertices, which are contracted when the edges or vertices are glued. We find that a set of the basic backgrounds is given by a tetrahedron of 1-region edges embedded into a 3-sphere (in the ribbon-disk case we also need a lattice with two disks connected to 3 ribbons embedded into the 3-sphere. The associated basis tensors correspond to the $F$-tensor (and the $R$-tensor) of a braided fusion category, and the CTL basic axioms correspond to the axioms of the braided fusion category. On the other hand we know that 3-dimensional 2-extended TQFTs are (modulo technical details) given by the Reshetikhin-Turaev construction [74] which takes a braided fusion category as input.

Fully extended (i.e., $n$-dimensional $n$-extended) TQFTs are said to correspond to microscopic physical theories. Indeed there is only one connected $(n-n)$-manifold, namely the point. So the corresponding CTLs has an $(n-1)$-region with a point as upper link, which is nothing but a boundary for the $n$-manifold. So the CTL corresponding to a fully extended TQFT has a CTL on manifolds with boundary, with gluing and indices at the boundary as sub-type. Such CTLs are exactly the ones that are contractions of TLs, i.e., microscopic local physical models.

Given a CTL type on $n$-manifolds with boundary there is a mapping to a CTL type corresponding to (arbitrarily) extended $n$-dimensional TQFTs: The lattice mapping cuts out a tubular neighbourhood around each $d$-region for $d \leq n-2$, and regards the remaining as a manifold with boundary. The tensor mapping uses the collection of the indices at the boundary of the tubular neighbourhood near a place in a $d$-region as the index associated to this place of the $d$-region.

### 6.1.9 3 + 1-dimensional phases

It is not difficult to find (C)TL types on 3-manifolds with boundary, that correspond to topologically ordered fixed point models in $3+1$ dimensions. The perhaps simplest possibility is to use 3SC(b)-lattices with one tensor for each 4-simplex that has one index for each of its 5 boundary 3-simplices. The Pachner moves will yield an axiom involving contractions of three copies of the basic tensor on each side (together with various technical axioms for inverting arrow orientatios and so on). In that sense the classification of 3-dimensional phases is already done (apart from the fact that we should use a universal TL type to get the most general phases).

In the literature a "classification" usually means finding an algebraic structure that classifies all possible phases. We do not really see the point in doing so, as an algebraic structure is nothing but a finite set of tensors fulfilling tensor-network axioms, which we can directly get from the Pachner moves.

There already exist some lattice models in $3+1$ dimensions, most notably the Walker-Wang models or Crane-Yetter state sum construction [75, 76] which take unitary braided fusion categories as input. Then there are also models by Kashaev [77] based on abelian groups. Both are unified in Ref. [78]. Those models are all captured in the simple TL type described above. We believe that this TL type also has more general solutions and that by going to a universal TL type we will find again more solutions, including ones without a topological (gapped) boundary (see Section (6.1.5)).

### 6.1.10 Other types of defects

There are lots of other interesting (C)TL types with lattices on higher order manifolds (with boundary) that we didn't consider yet. Let us sketch some of them.

**Defects inside a boundary**

Consider a (C)TL type on (boundary) 3/2-manifolds with the following (boundary) central link:

$$(6.5)$$

It describes defect lines within the $(1+1)$-dimensional physical boundary of a $(2+1)$-dimensional system, which we will call *boundary defects*. If the sub TL on 3-manifolds and the sub TL on 3-manifolds with boundary are irreducible, then the irreducible boundary defects correspond to the simple object labels of a fusion category whose Drinfel'd centre describes the anyon statistics of the model (so different boundaries correspond to Morita equivalent fusion categories). More precisely consider a CTL type on 3-manifolds with networks of lines and points embedded into the boundary, and the index lattice consisting of those points and lines. The algebraic structure corresponding to such a CTL type is this fusion category.

There is a fusion from physical boundaries times anyons to boundary defects. The lattice mapping takes the boundary defect lines and moves them into the interior by just a bit. Such a mapping corresponds to what is known as *anyon condensation* in the physics community: If the fusion of an anyon with a physical boundary is in the same phase as the trivial boundary defect, then the anyon is said to *condense* at the boundary. Otherwise it is said to be *confined*.

There is a mapping from boundary defects to anyons. The lattice mapping replaces every anyon world line by a thin tube of physical boundary carrying one boundary defect along.

We have already seen that physical boundaries correspond to tensor-network representations of the ground states. Now the boundary defects within a given physical boundary are exactly what is known as the *virtual symmetries* of the tensor network, i.e., in $2+1$ dimensions they correspond to the symmetry MPOs.

**Twists**

Consider a (C)TL type on (boundary) 3/2-manifolds with the following (boundary) central link:

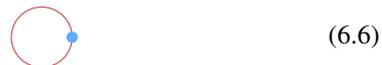 (6.6)

It describes lines in the $(2 + 1)$-dimensional bulk where codimension 1 defects end. In the physics community those are known as *twists* [79].

There is a fusion from anyons times twists to twists. The lattice mapping inserts an anyon world line next to the twist world line. The irrep coefficients of this fusion form an integer-valued representation of the anyon fusion ring.

## 6.1.11 The colour code as a (C)TL

In this section we sketch an alternative CTL type that contains the so-called colour-code [19, 80] as one specific solution. We will refer to CTLs of this type as 3**tcF-CTLs** where "tc" stands for "tri-coloured" referring to the decorations, "F" stands for "face", as this is where the indices are associated to.

**Backgrounds**

The backgrounds are 3-manifolds with boundary, with gluing at the boundary, just as for 3FE-CTLs and for 3E-CTLs.

**Lattices**

The lattices are 3CCb-lattices with only triangle faces and a tri-colouring of the vertices as decoration. That is, a map associating an element of the set {red, green, blue} to each vertex such that the vertices at the endpoints of each edge have different colours, or equivalently, each triangle has to be connected exactly to one red, one green and one blue vertex. The index lattice is given by the tri-coloured 2CC-lattice forming the boundary of the 3CCb-lattice. In order not to break our convention to use colour for the different regions of CTLs on higher order manifolds, we use 3 kinds of symbols to draw the vertices instead of different colours:

$$\text{red}: \bullet \qquad \text{green}: \circ \qquad \text{blue}: \odot \qquad (6.7)$$

Consider the following example for a patch of the index lattice

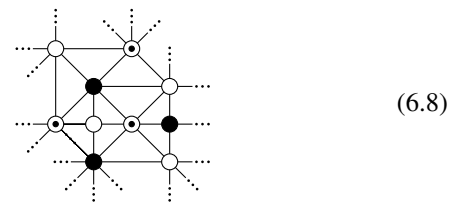 (6.8)

Two index faces can be glued just like for 3FE-CTLs, just that now the colours of the connected vertices have to match instead of the orientations of the surrounding vertices.

**Tensors**

The tensors are ordinary real tensors with one index type. One index is associated to every index face. When two faces are glued, the corresponding indices are contracted. One might also introduce normalizations when gluing two neighbouring triangles sharing the same face, but we won't go that much into detail.

**Basic tensors**

**Definition 215.** The red/green/blue **alternating $n$-double-pyramid** is the background consisting of $4n$ triangles and $2n+2$ vertices such that: There are two red/green/blue vertices (the two tips of the double-pyramid) which are each shared by $2n$ triangles. The rest of the vertices are half green/blue/red half blue/red/green (and are alternatingly placed at the equator of the double-pyramid). Its extended background is the 3-ball. E.g., the following red alternating 3-double-pyramid:

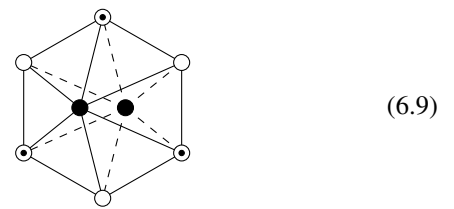 (6.9)

**Observation 145.** The alternating 2-double-pyramid can be seen as an octahedron with opposite edges having the same colour. So the red, green and blue alternating 2-double-pyramids are equal. We will refer to this background the **basic octahedron**.

**Observation 146.** Every red/green/blue alternating $n$-double-pyramid can be glued together from copies of the alternating 2-double-pyramid, e.g., for $n = 3$:

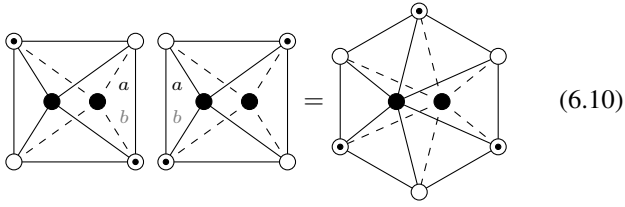

$$(6.10)$$

**Observation 147.** All lattices can be glued together from red alternating double-pyramids. A defining history for a given lattice can be obtained in the following way: Add one red vertex to the centre of every volume. For every red vertex in the lower link of the volume take one red alternating $n$-double-pyramid with $n$ equal to the number of blue/green vertices around the red vertex. The two tips of the red alternating double-pyramid are identified with the latter red vertex and the red vertex in the centre of the volume. Now glue all red alternating double-pyramids according to how they are located in the lattice. Of course the construction could be equally done with the blue/green vertices instead of red. In combination with Observation (146) we see that a set of basic lattices is given by basic octahedrons alone. The whole CTL is determined by the associated basic tensor.

**Properties and physical interpretation**

**Remark 149.** One can define 3 types of physical boundaries for 3tcF-CTLs, one for each colour. For each type there is the standard physical boundary by removing a vertex of that colour. Of course one fixed type of boundary can also represent the standard boundaries for the other two types of boundary. (Note that those boundaries do not have to be in different phases however.)

**Remark 150.** For every red/green/blue vertex in a 3tcF-CTL index lattice there is an alternating double-pyramid that can be glued with all of its back layer to the faces around this vertex. This leaves the tensor associated to this double-pyramid invariant. So the tensor associated to this double-pyramid when interpreted as a linear map from back to the front layer is a local ground space projector. The negative sum of those local ground space projectors for all vertices of a index lattice yields a commuting-projector Hamiltonian on this index lattice.

**The colour code as an example**

The following choice of basic tensors yields a solution to the CTL axioms. The basis set for the face indices is a two element set which we identify with the two elements of $\mathbb{Z}_2$.

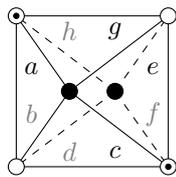

$$= \begin{cases} & a+b+c+d = 0 \land e+f+g+h = 0 \land \\ 1 \text{ if } & a+c+e+g = 0 \land b+d+f+h = 0 \land \\ & a+b+g+h = 0 \land d+c+e+f = 0 \\ 0 \text{ otherwise.} \end{cases} \qquad (6.11)$$

Here $+$ means $\mathbb{Z}_2$ multiplication. I.e., the tensor entries are only non-zero if, for every vertex, the parity of the indices of the faces connected to the vertex is even. The tensor associated to a general lattice (with the extended background of a 3-ball) is determined by the same rule.

Now the local Hamiltonian projector for this CTL, i.e., the tensor associated to an alternating double-pyramid, acts on a configuration of the indices of the faces around a vertex in the following way: First, it projects onto the parity even subspace for the indices. We can write this projector as $P_1 = (1 + z_1 z_2 z_3 \ldots)/2$ where $z_i$ is the according Pauli matrix acting on the $i$th index around the vertex. Second, it either leaves the index configuration invariant, or it flips all of them. In other words, it projects onto the symmetrized subspace under the action of the group $\mathbb{Z}_2$ that flips all the index values. We can write this projector as $P_2 = (1 + x_1 x_2 x_3 \ldots)/2$ where $x_i$ is the according Pauli matrix acting on the $i$th index around the vertex. In total the hamiltonian projector is given by

$$P_H = P_1 P_2 = (1 + z_1 z_2 z_3 \ldots)(1 + x_1 x_2 x_3 \ldots)/4. \quad (6.12)$$

Every state in the support of the projector has to be an eigenvector of both the operators $z_1 z_2 z_3 \ldots$ and $x_1 x_2 x_3 \ldots$ simultaneously. An according Hamiltonian is given by $H = \sum_{v \in V} -P_H^{(v)}$, where $P_H^{(v)}$ is the local ground state projector at the red (green/blue) vertex $v$. Thus, a ground state of the model has to be invariant under those $z$ and $x$ terms for the indices around every vertex.

If we go to the dual (index) lattice, then the faces are coloured red/green/blue and the indices/degrees of freedom live on the trivalent vertices. So the $x$ and $z$ terms above involve the vertices around every face. They are exactly the *stabilizers* of the *stabilizer code* known as the colour code.

# 6.2 General ideas

## 6.2.1 Robustness

For real tensors (or all tensor types which have a continuous set of tensors) there is a standard topology on the space of all tensors of fixed shape and bases, and a way to embed tensors with smaller basis into tensors with larger basis. So for each TL type with such a tensor type we can define a topology on the space of all TLs of this type. We can decompose this space of TLs into different equivalence classes in three different ways: Gauge families, TL phases and physical TL phases. Gauge families yield a finer decomposition than TL phases who in turn yield a finer decomposition of physical TL phases.

If a physical TL phase is an open set in the space of TLs, then we will call the phase *robust*. Robustness is a very important property in physics: As non-robust phases are unstable under arbitrarily small perturbations one expects to only observe robust phases in physics. However, robustness is not a defining property of a phase, and it is perfectly fine to talk about phases that are non-robust. This can also be important, as e.g., after a mapping, non-robust phases may become robust. Another question is about the cardinality of the set of gauge families, TL phases or physical TL phases: Those sets can be countable (i.e., discrete) or labeled by finitely many, or even infinitely many continuous parameters.

The general heuristics seems to be the following: The more powerful the basic moves of the lattice type are, the more robust are the gauge families, TL phases, and physical TL phases, and the fewer continuous parameters are needed to label them. E.g., the TL types from Section (5) were all based on topological lattice types. For most implementations in Section (5), gauge families are in one-to-one correspondence with the connected components of the space of TLs, and thus form a discrete set. E.g., for 2E-TLs the gauge families are labeled by three non-negative integers, each gauge family forming its own TL phase. However, e.g., all double-line 2E-TLs were in the same physical TL phase (the trivial one). The situation is slightly different for e.g., 2En-TLs: Their gauge families are parametrized by one continuous parameter corresponding to the euler characteristic 2En-TL. Now the stacking of a double-line 2En-TL with an euler characteristic 2En-TL is in the same TL phase as an euler characteristic 2En-TL, but the TL phases are still labeled by one continuous parameter. However, all euler characteristic 2En-TLs are in the same physical TL phase (the trivial one as they are all normalization TLs).

For 3FE-TLs the discreteness of the gauge families corresponds to what is known as *Ocneanou rigidity* [81] for unitary multi-fusion categories.

It would be interesting to study the TL space and the gauge families, TL phases and physical TL phases therein more systematically for different TL types. This should be quite straight forward in our formulation: The TL axioms are equations between tensor networks, and we can easily calculate the derivative of those axioms (i.e., left side minus right side, for fixed basis sets) with respect to the basic tensors, and the kernel of the corresponding Jacobi matrix. Second we can calculate the derivative of the basic tensors with respect to the gauge transformations, and the support of the corresponding Jacobi matrix. The latter support will be contained in the former kernel (for tensors fulfilling the TL axioms). If the two subspaces are equal then the set of gauge families is discrete. Otherwise the difference in dimension is the number of continuous parameters labeling the gauge families.

### 6.2.2 Real time and dissipative models

Our viewpoint allows comparing the real- and imaginary time evolution of a quantum system both as tensor networks. We saw that classifying (ground state) quantum phases corresponds to classifying the imaginary time evolution. So a very natural question is: In which physical TL phase is the real-time evolution? To which lattice type can it be extended? Since a (non-trivial) real-time evolution has a light-cone structure it cannot be extended to a topological lattice type. In contrast to topological field theory, conformal field theory does exist with euclidean and Lorentzian signature. Thus, one might speculate that if the imaginary time evolution extends to a euclidean-conformal lattice type, the real-time evolution extends to a Lorentzian-conformal lattice type. But then, if the imaginary time evolution is topological, does the real time evolution also extend to a lorenzian-conformal lattice type, or something simpler?

Another question is, when we slowly Wick rotate the tensor network, will there be a phase transition and at which point? It seems plausible to speculate that the imaginary time evolution stays topological after Wick rotation, until the real-time evolution is reached.

Another type of system that can be described as TL is a model with Liouvillian time evolution, and one can ask in which physical TL phase such a model is. One interesting question is, when we start with a doubled (ket plus bra layer) real-time evolution and add a dissipative noise, in which phase will we end up? Will it be the phase of the doubled imaginary-time evolution of the same model? This yields an interesting viewpoint on thermalisation.

It also yields another prespective on quantum error correction: In a topological code, the cycle of fixed point time evolution, dissipative noise, and decoding defines a (discrete) local dissipative time evolution (if the decoder is local what it better should be). Now if the decoder works it should preserve the logical qubits and discard all the noise. This is exactly what an imaginary time evolution does. Thus, it is very suggestive that the decoder works exactly if the corresponding dissipative TL is in the same physical TL phase as the (doubled) imaginary time evolution of the fixed point topological code model. So a working decoder can be seen as a (strong) perturbation of the real-time model that takes the model into the interior of the imaginary-time evolution.

### 6.2.3 Continuous extensions

One could think of adding continuous parameters to a lattice type (such that it is not a lattice type in the actual sense any more). Those parameters can have the interpretation of a length scale, such that one can take an actual continuum limit by choosing smaller and smaller length scales. This way one can obtain field (or continuum-time quantum) theory versions of discrete fixed point models.

### 6.2.4 Numerical algorithm for solving the (C)TL axioms

(Finitely axiomatized C)TLs are determined by a finite number of tensors fulfilling a finite number of tensor-network axioms. So for real tensors (or a similar tensor type) and a fixed basis set we have a finite set of real numbers (or similar) that fulfil a finite amount of polynomial equations. One can apply the Gauss-Netwon method to find the roots of those polynomial equations. This provides a systematic way of finding (C)TLs, even if they do not correspond to already well-known algebraic structures as e.g., for TL types in $3 + 1$ dimensions.

Of course such a numerical method of finding all (C)TLs for a fixed basis set cannot be efficient in the size of that basis set. In particular in generic cases already the number of different gauge families scales exponentially with the size of the basis set. However, the more complicated the fixed point models, i.e., the larger the basis set of the TL, the less likely the corresponding phases are to appear in realistic models. For models with practical importance, i.e., with small enough basis sets finding TLs with a Gauss-Newton method seems doable.

We have tested the Gauss-Newton method to numerically find 2E-CTLs up to basis size 5. We indeed found the zero CTL, the trivial CTL, the 2-dimensional double-line CTL, different delta CTLs, the complex numbers, the quaternions, and other direct sums. It turns out that the quaternion CTL (which is somewhat more complicated than the others) is less likely to be found depending on the starting point of the Gauss-Newton method.

### 6.2.5 Numerical algorithm for phase detection

Our definition of phases with tensor networks and tensor-network moves could be used to build a numerical algorithm for testing whether two models are in the same phase: First, one would represent both models as a tensor network (for field theories or continuum-time quantum mechanics this is done by choosing a fixed unit cell size, blocking and truncating as described in Sec. (4.1)). Then one variationally optimizes over a sequence of tensor-network moves transforming one tensor network into the other. As a first step, one might perform some blocking to get rid of the correlation length. The resulting algorithm would be quite similar in spirit to the tensor-entanglement-filtering approach [82]. Though our method does not (only) involve a blocking, and it would not have the problem of getting "unwanted" fixed points such as the "corner double-line (CDL) tensors" in Ref. [82].

## Acknowledgments

Special thanks goes to my ObstOffice comembers A. Nietner and M. Kesselring for lots of inspiring and fruitful discussions during the last two years. Thanks also to N. Tarantino for various discussions, and R. Sweke for reviewing the introduction. We thank the DFG (CRC 183, B01, and EI 519/15-1), the Templeton Foundation, the Studienstiftung des Deutschen Volkes, and the ERC (TAQ) for support.

# Bibliography

[1] X. Chen, Z.-C. Gu, and X.-G. Wen, "Local unitary transformation, long-range quantum entanglement, wave function renormalization, and topological order," *Phys. Rev. B* **82** (2010) 155138, arXiv:1004.3835.

[2] N. Schuch, D. Perez-Garcia, and I. Cirac, "Classifying quantum phases using matrix product states and PEPS," *Phys. Rev. B* **84** (2011) 165139, arXiv:1010.3732.

[3] N. Schuch, I. Cirac, and D. Perez-Garcia, "PEPS as ground states: degeneracy and topology," *Ann. Phys.* **325** (2010) 2153, arXiv:1001.3807.

[4] N. Bultinck, M. Marien, D. J. Williamson, M. B. Sahinoglu, J. Haegeman, and F. Verstraete, "Anyons and matrix product operator algebras," *Ann. Phys.* **378** (2017) 183–233, arXiv:1511.08090.

[5] M. B. Sahinoglu, D. Williamson, N. Bultinck, M. Marin, J. Haegeman, N. Schuch, and F. Verstraete, "Characterizing topological order with matrix product operators," (2014) , arXiv:1409.2150.

[6] F. Verstraete, J. I. Cirac, and V. Murg, "Matrix product states, projected entangled pair states, and variational renormalization group methods for quantum spin systems," *Adv. Phys.* **57** (2008) 143, arXiv:0907.2796.

[7] L. Landau, "On the theory of phase transitions," *Zh. Eksp. Teor. Fiz.* **7** (1937) 1932.

[8] A. Y. Kitaev, "Fault-tolerant quantum computation by anyons," *Ann. Phys.* **303** (2003) 2–30, arXiv:quant-ph/9707021.

[9] F. Pollmann, A. M. Turner, E. Berg, and M. Oshikawa, "Entanglement spectrum of a topological phase in one dimension," *Phys. Rev. B* **81** (2010) 064439, arXiv:0910.1811.

[10] C. Wille, O. Buerschaper, and J. Eisert, "Fermionic topological quantum states as tensor networks," *Phys. Rev. B* **95** (2017) 245127, arXiv:1609.02574.

[11] N. Bultinck, D. J. Williamson, J. Haegeman, and F. Verstraete, "Fermionic projected entangled-pair states and topological phases," *J. Phys. A* **51** (2017) 025202, arXiv:1707.00470.

[12] V. G. Turaev and O. Y. Viro, "State sum invariants of 3-manifolds and quantum 6j-symbols," *Topology* **31** (1992) 865–902.

[13] J. W. Barrett and B. W. Westbury, "Invariants of piecewise-linear 3-manifolds," *Trans. Amer. Math. Soc.* **348** (1996) 3997–4022, arXiv:hep-th/9311155.

[14] R. Dijkgraaf and E. Witten, "Topological gauge theories and group cohomology," *Commun. Math. Phys.* **129** (1990) 393–429.

[15] G. Kuperberg, "Involutory Hopf algebras and 3-manifold invariants," *Internat. J. Math.* **2** (1991) 41–66, arXiv:math/9201301.

[16] A. Kapustin and L. Fidkowski, "Local commuting projector Hamiltonians and the quantum Hall effect," (2018) , arXiv:1810.07756.

[17] M. A. Levin and X.-G. Wen, "String-net condensation: A physical mechanism for topological phases," *Phys. Rev. B* **71** (2005) 045110, arXiv:cond-mat/0404617.

[18] O. Buerschaper, J. M. Mombelli, M. Christandl, and M. Aguado, "A hierarchy of topological tensor network states," *J. Math. Phys.* **54** (2013) 012201, arXiv:1007.5283.

[19] H. Bombin and M. A. Martin-Delgado, "Topological quantum distillation," *Phys. Rev. Lett.* **97** (2006) 180501, arXiv:quant-ph/0605138.

[20] A. Kubica, B. Yoshida, and F. Pastawski, "Unfolding the color code," *New J. Phys.* **17** (2015) 83026, arXiv:1503.02065.

[21] C. Wille, R. Egger, J. Eisert, and A. Altland, "Simulating topological tensor networks with Majorana qubits," *Phys. Rev. B* (2018) , arXiv:1808.04529.

[22] M. Atiyah, "Topological quantum field theories," *Publications Mathématiques de lIHÉS* **68** (1989) 175186.

[23] M. B. Sahinoglu, *A tensor network study of topological quantum phases of matter*. PhD thesis, Universität Wien, 2016.

[24] M. Fukuma, S. Hosono, and H. Kawai, "Lattice topological field theory in two dimensions," *Commun. Math. Phys.* **161** (1994) 157–176, arXiv:hep-th/9212154.

[25] L. Chang, "Kitaev models based on unitary quantum groupoids," *J. Math. Phys.* **55** (2014) 041703, arXiv:arXiv:1309.4181.

[26] A. Kitaev and L. Kong, "Models for gapped boundaries and domain walls," *Commun. Math. Phys.* **313** (2012) 351–373, arXiv:1104.5047.

[27] V. Drinfel'd, "Quantum groups," *J. Math. Sci.* **41** (1988) 898.

[28] A. Altland and M. R. Zirnbauer, "Novel symmetry classes in mesoscopic normal-superconducting hybrid structures," *Phys. Rev. B* **55** (1996) 1142, `arXiv:cond-mat/9602137`.

[29] D. Nikshych, "Morita equivalence methods in classification of fusion categories," `arXiv:1208.0840`.

[30] G. Vidal, "Classical simulation of infinite-size quantum lattice systems in one spatial dimension," *Phys. Rev. Lett.* **98** (2007) 070201, `arXiv:cond-mat/0605597`.

[31] R. Orús and G. Vidal, "Infinite time-evolving block decimation algorithm beyond unitary evolution," *Phys. Rev. B* **78** (2008) 155117, `arXiv:0711.3960`.

[32] H. N. Phien, I. P. McCulloch, and G. Vidal, "Fast convergence of imaginary time evolution tensor network algorithms by recycling the environment," *Phys. Rev. B* **91** (2015) 115137, `arXiv:1411.0391`.

[33] A. J. Daley, C. Kollath, U. Schollwöck, and G. Vidal, "Time-dependent density-matrix renormalization-group using adaptive effective hilbert spaces," *J. Stat. Mech.* (2004) P04005, `arXiv:cond-mat/0403313`.

[34] M. Kliesch, C. Gogolin, and J. Eisert, "Lieb-Robinson Bounds and the simulation of time-evolution of local observables in lattice systems," in *Many-Electron Approaches in Physics, Chemistry and Mathematics*, V. Bach and L. Delle Site, eds., Mathematical Physics Studies, p. 301. Springer, 2014. `arXiv:1306.0716`.

[35] M. Levin and C. P. Nave, "Tensor renormalization group approach to two-dimensional classical lattice models," *Phys. Rev. Lett.* **99** (2007) 120601.

[36] V. Zauner-Stauber and F. Verstraete, "Symmetry breaking and convex set phase diagrams for the q-state potts model," *New J. Phys.* **18** (2016) 113033, `arXiv:1607.03492`.

[37] T. S. Cubitt, D. Perez-Garcia, and M. M. Wolf, "Undecidability of the spectral gap," *Nature* **528** (2015) 207–211, `arXiv:1502.04573`.

[38] T. Koma and B. Nachtergaele, "The spectral gap of the ferromagnetic XXZ-chain," *Lett. Math. Phys.* **40** (1997) 1–16, `arXiv:cond-mat/9512120`.

[39] W. L. Spitzer and S. Starr, "Improved bounds on the spectral gap above frustration free ground states of quantum spin chains," *Lett. Math. Phys.* **63** (2003) 165–177, `arXiv:math-ph/0212029`.

[40] M. B. Hastings and X.-G. Wen, "Quasi-adiabatic continuation of quantum states: The stability of topological ground state degeneracy and emergent gauge invariance," *Phys. Rev. B* **72** (2005) 045141, `arXiv:cond-mat/0503554`.

[41] B. Zeng and X.-G. Wen, "Gapped quantum liquids and topological order, stochastic local transformations and emergence of unitarity," *Phys. Rev. B* **91** (2015) 125121, `arXiv:1406.5090`.

[42] J. Eisert, M. Cramer, and M. B. Plenio, "Area laws for the entanglement entropy," *Rev. Mod. Phys.* **82** (2010) 277, `arXiv:0808.3773`.

[43] W. Shirley, K. Slagle, Z. Wang, and X. Chen, "Fracton models on general three-dimensional manifolds," *Phys. Rev. X* **8** (2018) 031051, `arXiv:1712.05892`.

[44] S. Bravyi and M. B. Hastings, "A short proof of stability of topological order under local perturbations," *Commun. Math. Phys.* **307** (2011) 609, `arXiv:1001.4363`.

[45] F. Verstraete and J. I. Cirac, "Renormalization algorithms for quantum-many body systems in two and higher dimensions," (2004) , `arXiv:cond-mat/0407066`.

[46] M. Fannes, B. Nachtergaele, and R. Werner, "Finitely correlated states on quantum spin chains," *Commun. Math. Phys.* **144** (1992) 443.

[47] B. Li, *Real operator algebras*. World Scientific Publishing Company, 2003.

[48] D. M. Greenberger, M. A. Horne, and A. Zeilinger, "Going beyond Bell's theorem," in *Bell's Theorem, Quantum Theory, and Conceptions of the Universe*, M. Kafatos, ed., pp. 69–72. Kluwer, 1989. `arXiv:0712.0921`.

[49] X. Chen, Z.-C. Gu, Z.-X. Liu, and X.-G. Wen, "Symmetry protected topological orders and the group cohomology of their symmetry group," *Phys. Rev. B* **87** (2013) 155114, `arXiv:1106.4772`.

[50] R. Verresen, R. Moessner, and F. Pollmann, "One-dimensional symmetry protected topological phases and their transitions," *Phys. Rev. B* **96** (2017) 165124, `arXiv:1707.05787`.

[51] D. Aasen, E. Lake, and K. Walker, "Fermion condensation and super pivotal categories," (2017) , `arXiv:1709.01941`.

[52] A. Kirillov and B. Balsam, "Turaev-Viro invariants as an extended TQFT," (2010) , `arXiv:1004.1533`.

[53] Z.-C. Gu, Z. Wang, and X.-G. Wen, "A classification of 2d fermionic and bosonic topological orders," *Phys. Rev. B* **91** (2015) 125149, `arXiv:1010.1517`.

[54] Y. Hu, Y. Wan, and Y.-S. Wu, "Twisted quantum double model of topological phases in two dimensions," *Phys. Rev. B* **87** (2013) 125114, `arXiv:1211.3695`.

[55] C.-H. Lin and M. Levin, "Generalizations and limitations of string-net models," *Phys. Rev. B* **89** (2014) 195130, `arXiv:1402.4081`.

[56] H. Bombin and M. A. Martin-Delgado, "A family of non-Abelian Kitaev models on a lattice: Topological confinement and condensation," *Phys. Rev. B* **78** (2008) 115421, `arXiv:0712.0190`.

[57] S. Majid, "Doubles of quasitriangular Hopf algebras," *Commun. Alg.* **19** (1991) 3061–3073.

[58] D. Gaiotto and A. Kapustin, "Spin TQFTs and fermionic phases of matter," *Int. J. Mod. Phys.* **A31** (2016) 1645044, `arXiv:1505.05856`.

[59] Q.-R. Wang and Z.-C. Gu, "Towards a complete classification of fermionic symmetry protected topological phases in 3d and a general group supercohomology theory," *Phys. Rev. X* **8** (2017) 011055, `arXiv:1703.10937`.

[60] C. V. Kraus, N. Schuch, F. Verstraete, and J. I. Cirac, "Fermionic projected entangled pair states," *Phys. Rev. A* **81** (2010) 052338, `arXiv:0904.4667`.

[61] L. Fidkowski and A. Kitaev, "Topological phases of fermions in one dimension," *Phys. Rev. B* **83** (2010) 075103, `arXiv:1008.4138`.

[62] A. Kitaev, "Unpaired Majorana fermions in quantum wires," *Physics-Uspekhi* **44** (2001) 131–136, `arXiv:cond-mat/0010440`.

[63] S. Ryu, A. Schnyder, A. Furusaki, and A. Ludwig, "Topological insulators and superconductors: ten-fold way and dimensional hierarchy," *New J. Phys.* **12** (2010) 065010, `arXiv:0912.2157`.

[64] S. Bravyi, "Contraction of matchgate tensor networks on non-planar graphs," *Contemp. Math.* **482** (2009) 179–211, `arXiv:0801.2989`.

[65] A. Jahn, M. Gluza, F. Pastawski, and J. Eisert, "Holography and criticality in matchgate tensor networks," (2017) , `arXiv:1711.03109`.

[66] A. Kitaev, "Anyons in an exactly solved model and beyond," *Ann. Phys.* **321** (2006) 2–111, `arXiv:cond-mat/0506438`.

[67] E. Lake and Y.-S. Wu, "Signatures of broken parity and time-reversal symmetry in generalized string-net models," *Phys. Rev. B* **94** (2016) 115139, `arXiv:1605.07194`.

[68] T. B. Wahl, H.-H. Tu, N. Schuch, and J. I. Cirac, "Projected entangled-pair states can describe chiral topological states," *Phys. Rev. Lett.* **111** (2013) 236805. `https://link.aps.org/doi/10.1103/PhysRevLett.111.236805`.

[69] D. Poilblanc, J. I. Cirac, and N. Schuch, "Chiral topological spin liquids with projected entangled pair states," *Phys. Rev. B* **91** (2015) 224431. `https://link.aps.org/doi/10.1103/PhysRevB.91.224431`.

[70] G. Vidal, "A class of quantum many-body states that can be efficiently simulated," *Phys. Rev. Lett.* **101** (2008) 110501, `arXiv:quant-ph/0610099`.

[71] R. J. Baxter and C. Domb, "Solvable eight-vertex model on an arbitrary planar lattice," *Philosophical Transactions of the Royal Society of London* 289.

[72] C. Boutillier and B. D. Tilire, "Statistical mechanics on isoradial graphs," `arXiv:1012.2955`.

[73] J. C. Baez and J. Dolan, "Higher-dimensional algebra and topological quantum field theory," *J. Math. Phys.* **36** (1995) 6073–6105, `arXiv:q-alg/9503002`.

[74] N. Reshetikhin and V. Turaev, "Invariants of 3-manifolds via link polynomials and quantum groups," *Invent. Math.* **103** (1991) 547–597.

[75] K. Walker and Z. Wang, "(3+1)-TQFTs and topological insulators," *Front. Phys.* **7** (2012) 150, `arXiv:1104.2632`.

[76] L. Crane and D. N. Yetter, "A categorical construction of 4D TQFTs," (1993) , `arXiv:hep-th/9301062`.

[77] R. Kashaev, "A simple model of 4d-TQFT," (2014) , `arXiv:1405.5763`.

[78] D. J. Williamson and Z. Wang, "Hamiltonian models for topological phases of matter in three spatial dimensions," *Ann. Phys.* **377** (2017) 311–344, `arXiv:1606.07144`.

[79] H. Bombin, "Topological order with a twist: Ising anyons from an abelian model," *Phys. Rev. Lett.* **105** (2010) 030403, `arXiv:1004.1838`.

[80] M. S. Kesselring, F. Pastawski, J. Eisert, and B. J. Brown, "The boundaries and twist defects of the color code and their applications to topological quantum computation," *Quantum* **2** (2018) 101, `arXiv:1806.02820`.

[81] P. Etingof, D. Nikshych, and V. Ostrik, "On fusion categories," *Ann. Phys.* **162** no. 2, (2005) 581–642, `arXiv:math/0203060`.

[82] Z.-C. Gu and X.-G. Wen, "Tensor-entanglement-filtering renormalization approach and symmetry protected topological order," *Phys. Rev. B* **80** (2009) 155131, `arXiv:0903.1069`.